\documentclass[12pt]{advphys_klm}
\usepackage{graphicx}
\usepackage{cite}
\usepackage{amssymb}
\usepackage{amsfonts}
 
\newcommand{\beqa}{\begin{eqnarray}}
\newcommand{\eeqa}{\end{eqnarray}}
\newcommand{\beq}{\begin{equation}}
\newcommand{\eeq}{\end{equation}}
\newcommand{\dg}{\dagger}
\newcommand{\sig}{\sigma}
\newcommand{\vektor}[1]{\mbox{\boldmath $#1$}}
\newcommand{\eff}{\mbox{\scriptsize{eff}}}
\newcommand{\Hub}{\mbox{\scriptsize{Hub}}}
\newcommand{\Hol}{\mbox{\scriptsize{Holstein}}}
\newcommand{\SSH}{\mbox{\scriptsize{SSH}}}
\newcommand{\crit}{\mbox{\scriptsize{crit}}}
\newcommand{\doex}{\mbox{\scriptsize{DE}}}
\newcommand{\KL}{\mbox{\scriptsize{KL}}}
\newcommand{\KLph}{\mbox{\scriptsize{PKL}}}
\newcommand{\DKL}{\mbox{\scriptsize{DKL}}}
\newcommand{\AKL}{\mbox{\scriptsize{AKL}}}
\newcommand{\RKKY}{\mbox{\scriptsize{RKKY}}}
\newcommand{\PAM}{\mbox{\scriptsize{PAM}}}

\newcommand{\tot}{\mbox{\scriptsize{tot}}}

\begin{document}

\pagenumbering{roman}
\setcounter{page}{1}
\null
\vskip 20pt
\begin{center}
\Large{\bf{The one dimensional Kondo lattice model at partial 
band filling}\footnote{Advances in Physics, Volume 53, Number 7, 
pp. 769 - 937, November 2004}}
\vskip 40pt
\normalsize
Mikl\'{o}s Gul\'{a}csi
\vskip 10pt
Department of Theoretical Physics,
Institute of Advanced Studies \\
The Australian National University,
Canberra, ACT 0200, Australia
\vskip 10pt
(July 14, 2003)
\vskip 60pt
\large{\bf{Abstract}}
\end{center}
\vskip 10pt
The Kondo lattice model introduced in 1977 describes 
a lattice of localized magnetic moments interacting 
with a sea of conduction electrons. It is one of the 
most important canonical models in the study of a class of 
rare earth compounds, called heavy fermion systems, 
and as such has been studied intensively by a wide 
variety of techniques for more than a quarter of
a century. This review focuses on the one dimensional
case at partial band filling, in which the number of 
conduction electrons is less than the number of 
localized moments. The theoretical understanding,
based on the bosonized solution, of the conventional 
Kondo lattice model is presented in great detail. 
This review divides naturally into two parts, the first 
relating to the description of the formalism, and the
second to its application. After an all-inclusive 
description of the bosonization technique, the 
bosonized form of the Kondo lattice hamiltonian
is constructed in detail. Next the double-exchange 
ordering, Kondo singlet formation, the RKKY interaction 
and spin polaron formation are described comprehensively. 
An in-depth analysis of the phase diagram follows, with
special emphasis on the destruction of the ferromagnetic 
phase by spin-flip disorder scattering, and of recent 
numerical results. The results are shown to hold for 
both antiferromagnetic and ferromagnetic Kondo lattice. 
The general exposition is pedagogic in tone. 
\tableofcontents
\cleardoublepage
\pagestyle{headings}
\pagenumbering{arabic}

%%%%%%%%%%%%%%%%%%%%%%%%%%%%%%%%%%%%%%%%%%%
%%  chapter 1
%%%%%%%%%%%%%%%%%%%%%%%%%%%%%%%%%%%%%%%%%%%%

\cleardoublepage
\chapter{\label{ch4}An Introduction to the Kondo Lattice}

The Kondo lattice is one of the most important canonical models 
used to study strongly correlated electron systems, and has been 
the subject of intensive study. Other canonical models for 
strongly correlated systems include the 
Hubbard model, which is discussed in section \ref{old3.2}, and the 
periodic Anderson model, which is introduced in section \ref{old4.2.2}
\footnote{A brief overview of the strongly correlated
electron systems in given in Appendix \ref{appd}.}. 
The Kondo lattice describes the interaction between 
a conduction band, containing Bloch-like delocalized electrons, 
and a lattice of localized magnetic moments. Its importance is 
due both to its relevance to several broad classes of real 
materials, and to the the fundamental theoretical challenge it 
presents; methods developed for the Kondo lattice are expected 
to aid in formulating methods for other strongly correlated electron 
systems.  Hereafter, follows a description of the Kondo lattice, and in 
particular the 1D Kondo lattice at partial band filling. 

This chapter provides a general introduction to the Kondo 
lattice model. The chapter is organized as follows: In section \ref{old4.1} 
the Kondo lattice is derived as a special case of a general 
two-band electron system. This serves to provide both a 
definition of the Kondo lattice model, and to provide a 
clear statement of the assumptions required 
to derive the model; the main assumption is that the Wannier 
states for one of the bands are atomic-like. 
Applications of the Kondo lattice to 
real materials are discussed in section \ref{old4.2}. The description 
of manganese oxide perovskites follows in straightforward fashion 
from the derivation of section \ref{old4.1}, and is discussed in 
section \ref{old4.2.1}. Section \ref{old4.2.2} considers the 
Kondo lattice description of 
rare earth and actinide compounds. The application in this case 
is indirect, and requires 
the derivation of the Kondo lattice from the 
more fundamental periodic Anderson model.

\section{\label{old4.1}Derivation of the Kondo lattice model}

The Kondo lattice is a special case of a general two-band 
electron system with interband interactions: The Kondo 
lattice specializes to the case in which the electrons 
in one of the bands remain localized at their lattice sites. 
In this section the Kondo lattice model is derived from a 
general two-band system with standard electron-electron 
interactions. The derivation serves two purposes. Firstly it 
defines the conventional Kondo lattice model, which in its 
1D form will be the primary focus of chapters \ref{ch5} and \ref{ch6}. 
Secondly, the derivation permits a clear statement  
of the assumptions required to derive the Kondo 
lattice from a general two-band system. This helps 
to identify the types of real materials which may be modelled  
by the Kondo lattice. The applications to 
real materials are discussed in section \ref{old4.2}. 

The starting point is a two-band  
electron system with band indices
$c$ and $f$. $c$ labels the conduction band, 
and $f$ labels the band of localized electrons; 
the $f$ is to suggest the localized $f$-electrons in rare 
earth and actinide compounds.\footnote{While this notation 
is fairly standard for the Kondo lattice, note that the 
localized electrons in the manganese oxides are 
in the $d$ band, cf.\ section \ref{old4.2.1} below.} 
In the simplest case, both $c$ and $f$ bands are 
assumed to have spin degeneracy only. (This point is 
discussed further in section \ref{old4.2}.)
The electron-electron interaction energy between  
two electrons at $x$ and $x'$ is given by 
$V_{\sig, \sig'}(x - x')$, as discussed in section \ref{old1.2.2}. 
The interaction is usually spin-isotropic, 
$V_{\sig, \sig'}(x - x') = V(x - x')$, and in this case 
represents Coulomb repulsion between the electrons. 
Since attention has been directed toward the spin-anisotropic 
Kondo lattice (Shibata, Ishii and Ueda 1995, 
Zachar, Kivelson and Emery 1996, Novais, {it et al.} 2002a,2002b),  
the Kondo lattice model will be derived here for 
general spin interactions. 
The total energy from electron-electron interactions 
in a lattice with two bands is given by 
\beqa
V = \frac{1}{2} \sum_{n_{1}, \ldots, n_{4}} 
\sum_{j_{1}, \ldots, j_{4}} \sum_{\sig, \sig'}
V_{\sig, \sig'}(n_{1}j_{1}, \ldots, n_{4}j_{4}) 
c^{\dg}_{n_{1}j_{1}\sig}c^{\dg}_{n_{2}j_{2}\sig'}
c^{}_{n_{3}j_{3}\sig'}c^{}_{n_{4}j_{4}\sig}
\label{4.1}
\eeqa
where $n_{i} = c, f$ labels the band and the $j_i$'s
labels the lattice site. This is a 
straightforward generalization of the single-band 
electron-electron interaction introduced in section 
\ref{old1.2.2}. From Eq.\ (\ref{1.2.9}),  
the matrix element in Eq.\ (\ref{4.1}) is given by 
\beqa
&& V_{\sig, \sig'}(n_{1}j_{1}, \ldots, n_{4}j_{4}) =  
\int_{L}dx \int_{L}dx'\, V_{\sig, \sig'}(x-x')   
\nonumber \\
&\times& \Phi^{*}_{n_{1}\sig}(x-j_{1}a)
\Phi^{*}_{n_{2}\sig'}(x'-j_{2}a)
\Phi^{}_{n_{3}\sig'}(x'-j_{3}a)
\Phi^{}_{n_{4}\sig}(x-j_{4}a)
\nonumber \\
\label{4.2}
\eeqa
in 1D, where $\Phi_{n\sig}(x - ja)$ is the wavefunction for the 
Wannier state $|nj\sig\rangle$ as in Eq.\ (\ref{1.6a}). The 
generalization to higher dimensions is straightforward. 

The Kondo lattice interaction $V_{\KL}$ may be derived from 
the full electron-electron interaction Eq.\ (\ref{4.1}) under 
the following assumptions, listed in order of importance: 
{\it i}) There is exactly one localized $f$-electron at each 
lattice site $j$. The set 
of lattice sites occupied by $f$-electrons is generally 
taken to be the entire lattice for the Kondo lattice model, 
but may be a small fraction of all lattice sites 
(modelling dilute Kondo impurities), 
and may even be a single lattice site 
(the original single impurity Kondo model
\footnote{A summary of the single impurity Kondo model results 
can be found in Appendix \ref{ch4kondo}.}). 
If this is the case, then the label $j$ in the Kondo lattice 
interaction (cf.\ Eq.\ (\ref{4.9}) below)
refers only to those lattice sites containing an 
$f$-electron. {\it ii}) The matrix elements 
$V_{\sig, \sig'}(n_{1}j_{1}, \ldots, n_{4}j_{4})$ of Eq.\ 
(\ref{4.2}) are assumed to be negligible, unless $j_{1} = 
j_{2} = j_{3} = j_{4}$. 
Thus only on-site interactions are considered, similar to the 
Hubbard model interaction. 
{\it iii}) Electron-electron interactions between electrons in the 
same band may be neglected. 
These three assumptions reduce the full electron-electron 
interaction of 
Eq.\ (\ref{4.1}) to on-site interactions between 
conduction electrons and $f$-electrons. Moreover, the 
interactions preserve the number of $f$-electrons at each  
site. The motivation for making assumptions {\it ii}) and 
{\it iii}) is 
mainly to distil the problem into the simplest form, but 
without losing the essential physics.
Assumption {\it iii}), neglecting intraband interactions, is made so 
as to focus on the interband interactions of central interest, 
and is often relaxed. On-site conduction electron-conduction 
electron interactions are considered for the 1D Kondo lattice 
in section \ref{old5.4}, and do not qualitatively alter the 
interband interactions. As for interactions between the 
localized electrons, note that on-site interactions are 
prohibited by assumption {\it i}). Nearest-neighbour interactions 
between the localized electrons in the 1D Kondo lattice are 
occasionally considered (White and Affleck 1996).
Indeed in real systems there will exist 
dipolar and exchange interactions between the localized 
electrons, as in the 1D Kondo lattice compound Cu(pc)I 
(Ogawa, {\it et al.} 1987). 
However, since for the Kondo lattice these 
interactions are nearest-neighbour, they will in general 
be far weaker than the on-site interactions
\footnote{Adding a large antiferromagnetic Heisenberg
intercation between the local moments produces
a spin-gapped metal (Sikkema, Affleck and White 1997) 
with unconventional pairing fluctuations 
(Coleman, Georges and Tsvelick 1997).}. A similar argument 
applies to nearest-neighbour conduction electron-conduction 
electron and conduction electron-localized electron 
interactions. This argument, which is identical to that made in 
the Hubbard model (cf.\ section \ref{old1.2.2} and see Hubbard (1963)), 
is the motivation for assumption {\it iii}). 
Assumption {\it i}) is the main assumption in the Kondo lattice, and 
distinguishes it from general two-band systems. The 
two-band materials for which assumption {\it i}) holds will be 
discussed in section \ref{old4.2}.

It is straightforward to verify that under 
assumptions {\it i}) - {\it iii}), Eq.\ (\ref{4.1}) reduces to
\beqa
V_{\KL} &=& V_{\rm dir} + V_{\rm ex}  \, , 
\nonumber \\
V_{\rm dir} &=& \sum_{j, \sig, \sig'}V^{\rm dir}_{\sig, \sig'} \, 
c^{\dg}_{cj\sig}c^{}_{cj\sig} n_{fj\sig'} \, , \quad
V^{\rm dir}_{\sig, \sig'} = V_{\sig, \sig'}(f0, c0, c0, f0)\, , 
\nonumber \\
V_{\rm ex} &=& -\sum_{j, \sig, \sig'}V^{\rm ex}_{\sig, \sig'}\,
c^{\dg}_{cj\sig}c^{}_{cj\sig'}c^{\dg}_{fj\sig'}c^{}_{fj\sig}\, , 
\quad
V^{\rm ex}_{\sig, \sig'} = V_{\sig, \sig'}(f0, c0, f0, c0)\, . 
\label{4.3} 
\eeqa
In Eq.\ (\ref{4.3}), 
$n_{fj\sig} = c^{\dg}_{fj\sig}c^{}_{fj\sig}$ is the $f$-electron 
number operator, and the matrix elements at right 
follow the notation of Eq.\ (\ref{4.2}). 

The Kondo lattice interaction $V_{\KL}$ may be written in 
conventional form by introducing pseudo-spin operators 
for the $f$-electrons: Since the $f$-electrons are confined to 
their lattice sites,  
the only degree of freedom available to the $f$-electrons 
is spin. Define the $f$-electron spin operator ${\bf S}_{j} = 
(S_{j}^{x}, S_{j}^{y}, S_{j}^{z})$ by 
\beqa
S^{x}_{j} &=& 
\frac{1}{2}\left( S^{+}_{j} + S^{-}_{j} \right) \, , 
\nonumber \\
S^{y}_{j} &=& 
\frac{1}{2i}\left( S^{+}_{j} - S^{-}_{j} \right) \, ,
\nonumber \\
S^{z}_{j} &=& 
\frac{1}{2}\left( n_{fj\uparrow} - n_{fj\downarrow}\right)\, . 
\label{4.4}
\eeqa
The spin raising and lowering operators are defined by
\beqa 
S^{+}_{j} = c^{\dg}_{fj\uparrow}c^{}_{fj\downarrow}\, , \quad
S^{-}_{j} = c^{\dg}_{fj\downarrow}c^{}_{fj\uparrow}\, .
\label{4.4a}
\eeqa 
A little algebra establishes the usual spin 
commutation relations for the 
Cartesian components of ${\bf S}_{j}$: 
\beqa
\left[ S^{\alpha}_{j}, S^{\beta}_{j'}\right] =
i\, \delta_{j, j'} \, \epsilon_{\alpha \beta \gamma}\, 
S^{\gamma}_{j} \, , 
\label{4.5}
\eeqa
where $\alpha, \beta, \gamma = x, y, z,$ and where 
$\epsilon_{\alpha \beta \gamma}$ is the third rank totally 
antisymmetric 
unit tensor. Using the localized spin operators, 
the Kondo lattice interaction 
$V_{\KL}$ of Eq.\ (\ref{4.3}) may be written in the form  
\beqa
V_{\KL} = \frac{J_{\parallel}}{2}\sum_{j}
\left(n_{cj\uparrow} - n_{cj\downarrow}\right) S^{z}_{j}
+ \frac{J_{\perp}}{2}
\sum_{j}\left(c^{\dg}_{cj\downarrow}c^{}_{cj\uparrow}S^{+}_{j} 
+ {\rm h.c.} \right) 
\label{4.6}
\eeqa 
to an additive constant depending on the dispersion of the
conduction electrons, where $n_{fj\sig} = c^{\dg}_{fj\sig}c^{}_{fj\sig}$
the $c$-electron number operator. 
The interaction parameters in Eq.\ (\ref{4.6}) are given by 
\beqa
J_{\parallel} &=& 2\left(V^{\rm dir}_{\parallel} 
- V^{\rm dir}_{\perp} - V^{\rm ex}_{\parallel} \right) \, ,
\nonumber \\
J_{\perp} &=& -2V^{\rm ex}_{\perp} \, 
\label{4.7}
\eeqa
where the direct and exchange integrals of Eq.\ (\ref{4.3}) 
have been decomposed into spin-parallel and spin-perpendicular 
components following Eq.\ (\ref{1.2.6}):
\beqa
V^{\rm dir}_{\sig, \sig'} 
&=& V^{\rm dir}_{\parallel}\delta_{\sig, \sig'} 
+ V^{\rm dir}_{\perp}\delta_{\sig, -\sig'}\, ,
\nonumber \\
V^{\rm ex}_{\sig, \sig'} 
&=& V^{\rm ex}_{\parallel}\delta_{\sig, \sig'} 
+ V^{\rm ex}_{\perp}\delta_{\sig, -\sig'}\, .
\label{4.8}
\eeqa
Note that $J_{\parallel} \neq J_{\perp}$ for a general  
spin-anisotropic interaction. For a spin-isotropic 
interaction, $V_{\sig, \sig'}(x - x') = 
V(x - x')$, the direct interaction between the conduction 
electrons and the $f$-electrons drops out of the problem, 
and the Kondo lattice interaction takes 
its conventional form 
\beqa
V_{\KL} = J\sum_{j} {\bf S}_{cj} {\bf \cdot} {\bf S}_{j}
\, , \quad \quad J = J_{\parallel} = J_{\perp} \, ,
\label{4.9}
\eeqa
where ${\bf S}_{cj}$ are pseudo-spin operators for the 
conduction electrons, defined as in Eqs.\ (\ref{4.4}) 
and (\ref{4.4a}) for the 
$f$-electrons. Eq.\ (\ref{4.9}) represents a Heisenberg-type 
interaction between an $f$-electron, and the conduction 
electron spin at the same site. The interaction parameter in 
the spin-isotropic case is given by the exchange integral 
\beqa
J = -2V^{\rm ex} = 
2\int_{L}dx \int_{L}dx'\, V(x-x') \Phi^{*}_{f}(x)\Phi^{*}_{c}(x')
\Phi^{}_{f}(x')\Phi^{}_{c}(x) \, ,
\label{4.10}
\eeqa
where $\Phi_{n}(x)$ is the Wannier wavefunction 
Eq.\ (\ref{1.6a}) at $j = 0$ without the Pauli spinor. 

The full Kondo lattice hamiltonian $H_{\KL}$ is obtained by 
adding the kinetic energy to the interaction $V_{\KL}$. 
Since the $f$-electrons are fixed at their lattice sites, 
the kinetic energy of the $f$-band is constant, and may be 
neglected. The full hamiltonian for the Kondo lattice then 
consists of $V_{\KL}$, together with the conduction electron 
kinetic energy. In the simplest case, the conduction electrons 
hop between nearest-neighbour sites $\langle ij \rangle$ only, 
as discussed in 
section \ref{old1.2.1} (cf. Eq.\ (\ref{1.2.4}) for 1D). 
The hamiltonian for the Kondo lattice is then
\beqa
H_{\KL} = 
-t\sum_{\langle ij \rangle}
\sum_{\sig}c^{\dg}_{ci\sig}c^{}_{cj\sig} 
+ J\sum_{j} {\bf S}_{cj} {\bf \cdot} {\bf S}_{j}  \, , 
\label{4.11}
\eeqa
for a spin-isotropic interaction. $H_{\KL}$ is the 
conventional Kondo lattice hamiltonian. It is 
straightforward to extend $H_{\KL}$ to  
include next-nearest-neighbour hopping, and so on, 
and spin-anisotropic interactions may be 
included by using Eq.\ (\ref{4.6}) for $V_{\KL}$
(see eg, Shibata, Ishii and Ueda (1995), Zachar, 
Kivelson and Emery (1996), Novais, {it et al.} (2002a,2002b)). 
While some of these 
(and other) variants of the Kondo lattice will be 
discussed at various stages later, the 
conventional Kondo lattice (in 1D) will be the {\it main} 
{\it focus} of this review.  
Unless otherwise noted, Eq.\ (\ref{4.11}) 
{\it defines} the Kondo lattice model as discussed here.

The Kondo lattice model contains two parameters. The first is 
the coupling $J$ (or the dimensionless parameter $J/t$). 
Both large and small values of $|J|/t$ are physically relevant, 
as will be discussed immediately below in section \ref{old4.2}. 
For large values the relevant sign of $J$ is negative, and 
this is called a ferromagnetic coupling since it favours an 
alignment of the conduction electron spin with the spin of the 
localized $f$-electron. For small values of $|J|/t$ the relevant 
sign of $J$ is positive, and this is called an antiferromagnetic 
coupling since it favours an opposite alignment of the conduction 
electron spin with the spin of the localized $f$-electron. 
The second 
parameter in the Kondo lattice is the number of conduction 
electrons. This is measured by the filling $n = N_{e}/N$, where 
$N_{e}$ is the number of conduction electrons and $N$ is the 
number of lattice sites. $N$ coincides with the number of 
$f$-electrons. A half-filled conduction band corresponds to 
$n = 1$, and $n<1$ is called a partially-filled band. 
It is usual to consider only fillings in the range 
$0 < n \leq 1$ in the Kondo lattice, and attention is restricted 
to this range.

\section{\label{old4.2}Relevance of the Kondo lattice to real materials}

The Kondo lattice describes materials in which the 
predominant interactions are between two distinct varieties 
of electrons; localized electrons possessed of a magnetic 
moment, and itinerant conduction electrons. The two varieties 
are described in the derivation of the previous section with 
the band indices $f$ and $c$, respectively. This situation 
is realised in two broad and important classes of materials: 
(i) Manganese oxide perovskites, in which there 
generally exists a 
mixture of ${\rm Mn}^{3+}$ and ${\rm Mn}^{4+}$ ions. 
(ii) Rare earth and actinide compounds, broadly classed as 
heavy fermion materials, in which atomic-like 
$f$-electrons interact with a conduction band. The Kondo 
lattice description of these materials is discussed in 
sections \ref{old4.2.1} and \ref{old4.2.2}, respectively. 
Before proceeding, 
however, it is useful to give a brief discussion of what 
the Kondo lattice model does {\it not} describe. 

In using the Kondo lattice to model real materials, first note 
that as an electron system the Kondo lattice model neglects  
electron-phonon coupling; it ignores interactions between the 
electrons and the vibrations of the underlying lattice of ions. 
This may result in the Kondo lattice model being unable to 
reproduce important properties of real materials. 
For example, Millis, Littlewood and Shraiman (1995)
have argued that this is indeed 
the case in the manganese oxide perovskites. The Kondo lattice 
is often used to model this class of materials 
(cf.\ section \ref{old4.2.1} below). Millis, {\it et al.} (1995) argue 
that the predictions of the Kondo lattice disagree with the 
experimental data by an 
order of magnitude or more. They propose that 
the discrepancy arises from the neglect of strong 
electron-phonon coupling coming from a Jahn-Teller splitting 
of the ${\rm Mn}^{3+}$ ion. If this is true (and the issue is not 
yet settled), then an electron-phonon 
interaction term must be added to $H_{\KL}$
\footnote{The effect of phonons on the 1D Kondo lattice model has been
studied by Gul\'{a}csi, Bussmann-Holder and Bishop (2003,2004) via bosonization.
Details can be found in section \ref{7.4}.}.

A second point to note is that the Kondo lattice model 
Eq.\ (\ref{4.11}) allows only two degrees of freedom for each 
band; either spin up or spin down. In the materials described 
by the Kondo lattice there is further degeneracy:
In the manganese oxides, the 
localized electrons are three Hund's rule coupled $d$-band 
electrons, and the localized band carries spin 3/2. In the 
heavy fermion 
materials, the localized electrons are $f$-electrons, and 
these carry an orbital degeneracy when the multiplet splitting 
is small. It follows that for a more detailed description of 
real materials, it would be necessary to include more than two 
degrees of freedom per band. A model which incorporates this 
is the Coqblin-Schrieffer model (Coqblin and Schrieffer 1969),
and may be
formally obtained from the standard Kondo lattice 
hamiltonian by generalizing to higher spin. 
While the use of Eq.\ (\ref{4.11})  
thus prevents a detailed description of the behaviour of 
individual compounds on a case by case basis, 
it does describe the interaction 
between extended and localized states which is the heart of the 
problem. Once the simplest non-degenerate case is understood, 
and much work remains to be done in this direction, then  
the effects of degeneracy may be gradually included. This may 
yield essentially the same behaviour, which is suggested for 
the ground-state phase diagram of the Kondo lattice. 
Dagotto, {\it et al.} (1998) observed   
the same basic phase diagram for both  
spin 3/2 or spin 1/2 localized moments in the 
ferromagnetic $J<0$ Kondo lattice. Alternatively, new effects 
may arise with the degeneracy, as for example the 
non-Fermi liquid behaviour found for the single impurity 
Kondo hamiltonian (Nozi\`{e}res and Blandin 1980)
\footnote{More accurately, the single impurity Kondo model 
turns out to be a {\it local} Fermi liquid, for details,
see Andrei, Furuya and Lowenstein (1983). The concept of
local a Fermi liquid was introduced by Newns and Hewson 
(1980) who used it to interpret experimental data
on rare earth compounds. It corresponds to a non-interacting
multi level resonant model.}. In either case an 
understanding of the simplest non-degenerate model is 
invaluable, and seems a prerequisite for understanding  
more detailed models.

\subsection{\label{old4.2.1}Manganese oxide perovskites}

The study of manganese oxide compounds with a perovskite 
crystal structure dates back almost half a century, 
and began with the work of Jonker and Van Santen (1950).
The compounds have the form 
${\rm R}_{1-x}{\rm A}_{x}{\rm MnO}_{3}$, where R = La, 
Nd, Pr is a trivalent rare earth, and A = Ca, Sr, Ba, Cd, Pb 
is divalent, and usually an alkaline earth. These materials 
present a rich phase diagram, with generic low-temperature 
phases as follows (Tokura, {\it et al.} 1996): At low doping 
$x \lesssim 0.2$, there is a spin-canted insulating state. 
This is often followed by a small doping region $\Delta x \sim 
0.05$ which presents a ferromagnetic insulator. For 
$0.2 \lesssim x \lesssim 0.5$ the Mn oxide perovskites 
are ferromagnetic metals. These materials have recently 
attracted much renewed interest due to the discovery of 
colossal magnetoresistance (Jin, {\it et al.} 1994):
The magnetoresistance 
\beqa 
\frac{\Delta \rho }{\rho (0, T)} = 
\frac{\rho (H, T) - \rho (0, T)}{\rho(0, T)} \, ,
\label{4.11a} 
\eeqa
where $\rho(H, T)$ is the resistivity in applied magnetic field 
$H$ at temperature $T$, is observed to vary
$\sim 100,000$ per cent in thin films of manganese oxide compounds 
near the Curie temperature in the metallic ferromagnetic phase. 
This has stimulated great interest because of the potential 
technological applications in magnetic recording heads. 

The relevance of the Kondo lattice to manganese oxide 
perovskites arises from the properties of the $3d$ shell 
electrons in Mn. In undoped compounds ${\rm RMnO}_{3}$, 
the manganese atoms are all triply ionized, and contain 
four $3d$ electrons in their outer shell. In the perovskite 
lattice the $3d$ band splits, and ${\rm Mn}^{3+}$ has the 
following configuration (Goodenough 1955): Three electrons 
occupy the lower three-fold degenerate localized $t_{2g}$ 
orbitals, and there is one electron in an upper two-fold 
degenerate delocalized $e_{g}$ orbital. Upon doping the 
trivalent rare earth R with a divalent atom A, such as an 
alkaline earth, extra electrons are stripped from the Mn 
atoms, and there exists a mixture of ${\rm Mn}^{3+}$ and 
${\rm Mn}^{4+}$ ions. The latter are missing the outer 
$e_{g}$ electron. 

The Kondo lattice describes the electrons in the $3d$ Mn 
shell as follows: A very strong 
Hund's rule coupling forces the alignment of the spins of 
the localized $t_{2g}$ electrons, and these act as a localized 
spin 3/2. (See Landau and Lifshitz (1965), for example, for a 
discussion of Hund's rules.) 
The $t_{2g}$ electrons in the manganese oxides form 
the localized band in the Kondo lattice model, and as noted 
above are approximated by spins 1/2 in the simplest case. 
The Kondo lattice conduction band models the the delocalized 
$e_{g}$ orbitals. These are coupled to the localized electrons 
by a Hund's rule coupling; as for the Hund's rule coupling 
between the $t_{2g}$ electrons, the coupling is strong and 
favours an alignment of the $e_{g}$ electron spin with that of 
the localized $t_{2g}$ spins. Thus the parameter regime of the 
Kondo lattice which is appropriate for the manganese oxides 
is one with a large ferromagnetic coupling: $J<0$, 
$|J|/t > 1$. 
The conduction band filling appropriate for these materials 
is variable. In undoped manganese oxide compounds the conduction 
band is half-filled, with one $e_{g}$ conduction electron per 
localized spin. Upon doping, one obtains a Kondo lattice with 
a partially-filled conduction band having less than one 
conduction electron per localized spin. 

\subsection{\label{old4.2.2}Rare earth and actinide compounds}

Many intermetallic compounds involving rare earth or actinide 
elements exhibit complex and intriguing properties, and have 
fascinated both experimentalists and theorists for decades. 
One broad class of these compounds are the heavy fermion 
materials; a characteristic feature is that the linear 
coefficient of the specific heat is two or three orders of 
magnitude greater in these compounds than in normal 
metals.\footnote{A similar enhancement is observed for the 
spin susceptibility.}
This enhancement may be accounted for by supposing that the 
quasiparticle effective mass $m^{*}$ (Nozi\`{e}res 1964) 
is two or three orders of magnitude greater than the bare 
electron mass, and thus the heavy fermions. Heavy fermion 
materials exhibit a striking diversity of ground states, 
including magnetically ordered states  
(${\rm CeAl}_{2}$ and ${\rm U}_{2}{\rm Zn}_{17}$), 
novel (non-BCS) superconducting states  
(${\rm CeCu}_{2}{\rm Si}_{2}$ and ${\rm UBe}_{13}$), and 
ground states which are neither magnetically ordered nor 
superconducting (${\rm CeAl}_{3}$ and ${\rm UAl}_{2}$).  
A detailed discussion of the early heavy 
fermion compounds and their experimental properties  
may be found in the review by Stewart (1984). 
A review discussion of effective theoretical models for 
heavy fermion materials is given by Lee, {\it et al.} (1986). 

Along with the heavy fermion compounds, there are several 
related classes of compounds containing rare earth and actinide 
elements which have attracted great interest in the last 
decade. One is the class of Kondo insulators, and includes 
CeNiSn, ${\rm Ce}_{3}{\rm Bi}_{4}{\rm Pt}_{3}$, 
TmSe, and UNiSn. These are 
small gap semiconductors in which the gap, of only a few 
meV, arises from hybridization between the $f$-electrons 
in the rare earth and actinide ions, and a conduction band. 
The Kondo insulators show very different behaviour from normal 
band insulators which have a gap at least of a few tenths of an 
eV. The Kondo insulators are reviewed by Aeppli, {\it et al.} 
(1992) and Fisk, {\it et al.} (1995).  
Another class is the low-carrier-density Kondo systems 
which include the trivalent cerium monopnictides CeX (X = P, 
As, Sb, Bi), and uranium and ytterbium analogues USb, YbAs, 
among others. (See Suzuki (1993) for a brief review.) 
These systems have carriers (either conduction electrons or 
holes) whose densities are much less than those of the 
magnetic rare earth or actinide ions. The low-carrier-density 
Kondo systems show heavy fermion behaviour including the extreme 
case of insulating states, and exhibit very interesting and 
complex magnetic properties. These include ferromagnetic 
ordering in one plane, and a complex stacking of 
the ferromagnetic planes through the crystal. 

The Kondo lattice model is relevant to all of the  
compounds mentioned above; heavy fermion materials, 
Kondo insulators, 
and low-carrier-density Kondo systems. This is due 
to the $f$-electrons from the rare-earth and actinide elements, 
which remain essentially atomic-like in the compounds. 
The rare earth elements from cerium to thulium have atomic 
configurations $[{\rm Xe}]4f^{n}5d^{0 \: {\rm or} \: 1}6s^{2}$ 
with partially filled $4f$ shells $2 \leq n \leq 13$. 
There is a similar progression in the actinide series as the 
$5f$ shell is filled. The partially filled $f$-shells lead to a 
variety of magnetic states for these elements. In compounds 
involving these elements the $f$ orbitals remain strongly 
localized, and essentially retain their atomic character: The 
magnetic effects present in the isolated atoms persist in many  
rare earth and actinide compounds. Thus the sites containing 
rare earth or actinide atoms often possess 
a magnetic moment corresponding to the Hund's rule  
maximization of the total $f$-electron spin. 
In the compounds discussed above the $f$-electrons 
interact with electrons in the conducting $d$ (or hybridized 
$s$-$d$) band. The conduction band filling depends on the 
particular compound structure, and varies from small fillings 
in the low-carrier-density Kondo systems to half-filling for the 
Kondo insulators. The interaction between the conducting electrons 
and the localized electrons in the $f$ orbitals is the basis 
of the relevance of the Kondo lattice model to these materials. 

The description of heavy fermion, Kondo insulating, and 
low-carrier-density Kondo systems using the Kondo lattice model 
is indirect, and proceeds via the more fundamental periodic 
Anderson model. To introduce this model, it is best to  
consider a concrete example (Varma 1976). 
SmS is an ionic semiconductor at atmospheric pressure, and 
contains ${\rm Sm}^{2+}$ and inert ${\rm S}^{2-}$ ions. 
The $4f^{6}$ 
band level\footnote{The bands corresponding to the $4f$ orbitals 
are energetically narrow, or atomic-like, with a 
delta-function density of states $\rho(\varepsilon) 
\approx \delta(\varepsilon - \varepsilon_{f})$, where 
$\varepsilon_{f}$ is the $f$ level. Note that these are not 
bands in the usual sense; band theory 
begins from a basis of non-interacting 
delocalized Bloch states (cf.\ section \ref{old1.1.1}), and provides a 
poor description of the properties of partially-filled $4f$ 
shells, in which the $f$ electrons are localized and interact 
strongly with each other. For comparison with the energies of 
well-defined conduction bands, it is however 
convenient and conventional to consider a $4f$ `band' for a 
given compound with a $4f^{n}$ nominal occupation on 
constituent rare earth atoms. The $f$ level $\varepsilon_{f}$ 
is then the energy for the process 
$4f^{n} \rightarrow 4f^{n-1}$ of removing an $f$ electron. 
For example, in SmS $\varepsilon_{f}$ is the energy for the 
process $4f^{6} \rightarrow 4f^{5}$ (Varma 1976, 
Hewson 1993).}
 $\varepsilon_{f}$ lies just below the (unoccupied) 
$5d$ and $6s$ conduction bands in Sm. If pressure is applied,  
the lower of the crystal-field split $5d$ bands of Sm 
broadens, due to the increased overlap of its Wannier states, 
and moves down in energy relative to $\varepsilon_{f}$. 
Ultimately the $5d$ band crosses the level of the $4f^{6}$ band.
When this occurs, there are valence fluctuations 
on each Sm site, as $f$ electrons enter $5d$ conduction 
band states, and SmS becomes metallic. The metal-insulator 
transition is accompanied by a change in the colour and 
volume of SmS. The central lesson of this transition is that 
in order to understand the metallic state, 
it is necessary to understand excitations from 
the $4f$ localized band into the $5d$ conduction band. This 
situation of fluctuating valence is described by the periodic 
Anderson model, which considers interactions in which localized 
electrons may be 
excited to the conduction band, and vice versa, in interband  
interactions. 

The periodic Anderson model is the formal 
extension to the lattice of the 
single impurity Anderson model (Anderson 1961). 
It describes a band of conduction electrons together 
with localized $f$ orbitals at each lattice site. The 
interaction between the localized and extended states
is distinct from that in the Kondo lattice model, and 
describes excitations into and out of the localized 
$f$ orbitals. On a lattice 
with $N$ sites, the periodic Anderson hamiltonian is given by 
\beqa
H_{\PAM} &=& \sum_{k, \sig} 
\varepsilon (k) c^{\dg}_{ck\sig} c^{}_{ck\sig} + 
\varepsilon_{f}\sum_{j, \sig} n_{fj\sig} 
+ U\sum_{j} n_{fj\uparrow}n_{fj\downarrow}  
\nonumber \\
&+& N^{-1/2}\sum_{k,j, \sig}\left( 
V_{k}e^{ikja}c^{\dg}_{fj\sig}c^{}_{ck\sig} 
+ {\rm h.c.}\right) \, ,
\label{4.12a}
\eeqa
where the conduction electrons are written in terms of Bloch 
states with dispersion $\varepsilon(k)$. 
$\varepsilon_{f}$ is the level of the flat band of 
localized $f$ orbitals. The hybridization $V_{k}$ gives the 
amplitude for a localized $f$-electron to be excited to a 
conduction band Bloch state with crystal momentum 
$k$.\footnote{According to standard band theory, $V_{k}$ would 
be zero, since the states in different bands are orthogonal 
(cf.\ section \ref{old1.1.1}). However, as noted above, the $f$ orbitals 
are strongly correlated, and do not constitute a band in the 
rigorous sense.} An important element in $H_{\PAM}$ is the 
Coulomb repulsion $U$ between on-site $f$-electrons. This is the 
by far the strongest interaction in real materials,  
$U \sim 5\, {\rm eV}$ (Varma 1976), and large $U$ 
will be assumed in the following. 

The hybridization between the conduction band and the localized 
orbitals can be an important element in rare earth and actinide 
compounds, as is clear from the example of SmS above. 
Hybridization makes the periodic Anderson model 
superficially different from the Kondo lattice 
model, in which the $f$-electrons are fixed at their 
lattice sites, and excitations from localized orbitals to 
the conduction band are forbidden. Schrieffer and Wolff (1966)  
showed that this difference is in fact superficial only, and 
that the periodic Anderson model reduces to the Kondo lattice 
model in the local moment regime. (The exact canonical transformation
given later on in section \ref{sectionSW} is valid for any 
value of the periodic Anderson model parameters.) 
The local moment regime has 
the $f$ level $\varepsilon_{f}$ below the conduction electron 
chemical potential. In this case, for zero hybridization 
$V_{k} = 0$, the ground-state consists of one localized 
electron in each $f$-orbital, and a non-interacting Fermi sea 
of conduction electrons with chemical potential 
$\varepsilon(k_{F})$ at zero temperature. 
($k_{F}$ is the conduction electron Fermi momentum.) 
The local moment  
regime thus has $\varepsilon_{f} - \varepsilon(k_{F}) < 0$ and 
$\varepsilon_{f} - \varepsilon(k_{F}) + U > 0$. 
(Large $U$ is assumed.) The Schrieffer-Wolff transformation 
(Schrieffer and Wolff 1966) 
consists of perturbing $H_{\PAM}$ out of the $V_{k}=0$ 
ground-state to second-order in the hybridization, and 
yields the hamiltonian 
\beqa
H = \sum_{k, \sig} 
\varepsilon (k) c^{\dg}_{ck\sig} c^{}_{ck\sig} &+& 
\frac{1}{2N}\sum_{k, k', j} J_{k, k'} e^{i(k'-k)ja} \left\{ 
\left(c^{\dg}_{ck\uparrow}c^{}_{ck'\uparrow} -
c^{\dg}_{ck\downarrow}c^{}_{ck'\downarrow}\right)S^{z}_{j}
\right. \nonumber \\
&+& \left. c^{\dg}_{ck\downarrow}c^{}_{ck'\uparrow}\, 
S^{+}_{j} + c^{\dg}_{ck\uparrow}c^{}_{ck'\downarrow}\, 
S^{-}_{j} \right\} \, , 
\label{4.13}
\eeqa
to an unimportant pure potential scattering term involving the 
conduction electrons (see Eq. (\ref{sw5}) and Appendix \ref{ch4sw}). 
Eq.\ (\ref{4.13}) is a Kondo lattice hamiltonian, as in 
Eq.\ (\ref{4.11}), but written in a basis of Bloch states
\footnote{
The same notation is pertained as in Eqs.\ (\ref{4.4}) and
(\ref{4.4a}) i.e., ${\bf S}_{j}$ represent the pseudo-spin 
operators for the $f$-electrons.}. 
It is a valid effective hamiltonian for the periodic 
Anderson model provided the hybridization $|V_{k}|$ is small 
compared to $|\varepsilon_{f} - \varepsilon(k_{F})|$ and to
$|\varepsilon_{f} - \varepsilon(k_{F}) + U|$, so that the 
terms neglected in the perturbation expansion are negligible.
The Schrieffer-Wolff transformation gives a weak 
antiferromagnetic Kondo coupling for large $U$ in the local 
moment regime (Schrieffer and Wolff 1966):
\beqa
J_{k, k'} = V^{*}_{k}V^{}_{k'}\left\{ 
\frac{1}{U + \varepsilon_{f} - \varepsilon(k')} +
\frac{1}{\varepsilon (k') - \varepsilon_{f}}  \right\} \geq 0\, .
\label{4.14}
\eeqa
The Kondo lattice hamiltonian $H_{\KL}$ with a weak 
antiferromagnetic coupling ($J>0$, $J/t < 1$) thus describes 
rare earth and actinide compounds with weak valence 
fluctuations in the local moment regime.

\subsection{\label{sectionSW}The Exact Schrieffer-Wolff transformation}

As presented in the previous section, more than thirty years ago, 
Schrieffer and Wolff (1966) established the relationship between 
the single impurity Anderson and the Kondo hamiltonian's 
using a canonical transformation. 
The transformation was originally carried out, up to first 
order, and later on extended to the third order (Zaanen and 
Ole\'{s} 1988, Zhou and Zheng 1992, Kolley, Kolley and Tietz 1992, 
Kehrein and Mielke 1996). These results were later reobtained  
by other numerous (fourth and sixth order) perturbation 
calculations (Yosida and Yoshimori 1973). 

Scrieffer and Wolff (1966) originally used a canonical transformation:
their method is followed here. An equivalent transformation can be
accomplished via perturbation theory. 
In this case, see Eq.\  (\ref{egy}), 
transformed Hamiltonian terms $H_{n}$ are 
proportional to $V^{n+1}$, hence correspond to the $2 n$ perturbation 
expansion in $H_{V}$ (Yosida and Yoshimori 1973), e.g., 
the first transformed Hamiltonian term $H_1$ correspond 
to the second order perturbation result, the third transformed 
Hamiltonian term to the sixth order perturbation result, etc. 

The starting Hamiltonian is Eq.\ (\ref{4.12a}) which for a
single impurity Anderson model is re-written as
$H_{\rm SIAM} =    H_{0} + H_{V}$, where 
\beqa
H_{0} = \sum_{k, \sig} 
\varepsilon (k) c^{\dg}_{ck\sig} c^{}_{ck\sig} + 
\varepsilon_{f} \sum_{\sig} n_{f \sig} 
+ U n_{f j \uparrow} n_{f j \downarrow} \, , 
\label{sw1}
\eeqa
and 
\beqa
H_{V} = \sum_{k, \sig} \left( 
V_{k} c^{\dg}_{f \sig} c^{}_{c k \sig} + {\rm h.c.}\right) \, .
\label{sw2}
\eeqa
There is no constraint on the value of $\varepsilon_{f}$, and so the 
obtained results are valid for the general asymmetric Anderson model. 

The transformed Hamiltonian $e^{{\cal S}} ( H_{0} + H_{V} ) 
e^{-{\cal S}}$ can be written in the form $H_{0} + 
\sum^{\infty}_{n = 1} [ 1 / n! - 1 / (n+1)!] \: H_{n}$, 
with 
\beqa
{\cal S} = \sum_{{k}, \sigma} 
( A_{{k}} + Z_{{k}} 
c^{\dagger}_{f -\sigma} c^{}_{f -\sigma} ) 
( V_{{k} } c^{}_{c {k} \sigma} 
c^{\dagger}_{f \sigma} - {\rm h.c.}) \; ,
\label{sw3}
\eeqa
and 
\begin{equation}
H_{n} \: = \: {\buildrel{n \: {\rm times}}\over
{\overbrace{
[{\cal S}, [{\cal S}, [{\cal S}, \: \cdots \: [{\cal S}}}},H_{V}]
\: \cdots \: ]]] \; , 
\label{egy}
\end{equation}
where 
\beqa
A_{k} &=& 1/(-\varepsilon ({k}) + \varepsilon_{f})  
\nonumber \\
Z_{k} &=& 1/(-\varepsilon ({k}) + \varepsilon_{f} + U) - 
1/(-\varepsilon ({k})  + \varepsilon_{f}) \; . 
\label{sw4}
\eeqa
This satisfies the equation $H_{V} + [{\cal S}, H_{0}] = 0$, see
also Appendix \ref{ch4sw}. 

The first term of the transformation is the celebrated 
Schrieffer and Wolff (1966) result: 
\begin{eqnarray}
H_{1} &=& \sum_{{k}, {k}^{\prime}, \sigma} 
V_{{k}} V_{{k}^{\prime}} \: \biggl\{ \Big[
(A_{{k}} + A_{{k}^{\prime}}) + 
(Z_{{k}} + Z_{{k}^{\prime}}) n^{}_{f} \Big] 
 (n^{}_{f \sigma} \delta_{{k}, {k}^{\prime}} 
- c^{\dagger}_{c {k} \sigma} c^{}_{c {k}^{\prime} \sigma}) 
\nonumber \\
&& +(Z_{{k}} + Z_{{k}^{\prime}}) 
 \: \Big[ c^{\dagger}_{c {k} - \sigma}
c^{}_{c {k}^{\prime} \sigma} c^{\dagger}_{f \sigma}
c^{}_{f - \sigma} 
\nonumber \\
&& - \frac{1}{2} (c^{\dagger}_{c {k} - \sigma} 
c^{\dagger}_{c {k}^{\prime} \sigma} c^{\dagger}_{f - \sigma} 
c^{}_{f \sigma} 
+ c^{\dagger}_{f - \sigma} c^{}_{f - \sigma} 
c^{}_{c {k} - \sigma} c^{}_{c {k}^{\prime} \sigma}) 
\Big] \biggr\} \; . 
\label{sw5}
\end{eqnarray} 
The coefficient of the term $c^{\dagger}_{c {k} \sigma} 
c^{}_{c {k}^{\prime} - \sigma} c^{\dagger}_{f - \sigma} 
c^{}_{f \sigma}$ gives the strength of the Kondo exchange 
interaction, $J_{1}$ given in Eq.\ (\ref{4.14}). To simplify
the notation,  only the low energy contributions will be
kept (Schrieffer and Wolff 1966) hence 
$\varepsilon ({k}) \equiv \varepsilon ({\bf k})$ will depend 
on $\vert {\bf k} \vert$ and all the $\vert {\bf k} \vert$'s 
$\approx$ $k_{F}$. 
In this, Eq.\ (\ref{4.14}) can be simply written as
\beqa
J_{1} \: = \: 2 V^2 Z  \, , 
\label{sw6}
\eeqa
where $V \equiv V_{k_F d}$,  $A \equiv A_{k_F}$ and $Z \equiv Z_{k_F}$.

Continuing the transformation to second order, it is found
(Chan and Gul\'{a}csi 2001a,2004):  
\begin{equation}
H_{2} = \sum_{{k}, \sigma} \: 
({\widetilde{V^{}_{2}}} + {\widetilde{V^{c}_{2}}} 
n_{c {k} \sigma} + {\widetilde{V^{f}_{2}}} n_{f \sigma} 
+ {\widetilde{V^{cf}_{2}}} n_{c {k} \sigma} n_{f \sigma})
(c^{\dagger}_{c {k} - \sigma} c^{}_{f - \sigma} + 
{\rm h.c.}) \; ,
\label{ketto}
\end{equation}
where ${\widetilde{V^{}_{2}}} = - 4 A^2 V^3$, 
${\widetilde{V^{c}_{2}}} = 2 Z V^3 (A - Z)$, 
${\widetilde{V^{f}_{2}}} = -2 Z V^3 (5 A - 4Z)$ 
and ${\widetilde{V^{cf}_{2}}} = 6 Z^2 V^3$.
Usually these (and subsequently all even order) terms are   
neglected on the ground that they only renormalize the 
starting Hamiltonian. However, this is not entirely correct
as it was pointed out by Chan and Gul\'{a}csi (2001a,2004). 
A close inspection of Eq.\ \ref{ketto} reveals that 
the second (and all even) order terms 
take care of all possible single particle 
(and two particles starting from $n \ge 3$) 
virtual processes which can occur between 
the energy shells of the impurity site. The presence 
of the new terms in Eq.\ \ref{ketto} will have radical 
consequences to the behaviour of the Kondo exchange coupling 
constant, which in the third order (Chan and Gul\'{a}csi 2001a,2004) 
can be expressed as: 
\begin{equation}
J_{3} = 2 V [ Z({\widetilde{V^{}_{2}}} 
+ {\widetilde{V^{f}_{2}}}) - A({\widetilde{V^{c}_{2}}} - 
{\widetilde{V^{f}_{2}}})] \; , 
\label{harom}
\end{equation}
or explicitly: 
\beqa
J_{3} \: = \: - 2^{4} V^{4} (2 A^2 Z + 2 A Z^2 + Z^3) \; . 
\label{sw7}
\eeqa

\begin{figure}[ht]
\centering
\includegraphics[scale=0.7]{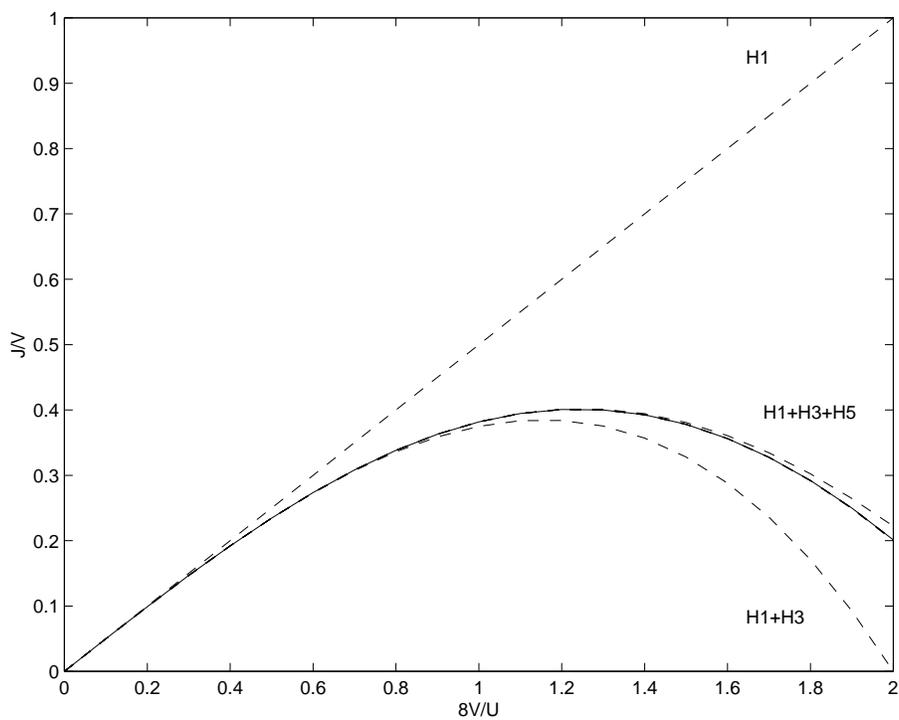}
\vspace{3.0mm}
\caption{\label{fig:kondoresult}
The spin coupling constant $J$ up to different order of 
transformation. H1 is the Schrieffer Wolff result, H1+H3 is the result
up to the third order of the transformation, and similarly for H1+H3+H5. 
Solid line shows the result up to infinite order.} 
\end{figure}

The contributions of $H_{1} + H_{3}$ are plotted in Fig.\ 
\ref{fig:kondoresult}, 
which shows that the Kondo exchange can reverse sign. This 
sign oscillation is due to the ${\widetilde{V^{c}_{2}}}$ 
and ${\widetilde{V^{f}_{2}}}$ new terms (which are equal 
for the symmetric case) of Eq.\ (\ref{ketto}): 
take for example the case where two electrons of opposite 
spin occupy the $f$ shell as an intermediate state. 
Instead of the the ``down'' electron, which caused the excitation, 
returning to the Fermi sea, the ``up''
electron did so, it picks up a change of sign because of
the Pauli principle (the ordering of the two electrons has
been interchanged). In perturbation theory, the $J_{3}$
term corresponds to a sixth order process (as mentioned earlier)
of two exchanged electrons. This phenomena is 
the Friedel (i.e., Ruderman-Kittel) oscillations, 
where the polarisation periodically reverses as one proceeds 
to higher orders. Similar oscillations can also be observed  
in the renormalization group approach (Wilson 1975) 
in terms of energy shells. This sign oscillation 
is important for our analysis, as this will guarantee 
that the  $J_{n}$ series is convergent. 
In this respect, it is interesting to mention that the 
$n$-th order canonical transformation corresponds to the 
$m$-th shell ($2 m + 1 = n$) calculation in the 
renormalization group approach (Wilson 1975). 
For asymmetric cases $\varepsilon_{f}$ can be chosen 
in such a way to get rid of this fluctuation, for some 
examples see Fig.\ \ref{fig:andersonasymetric}. 

The canonical transformation can be continued up to eleventh order
(Chan and Gul\'{a}csi 2001a,2004) however already in fifth order
a pattern in the $J$ coefficients is observed, see 
Table \ref{tab:1-3OrderResult} of Appendix \ref{ch4sw}, or 
Chan and Gul\'{a}csi (2001a,2003,2004): 
\begin{equation}
J_n \: = \: (-1)^{n-1} 2^{3(n-1)/2} \: 
\biggl[ \frac{1}{n!} - \frac{1}{(n + 1)!} \biggr] \: 
Z V^{n+1} \Big[ (Z+A)^{2} + A^{2} \Big]^{(n - 1)/2} \; . 
\label{negy}
\end{equation}
This pattern has been
proven to be valid for any order by mathematical induction.
The proof is presented in detail in Appendix \ref{ch4sw}, the derivation 
is based on Chan and Gul\'{a}csi (2003). 

Summing up the coefficients from Eq.\ (\ref{negy}) up to infinity 
a remarkably simple result is obtained (Chan and Gul\'{a}csi 2001a,2004): 
\begin{eqnarray}
J \: = \: 2 Z V^{2} \: \biggl( \frac{\sin\beta}{\beta} 
+ \frac{\cos\beta -1}{\beta^{2}} \biggr) \; , 
\label{ot}
\end{eqnarray}
where
\begin{equation}
\beta \: = \: \sqrt{2^{3}V^{2} [ (Z+A)^{2}+A^{2}]} \; . 
\end{equation}

\begin{figure}[ht]
\centering
\includegraphics[scale=0.55]{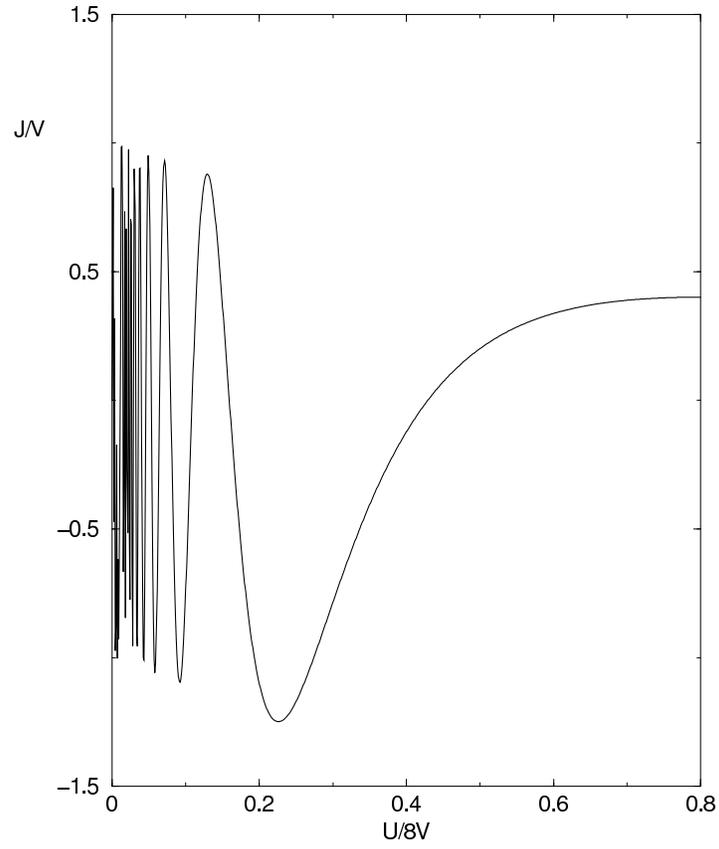}
\vspace{3.0mm}
\caption{\label{fig:kondoJresult}
The exact Kondo exchange interaction, Eq.\ (\ref{ot}) is 
plotted as a function of $U / 8 V$. This way of plotting Eq.\ (\ref{ot})  
shows clearly the two different scaling limits:
local moment regime (Kondo limit) for $U \gg V^2$ 
and the oscillatory mixed valence regime, $\varepsilon_{f} 
\approx \epsilon (k_F)$.} 
\end{figure}

\begin{figure}[ht]
\centering
\includegraphics[scale=0.60]{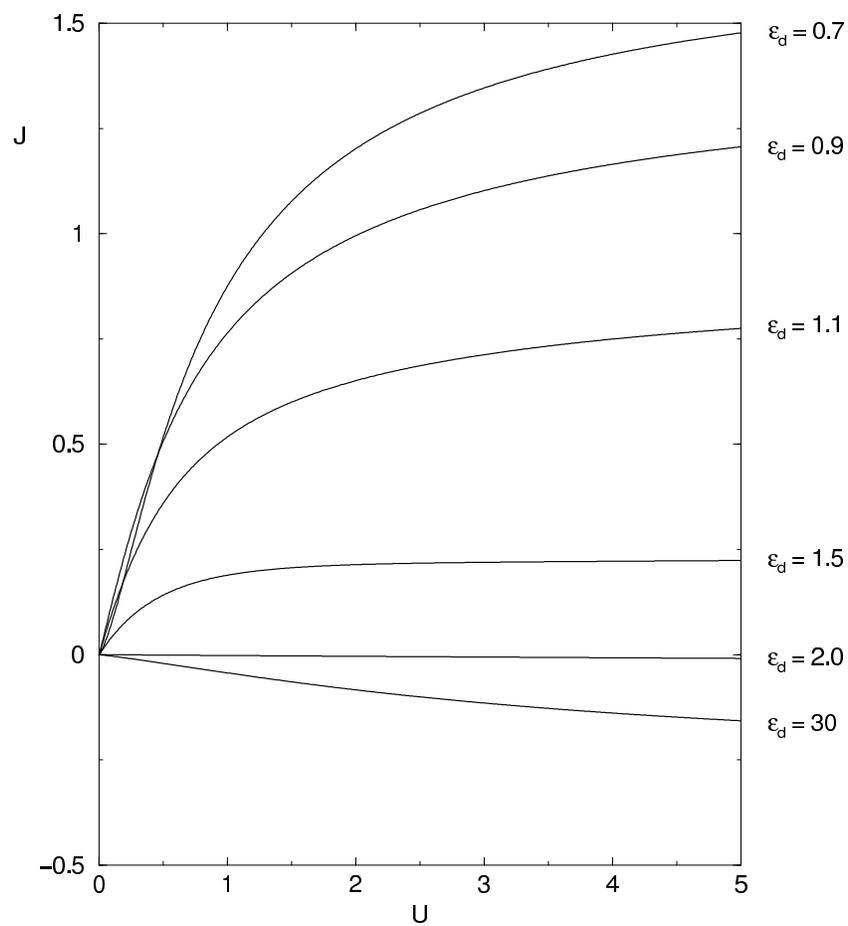}
\vspace{3.0mm}
\caption{\label{fig:andersonasymetric}
Several asymmetric cases are presented for the simplified
case of $V = 1$ and $\epsilon (k_F) = 0$.} 
\end{figure}

For the symmetric case, where $\varepsilon_{f} = - U/2$, $J$ 
becomes: 
\begin{equation}
J_{\rm sym} \: = \: V \sin(2VZ) \: + \: \frac{1}{2Z}[\cos(2VZ) - 1] 
\label{hat}
\end{equation}
For simplicity we consider $\epsilon (k_F) = 0$ hereafter. 
For this case the results for the symmetric Anderson model 
are plotted in Fig.\ \ref{fig:kondoJresult} as a function of 
$U/8V$. 

The two fixed points of the symmetric Anderson model can be 
seen clearly in Figs.\ \ref{fig:kondoJresult}. For $U \gg V^2$, 
Eq.\ (\ref{hat}) reduces to:
\begin{equation}
J_{\rm sym} \: \approx \: {\frac{U}{8}} \biggl( \cos {\frac{8 V}{U}} 
- 1 \biggl) \; .
\end{equation}
In the limit $U \rightarrow \infty$, 
this equation reduces to the well-known Schrieffer-Wolff 
result of the local moment or Kondo regime. 
However, this is not the case for smaller values of $U$. 

As $U$ decreases, $V^2 \gg U$ and $\varepsilon_{f}$ approaches 
$\epsilon (k_F)$, we cross over to the mixed valence regime. 
Here the $s$ and $f$ bands overlap which causes  
the virtual excitations of the impurity energy levels (shells) 
and the $s-f$ multiple scattering processes to dominate.
These processes cause most of the perturbative approaches
to fail. However, the transformation of Chan and Gul\'{a}csi 
(2001a,2003,2004) is still convergent. 
For $U \rightarrow 0$ Eq.\ (\ref{hat}) reduces to
\begin{equation}
J_{\rm sym} \: \approx \: V \sin {\frac{8 V}{U}} \; ,
\end {equation}
The strong oscillations which appear at very small values of $U$
(clearly seen in Fig.\ \ref{fig:kondoJresult}) are due to the multiple 
scattering processes.

These fluctuations are not present in the asymmetric Anderson
model, as shown in Fig.\ \ref{fig:andersonasymetric} where 
Eq.\ (\ref{ot}) is plotted 
for different values of $\varepsilon_{f}$. When there is no 
$s-f$ mixing, multiple scattering processes are absent and 
the oscillations in $J$ disappear completely. 

{\bf The 1D periodic Anderson model}: In the case of the
1D periodic Anderson model, all the terms of the infinite order
canonical transformation can be calculated exactly 
(Chan and Gul\'{a}csi 2003), details are presented in Appedix A. 

The starting hamiltonian  is the 
1D periodic Anderson model from Eq.\ (\ref{4.12a}) 
re-written in real space: 
\begin{equation}
H_{\PAM} = H_{0} + H_{V} \, ,
\label{sw8}
\end{equation}
with
\begin{eqnarray}
H_{0} & = & t \sum_{i, \sigma} (c^{\dagger}_{c i+1 \sigma}
c^{}_{c i \sigma} +  {\rm h.c.}) 
- \mu \sum_{i, \sigma} c^{\dagger}_{c i \sigma} c^{}_{c i \sigma}
\nonumber \\
&& +  \varepsilon_{f} \sum_{i, \sigma}
c^{\dagger}_{f i \sigma} c^{}_{f i \sigma} 
+  U \sum_{i} n^{}_{f i -\sigma} n^{}_{f i \sigma} \, , 
\label{sw9} \\
H_{V} & = & V \sum_{i \sigma} ( c^{\dagger}_{f i \sigma}
c^{}_{c i \sigma} + {\rm h.c} ) \, .
\label{sw10}
\end{eqnarray}
 
The transformed Hamiltonian 
\beqa
\tilde{H}{}_{\PAM} &=& e^{{\cal S}} ( H_{0} + H_{V} ) e^{-{\cal S}} 
\nonumber \\
&=& H_{0} + \sum^{\infty}_{n = 1} [ 1 / n! - 1 / (n+1)!] \: H_{n}  \, ,
\eeqa
can be written as:
\begin{equation}
\tilde{H}{}_{\PAM} = H_{0} + H_{odd} + H_{even} \, . 
\label{pamtransham}
\end{equation}

The odd hamiltonian terms are given in Eq.\ (\ref{eq:sumoddH}),
with the exact coefficients in Eqs.\ (\ref{eq:J}) - (\ref{eq:M}). 
It can be observed that the canonical transformation generates, 
besides the terms which renormalize the starting Hamiltonian, three 
new effective interactions, $J$, $P$, and $K$, and a higher order  
triplet-creating term, $M$. $J$ is the Kondo coupling, 
and is identical to Eq.\ (\ref{ot}), the single
impurity Anderson model result. 
$P$ a Josephson type two particle intersite
tunnelling and $K$ an effective on-site Coulomb repulsion
for the conduction electrons.

Eq.\ (\ref{eq:sumeven}) is the final result for the even hamiltonian 
terms, with the exact coefficients given in Eqs.\ (\ref{eq:R}) - (\ref{eq:Q}).
It is interesting to remark that the hybridization term still exists
after the transformation in the final result. However, 
its magnitude has been reduced from $V$ to
$ \approx AV^{3} / 3$ (see, Eq.\ \ref{eq:R}) for
$V$ and $\beta$ less than one. In a real lattice, these
conditions are true as the kinetic energy of the free
electrons is much larger than the hybridization energy, ie
$V/\mu << 1$. 

We conclude this section by quoting the results of the 1D 
canonical transformation: The 1D periodic Anderson model,
$H_{\PAM}$ defined in Eqs.\ (\ref{sw8}), (\ref{sw9}) and (\ref{sw10})
allows an exact mapping to an effective hamiltonian, $\tilde{H}{}_{\PAM}$ 
of the form presented in Eq.\ (\ref{pamtransham}). The transformed
hamiltonian contains nine new terms, from which $J$ represents
the only effective magnetic interaction. One may neglect
the rest of the terms if only interested in magnetism. What remains is 
the exact 1D Kondo lattice hamiltonian, see Eq. (\ref{4.11}): 
\beqa
H_{\KL} = 
-t \sum_{i \sig} (c^{\dg}_{c i \sig}c^{}_{c i+1 \sig} + {\rm h.c.}) 
+ J \sum_{i} {\bf S}_{c i} {\bf \cdot} {\bf S}_{i}  \, , 
\label{4.11new}
\eeqa
where $J$ is given by Eq.\ (\ref{eq:J}) or equivalent Eq.\ (\ref{ot}). 

It is interesting to mention that such an infinite order 
canonical transformation 
does work for other model hamiltonians which have a similar
structure to the Anderson model (so-called cylindrical quantum 
symmetry (Wagner 1986)). The method has been
successfully applied to the 2D two band Hubbard model by Chan and
Gul\'{a}csi (2000,2001b,2002). This infinite order canonical
transformation method resembles the {\it projective} 
{\it renormalization} method 
introduced by Glazek and Wilson (1993).   

\subsection{\label{old4.2.3}An aside on notation}

Several labels have arisen over the years to distinguish 
various aspects of the behaviour of the materials described by 
the Kondo lattice and periodic Anderson models. 
It is perhaps useful to list these various labels, and 
to briefly discuss their physical significance. 
The basic physical distinction at the heart of all the 
different labels may be understood in terms of the 
different physically relevant parameter regimes of the Kondo 
lattice model. From section \ref{old4.2.1}, one of these regimes has 
strong ferromagnetic coupling: $J < 0$ and $|J|/t > 1$. This 
arises from an on-site Hund's rule coupling between the 
localized electron spin and the spin of the the conduction 
electrons. From section \ref{old4.2.2}, 
the second of the regimes of the Kondo lattice has 
weak antiferromagnetic coupling: $J>0$ and $J/t <1$.
This arises from the Schrieffer-Wolff transformation of the 
periodic Anderson model as an effective hybridization. 

A summary of the original labels used to distinguish these 
two regimes is presented by Varma (1976).  
The first regime, with strong ferromagnetic coupling, is 
termed the `mixed valence' regime, and refers to the mixtures 
of ${\rm Mn}^{3+}$ and ${\rm Mn}^{4+}$ ions in the doped 
manganese oxide perovskites 
${\rm R}_{1-x}{\rm A}_{x}{\rm MnO}_{3}$. The second regime, 
with weak antiferromagnetic Kondo coupling, is termed the 
`fluctuating valence' regime, and refers to the continuous 
non-integral valence fluctuations at each site due to 
hybridization with the conduction band. 
Since he found these original labels somewhat mysterious, 
Varma (1976) proposed that a better label for the 
mixed valence materials would be `inhomogeneously 
mixed-valent' to emphasize that the different ions existed 
as separate entities. They were then distinguished from the 
fluctuating valence materials, which Varma proposed to 
call `homogeneous mixed-valent' to signify that the 
valence mixing occurred at each site continuously and 
homogeneously from site to site. These labels are now 
little used. 

With the discovery of heavy fermion compounds, another 
set of labels came into use. `Anomalous rare earth 
compound' signified the heavy fermion materials, which 
have hybridization between localized and extended states
and are described by a weak antiferromagnetic Kondo coupling. 
`Normal rare earth compounds' was reserved for those compounds 
in which the hybridization was negligible (Hewson 1993). 
This last labelling is now becoming less used, although it has 
a physical basis that further clarifies the distinction. 
In anomalous rare earth compounds, the level of the 
localized $f$ orbitals is close to the conduction electron 
Fermi level, and hybridization is important. The corresponding 
Kondo lattice has a weak antiferromagnetic coupling. In the 
normal rare earth compounds the $f$ level is well below that 
of the conduction electron Fermi level. In this case 
hybridization is unimportant and there is only Coulomb 
repulsion between localized and conduction electrons. This leads 
to a Kondo lattice with a ferromagnetic coupling of variable 
magnitude. 

For ease of expression, the localized electrons will be 
referred to as the localized spins from now on. 
This describes their degrees of freedom in the Kondo 
lattice, and permits easy reference to all the parameter 
regimes of interest;  it avoids the ambiguity of having 
`$f$-electrons' occupying $d$ band orbitals when referring 
to Mn oxide perovskites. 
It also distinguishes them clearly from the conduction 
electrons, and the adjective `conduction' may occasionally be 
dropped without confusion.

%%%%%%%%%%%%%%%%%%%%%%%%%%%%%%%%%%%%%%%%%%%
%% chapter 2
%%%%%%%%%%%%%%%%%%%%%%%%%%%%%%%%%%%%%%%%%%%%

\cleardoublepage
\chapter{\label{old4.3}Interactions in the Kondo Lattice Model}

This chapter
discusses the interactions present in the Kondo lattice in 
its various parameter regimes. Section \ref{old4.3.1} derives the RKKY 
interaction, which operates at weak-coupling. In section \ref{old4.3.2} 
the formation of Kondo singlets in the single impurity Kondo 
model is summarized, and related where possible to singlet 
formation in the lattice case. Section \ref{old4.3.3} discusses 
double-exchange ordering, which occurs in the Kondo lattice 
when the conduction band 
is under half-filled. Double-exchange has long been known to 
be an important mechanism in manganese oxide compounds 
(Zener 1951, Anderson 1955), and it is usual to consider 
double-exchange ordering for the Kondo lattice with a 
ferromagnetic coupling $J<0$. It is far less common to consider 
double-exchange for the antiferromagnetic $J>0$ Kondo 
lattice.\footnote{This in spite of the fact that double-exchange 
ferromagnetic ordering has been observed in rare earth compounds 
as far back as 1979. See Varma (1979) for theory, and 
Batlogg, Ott and Wachter (1979) for experiments 
on the thulium compound ${\rm TmSe}_{0.83}{\rm Te}_{0.17}$.} 
Nevertheless, double-exchange ordering occurs also for  
antiferromagnetic coupling, and in section \ref{old4.3.3} a  
microscopic derivation is given. Since this work is new, it is 
discussed in some detail and forms the longest section of this 
chapter. Previously known results 
for the 1D Kondo lattice are summarized in section \ref{old4.4}. Results 
for the lattice with a half-filled conduction band are given 
briefly in section \ref{old4.4.1}. A more 
detailed discussion of results for the partially-filled 
case is given in section \ref{old4.4.2}. The latter results provide an 
important test for the theory of the 1D Kondo 
lattice which is presented in chapters \ref{ch5} and 
\ref{ch6}. 

There are several parameter regimes of the Kondo lattice 
in which the dominant interaction processes may be 
identified. In this section these interactions are 
discussed. At weak-coupling $|J|/t \ll 1$, second order 
perturbation theory gives the Ruderman-Kittel-Kasuya-Yosida 
(RKKY) interaction (Ruderman and Kittel 1954, Kasuya 1956,
Yosida 1957)\footnote{For a different derivation, see
Van Vleck (1962).} This is an 
effective interaction between the localized spins which 
is mediated by the conduction electrons. The derivation 
of the RKKY interaction is given in section
\ref{old4.3.1}, and its divergence in 1D is discussed. 
Section \ref{old4.3.2} discusses Kondo singlet formation, 
which has a long history in the theory of 
systems with dilute magnetic impurities. Some of the results 
for dilute systems are summarized, before passing to 
singlet formation in the $J>0$ Kondo lattice. 
The RKKY interaction and Kondo singlet formation operate 
both in the Kondo lattice, and 
in systems with a very small fraction of localized spins, 
which model systems with dilute magnetic impurities. This 
is not the case for the double-exchange interaction, which 
operates only in the opposite case where the localized spins 
outnumber the conduction electrons, and occurs in the Kondo 
lattice with a partially-filled conduction band $n < 1$.
Section \ref{old4.3.3} discusses the double-exchange interaction, 
and determines some of its properties. 
Although recognized as early as 1951 as an important 
interaction in the perovskite manganese oxides 
(Zener 1951), double-exchange has only 
recently been discussed in relation to the $J > 0$ Kondo 
lattice (Yanagisawa and Shimoi 1996, Honner and Gul\'{a}csi 1998b).

\section{\label{old4.3.1}The RKKY interaction} 

For $J = 0$, the conduction electrons in the Kondo 
lattice are in their non-interacting ground-state 
$|0\rangle$ (assumed non-degenerate), 
as described in section \ref{old1.2.1}. The ground-state 
energy of the conduction electrons, with the 
Kondo lattice interaction $V_{\KL}$ considered as a 
perturbation, is given by Raleigh-Schr\"{o}dinger 
perturbation theory (Gross, Runge and Heinonen 1991) 
as follows:
\beqa
E = \langle 0|V_{\KL}|0\rangle 
- \sum_{n \neq 0} \frac{|\langle n|V_{\KL}
|0\rangle|^{2}}{E_{n}} 
+ \cdots 
\label{4.2.1} 
\eeqa
where the zero of energy is chosen so that 
$H_{0}|0\rangle$ vanishes, and  
where $|n\rangle$ are non-interacting excited 
states with excitation energies $E_{n} > 0$. 
The only excited states giving non-vanishing 
matrix elements in Eq.\ (\ref{4.2.1}) are those 
of the form 
$|n\rangle = c^{\dg}_{ck\sig}c^{}_{ck'\sig'}|0\rangle$ 
with $|k| > k_{F} > |k'|$. To second order, 
straightforward computation gives 
\beqa
E &=& \sum_{j, l} {\cal J}_{\RKKY}
(j-l)\,{\bf S}_{j} {\bf \cdot} 
{\bf S}_{l} \, , 
\nonumber \\
{\cal J}_{\RKKY}(j-l) &=& -\frac{J^{2}}{2N^{2}}
\sum_{|k'| < k_{F} < |k| \leq \pi/a} 
\frac{e^{i(k'-k)(j-l)a}}{\varepsilon(k) 
- \varepsilon(k')} \, ,
\label{4.2.2}
\eeqa
where $N$ is the number of lattice sites.
Eq.\ (\ref{4.2.2}) implies that 
the complete $2^{N}$-fold spin-degeneracy of the 
localized spins at $J = 0$ is 
lifted by perturbation in the conduction 
electrons, so that the Kondo lattice is in its 
ground-state if the localized spins order 
so as to minimize $E$. Perturbation 
theory thus generates an effective interaction 
${\cal J}_{\RKKY}(j-l)$ between the localized spins, 
called the RKKY interaction, which is mediated by the 
conduction electrons. 

The form of Eq.\ (\ref{4.2.2}) is generic to any 
dimension. What differs in different dimensions 
is the evaluation of the summation over 
$k$ and $k'$. This gives different 
RKKY interactions ${\cal J}_{\RKKY}$ 
depending on dimensionality. The integrals  
may be evaluated in closed form by approximating the 
conduction electrons by free electrons (valid for 
weak-interactions for which the perturbation 
expansion is accurate). The calculations are 
carried out by Aristov (1997)\footnote{see also 
Yafet (1987) for 1D} with the results  
\beqa
{\cal J}_{\RKKY}(r) = \left\{
\begin{array}{ll}  
\frac{m_{e}J^{2}}{2\pi}\left[
{\rm Si}(2k_{F}r) - \frac{\pi}{2} \right] 
 & {\rm 1D}\, ,
\nonumber \\
\frac{m_{e}k_{F}^{2}J^{2}}{8\pi}\left[
J_{0}(k_{F}r)Y_{0}(k_{F}r)
+J_{1}(k_{F}r)Y_{1}(k_{F}r)\right] 
& {\rm 2D}\, , 
\nonumber \\
\frac{m_{e}k_{F}J^{2}}{16\pi^{3}r^{3}}
\left[\cos(2k_{F}r) 
- \frac{\sin(2k_{F}r)}{2k_{F}r}\right]
& {\rm 3D}\, , \nonumber 
\end{array}\right.
\nonumber \\
\label{4.2.3}
\eeqa
where $r = |(j-l)a|$ is the distance between the lattice 
sites $j$ and $l$, and $m_{e}$ is the bare conduction 
electron mass. The special functions in Eqs.\ 
(\ref{4.2.3}) are the sine integral Si 
and the order $n$ 
Bessel functions of the first and second kind, 
$J_{n}$ and $Y_{n}$ respectively. 

In 3D, the RKKY interaction decreases at large distances 
with $r^{-3}$, and oscillates with wave vector $2k_{F}$ 
(i.e.\ with period $\pi /k_{F}$ in real space). This 
favours an oscillatory alignment of the localized spins with the 
same periodicity. Ordering of the localized moments with 
wave vector $2k_{F}$ is characteristic of the RKKY interaction 
in any dimension, and a similar alignment is favoured in 
2D.\footnote{Note that more complicated magnetic structures 
can arise with the stacking of 2D planes in a crystal
(Aristov 1997).}
In 1D the RKKY interaction leads to a particularly strong 
$2k_{F}$ ordering of the localized spins. Indeed the  
RKKY interaction in this case is divergent: 
The Fourier component of 
${\cal J}_{\RKKY}$ at wave vector $k$ is defined through  
the Fourier transform
\beqa
{\cal J}_{\RKKY}(j-l) = 
\frac{1}{2\pi}\int_{-\infty}^{\infty}dk\, 
{\cal J}_{\RKKY}(k)\, e^{ik(j-l)a} \, .
\label{4.2.3a} 
\eeqa
Using the 1D form of the RKKY interaction from 
Eq.\ (\ref{4.2.3}), it may be shown that (Yafet 1987)
\beqa
{\cal J}_{\RKKY}(k) \propto 
\frac{1}{k}\ln \left| \frac{2k_{F} + k}{2k_{F} - k} 
\right| \, , 
\label{4.2.4}
\eeqa
which has a logarithmic divergence at $2k_{F}$. 
${\cal J}_{\RKKY}(k)$ is discussed by Kittel (1968),
together with the corresponding 
2D and 3D forms, which do not diverge. The 
divergence of ${\cal J}_{\RKKY}(k)$ in 1D 
is typical of the results of perturbation 
theory for 1D systems. The significance of the 
divergence is as follows: Ordering 
of the localized spins with wave vector $2k_{F}$ is 
still expected to occur in the 1D Kondo lattice at 
weak-coupling, much as in 3D. (This expectation is 
confirmed by the results of numerical simulations, 
cf.\ section \ref{old4.4} below). 
However, it is not possible to use the RKKY interaction 
itself to describe the $2k_{F}$ ordering in 1D, for 
since the interaction diverges there is no lower 
bound on the ground-state energy of the 1D Kondo lattice 
if it is described using the RKKY interaction. It is 
necessary to go beyond perturbation theory to describe 
the $2k_{F}$ correlations in the localized spins. 
In chapter \ref{ch5} a $2k_{F}$ ordering of the 
localized spins is obtained at weak-coupling by using 
a bosonization description of the conduction electrons.

\section{\label{old4.3.2}Kondo singlet formation}

The Kondo lattice model can be considered to be the formal 
extension to the lattice of the single impurity Kondo hamiltonian, 
which has a single localized spin in a sea of conduction 
electrons. The single impurity Kondo hamiltonian is given by 
\beqa
H_{\rm imp} = 
-t\sum_{<ij>}\sum_{\sig} c^{\dg}_{ci\sig}c^{}_{cj\sig} 
+ J{\bf S}_{c0} {\bf \cdot} {\bf S}_{0}  \, , 
\label{4.2.5}
\eeqa
and describes a system with a magnetic impurity, modelled 
by a single localized spin at the 
origin. An antiferromagnetic Kondo coupling $J > 0$ is  
assumed for $H_{\rm imp}$. Historically, intensive study of the 
single impurity Kondo model preceded that of the Kondo 
lattice.\footnote{Note, however, that the lattice problem  
in a certain sense preceeds the work on the impurity 
Kondo hamiltonian: Fr\"{o}hlich and Nabarro (1940)
considered the lattice case in 1940 in their work on the magnetic 
ordering of nuclear spins.} As a result $H_{\rm imp}$ 
is now well-understood, and has even been solved exactly.  
In the following some of the results for the single impurity 
Kondo hamiltonian are summarized. An emphasis is placed on 
results which have relevance for the Kondo lattice, 
and no more than a skeletal outline of the 
huge body of work devoted to $H_{\rm imp}$ is intended. 
A thorough discussion of the numerous solution methods 
may be found in the book by Hewson (1993), 
together with references. 

Interest in the single impurity Kondo model arose from the 
observation of anomalous behaviour in the conduction electron 
resistivity in dilute magnetic alloys. In simple metals, 
and in alloys with non-magnetic impurities,
the resistivity drops monotonically as the temperature $T$
decreases to $T=0$. This is because the main 
contribution to the resistivity at low temperatures is from 
electron-phonon scattering, and decreases as $T^{5}$ at 
low $T$. In metals with dilute magnetic impurities, such as 
iron in gold, the resistivity is not monotonic with temperature, 
but passes through a minimum before rising again as 
$T \rightarrow 0$. 
A breakthrough in understanding the occurrence of 
the resistance minimum was achieved by Kondo (1964), 
who calculated the resistivity of $H_{\rm imp}$ to third 
order in the coupling $J$ by standard diagrammatic perturbation 
theory. He found that at third order the interaction leads to 
singular spin scattering of the conduction electrons with the 
magnetic impurity, and gives a $-\log T$ contribution to the 
resistivity. Since $-\log T$ increases as $T \rightarrow 0$, 
this explained the occurrence of the resistance minimum. 

Since $\log T$ diverges as $T \rightarrow 0$, it was clear that 
perturbation theory fails at low temperatures below the 
resistance minimum. Thus, while Kondo's calculation provided 
a first understanding of the effect of dilute magnetic 
impurities, the method could not access the very low temperature 
regime. The problem of finding a solution valid as $T 
\rightarrow 0$ became known as the {\it Kondo} {\it problem}. 
This was essentially solved in the 1970's by scaling arguments
\footnote{For more details about the Kondo model, see Appendix \ref{ch4kondo}}; 
Anderson's poor man's scaling (Anderson 1970), and the 
numerical renormalization group of Wilson (1975). The 
basic idea is that as the temperature is lowered, the high-energy 
states of the conduction electrons (those far from the 
Fermi surface) may be eliminated. This yields an effective 
hamiltonian, defined for a conduction band with a reduced 
bandwidth, and containing a modified or `renormalized' coupling 
between the electrons and the localized spin. The details of 
the scaling procedure may be found in the original papers, 
and are also reproduced in great detail by Hewson (1993). 
The results show that when the scaling reaches a characteristic 
Kondo temperature $T_{K}$, an initially small antiferromagnetic 
coupling $J>0$ becomes large, and the conduction electrons 
form a magnetically neutral spin-singlet with the localized spin; 
the magnetic impurity is quenched. The resistance minimum 
in dilute magnetic alloys then reflects the formation of  
strongly-coupled screening clouds of 
conduction electrons around the dilute sites which 
contain impurities. 

The correctness of the scaling approach was confirmed in 1980 
with the discovery of an exact solution to $H_{\rm imp}$ by 
Andrei (1980) and Wiegmann (1980)\footnote{
For a review, see Andrei, Furuya and Lowenstein (1983). 
The exact solution of the single impurity Anderson model,
is reviewed by Tsvelick and Wiegmann (1983).}, 
using the Bethe ansatz. The exact solution 
verified that the single impurity 
model contains one energy scale below the conduction 
electron bandwidth, called the Kondo temperature, which 
measures the energy for the quenching of the localized spin 
via singlet formation with the conduction electrons. 
The form of the Kondo temperature (see Eq.\ (\ref{a1}) 
of Appendix \ref{ch4kondo}) is 
\beqa
T_{K} \propto \exp(-1/ 2 \rho(\varepsilon_{F}) J) \, ,
\label{4.2.6} 
\eeqa
where the density of conduction electron states at the 
Fermi level is 
\beqa
\rho(\varepsilon_{F}) = 2 \sum_{k}\delta \left(\varepsilon(k) 
-\varepsilon(k_{F})\right)\, .
\label{4.2.7}
\eeqa
As a result of the existence of one energy scale, the 
low temperature thermodynamic properties of the model are 
universal functions of $T/T_{K}$. These properties are 
summarized in Appendix \ref{ch4kondo}. 

At present it is largely unclear whether, and to what extent, 
the results for the 
single impurity Kondo hamiltonian apply to the Kondo lattice. 
Certainly there are aspects of the single impurity 
solution which have no analogue in the lattice case. The most 
important of these is the extent of the Kondo screening cloud: 
When the conduction electrons screen the localized spin, 
forming a singlet with it, renormalization group 
arguments suggest that the screening cloud extends in real-space 
over a scale $\xi_{K} \approx v_{F}/k_{B}T_{K}$, where $v_{F}$ is 
the Fermi velocity of the electrons (cf.\ Eq.\ (\ref{Vf}))
(S{\o}rensen and Affleck 1996). 
Since $T_{K}$ is generally of the order 
of tens of degrees, the screening cloud extends over thousands 
of lattice spacings. This cannot occur in the lattice case, 
for which the extent of the screening cloud 
per localized spin can extend to the order of only one  
lattice spacing. Moreover, any screening in the lattice must be 
qualitatively very different from that in the single impurity 
problem. In the lattice the number of conduction electrons is 
less than, or of the order of the number of localized spins. 
Thus the screening of each localized spin by a large number of 
conduction electrons, as occurs at low temperatures in the 
impurity problem, cannot occur in the lattice case. 
Notwithstanding these problems, it is common in the Kondo 
lattice to propose a Kondo temperature which, analogous to the 
single impurity case, measures the energy scale for the 
formation of spin singlets around the localized spins. 
To understand the basis for this, it is necessary to consider 
the Kondo lattice at strong-coupling. 

\begin{table}
\centering
\begin{tabular}{|c|c|c|c|c|} 
\hline 
State & Energy & $S_{\tot}$ & $S^{z}_{\tot}$ & $n_{c}$ \\ 
\hline \hline
$(|\uparrow 0 1\rangle - |\downarrow 1 0 \rangle)/\sqrt{2}$ 
& $-3J/4$ & 0 & 0 & 1 \\ \hline 
$|\uparrow 0 0\rangle $ 
& 0 & 1/2 & 1/2 & 0 \\  
$|\downarrow 0 0\rangle $ 
& 0 & 1/2 & -1/2 & 0 \\  
$|\uparrow 1 1\rangle $ 
& 0 & 1/2 & 1/2 & 2 \\  
$|\downarrow 1 1\rangle $ 
& 0 & 1/2 & -1/2 & 2 \\ \hline 
$|\uparrow 1 0\rangle $ 
& $J/4$ & 1 & 1 & 1 \\  
$(|\uparrow 0 1\rangle + |\downarrow 1 0 \rangle)/\sqrt{2}$ 
& $J/4$ & 1 & 0 & 1 \\  
$|\downarrow 0 1\rangle $ 
& $J/4$ & 1 & -1 & 1 \\ \hline 
\end{tabular}
\caption{Eigenbasis of $V_{\KL}$ for each lattice site $j$. The 
notation for the states is $|S^{z}_{j}, n_{cj\uparrow}, 
n_{cj\downarrow}\rangle$, where $S^{z}_{j} = \uparrow, 
\downarrow$ is the $z$ component of the localized spin at $j$, 
and $n_{cj\sig} = 0, 1$ is the number of conduction electrons of 
spin $\sig$ at $j$. ${\bf S}_{\tot} 
= {\bf S}_{cj} + {\bf S}_{j}$ is the total spin, and 
$n_{c} = n_{cj\uparrow} + n_{cj\downarrow}$ is the total number 
of conduction electrons. The basis consists of a singlet state, 
two doublets, and a triplet. The singlet state has 
the lowest energy for $J>0$, and the triplet states have   
the lowest energy for $J<0$.}
\end{table}

At infinitely strong-coupling $|J| = \infty$, the conduction 
electron hopping is ineffective and the Kondo lattice 
hamiltonian $H_{\KL}$ reduces to the interaction 
$V_{\KL}$ of Eq.\ (\ref{4.9}). It is straightforward to 
diagonalize $V_{\KL}$, and the eight eigenstates per site 
are listed in Table 4.1. For infinite antiferromagnetic 
coupling $J = \infty$, the lowest energy state for each site 
is an on-site singlet, involving a single 
conduction electron pairing 
with a localized spin at the same site. This pairing quenches 
the localized spin, yielding a site with total spin zero, and 
this is the analogue in the lattice of the screening in the 
single impurity Kondo hamiltonian. The Kondo temperature is 
the binding energy of the singlet; the lowest excited state with 
the same number of conduction electrons is a triplet state, so 
that $k_{B}T_{K} = J$ as $J \rightarrow \infty$ (Lacroix 1985).
At finite coupling 
the Kondo temperature is more difficult to define. In fact 
when the conduction band is partially-filled there is no such 
generally accepted definition to date.
For a half-filled conduction band, there is 
a gap for the spin excitations, extending from infinite $J$ 
down to small $J$. Moreover, the ground-state of the 
Kondo lattice at half-filling 
is a total spin-singlet. (These properties are discussed 
further in section \ref{old4.4.1}). This permits an identification of the 
size of the spin gap 
with the Kondo temperature, and at weak-coupling 
numerical simulations give a form identical to that for the 
single impurity case, but enhanced by a factor of 1.4 
(Shibata, {\it et al.} 1996): 
\beqa
T_{K} \propto \exp(-1/1.4\, \rho(\varepsilon_{F})J).
\label{4.2.8} 
\eeqa
It should be stressed that the nature of the screening in the 
lattice at weak-coupling is unlike that in the single impurity 
case.

\section{\label{old4.3.3}Double-exchange ordering}

Double-exchange was introduced in 1951 by Zener (1951)
to describe ferromagnetism in the 
manganese oxide perovskites. Zener considered Mn oxide  
compounds ${\rm La}_{1-x}{\rm A}_{x}{\rm Mn}{\rm O}_{3}$, with
$0 < x < 1$ and A $=$ Ca, Sr or Ba. The compounds 
contain both ${\rm Mn}^{3+}$ and ${\rm Mn}^{4+}$ ions, 
in concentrations $1-x$ and $x$ respectively. For 
$x = 0$ the compounds are insulating, while at moderate 
doping $x \gtrsim 0.2$ they are ferromagnetic and conducting. 
Zener proposed that the close connection between 
ferromagnetism and conduction in these materials could be 
accounted for by supposing that the $e_{g}$ electrons on  
${\rm Mn}^{3+}$ ions could hop to vacant $e_{g}$ orbitals on 
neighboring ${\rm Mn}^{4+}$ ions. Since hopping electrons 
tend to preserve their spin, and since Hund's rule coupling 
strongly favours an alignment of the $e_{g}$ spin with that of 
the localized $t_{2g}$ electrons (cf.\ section \ref{old4.2.1}), this 
hopping should favour a ferromagnetic alignment of the spins of 
the $t_{2g}$ electrons on neighboring Mn ions.  
Since the hopping of the $e_{g}$ electrons occurs through an 
intermediate ${\rm O}^{2-}$ ion, Zener called the ferromagnetic 
alignment induced by the hopping the double-exchange 
interaction. The name is somewhat unfortunate, since the 
interaction is not an exchange interaction in the usual sense, 
but simply reflects the tendency of hopping electrons to 
preserve their spin. 

A microscopic derivation of the double-exchange interaction 
was given by Anderson and Hasegawa (1955) on a 
two-site Kondo lattice with a ferromagnetic coupling $J<0$, 
which models the Hund's rule coupling in the Mn oxides. 
However, double-exchange operates regardless of the sign 
of the coupling, and requires only that there be more 
localized spins than conduction electrons; the fact that the 
hopping electron aligns opposite or parallel to the localized 
spins at each site is irrelevant 
to the preservation of spin while hopping. It is the latter 
which forces the localized spins to align. The first hints of 
this are in Anderson and Hasegawa's (1955) original work;
they noticed that 
the sign of the coupling was largely irrelevant to the 
ferromagnetic ordering within a semiclassical approximation 
for the localized spins. More recently, in the 1990s, a 
succession of rigorous results and numerical simulations have 
established a large region of ferromagnetism at partial 
conduction band fillings in the $J>0$ Kondo lattice (reviewed 
in section \ref{old4.4.2}). The ferromagnetism has been attributed to 
double-exchange (Yanagisawa and Shimoi 1996).
Not withstanding this, double-exchange 
is largely neglected in discussions of the $J>0$ Kondo lattice, 
with most papers focusing instead on the competition between 
RKKY interactions and Kondo singlet formation. Given this 
neglect, it seems worthwhile to present a detailed microscopic 
derivation of double-exchange for $J>0$, analogous to that for 
the $J<0$ Kondo lattice. 

To establish double-exchange in the Kondo lattice for either 
sign of the coupling $J$, consider a Kondo lattice with two 
sites and one conduction electron. This system is equivalent 
to the system studied by Anderson and Hasegawa (1955),
in which double-exchange ordering 
was shown to occur within a semiclassical approximation for the 
localized spins. Apart from minor differences in some numerical 
factors, double-exchange was found to operate independent of 
the sign of the coupling $J$ in the semiclassical case. 
Double-exchange ordering was also demonstrated by Anderson and 
Hasegawa for quantum localized spins, but in a reduced 
(4-dimensional) Hilbert space. 
The reduced space is sufficient if 
the coupling $J<0$ is ferromagnetic, and this was the case of 
interest to Anderson and Hasegawa who considered the Mn 
oxides. For antiferromagnetic 
Kondo coupling $J>0$, it is no longer sufficient 
to operate in a 4-dimensional Hilbert space. In the 
following the complete 16-dimensional Hilbert space of the 
system is considered. 

States of the two site Kondo lattice   
may be written 
$$|S^{z}_{1}, n_{c1\uparrow}, n_{c1\downarrow}
: S^{z}_{2}, n_{c2\uparrow}, n_{c2\downarrow} \rangle \; ,$$ 
in a straightforward generalization of the single-site 
states used in Table 4.1; 
$S^{z}_{j} = \uparrow, \downarrow$ is the $z$ component of the 
localized spin at site $j =1,2$, and $n_{cj\sig} = 0,1$ 
is the number of conduction electrons of spin $\sig$ at site 
$j$.  A single conduction electron can be in one of the 4 
states $n_{cj\sig}$, and for each of these  configurations 
there are 4 possible configurations for the localized 
spins $S^{z}_{1} S^{z}_{2}$: $\uparrow \uparrow$, 
$\uparrow \downarrow$, $\downarrow \uparrow$, 
$\downarrow \downarrow$.  Thus a basis for the two site Kondo 
lattice with one conduction electron has 16 elements. 
The $16 \times 16$ matrix of the Kondo lattice hamiltonian 
Eq.\ (\ref{4.11}) operating on basis elements 
$|S^{z}_{1}, n_{c1\uparrow}, n_{c1\downarrow}
: S^{z}_{2}, n_{c2\uparrow}, n_{c2\downarrow} \rangle$ is 
easily constructed, and the eigenvalues and eigenvectors 
are obtained in analytic form. 
This gives a complete solution of the system, and may be used 
to analyse the ground-state properties as a function of $t/J$.  

For ferromagnetic coupling $J<0$ the ground-state properties 
are relatively simple, and are well-described within  
a semiclassical approximation (Anderson and Hasegawa 1955). 
This case is included in the discussion here because the 
analysis for $J<0$ aids in 
identifying similar behaviour in the more complicated 
case of antiferromagnetic coupling. 
The ground-state energy of the 
system for infinite coupling  
$J \rightarrow -\infty$ is $-|J|/4$. This corresponds to a 
triplet pairing between the conduction electron and a 
localized spin at the same site, together with an unpaired 
localized spin at the other site. There are 12 distinct   
configurations of states of this form, and the ground-state 
is 12-fold degenerate at $J = -\infty$. 
Conduction electron hopping partially 
lifts this degeneracy. For example, the state 
$|\uparrow 1 0 : \uparrow 0 0 \rangle$ is preferred over the 
state $|\uparrow 1 0 : \downarrow 0 0 \rangle$ because the 
conduction electron can hop to site 2 while maintaining its 
spin parallel to the localized spin at the new site. 
The ground-state energy for finite ferromagnetic coupling is 
\beqa
E_{0} = -\frac{|J|}{4} - t\, , \quad \quad J<0\, ,
\label{fmgse}
\eeqa
and the ground-state is only 4-fold degenerate. This 
degeneracy may be understood by considering the total 
spin operator 
${\bf S}_{\tot} = \sum_{j} {\bf S}_{j} + {\bf S}_{cj}$. 
Each of the four ground-states is an eigenvector of 
$S_{\tot}^{2}$ with total spin $3/2$, and the ground-state 
degeneracy simply reflects the 4 choices 
of the $z$ component $S^{z}_{\tot}$ of the total spin. 
The Kondo lattice hamiltonian Eq.\ (\ref{4.11}) does not mix 
states with different values of $S^{z}_{\tot}$, and it suffices 
to consider a subspace in which $S^{z}_{\tot}$ is fixed. 
In the subspace with $S^{z}_{\tot} = 3/2$,
the ground-state is unique and takes a particularly 
simple form:
\beqa
|\psi_0 \rangle = \frac{1}{\sqrt{2}}\left(
|\uparrow 1 0: \, \uparrow 0 0\rangle + 
|\uparrow 0 0: \, \uparrow 1 0\rangle \right)
\quad \, \, J<0\, .
\label{fmgs}
\eeqa
This state is the prototype for double-exchange ordering. 
For finite ferromagnetic coupling the system minimizes energy 
by conduction electron hopping, and by having the conduction 
electron spin aligned parallel to the localized spin at each 
site. Since hopping electrons tend to 
conserve their spin, called coherent electron hopping, 
the minimal 
energy is obtained if all the localized spins align with the  
conduction electron spin.  The ferromagnetic ordering induced 
on the localized spins by coherent conduction electron 
hopping is the double-exchange interaction.  

Consider now an antiferromagnetic Kondo coupling $J>0$. In 
this case the conduction electron spin tends to align opposite 
to the localized spin. At infinite coupling      
$J = \infty$, the ground-state energy is $-3J/4$ and the 
ground-state consists of a singlet pairing between the 
conduction electron and a localized spin at the same site 
(i.e.\ a localized Kondo singlet), together with 
an unpaired localized spin at the other site. The ground-state 
is 4-fold degenerate.  Conduction electron hopping partially 
lifts this 
degeneracy, similar to the case for ferromagnetic coupling. 
For finite antiferromagnetic coupling the ground-state energy is
given by the solution of the $16 \times 16$ matrix as
\beqa
E_{0} = -\frac{J}{4} - \frac{1}{2}\sqrt{J^{2} + 2Jt + 4t^{2}}\, ,
\quad \quad J>0\, ,
\label{afmgse}
\eeqa
and the ground-state is two-fold degenerate.  Both of the 
ground-states are eigenstates of $S^{2}_{\tot}$ with total 
spin 1/2. The degeneracy is due to the two choices of 
$S^{z}_{\tot}$, and as for $J<0$  
it suffices to consider a subspace with fixed $S^{z}_{\tot}$. 
The unique ground-state with $S^{z}_{\tot}=1/2$ is given by  
\beqa
|\psi_{0}\rangle &\propto & |\, s_{1} \uparrow \rangle + 
|\uparrow s_{2} \rangle 
\nonumber \\
&+& \frac{t}{J/4 - E_{0}}\left\{
|\uparrow 01:\,  \uparrow 00\rangle + |\uparrow 00: \, 
\uparrow 01\rangle  \right.  
\nonumber \\
&& \left.  \quad - |\uparrow 10 :\, \downarrow 00\rangle 
-|\downarrow 00: \, \uparrow 1 0\rangle \right\}\, 
\quad \, \,  J>0\, .
\label{afmgs}
\eeqa
The proportionality constant is the normalization, and the 
$J = \infty$ behaviour has been separated out using the localized 
Kondo singlet states:  
\beqa
| \, s_{1}\uparrow \rangle &=& |\uparrow 01:\, 
\uparrow 00\rangle 
- |\downarrow 10:\, \uparrow 00\rangle  \, , 
\nonumber \\
|\uparrow s_{2} \rangle &=& |\uparrow 00:\, \uparrow 01\rangle 
- |\uparrow 00:\, \downarrow 10\rangle  \, .  
\label{ss}
\eeqa
The ground-state Eq.\ (\ref{afmgs}) involves six basis 
elements, and falls 
outside the 4-dimensional space used by Anderson and 
Hasegawa (1955) to establish double-exchange for 
ferromagnetic coupling $J<0$. 
Although somewhat obscured, double-exchange ferromagnetic 
ordering of the localized spins is 
present also in the ground-state of the 
$J>0$ Kondo lattice, as can be shown in two steps. 
First, the correlation between the localized spins 
in the ground-state Eq.\ (\ref{afmgs}) is calculated. Second  
the $J>0$ ground-state is rewritten in a form in which 
double-exchange is made manifest, i.e.\ similar in form to 
Eq.\ (\ref{fmgs}), but with the conduction 
electron spin aligning opposite to the localized spins.

\begin{figure}[ht]
\centering
\includegraphics[scale=0.7]{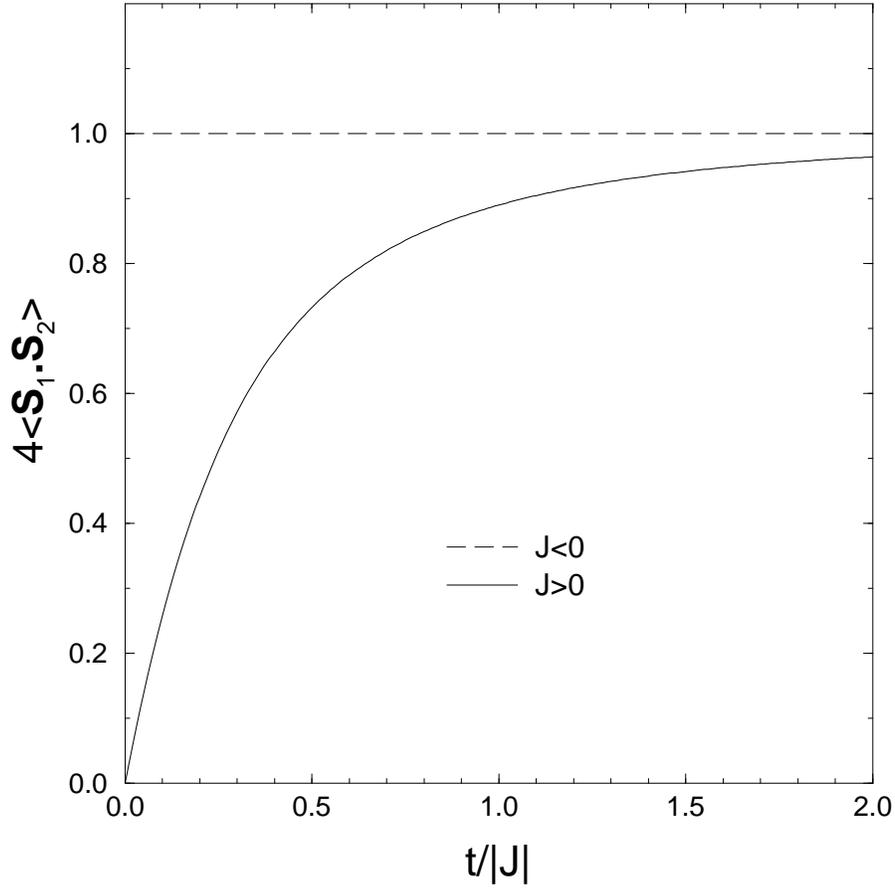}
\vspace{3.0mm}
\caption{\label{oldfig4.1}The ground-state correlation    
$\langle \psi_{0}|{\bf S}_{1} {\bf \cdot} 
{\bf S}_{2}|\psi_{0}\rangle / \langle \psi_{0}|\psi_{0}\rangle$
between the localized spins in the Kondo lattice with two 
sites and one conduction electron, from Eq.\ (\ref{afmcor}). 
$t \geq 0$ is the conduction electron hopping, 
and $J$ is the coupling 
between the electron and the localized spins. 
For antiferromagnetic coupling $J>0$, the contribution to the 
correlation from each spin component $x, y, z$ is identical.} 
\end{figure}

The ground-state correlation between 
the localized spins is given by 
\beqa
\frac{\langle \psi_{0}|S^{\alpha}_{1}
S^{\alpha}_{2}|\psi_{0}\rangle}{\langle \psi_{0}|\psi_{0}\rangle}
= \frac{1}{4} \left[ 1+ \frac{t}{J/4 -E_{0}} + 
\frac{J/4 -E_{0}}{t}\right]^{-1} 
\label{afmcor}
\eeqa
where $\alpha = x,y,$ or $z$.
The correlation is ferromagnetic for each component of the 
spin, and the contribution from each component is identical.
The full spin-spin correlation between the localized spins is 
plotted in Fig.\ \ref{oldfig4.1} as a function of $t/J$. 
The ferromagnetic correlation between the localized spins
grows quickly from zero at infinite coupling, and when $t = J$ 
is already at 90\% of the possible maximum of 1/4. 
For comparison, Fig.\ \ref{oldfig4.1} also includes 
the correlation between the localized spins in the 
ground-state for $J<0$. The result for $J<0$  
coincides with the correlation for any sign of the 
coupling within a semiclassical approximation for the 
localized spins (Anderson and Hasegawa 1955).

It remains to show that  
the ferromagnetic ordering between the localized spins 
for $J>0$ is due to the double-exchange interaction. 
Double-exchange in the $z$ direction of the spin is manifest in 
Eq.\ (\ref{afmgs}) in the third and fourth states: 
\beqa
|{\rm de}_{z}\rangle = |\uparrow 0 1 : \, \uparrow 0 0\rangle 
+|\uparrow 0 0 : \, \uparrow 0 1 \rangle  \, .
\label{afmdez}
\eeqa
This has the form of Eq.\ (\ref{fmgs}), 
and differs only in that for antiferromagnetic 
coupling $J>0$, the conduction electron spin aligns 
opposite to the localized spins. The 
remaining terms in the $J>0$ ground-state, 
Eq.\ (\ref{afmgs}), represent double-exchange ordering 
along the $x$ and $y$ spin axes, together with 
residual localized Kondo singlets. To see this, 
introduce the eigenstates of $S^{x} =(S^{+} + S^{-})/2$ 
and $S^{y} =(S^{+} - S^{-})/2i$. These are defined in terms  
of the spin $z$ basis by:
\beqa
|\uparrow_{x}\rangle &=& 
(|\downarrow \rangle + | \uparrow \rangle)/\sqrt{2}\, , 
\quad 
|\uparrow_{y}\rangle = 
i(i|\downarrow \rangle + | \uparrow \rangle)/\sqrt{2}\, , 
\nonumber \\
|\downarrow_{x}\rangle &=& 
(|\downarrow \rangle - | \uparrow \rangle)/\sqrt{2}\, , 
\quad 
|\downarrow_{y}\rangle = 
i(i|\downarrow \rangle - | \uparrow \rangle)/\sqrt{2}\, . 
\nonumber
\eeqa
(The convention for the notation is that spins without a 
subscript always refer to the $z$ direction.) 
The arbitrary phases in these states have been chosen because 
they are convenient for working in the $S^{z}_{\tot}=1/2$ 
subspace. Double-exchange in the $\alpha = x,y$ spin 
directions,    
and with an antiferromagnetic coupling $J>0$, is described by 
\beqa
|{\rm de}_{\alpha}\rangle &=&
([\, |\uparrow_{\alpha} 0 1 : \, \uparrow_{\alpha} 0 0\rangle
+ |\uparrow_{\alpha} 0 0 : \, \uparrow_{\alpha} 0 1\rangle]
\nonumber \\
&+& [\, |\downarrow_{\alpha} 1 0 : \, \downarrow_{\alpha} 0 
0\rangle
+ |\downarrow_{\alpha} 0 0 : 
\, \downarrow_{\alpha} 1 0\rangle])/\sqrt{2}\, . 
\label{afmdealpha}
\eeqa
In these states the conduction electron occupation numbers 
refer to the spin direction $\alpha$: 
$n_{cj\sig_{\alpha}} = 0,1$ 
gives the number of conduction electrons with spin 
$\uparrow_{\alpha}$ or $\downarrow_{\alpha}$ at site $j$. The 
states $|{\rm de}_{\alpha}\rangle$ for $\alpha = x, y$ have the 
same form, and the same normalization, as that of 
Eq.\ (\ref{afmdez}) for double-exchange in the $z$ direction. 
The only difference is that the 
$S^{\alpha}_{\tot} = 1/2$ and $-1/2$ 
double-exchange forms (enclosed 
by the square brackets in Eq.\ (\ref{afmdealpha})) have been 
summed, which is the appropriate combination for the 
$S^{z}_{\tot} = 1/2$ subpace. The 
ground-state Eq.\ (\ref{afmgs}) for $J>0$ may now be written as
\beqa
|\psi_{0}\rangle & \propto & \left(1 - \frac{t}{J/4-E_{0}} 
\right)
(|\, s_{1} \uparrow \rangle + |\uparrow s_{2}\rangle) 
\nonumber \\
&+& \frac{t}{J/4 - E_{0}}(|{\rm de}_{x}\rangle
+ |{\rm de}_{y}\rangle + |{\rm de}_{z}\rangle ) \, . 
\label{afmgsdeform}
\eeqa
The proportionality constant is the normalization, and coincides 
with that in Eq.\ (\ref{afmgs}). As $t/J$ increases, 
$|\psi_{0}\rangle$ describes weight being taken 
away from the localized Kondo singlet states 
$|\, s_{1} \uparrow \rangle$ and $|\uparrow s_{2}\rangle $,     
which define the non-magnetic $J = \infty$ ground-state. 
The weight 
is shifted in equal proportion to the double-exchange terms 
$|{\rm de}_{\alpha}\rangle $ along the three spin axes 
$\alpha = x, y, z$. This produces 
the spin-isotropic ferromagnetic ordering between the localized 
spins, as in Fig.\ \ref{oldfig4.1}. Eq.\ (\ref{afmgsdeform}) makes it clear 
that the ferromagnetic ordering observed in Fig.\ \ref{oldfig4.1} 
for $J>0$ is due to the double-exchange interaction, 
as for the $J<0$ case. 

It is useful to comment on the differences between the 
ground-states 
for different signs of the coupling $J$. For ferromagnetic 
coupling the double-exchange ordering can be described 
using just 
the spin $z$ direction, at least in the subspace with maximal 
$S^{z}_{\tot}$. For antiferromagnetic coupling  
all spin directions were required. 
This difference is superficial, and 
while it increases the complexity of the description for 
$J>0$, there is no difference in the physics. The reason   
the simple form Eq.\ (\ref{fmgs}) can be obtained for $J<0$ is 
because the ground-state has maximal $S^{2}_{\tot}$ 
corresponding to fully saturated ferromagnetism. Thus, by 
describing the system 
in the subspace with maximal $S^{z}_{\tot}$, 
all the spins in the 
system are aligned, and the spin-flip part 
$J(S^{-}_{cj}S^{+}_{j} + S^{+}_{cj}S^{-}_{j})/2$ of the 
hamiltonian Eq.\ (\ref{4.11}) becomes ineffective. For 
antiferromagnetic coupling $J<0$ the ferromagnetism is 
not saturated. The localized 
spins align, but the conduction electron spin is opposite to 
that of the localized spins. In this case it is not possible to 
choose a subspace of $S^{z}_{\tot}$ in which the spin-flip 
part of the hamiltonian becomes ineffective: All spin 
directions are required to extract the double-exchange 
ordering in the ground-state for $J>0$. Nonetheless 
the physics of the ordering remains unaffected. 
For example, similar complications ensue for $J<0$ if one 
chooses the subspace with $S^{z}_{\tot} =1/2$. 

An intrinsic difference between the ground-states for different 
signs of the coupling is the residual weight attached to the 
localized Kondo singlets when $J>0$. This is due to the 
increased gain in energy $-3J/4$ for a Kondo singlet, over the 
energy gain $-|J|/4$ for a triplet state when $J<0$. 
The energy gain per 
site due to double exchange is $-|J|/4$ for either sign of the 
coupling. Thus, when $J<0$ the system gains just as much energy 
by triplet formation as by double-exchange, and there is no 
competition between the two. This is clear in the ground-state 
Eq.\ (\ref{fmgs}), which is consistent both with double-exchange 
ordering, and with a superposition of triplet formation at each 
site. As a result of this, the ground-states for $J<0$ do not 
evolve as $t/J$ increases, and the maximum ferromagnetic 
ordering between the localized spins sets in with $t/J$ 
arbitrarily small, as is clear from Fig.\ \ref{oldfig4.1}. 
For $J>0$ there is a competition between singlet 
formation and double-exchange ordering. When $t/J$ is small 
Kondo singlet formation dominates, and the ferromagnetic 
correlation 
between the localized spins vanishes as $t/J \rightarrow 0$. 
As $t/J$ increases, there is a larger energy gain for conduction 
electron hopping and this favours double-exchange. 
At large $t/J$ double-exchange dominates, and no residual Kondo 
singlets remain. 
The localized spins are then strongly ferromagnetically ordered. 
The crossover between the limiting behaviours is shown in 
Fig.\ \ref{oldfig4.1}, and is characterized by $t/(J/4 - E_{0})$ with 
$E_{0}$ given by Eq.\ (\ref{afmgse}).

\section{\label{old4.4}1D Kondo lattice results} 

The Kondo lattice model has been studied intensively for 
over two decades.  Notwithstanding this effort, reliable 
results on the Kondo lattice are few. The extension to the 
lattice case of methods developed for the single impurity Kondo 
model are either impossible, as in the Bethe ansatz solution 
of Andrei (1980) and Wiegmann (1980), or involve 
uncontrolled approximations. Primary examples of the latter are 
$1/N$ expansions,\footnote{Following convention, $N$ here denotes 
the degeneracy of the localized spin orbitals. Elsewhere 
$N$ denotes the number of lattice sites.} 
the slave boson method, and Gutzwiller approximations. 
(See Tsunetsugu, Sigrist and Ueda (1997) for references 
and discussion.) These methods have been quite successful 
in describing the formation of a coherent band of heavy 
quasiparticles, as is observed in the heavy fermion compounds. 
However, the methods taken over from the single impurity Kondo 
model focus on local correlations only, and have offered no 
consensus on the ground-state phases of the Kondo lattice. 
Thus, while the various methods developed on the basis of the 
single impurity model appear to capture some 
of the physics of the lattice problem, it is {\it a priori} 
unclear as to which aspects of the various solutions 
are reliable, and which are not. 

For the 1D Kondo lattice, some rigorous results 
have appeared in the 1990s, and have been 
supplemented by the results of a variety of numerical 
simulations. All these results are in substantial agreement, and 
give support to the view that the broad features of the 
ground-state phase diagram of the 1D Kondo lattice are now 
known. It is useful for later reference to summarize 
these results, and to provide a picture of the properties of the 
1D Kondo lattice as it was understood in the early 1990s. 

The main focus of this review is the conventional 1D Kondo 
lattice model as defined in Eq.\ (\ref{4.11}). 
Variants of the conventional spin-isotropic model, such as the 
inclusion of nearest-neighbour (White and Affleck 1996, Sikkema, 
Affleck and White 1997, Coleman, Georges and Tsvelick 1997)
or considering spin-anisotropic interaction (Shibata, Ishii and Ueda 1995, 
Zachar, Kivelson and Emery 1996, Novais, {it et al.} 2002a,2002b),
do not constitute the main topic of discussion of this review
Even though some of these extensions of the conventional Kondo
lattice will be discussed at later stages of this review. 

Hence, section \ref{old4.4.1} only 
contains a brief discussion of results for  
the 1D Kondo lattice with a half-filled conduction band.
Since the lattice with a partially-filled (ie\ less than  
half-filled) conduction band is the focus of interest in 
chapters \ref{ch5} and \ref{ch6}, section \ref{old4.4.2} contains a detailed  
discussion of previously known results for the 1D Kondo lattice 
at partial conduction band filling.

\subsection{\label{old4.4.1}Half-filled conduction band}

The Kondo lattice with a half-filled conduction band is 
thought by some (Yu and White 1993, Wang, Li and Lee 1993,
Guerrero and Yu 1995)
to be an effective model for the class of Kondo 
insulators. As discussed in section \ref{old4.2.2}, the
Kondo insulators are small gap semiconductors, 
in which the gap derives from a hybridization between 
singly-occupied localized $f$-orbitals, and a half-filled 
conduction band (Aeppli and Fisk 1992, Fisk, {\it et al.} 1995).
There are some doubts as to whether the Kondo insulators 
are in the local moment regime (Varma 1994),
and therefore some question as to whether 
the Kondo lattice is applicable, or whether the more 
fundamental periodic Anderson model is required 
(cf.\ section \ref{old4.2.2}). This issue will not be addressed here.

Half-filling is defined by $n = N_{e}/N =1$, where $N_{e}$ is the 
number of conduction electrons and $N$ the number of lattice 
sites (i.e.\ the number of localized spins). A rigorous theorem 
holds for the half-filled Kondo lattice in any dimension on 
a bipartite lattice (Tsunetsugu, {\it et al.} 1992, Shen 1996, 
Tsunetsugu, Sigrist and Ueda 1997): For antiferromagnetic coupling 
$J>0$ the ground-state is unique and has 
zero total spin $S_{\tot} = 0$ (i.e.\ the ground-state is a 
total spin singlet). The same  conclusion holds 
for ferromagnetic coupling $J<0$ provided the two sublattices 
have the same number of sites. 
Beyond the rigorous proof that the ground-state is a total spin
singlet in any dimension, there is substantial evidence that the 
ground-state also has a spin gap at least in 1D. 
The ground-state of the 
half-filled 1D Kondo lattice thus forms a spin-liquid. To 
discuss the evidence for this, it is convenient to consider the 
$J>0$ and $J<0$ cases separately.

{\bf Antiferromagnetic coupling $J>0$}: At infinite coupling 
$J = \infty$ the ground-state at half-filling consists of $N$ 
on-site Kondo 
singlets (cf.\ Table 4.1). There is a spin gap of size 
$J$ to an on-site triplet state, and a larger charge gap 
of $3J/2$, corresponding to the hopping of a conduction electron 
to a neighboring site. The persistence of the spin gap (and a 
larger charge gap) down to arbitrarily small coupling strengths 
$J>0$ has been established by the following numerical 
simulations: quantum Monte Carlo at $J/t = 1.6$ 
(Fye and Scalapino 1990); 
exact diagonalization on systems of up to 10 sites, 
and over the full range of couplings $J>0$ 
(Tsunetsugu, {\it et al.} 1992); 
density-matrix renormalization group studies on lattices of up to 
24 sites and over a wide range of coupling strengths 
(Yu and White 1993). 
The numerical results are further supported by approximate 
analytic techniques; Gutzwiller-projected mean-field solutions 
(Wang, Li and Lee 1993),
and a mapping of the Kondo lattice to a 
nonlinear sigma model at weak-coupling, within a semiclassical 
approximation for the localized spins (Tsvelik 1994).
\footnote{Spin gaps are also observed in 
bosonization treatments of the 1D half-filled Kondo lattice. 
See Fujimoto and Kawakami (1997) and Le Hur (1998) for $J>0$ and 
Le Hur (1997) for $J<0$. These treatments are not for the 
pure Kondo lattice of Eq.\ (\ref{4.11}), but include also a 
direct interaction between the localized spins.} 
As noted in section \ref{old4.3.2}, the singlet ground-state and spin gap 
in the half-filled Kondo lattice permits a formal identification 
of the Kondo temperature as the energy of the spin gap. At 
strong-coupling this is linear in $J$ and at weak-coupling takes 
an exponential form Eq.\ (\ref{4.2.8}) which is similar to, 
but enhanced over, the single impurity result Eq.\ (\ref{4.2.6}). 

{\bf Ferromagnetic coupling $J<0$}: Substantially less work has 
been done on the half-filled Kondo lattice with $J<0$. 
Exact-diagonalization results in 1D, together with finite size 
scaling (Tsunetsugu, {\it et al.} 1992, Shibata, {\it et al.} 1996),
show that there exists a spin gap for 
$J<0$ as for the half-filled $J>0$ Kondo lattice. Nevertheless 
the nature of the gap for $J<0$ is different. For $J>0$, 
the strong-coupling behaviour is that of the Kondo spin liquid, 
while for $J<0$ it scales to a spin 1 
chain, and reduces to a Haldane gap state at strong-coupling; 
instead of the spin gap increasing as $J \rightarrow -\infty$, 
it decreases at large $J$ (Tsunetsugu, {\it et al.} 1992). 

{\bf Correlations}: The correlations between nearest-neighbour 
localized spins in the half-filled 1D KLM are antiferromagnetic 
for any sign of the coupling. This may be understood 
beginning from the $|J| = \infty$ limits where there is one 
conduction electron localized at each site. At strong but 
finite coupling there is weak virtual conduction electron hopping 
to neighboring sites. By the Pauli principle, this is 
possible only if the conduction electron on the neighboring site 
has opposite spin. For strong but finite ferromagnetic or 
antiferromagnetic coupling, the weak antiferromagnetism of the 
localized conduction electrons induces a similar 
antiferromagnetic ordering on the underlying localized spins. 
Strong antiferromagnetism at weak-coupling is expected on the 
basis of the RKKY interaction which  oscillates in sign with   
wave vector $2k_{F}$ (cf.\ section \ref{old4.3.1}). At half-filling 
$k_{F} = \pi /2a$, and so the RKKY interaction changes sign at 
neighboring lattice sites. 
These expectations are supported by the results of numerical 
simulations for $J>0$ (Fye and Scalapino 1990, Yu and White 1993),
and by perturbation theory at large ferromagnetic couplings  
(Tsunetsugu, {\it et al.} 1992). 

{\bf Higher dimensions}: As mentioned earlier, a great deal of 
work has been carried out on the one dimensional Kondo lattice model
for over two decades. However, in higher dimensions results are
few, and most of them are numerical studies for the half filled band, 
spin liquid phase. These studies show that the spin liquid
phase, at large $J$, 
becomes unstable against antiferromagnetic long-range order
formation at small $J$, at a finite $J_c$. 
Above the transition, a popular 
scenario (P\'{e}pin and Lavagna 1999, Lavagna and P\'{e}pin 2000) 
involves antiferromagnetic spin fluctuations
coupling to the Fermi liquid properties. The low temperature
physics of this is controlled by a collective mode that softens
at the antiferromagnetic transition (Lavagna and P\'{e}pin 2000). 
The transition occurs at $J_c \approx 1.4$ for the two dimensional
case and around $J_c \approx 1.8 - 2.0$ for the three dimensional one.
The estimates for the critical coupling in two dimensions have been
confirmed by several different calculations, such as mean-field
approaches (Lacroix and Cyrot 1979, Zhang and Yu 2000, Jurecka
and Brening 2001), quantum Monte Carlo simulations (Wang, Li and
Lee 1994, Assad 1999, Capponi and Assad 2001)
and series expansions (Shi {\it et al.} 1995, Zheng and Oitmaa 2003). 
In the three dimensional case only series expansions have been 
implemented until now (Shi {\it et al.} 1995, Zheng and Oitmaa 2003). 
It is interesting to remark that the series expansion methods,
both in two and three dimensions, suggest that the spin excitations
becomes gapless at the quantum critical point, while the charge
excitations ramain gapped. This may suggest that similarly to the
one dimensional case, the nonlinear
sigma model is the appropriate low energy theory also
in hiher dimensions.

\subsection{\label{old4.4.2}Partially-filled conduction band}

Results for the 1D Kondo lattice with a partially-filled 
conduction band $n = N_{e}/N <1$ fall into two categories; 
rigorous results proved using the Perron-Frobenius theorem, 
and results obtained by numerical simulations. These results are 
summarized in turn. Most apply to the case of antiferromagnetic 
coupling $J>0$. 

{\bf Rigorous results in limiting cases}: 
Two rigorous results have been proven for the 1D Kondo lattice 
hamiltonian, Eq.\ (\ref{4.11}). 

(i) {\it The Kondo lattice with one conduction electron}. 
Sigrist, Tsunetsugu and Ueda (1991) proved that the 
ground-state of the Kondo lattice with one conduction electron 
forms an incompletely saturated ferromagnet, with total spin 
$S_{\tot} = (N - 1)/2$, for $J>0$, and forms 
a fully saturated ferromagnet, $S_{\tot} = (N + 1)/2$, for 
$J<0$. For $J<0$, the theorem is trivial, since the spin-flip 
part of the Kondo lattice interaction is ineffective, much as 
for the two-site lattice discussed in section \ref{old4.3.3}; the 
ground-state consists of an alignment of all the localized 
spins, together with the conduction electron in a zero 
momentum Bloch state and with spin opposite to that of the 
localized spins. For antiferromagnetic coupling $J>0$ the 
theorem is non-trivial, and is proved by 
adopting a special phase convention for the conduction 
electron states, and then applying the Perron-Frobenius theorem 
of matrix theory (Sigrist, Tsunetsugu and Ueda 1991).  
The Perron-Frobenius theorem also proves that the ground-state 
is unique, except for the usual 
$2S_{\tot} + 1$ choices for the $z$ component 
$S^{z}_{\tot}$ of the total spin. 
A detailed presentation of the proof, together with the required  
matrix theory, is given by Yanagisawa and Harigaya (1994).  
The theorem holds in any dimension, and for all finite coupling 
strengths $0 < J < \infty$. It applies also if the conduction 
electron hopping goes beyond nearest-neighbour, provided that 
the hopping amplitudes remain negative.

Sigrist {\it et al.} (1991) also constructed 
the maximal $S_{\tot}^{z}$ ground-state of the $J>0$ Kondo 
lattice with one conduction electron, and used it to 
investigate some of the ground-state properties.   
The conduction electron in the lattice of localized spins is a 
spin polaron. It moves dressed by a polarization cloud of 
localized spins, which tend to align antiparallel to the 
conduction electron spin. The effective 
mass $m^{*}$ of the electron is thereby slightly enhanced, and 
varies from $m^{*} = m_{e}$ at $J = 0$ to $m^{*} = 2m_{e}$ as 
$J \rightarrow \infty$, where $m_{e}$ is the bare electron mass. 
The spatial extension of the polarization cloud is given by the 
correlation between the conduction electron spin and nearby 
localized spins. As $J \rightarrow \infty$ the extension shrinks 
down to on-site correlation only, as the conduction electron 
localizes in an on-site Kondo singlet. For small $J$, the 
correlation decays exponentially with a coherence length 
$\sqrt{2t/J}$, which characterizes the extension of the 
polarization cloud.  

Using exact diagonalization on systems of up to 14 sites, 
Sigrist, Ueda, and Tsunetsugu (1992a) later 
investigated the interactions between spin polarons in the 
$J>0$ 1D Kondo lattice with two conduction electrons. As $J$ 
is increased, the electrons separate, forming separate spin 
polarization clouds, and form strongly 
repulsive on-site Kondo singlets as $J \rightarrow \infty$

(ii) {\it Strong-coupling expansion}. For partial 
conduction band filling $n<1$, the $J = \infty$ ground-state 
consists of $N_{e}$ on-site Kondo singlets, and $N-N_{e}$ 
free unpaired localized spins (cf.\ Table 4.1). The ground-state 
has a $2^{N-N_{e}}$--fold spin degeneracy.
Introducing corrections of order $t/J$, the on-site Kondo 
singlets hop to nearest-neighbour unpaired sites 
with amplitude $-t/2$ (Hirsch 1984).\footnote{The fact that the 
hopping amplitude is renormalized by a factor 2 is the reason 
that the effective mass $m^{*} = 2m_{e}$ as $J \rightarrow 
\infty$ for the spin polaron in the Kondo lattice with one 
conduction electron.} 
Hopping Kondo singlets alter the charge configuration, 
but not the spin configuration: The unpaired 
localized spin and the Kondo singlet simply exchange sites, 
with no flipping of the localized spin. Thus at ${\cal O}(t/J)$ 
the charge and spin degrees of freedom are decoupled,   
and the picture formed by focusing on mobile Kondo singlets up to 
${\cal O}(t/J)^{2}$ is given by Hirsch (1984). 

A qualitatively different picture emerges by considering 
${\cal O}(t/J)^{2}$ corrections which alter the spin 
configuration. Sigrist, Tsunetsugu, Ueda and Rice (1992b)
performed a systematic expansion in powers of 
$t/J$ out of the $J = \infty$ ground-state. They neglected the 
higher-energy excitations to on-site triplets,  which cost an 
energy $J$, and neglected the 
states with two conduction electrons at the same site, which cost 
an energy $3J/2$ corresponding to the destruction of two Kondo 
singlets. At order $(t/J)^{2}$, Sigrist {\it et al.} obtain a 
series of complicated hopping processes, one of which alters 
the spin configuration. Using the Perron-Frobenius theorem, 
it is then proved that the ground-state spin degeneracy is 
lifted at 
order $(t/J)^{2}$ to give an incompletely saturated ferromagnet 
with total spin $(N - N_{e})/2$ corresponding to the 
alignment of all unpaired localized spins (Sigrist, {\it et al.} 1992b). 
The theorem holds for all partial conduction band fillings $n<1$.

The strong-coupling result was later extended by Yanagisawa 
and Harigaya (1994), who used the Perron-Frobenius theorem 
to prove a similar result for the partially-filled 1D 
Kondo lattice with a strong conduction electron Hubbard repulsion 
$U\sum_{j}n_{cj\uparrow}n_{cj\downarrow}$ (cf.\ Eq.\ 
(\ref{1.2.10})) 
added to the standard hamiltonian Eq.\ (\ref{4.11}): 
As $U \rightarrow \infty$, the ground-state of the Kondo 
lattice for $n<1$ has total spin $S_{\tot}= (N - N_{e})/2$ for 
all antiferromagnetic couplings $J>0$, and 
$S_{\tot} = (N + N_{e})/2$ for all ferromagnetic couplings 
$J<0$. Thus 
ferromagnetism occupies the whole of the phase diagram 
($0 < n < 1$, $J \neq 0$) if there is a strong Hubbard repulsion 
between the conduction electrons.

{\bf Results by numerical simulations}: The rigorous results for 
the Kondo lattice apply only in limiting parameter regimes; 
vanishing conduction band filling $n \rightarrow 0$, and the 
strong-coupling 
limit $t/J \rightarrow 0$. In both these limits a ferromagnetic 
ordering is favoured; incompletely saturated ferromagnetism for 
$J>0$, and fully saturated ferromagnetism for $J<0$. 
The remainder of the ground-state phase diagram was tackled 
using numerical simulations.

{\it Antiferromagnetic coupling}:  
Several numerical studies by a variety of different methods 
have been carried out on the partially-filled 1D Kondo lattice 
in the early 1990s. For antiferromagnetic couplings 
$J>0$, the earlier studies have been completed by: (i) Troyer and W\"{u}rtz 
(1993), who used quantum 
Monte Carlo on systems of up to 24 sites, and at conduction 
band fillings $n = 1/3$ and $2/3$. 
(ii) Tsunetsugu, Sigrist and Ueda (1993), 
who used exact diagonalization on systems 
of up to 9 sites and over the full range of fillings 
$0 < n <1$. (iii) Moukouri and Caron (1995), 
who used the density-matrix renormalization group 
on systems of up to 75 sites at $n = 0.7$. 
(iv) Caprara and Rosengren (1997), who used the 
infinite-size version of the density-matrix renormalization group 
on systems of up to 202 sites, and over a range of small 
fillings $0 < n \leq 1/2$. 

\begin{figure}[ht]
\centering
\includegraphics[scale=0.7]{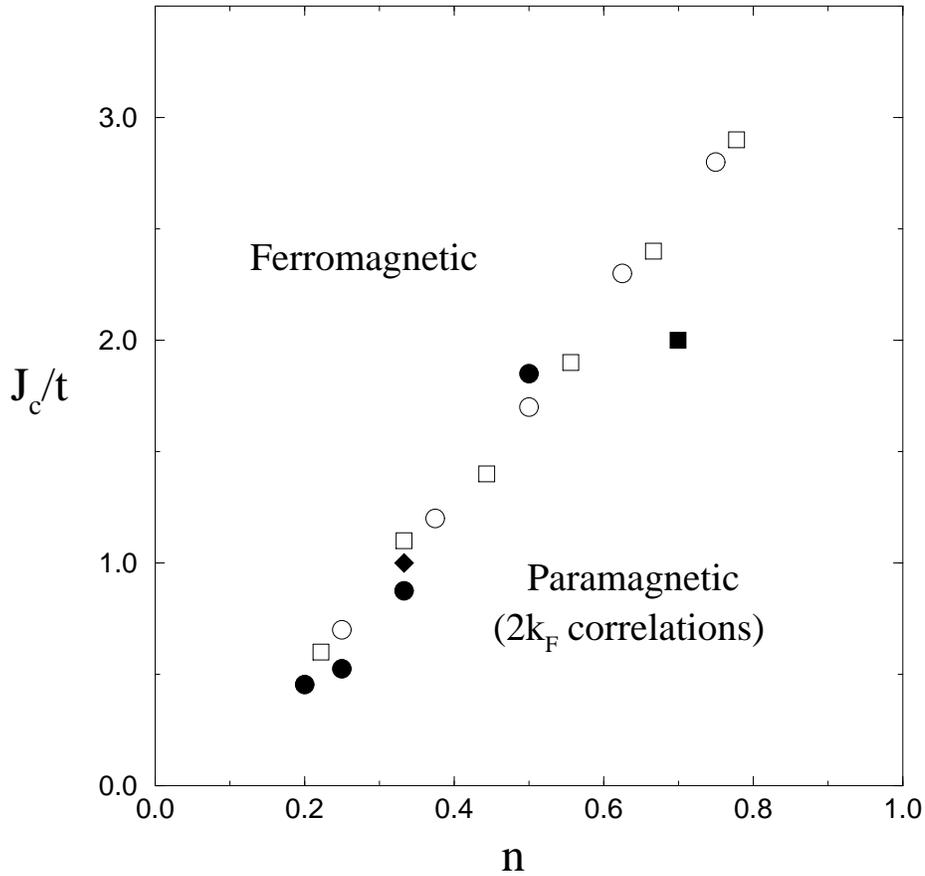}
\vspace{3.0mm}
\caption{\label{oldfig4.2}Phase diagram of the 1D Kondo lattice 
with antiferromagnetic coupling $J>0$, as determined by the
first numerical simulations. Critical points $J_{c}$ 
for the ferromagnetic-paramagnetic transition are shown: 
the filled diamond is the quantum Monte Carlo point 
(Troyer and W\"{u}rtz 1993);  
open circles and squares are exact 
diagonalization results (Tsunetsugu, Sigrist and Ueda 1993); 
the filled square is the density-matrix renormalization 
group result (Moukouri and Caron 1995); the filled circles 
are the infinite size density-matrix renormalization group 
results (Caprara and Rosengren 1995).}
\end{figure}

The results of all these studies are in very good agreement 
with each other, as well as with the rigorous results, and the 
basic picture they present is now generally accepted. 
From the numerical results on larger systems ((i), (iii), (iv)), 
the 1D Kondo lattice with $n < 1$ and 
$J > 0$ shows a ferromagnetic ordering of the localized spins at 
stronger couplings. This is identified by the structure factor 
of the localized spins (i.e.\ the Fourier transform of the 
real space correlation function for the localized spins) 
(Troyer and W\"{u}rtz 1993):
\beqa
C(k) = \frac{1}{N}\sum_{j,l}^{N} e^{ik(j-l)a}\, \langle 
S^{z}_{j} S^{z}_{l} \rangle \, .
\label{lsss}
\eeqa
The simulations on the larger systems show a peak in 
$C(k)$ at $k = 0$, which indicates an alignment of the localized 
spins. This is accompanied by a very weak ferromagnetic 
correlation in the spin structure factor for the conduction 
electrons. 
As the coupling is lowered, the Kondo lattice undergoes a 
transition to a paramagnetic phase at a critical coupling 
$J_{c}$ in the weak to intermediate range ($0 < J_{c}/t \lesssim 
4$). The paramagnetic phase is characterized by a peak in the 
localized spin structure factor at $2k_{F}$, where $k_{F}$ is 
the conduction electron Fermi momentum. The correlations between 
the localized spins in the paramagnetic phase are characteristic 
of the RKKY interaction, though recall 
that the RKKY interaction diverges in 1D, as is typical of 
perturbative expansions in low dimensions (cf.\ section \ref{old4.3.1} 
and the discussion given there). In the paramagnetic phase 
the correlations in the conduction electron spins again 
weakly track those of the localized spins, so there is a very 
weak $2k_{F}$ peak in the structure factor of the conduction 
electron spins. The critical coupling $J_{c}$ marking the 
ferromagnetic-paramagnetic transition depends on the filling, 
$J_{c} = J_{c}(n)$, and is identified in the numerical studies 
(i), (iii) and (iv) as the point where the intensity of the 
$k = 0$ peak in $C(k)$ is equal to the intensity of the 
$k = 2k_{F}$ peak; at stronger couplings the $k = 0$ peak 
dominates (ferromagnetism), and at weaker couplings the 
RKKY-like $k = 2k_{F}$ peak dominates (paramagnetism). 
The obtained transition points are shown in Fig.\ \ref{oldfig4.2}.

The exact diagonalization study (ii) of Tsunetsugu 
{\it et al.} (1993) 
presents the same phase diagram as that obtained by 
the studies on larger systems; some representative phase 
transition points obtained by exact digonalization are shown 
in Fig.\ \ref{oldfig4.2}. Contrary to the other studies, the transition 
points in this case are obtained by calculating the total 
spin of the system. The transition points mark the values at 
which the total spin changes from incompletely saturated  
ferromagnetism $S_{\tot} = (N-N_{e})/2$ to the 
paramagnetic value of minimal total spin $S_{\tot} = 0$ 
or $1/2$ depending on whether $N-N_{e}$ is even or odd. 
At smaller fillings, there is a small crossover 
regime between these extremities which signifies a 
continuous phase transition. The analytic work of 
chapter \ref{ch5} predicts this type of ferromagnetic-paramagnetic 
transition for the 1D Kondo lattice. At larger fillings the 
crossover is sharp, and signifies a first-order transition. 
The results of chapter \ref{ch5} dispute this as a property in a
thermodynamically large Kondo lattice, and the sharp crossover 
must be due to the small system sizes $N \leq 9$ for which the
exact diagonalization method is practicable. Indeed, 
in their review Tsunetsugu, Sigrist and Ueda (1997), 
the authors of the exact diagonalization study   
state that the small size of the systems makes it 
difficult to judge the type of transition on the basis of their 
results.

The recent numerical results (McCulloch, {\it et al.} 1999, 2001,
2002) confirmed the above results. The high
accuracy of the modern numerical algorithms allowed 
the whole phase diagram to be mapped out for all values 
of electron doping levels, see section \ref{6.5}. 

{\it Ferromagnetic coupling.} For ferromagnetic couplings 
$J<0$, the Kondo lattice has been studied by Dagotto's group 
(Dagotto, {\it et al.} 1998, Yunoki, {\it et al.} 1998)
using a variety of numerical methods, 
including quantum Monte Carlo, the density-matrix renormalization 
group, and Lanczos methods. The general phase diagram they 
obtain, from studies of the Kondo lattice in one-, two- and 
infinite-dimensions, and for both quantum and semiclassical 
localized spins, is as follows: At stronger coupling the 
ground-state is ferromagnetic, as identified by a $k=0$ peak 
in the structure factor $C(k)$ of the localized spins 
(cf.\ Eq.\ (\ref{lsss})). At weaker coupling the ground-state 
is a paramagnet, with incommensurate correlations in $C(k)$ 
which usually correspond to $2k_{F}$, as expected on the 
basis of the RKKY interaction. 
The crossover between the ferromagnetic and paramagnetic 
phases occurs at a critical coupling dependent on filling. This 
transition line has the same general shape as that for $J>0$ at 
small fillings (cf.\ Fig.\ \ref{oldfig4.2}), but turns upward as $n 
\rightarrow 1$ and approaches the half-filling line 
asymptotically (cf.\ Fig.\ \ref{oldfig6.5} for some representative points). 
Dagotto's group also find a phase separated region close to 
half-filling at strong coupling. 

The status of the results of Dagotto's group 
(Dagotto, {\it et al.} 1998, Yunoki, {\it et al.} 1998)
is at present unclear. There are problems both in the 
great variety of scales of the transition line in different 
simulations, and in the appearance of some 
non-$2k_{F}$ correlations at weak-coupling even in higher 
dimensions where the RKKY interaction operates without problems.
Finally, the appearance of phase separation has been questioned
recently. Several new simulations
(Horsch, Jaklic and Mack 1999, Batista, {\it et al.} 1998, 2000,
Garcia, {\it et al.} 2002, Koller, {\it et al.} 2003) suggests 
that the phase separated region close to half filling is rather 
a phase dominated by ferromagnetic polarons with one single trapped 
charge carrier. This is in perfect agreement with the polaronic
picture first proposed by Honner and Gul\'{a}csi (1998b), for more details
also see section \ref{old6.1}. It may even be that the previously
attributed phase separation regime is actually a polaronic liquid,
as it appears (see Fig.\ \ref{Ian-kondo-PRB-fig1}) for $J>0$. 

%%%%%%%%%%%%%%%%%%%%%%%%%%%%%%%%%%%%%%%%%%%
%% chapter 3
%%%%%%%%%%%%%%%%%%%%%%%%%%%%%%%%%%%%%%%%%%%%

\cleardoublepage
\chapter{\label{ch2}Bosonization Formalism}

Landau Fermi liquid theory fails in 1D due to a 
peculiarity of the low-energy excitations of 
interacting 1D many-electron systems. For 
interacting 3D systems, the low-energy excitations 
are particle-like, and lead to Landau's one-to-one 
correspondence between interacting and 
non-interacting eigenstates. This leads  
in turn to the concept of fermionic quasiparticle 
excitations. For interacting 1D systems,  
the low-energy excitations are collective density 
fluctuations, and involve large numbers of excited 
electrons acting as a coherent whole. 
This destroys the one-to-one correspondence 
between non-interacting and interacting eigenstates: The 
non-interacting eigenstates of course remain particle-like 
in 1D, and satisfy Fermi statistics, but the interacting 
eigenstates are collective density fluctuations, and  
satisfy Bose statistics. 

The special properties of interacting 1D systems lead 
naturally to a description in terms of bosonic excitations, 
which are the collective density fluctuations. 
The description in terms of bosons  
is far simpler than the standard description in 
terms of non-interacting electron states, because the number of 
bosons required to describe the system is far less than the 
number of excited electrons. In some cases, the description 
in terms of bosons is so simple that the  
problem may be solved exactly. This occurs, for example, in the 
Tomonaga-Luttinger model (Tomonaga 1950, Luttinger 1963, Mattis
and Lieb 1965),  
which has small momentum transfer or forward scattering 
interactions between the electrons. 

The description of 1D many-electron systems in terms of 
bosonic density fluctuations is called bosonization. 
The central elements in the formalism of bosonization 
are Bose 
representations; fermionic operators, such as the electron 
fields $\psi^{\dg}_{\sig}(x)$ (cf.\ Eq.\ (\ref{1.22})), 
may be written in terms of bosonic density fluctuations. 
This gives a Bose representation for 
$\psi^{\dg}_{\sig}(x)$ which may be substituted back 
into the hamiltonian to give a more tractable 
description of the system. Bose representations may also 
be used to simplify the calculation of physical properties,  
such as correlation functions. A complete 
bosonization formalism consists of deriving Bose 
representations for all the fermionic operators appearing 
in a system's hamiltonian. In this chapter, the 
bosonization formalism is derived for 1D many-electron 
systems, defined both in a continuum and on a lattice. 

The derivation of bosonization given in this chapter is 
based on Honner and Gul\'{a}csi (1997a,1998b), 
but the resulting Bose representations are variations on a 
well worn theme. 
Bosonization techniques have a long history in condensed 
matter physics, and first appear in recognizable form in 1950 
in a paper by Tomonaga (1950).\footnote{Some of the 
basic concepts of bosonization can be traced back further; 
an approximate 3D treatment was given by Bloch in 1934
(Stone 1994).} 

In general terms bosonization, as applied to 1D many-electron
systems, historically has developed along two directions: the
field theoretical bosonization and the so-called constructive 
bosonization. The field theoretical bosonization evolved out 
of solutions to models in the area of high energy physics. 
Introduced by Tomonaga (1950), and developed to its current 
form by Mattis and Lieb (1965), Schotte and Schotte (1969), 
Coleman (1975), Luther and Peschel (1975), Mandelstam (1975) 
and Heidenreich, {\it et al.} (1975). For good reviews 
on the field theory bosonization, see Ha (1984), 
Fradkin (1991), Shankar (1995), Gul\'{a}csi (1997a). 
Constructive bosonization, introduced by Emery (1979) and later
on by Haldane (1981) in terms of a Fock space identification 
of operators, gives a simpler, more transparent derivation of the boson
representation of fermion operators.  An explicit picture of the physics
corresponding to the boson representation may be obtained directly from 
the formalism, see the pedagogical articles of 
Sch\"{o}nhammer and Meden (1996), Delft and Sch\"{o}ller (1998).
and the finite temperature extension in Bowen and Gul\'{a}csi (2001). 

However, bosonization is usually derived beginning 
from a field theory approximation to the condensed matter system 
of interest; the Fermi sea is replaced by two Dirac seas 
(cf.\ section \ref{old1.2.3}).
In this chapter, the bosonization formalism is derived 
beginning from the original system with a finite Fermi sea,
{\it \'{a} la} constructive bosonization.  
This requires different methods and considerably more 
work than the usual derivation, but the resulting 
Bose representations have similar forms. 
There are two main rewards for the extra effort. 
Firstly, a solution is obtained for a long-standing problem 
regarding the interpretation of terms in Bose representations.
Secondly, the Bose representations are better
suited to describing two-component systems
\footnote{The common examples of two-component systems are 
the two band models (Gul\'{a}csi and Bedell 1994),  
and the two chain problem (Finkel'stein and Larkin 1993). Both
of these cases were solved using a field theoretical 
bosonization.}, in 
particular the Kondo lattice model, in which only one of the 
components is bosonized. Before discussing these points in more 
detail, it is useful to establish some 
terminology. For brevity, 
the model with the two Dirac seas replacing the Fermi sea  
is called a Luttinger model,\footnote{In the literature 
the Luttinger model is sometimes understood to imply also 
certain interactions, specifically forward scattering 
interactions as originally considered by Tomonaga (195)
and by Luttinger (1963). This is not implied here; 
the Luttinger model simply refers to the introduction of the 
Dirac seas. Following modern convention (Voit 1994), 
the Luttinger model with forward scattering 
interactions is called the Tomonaga-Luttinger model.} and 
the original system with the finite Fermi sea is called a 
realistic system.

In deriving Bose representations directly for a realistic 
system, a long-standing problem of interpretation is solved. 
The usual bosonization procedure begins from a Luttinger model, 
and then maps the results back to the realistic system of 
interest. This mapping sometimes runs into problems. 
The Luttinger model bosonization involves a length $\alpha$, 
which of necessity is infinitesimal, and is to be taken to zero 
at the end of a calculation (Voit 1994). (The Luttinger 
model bosonization is reviewed in Appendix \ref{appa}.)
In a realistic system, with a Fermi sea instead of Dirac seas, 
sending $\alpha$ to zero can give non-physical results.
For example, in the 1D attractive backscattering electron gas, 
the spin excitations are gapped, and bosonization gives 
a gap size proportional to $\alpha^{-1}$ (Luther and Emery 1974). 
If $\alpha \rightarrow 0$, then the gap size is infinite and 
the energy spectrum exhibits a non-physical divergence.
In these situations, it becomes necessary to reinterpret $\alpha$ 
as a finite physical quantity relating to the realistic 
system. The reinterpretation is highly problematic, 
as no consensus has been reached on the general meaning of 
$\alpha$ in realistic systems, and finite $\alpha$ is in any case 
inconsistent with the Luttinger model bosonization, where 
$\alpha$ must of necessity be taken to zero  
in order to get the correct representation for 
the density operator 
$\rho_{r\sig}(x) = \psi^{\dg}_{r\sig}(x)\psi^{}_{r\sig}(x)$ 
(cf.\ Eq.\ (\ref{A.9})). 
In the derivation given in this chapter, it becomes clear that 
$\alpha$ measures the minimum wavelength of the density 
fluctuations which satisfy bosonic commutation relations. 
This is discussed further in section \ref{old2.2.2} below, and is related 
to previous interpretations in section \ref{old3.1.3}. 

The second main benefit for deriving bosonization directly 
for realistic systems is for applications in the 1D Kondo 
lattice model. In general, the Luttinger model bosonization 
provides at least a qualitatively complete description of 
the low-energy physical properties of simpler one-component 
systems, such as the 1D Hubbard model (cf.\ section \ref{old3.2}). 
This is because 
the density fluctuations generally remain  
bosonic, at least qualitatively, down to wavelengths of the 
order of the average inter-electron spacing. This spacing is 
the minimum physically meaningful length scale in one-component 
systems, and bosonization thus provides a qualitatively complete 
description of all physical processes; $\alpha$ 
acts in these cases as a harmless short-distance cut-off, 
much as in field theory,
and may have to be kept finite in order to avoid divergences,  
but otherwise is largely irrelevant 
in the description of the physical 
properties of the system.\footnote{This is the physical 
content of Emery's identification (Emery 1979) 
of $\alpha$ with the lattice spacing $a$. Emery's 
interpretation, which has wide currency, 
is discussed further in section \ref{old3.1.3}.}
In the Kondo lattice model, which is a two-component system 
containing conduction electrons and localized spins, 
these arguments do not apply. To be precise, the average 
inter-conduction electron spacing in the partially-filled 
Kondo lattice is greater than the minimum physically meaningful 
length scale in the system. In the Kondo lattice  
the minimum length scale is set  
by the lattice spacing $a$ between the localized spins. 
This is smaller than the average inter-conduction electron 
spacing when the conduction band filling is below half. Thus, in 
the Kondo lattice with a bosonized conduction band, 
it may happen that important physical 
processes between neighbouring localized spins are missed if 
$\alpha$ for the conduction electrons is not treated correctly,  
and is made arbitrarily small as in the Luttinger model. 
This does in fact occur; a Luttinger model bosonization misses  
the double-exchange ordering of the localized spins, whereas 
the bosonization developed for realistic systems does 
not (Honner and Gul\'{a}csi 1997a,1998b). 
A comprehensive description of the 1D Kondo lattice via  
bosonization is contained in chapters \ref{ch5} and 
\ref{ch6}. 

The organization of this chapter is as follows: 
Section \ref{old2.1} defines the systems for which a bosonization 
description can be rigorously derived. 
To include systems with interactions of maximal generality, 
it is assumed only that the low-energy properties of 
the interacting system involve states within a momentum range 
$\pm k_{0}$ of the Fermi surface at $\pm k_{F}$. 
Section \ref{old2.2}  states and 
proves two theorems, which follow from the structure of the 
1D state space. These provide the foundation for bosonization.
The first theorem, proved in section \ref{old2.2.1}, establishes 
that certain density fluctuation components satisfy a bosonic 
algebra. The length $\alpha$ is 
introduced here as the wavelength below which   
the density fluctuations cease to satisfy the bosonic algebra. 
Section \ref{old2.2.2} contains a discussion of $\alpha$. 
The second theorem, proved in section \ref{old2.2.3}, 
establishes that the density 
fluctuations generate all the states in the subspace defined  
by the bandwidth cut-off $k_{0}$, provided that $k_{0} \ll k_{F}$.
The theorems of section \ref{old2.2} lead to a rigorous but simple 
prescription for deriving Bose representations, 
and allow the representations to be derived without direct 
reference to interactions. The prescription is stated at the 
beginning of section \ref{old2.3}, and is used 
to derive Bose representations for the Fermi operator 
forms of interest; section \ref{old2.3.1} derives the Bose 
representation for the non-interacting 
hamiltonian $H_{0}$, section \ref{old2.3.2} the representation for 
Fermi field operators, and section \ref{old2.3.3} derives the 
representations for operators bilinear in the Fermi fields. 
The results of section \ref{old2.3}, which apply to 1D continuum systems, 
are generalized to 1D lattice systems in section \ref{old2.4}.

\section{\label{old2.1}System Description and Notations}

{\bf Non-interacting system}: Consider a 1D continuum system, 
as described in section \ref{old1.1.1}, with an average number of 
$N_{e}$ electrons. The non-interacting hamiltonian is given by 
\beqa
H_{0} = \sum_{k, \sig}\left(\varepsilon(k) - \mu \right)
c^{\dg}_{k\sig}c^{}_{k\sig} 
\label{2.1.1}
\eeqa
where the dispersion $\varepsilon(k) = k^{2}/2m_{e}$ and $m_{e}$ 
is the bare electron mass. This is the same as in section 
\ref{old1.2.1}, Eq.\ (\ref{1.2.1}), but with a chemical potential 
$\mu$ subtracted to facilitate dealing with systems in which the 
electron number may vary slightly about $N_{e}$. For the 
non-interacting zero 
temperature systems considered here $\mu = \varepsilon(k_{F})$. 
The non-interacting ground-state $|0\rangle$, assumed 
non-degenerate, has all momentum/spin 
states $|k\sig\rangle$ with $|k| < k_{F}$ occupied, 
and all other states empty, where the Fermi momentum 
$k_{F} = N_{e}\pi/2L$ (cf.\ section \ref{old1.2.1}). 
It is convenient to choose the zero of energy so that 
$H_{0}|0\rangle$ vanishes. 

{\bf Notations}: The operator 
$\rho_{\sig}(x) = \psi^{\dg}_{\sig}(x)\psi^{}_{\sig}(x)$ for the 
density of electrons of spin $\sig$ at $x$ may be Fourier  
decomposed into plane wave components using Eq.\ (\ref{1.22}):
\beqa
\rho_{\sig}(x) = L^{-1}\sum_{k} \rho_{\sig}(k)\, 
e^{ikx}\, .
\label{2.1.2}
\eeqa
$\rho_{\sig}(k)$ is a density fluctuation operator, and measures 
the weight of the density wave with wave vector $k$ contributing 
to the electron density at $x$. It is a collective excitation 
involving the coherent superposition of a large number of 
electron-hole pairs: 
\beqa
\rho_{\sig}(k) = \sum_{k'} c^{\dg}_{k'\sig}
c^{}_{k'+k\sig} \, .
\label{2.1.3}
\eeqa
Density fluctuations $\rho_{\sig}(k)$ play a central role in 
bosonization. The bosons in bosonization are not quite the 
density fluctuations, however, but are chiral 
components of the fluctuations describing 
collective excitations
about the Fermi points at $+k_{F}$ and $-k_{F}$ separately.
Consequently, $\rho_{\sig}(k)$ is decomposed into  
right-moving ($r = +$) and left-moving ($r = -$) components
\beqa
\rho_{\sig}(k) &=& \sum_{r} \rho_{r\sig}(k)\, ,
\nonumber \\
\rho_{+ \sig}(k) &=& \sum_{\overline{k} > 0} 
c^{\dg}_{\overline{k}-\frac{k}{2}\sig}
c^{}_{\overline{k}+\frac{k}{2}\sig} \, , \nonumber \\
\rho_{- \sig}(k) &=& \sum_{\overline{k} \leq 0} 
c^{\dg}_{\overline{k}-\frac{k}{2}\sig}
c^{}_{\overline{k}+\frac{k}{2}\sig} \, ,  
\label{2.1.4}
\eeqa
where $\overline{k} = 2\pi n/L$ or $2\pi(n+1/2)/L$, $n$ an 
integer, as $k=2\pi m/L$ with $m$ even or odd, respectively. 
The decomposition of Eq.\ (\ref{2.1.4}) is not 
unique,\footnote{An alternative decomposition, designed to 
investigate short-wavelength fluctuations with 
wavevectors $|k| \approx 2k_{F}$, is used in section 
\ref{old2.2.2}, cf.\ Eqs.\ (\ref{B.1}).} and is chosen  
so as to maximize the domain of validity of Tomonaga's result, 
Eq.\ (\ref{tom}) below. 

When $k = 0$, $\rho_{\sig}(k)$ counts the number of electrons 
of spin $\sig$. It is conventional in bosonization 
to give the number operator components 
$\rho_{r\sig}(0)$ a separate notation, normal-ordered with 
respect to the non-interacting ground-state $|0\rangle$:
\beqa
N_{r\sig} = \rho_{r\sig}(0) - \langle 0| \rho_{r\sig}(0) 
| 0 \rangle \, .
\label{2.1.5}
\eeqa
It is convenient also to define `generalized' non-interacting 
ground-states $|\{N_{r\sig}\}\rangle 
= |N_{+\uparrow}, N_{+\downarrow}, N_{-\uparrow}, 
N_{-\downarrow} \rangle$. These states have all momentum states 
$|k\sig\rangle$ occupied for $0 \leq rk < 
k_{F} + 2\pi N_{r\sig}/L$. In this notation the  
true non-interacting
ground-state is just $|0\rangle \equiv |0,0,0,0\rangle$.  
For a fixed total number $N_{e}$ of electrons, 
$\sum_{r\sig}N_{r\sig}$ vanishes.
Note that normal-ordering as in Eq.\ (\ref{2.1.5}) is a 
non-trivial operation in the Luttinger model (cf.\ Appendix \ref{appa}), 
since the number of $r$-electrons is infinite. 
However, normal-ordering is always trivial in a realistic 
condensed matter system because the number of 
electrons is finite.

{\bf Interactions and the subspace ${\cal H}_{k_{0}}$}:
Consider adding to $H_{0}$ an interaction term $V$ as in 
Eq.\ (\ref{1.2.7}). The 
interaction may be Coulomb electron-electron repulsion, 
for example, but it is useful for the general application of the 
bosonization formalism to leave $V$ as unspecified as possible: 
It will become clear below that bosonization describes 
the ground-state and low-energy excited-state properties of 
any interacting system provided the 
interactions considered are not `too strong'. A measure of 
interaction strength appropriate for bosonization is the 
deformation of the interacting ground-state momentum 
distribution 
$\langle \psi_{0}|n_{\sig}(k)|\psi_{0}\rangle$ 
from its non-interacting step-function form $n^{0}_{\sig}(k)
= \langle 0|n_{\sig}(k)|0\rangle = \theta(k_{F} - |k|)$. 
($n_{\sig}(k) = c^{\dg}_{k\sig}c^{}_{k\sig}$, and 
$|\psi_{0}\rangle$ denotes the interacting ground-state.)
The deformation of the momentum distribution 
may be quantified by introducing 
a bandwidth cut-off $k_{0} \geq 0$ (S\'{o}lyom 1979), which limits 
the range of states about $\pm k_{F}$ which are affected when 
$V$ is turned on. A bandwidth cut-off $k_{0}$ 
defines a subspace ${\cal H}_{k_{0}}$ of the full  
Hilbert space as follows: ${\cal H}_{k_{0}}$ is spanned by those 
many-electron states in which the single-electron  
states $|k\sig\rangle$ with $|k| \leq k_{F} - k_{0}$ remain 
occupied, and those with $|k| \geq k_{F} + k_{0}$ remain empty.
For a given interaction term $V$, $k_{0}$ is chosen so that 
the ground-state $|\psi_{0}\rangle$ of the interacting 
hamiltonian 
$H = H_{0} + V$ belongs to the subspace ${\cal H}_{k_{0}}$. 
This does not define the bandwidth cut-off $k_{0}$ uniquely, 
but does provide a lower bound on possible choices for $k_{0}$. 

It will become clear in section \ref{old2.2.3} that bosonization 
provides a complete description for 
systems with weak interactions $V$, for which a  
bandwidth cut-off satisfying $k_{0} \ll k_{F}$ 
may be chosen. Bose representations for Fermi operators are 
defined on the subspace ${\cal H}_{k_{0}}$, and provide 
a simpler description of interacting 1D many-electron systems.

{\bf Linear dispersion approximation}:
For bosonization to provide a complete description 
of an interacting system, it is necessary to make   
an approximation related to, but going somewhat 
beyond, the restriction to the subspace ${\cal H}_{k_{0}}$:
The dispersion $\varepsilon (k)$ must be approximated 
by its expansion to first order in $|k| - k_{F}$. Thus
\beqa
\varepsilon(k) = \varepsilon(k_{F}) + v_{F}(|k|-k_{F}) 
\, , \label{2.1.6} 
\eeqa
where the Fermi velocity 
$v_{F} = d\varepsilon(k_{F})/dk = k_{F}/m_{e}$. 
Eq.\ (\ref{2.1.6}) will be a good approximation provided 
that $k_{0} \ll k_{F}$, so that only states close to 
$\pm k_{F}$ are affected when the interactions are turned on. 

The linear dispersion approximation 
is less valid in bosonization than it is in Landau 
Fermi liquid theory. In Landau's theory, the linear dispersion 
approximation (Baym and Pethick 1978) is made for the 
quasiparticle dispersion, and since quasiparticles only 
exist very 
close to the Fermi surface, the approximation is always a very 
good one. In bosonization the linear dispersion approximation 
is made on the bare electron dispersion. For any finite 
interaction strength, bare electrons will be excited from 
momentum states a finite distance from the non-interacting 
Fermi surface. An approximation of the energy of these 
excitations with a linear dispersion may therefore introduce 
a small but finite error. 

The corrections due to deviations from a linear dispersion are 
not entirely trivial and destroy some of the more 
attractive features of bosonization, in particular exact 
solvability in simple models (Haldane 1981) and 
absolute spin-charge separation (Matveenko and Brazovskii 1994).  
Beyond the loss of these features, which are not essential 
to bosonization, the effects of a non-linear dispersion are 
weak and relatively harmless (Haldane 1981, Voit 1994). 
For completeness, the changes due to a non-linear dispersion on the
Bose representations derived in section \ref{old2.3} will be noted as 
they arise.

\section{\label{old2.2}Two Theorems on the Subspace ${\cal H}_{k_{0}}$}

The 1D state space is highly restrictive, and is distinguished 
from its higher dimensional cousins by possessing low-energy 
particle-hole excitations out of the non-interacting 
ground-state $|0\rangle$ with momenta close 
to $0$ and $\pm 2k_{F}$ only: 
In higher dimensions, the low-energy excitations possess  
momenta through all intermediate values as well.
The special low-energy properties of the 1D state space, 
together with the degeneracy introduced by the linear 
dispersion approximation, give rise to two theorems which 
provide a rigorous basis for bosonization. In this 
section the two theorems are stated and proved, and the 
conditions under which they hold is discussed.

\subsection{\label{old2.2.1}Tomonaga's bosons} 

Tomonaga's 1950 paper (Tomonaga 1950) marks the beginning of 
systematic bosonization techniques in condensed matter. Although 
now almost half a century old, the paper remains highly 
relevant. Indeed the development of bosonization 
given in this chapter is very much in Tomonaga's spirit, 
and may be viewed as something like an updated version of 
his work: The development given here includes 
results translated from field theory, in particular the 
representation of Fermi operators in terms of bosons,
which were not discovered until around the early 1970's 
(Schotte and Schotte 1969, Mattis 1974, Luther and Peschel 1974). 
Tomonaga's central insight may be formulated as a theorem. \\ \\
{\bf Theorem}: On ${\cal H}_{k_{0}}$ 
the right- and left-moving density fluctuation components 
$\rho_{r\sig}(k)$ satisfy the bosonic algebra 
\beqa
[\rho_{r\sig}(k), \rho_{r'\sig '}(k')]
= \delta_{r,r'}\,  \delta_{k,-k'}\, \delta_{\sig,\sig '}\,
\frac{rkL}{2\pi}
\label{tom}
\eeqa
provided that $|k|, |k'| < 2(k_{F} - k_{0})/3$. \\ \\
The proof is straightforward beginning from the definitions  
Eqs.\ (\ref{2.1.4}) of the density fluctuation components 
$\rho_{r\sig}(k)$, but is somewhat tedious due to 
the many different cases to be considered. Tomonaga (195)
considers each case in detail, and it will 
suffice here to give a few representative examples.
The density fluctuation commutators yield a difference 
of sums over $\overline{k}$ of operator products. For example 
\beqa
[\rho_{+\sig}(k), \rho_{+\sig'}(k')] = 
\delta_{\sig, \sig'}\left\{
\sum_{\stackrel{\scriptstyle{\overline{k} 
> -k/2}}{\scriptstyle{\overline{k} > k'/2}}} 
- \sum_{\stackrel{\scriptstyle{\overline{k} 
> k/2}}{\scriptstyle{\overline{k} > -k'/2}}}
\right\}c^{\dg}_{{\overline k}-\frac{k}{2}-\frac{k'}{2} \sig}
c^{}_{{\overline k}+\frac{k}{2}+\frac{k'}{2} \sig '}\, .
\label{2.2.2}
\eeqa
Under the stated conditions, it is easily verified that 
the argument of $c^{\dg}$ falls in the range of 
occupied states in ${\cal H}_{k_{0}}$, or that the argument of 
$c^{}$ falls in the range of empty states, and so the 
commutator vanishes if the arguments 
of $c^{\dg}$ and $c$ are distinct.
For example, for $k \geq 0$ and 
$k' < -k$, Eq.\ (\ref{2.2.2}) reduces to 
\beqa
[\rho_{+\sig}(k), \rho_{+\sig'}(k')] = 
\delta_{\sig, \sig'}
\sum_{-k/2 < \overline{k} \leq -k'/2}
c^{\dg}_{{\overline k}-\frac{k}{2}-\frac{k'}{2} \sig}
c^{}_{{\overline k}+\frac{k}{2}+\frac{k'}{2} \sig '}\, .
\label{2.2.3}
\eeqa
Given the restrictions on $|k|$ and $|k'|$ in the statement 
of the theorem, the argument of $c^{\dg}$ in Eq.\ (\ref{2.2.3}) 
always falls in the range of states between 
$-(k_{F} - k_{0})$ and $k_{F} - k_{0}$, which are occupied 
for all states in the subspace ${\cal H}_{k_{0}}$. Consequently  
the commutator Eq.\ (\ref{2.2.3}) vanishes. Other cases are 
similar. The exception to this rule is when $k = -k'$ for 
$r = r'$, in which case the arguments of $c^{\dg}$ and $c^{}$ 
coincide: The operator product in Eq.\ (\ref{2.2.2}) is then a 
number operator, and counts the core states between $-k/2$ and 
$k/2$ which are occupied for all states in ${\cal H}_{k_{0}}$. 
This gives the non-zero component in Eq.\ (\ref{tom}).

Instead of $\rho_{r\sig}(k)$, it will be useful occasionally 
to consider canonical Bose operators. These take the form of 
weighted density fluctuation components:
\beqa
b_{k\sig} = \left\{
\begin{array}{lc}\sqrt{\frac{2\pi}{|k|L}}\,   
\rho_{+\sig}(k)\, ,
& 0 < k < 2(k_{F} - k_{0})/3\, , \\
\sqrt{\frac{2\pi}{|k|L}}\,   \rho_{-\sig}(k)\, ,
& -2(k_{F} - k_{0})/3 < k < 0\, . 
\end{array} \right. 
\label{2.2.4}
\eeqa
Note that the case $k=0$ is excluded. From Eq.\ (\ref{tom}), 
$b^{}_{k\sig}, b^{\dg}_{k'\sig'}$ satisfy canonical Bose 
commutation relations (cf.\ Eqs.\ (\ref{1.17})) on 
${\cal H}_{k_{0}}$:
\beqa
\left[b^{}_{k\sig}, b^{}_{k'\sig'}\right] & = &
\left[b^{\dg}_{k\sig}, b^{\dg}_{k'\sig'}\right] = 0 \, , 
\nonumber \\
\left[b^{}_{k\sig}, b^{\dg}_{k'\sig'}\right] & = &
\delta_{k, k'} \delta_{\sig, \sig'}\, .
\label{2.2.5} 
\eeqa
The boson destruction operators $b_{k\sig}$ destroy the 
`generalized' ground-states: 
\beqa
b_{k\sig}|\{N_{r\sig}\}\rangle = 0
\label{2.2.6} 
\eeqa
for all $k < 2(k_{F} - k_{0})/3$.

In the Luttinger model formulation of bosonization 
(cf. Appendix \ref{appa}), there is a result analogous to 
Eq.\ (\ref{tom}), Eq.\ (\ref{A.4}), which is proved in an 
entirely different manner. Indeed the way in which the non-zero 
commutators are obtained in the Luttinger model relies crucially 
on the existence of an infinite number of particles, and 
thus has no analogue in a realistic condensed matter system. 
The Luttinger model result analogous to Eq.\ (\ref{tom}) 
applies for density 
fluctuations $\rho_{r\sig}(k)$ with $|k|$ arbitrarily 
large. For the realistic system considered here, 
there is a finite limit on the wave vectors $|k|$ for  
$\rho_{r\sig}(k)$ to satisfy Eq.\ (\ref{tom}). This is a 
significant point, and the following section discusses the 
finite limit in some detail. 

\subsection{\label{old2.2.2}Wavelength limit for bosonic density fluctuations}

The proof of Eq.\ (\ref{tom}) gives the  
restriction $|k| < 2(k_{F} - k_{0})/3$ on the wave vectors 
for density fluctuations $\rho_{r\sig}(k)$ to be bosonic. The 
precise value $2(k_{F} - k_{0})/3$ is not of particular  
physical significance; it simply denotes the value for which the 
commutation relations Eq.\ (\ref{tom}) are rigorously obeyed. 
The rigorous result will be 
assumed for the development of the bosonization formalism, but in 
applications to specific systems it will become clear 
(cf.\ section \ref{old3.2.3}) that the 
fluctuations are approximately bosonic over a wider 
range. Indeed by retracing the proof of Eq.\ (\ref{tom}), it may 
be verified that fluctuations with slightly shorter wavelengths 
are bosonic to a very good approximation. 

The central question in applications to specific systems 
is then the following: Up to which wave vector 
do the density fluctuations $\rho_{r\sig}(k)$ satisfy bosonic  
commutation relations approximately? Equivalently, at which 
wavelength, if any, does the character of the density 
fluctuations change, so that longer wavelength fluctuations are 
at least qualitatively bosonic in character, but not shorter 
wavelength fluctuations? As already noted, in the Luttinger 
model the fluctuations are bosonic over all finite wavelengths. 
In a realistic system it turns out that this is false; 
fluctuations with wavevectors $\approx \pm 2k_{F}$ (i.e.\ with 
wavelengths of the order of the average inter-electron spacing) 
are certainly not bosonic. This is established in the next 
paragraph.

A lower limit on the wavelengths for bosonic density 
fluctuations, at the order of the inter-electron spacing, is 
plausible on basic physical grounds: 
The system is composed of fermions, and it is therefore  
expected that density fluctuations 
over the scale of the average inter-electron spacing would have 
single-electron characteristics, and not the bosonic character 
of the longer wavelength collective fluctuations. 
This conclusion may be 
established formally by two methods; either 
the decomposition of $\rho_{\sig}(k)$ as in Eqs.\ (\ref{2.1.4}) 
may be kept, or the decomposition may be altered to 
exhibit the $2k_{F}$ fluctuations explicitly.
In the first case, when $|k| \approx 2k_{F}$, 
the components $\rho_{r\sig}(k)$ describe excitations from one 
Fermi point to the other. Following the same method used to 
prove Eq.\ (\ref{tom}), it may be verified that the commutation 
relations are operator-valued, and have a complicated 
dependence on the electron momentum distribution 
arbitrarily close to the Fermi points. The commutators 
do not simplify to $c$ numbers, even for arbitrarily weak 
interactions. To further clarify this point, 
and in particular to exhibit explicitly the type 
of algebra that the $2k_{F}$ fluctuations do satisfy, 
it is convenient to generalize the 
decomposition Eqs.\ (\ref{2.1.4}) of $\rho_{\sig}(k)$ to include 
short-wavelength components $\rho^{s}_{r\sig}(k)$ as follows:
\beqa
\rho_{\sig}(k) &=& \sum_{r = \pm} \rho_{r\sig}(k) 
+ \rho^{s}_{\sig}(k) \, ,
\nonumber \\
\rho_{r\sig}(k) &=& \sum_{r\overline{k} > k_{F}/2}
c^{\dg}_{\overline{k} - \frac{k}{2} \sig}
c^{}_{\overline{k} + \frac{k}{2} \sig}  \, ,  
\nonumber \\
\rho^{s}_{\sig}(k) &=& \sum_{-k_{F}/2 \leq \overline{k} 
\leq k_{F}/2}
c^{\dg}_{\overline{k} - \frac{k}{2} \sig}
c^{}_{\overline{k} + \frac{k}{2} \sig}  \, .  
\label{B.1}
\eeqa
On the subspace ${\cal H}_{k_{0}}$ (with $k_{0} \ll k_{F}$), 
$\rho^{s}_{\sig}(k)$ describes density fluctuations with 
wavevectors $k \approx \pm 2k_{F}$. 
These correspond to the transfer of  
electrons from the vicinity of $k_{F}$ to the vicinity of 
$-k_{F}$ for $k \approx 2k_{F}$, and from $-k_{F}$ to $k_{F}$ 
for $k \approx -2k_{F}$. 
Corresponding to these cases, it is convenient to rewrite the 
short-wavelength density fluctuations as 
\beqa
\rho_{r2k_{F}\sig}(k_{r}) &=& \rho^{s}_{\sig}(r2k_{F} + k_{r})
\nonumber \\
&=& \sum_{-k_{F}/2 \leq \overline{k} \leq k_{F}/2}
c^{\dg}_{\overline{k} - rk_{F} - \frac{k_{r}}{2} \sig}
c^{}_{\overline{k} + rk_{F} + \frac{k_{r}}{2} \sig}  \, ,   
\label{B.2}
\eeqa
where the reduced wave vector $k_{r}$ is small. 
Proceeding much as in the 
proof of Tomonaga's result, Eq.\ (\ref{tom}), it is 
straightforward to 
verify that on ${\cal H}_{k_{0}}$, and for 
$|k|, |k'|, |k_{r}|, |k_{r}'| \ll k_{F}$, the following algebra 
is satisfied: 
\beqa
\left[\rho_{r\sig}(k), \rho_{r'\sig'}(k')\right] &=& 
\delta_{r,r'}\,\delta_{k,-k'}\,\delta_{\sig,\sig'}\, 
\frac{rkL}{2\pi}\, ,
\nonumber \\
\left[\rho_{r2k_{F}\sig}(k_{r}), 
\rho_{r'2k_{F}\sig'}(k_{r}')\right] &=& 
r \delta_{\sig,\sig'}(1-\delta_{r,r'})
\left\{\rho_{-\sig}(k_{r} + k_{r}') 
- \rho_{+\sig}(k_{r} + k_{r}')\right\}\, ,
\nonumber \\
\left[\rho_{r2k_{F}\sig}(k_{r}), \rho_{r'\sig'}(k')\right] &=& 
rr'\delta_{\sig,\sig'}\, 
\rho_{r2k_{F}\sig}(k_{r} + k') \, .
\label{B.3}
\eeqa
This algebra is bosonic only for the longer wavelength density 
fluctuations $\rho_{r\sig}(k)$. The shorter wavelength 
fluctuations $\rho_{r2k_{F}\sig}(k_{r})$, with wave vectors 
$\approx \pm 2k_{F}$, satisfy an operator-valued algebra 
on ${\cal H}_{k_{0}}$, which is qualitatively non-bosonic.

The above considerations make it clear that 
in realistic condensed matter systems there is 
always a qualitative change in the behaviour of the density 
fluctuations as the wavelength decreases. This change occurs at 
least at wavelengths of the order of the average inter-electron 
spacing, but more generally may occur at longer wavelengths, 
$\approx (k_{F} - k_{0})^{-1}$ for example, depending on the 
particular type of interaction. The inverse length 
$\alpha^{-1}$ will be used to denote the wave vector 
at which density fluctuations cease to exhibit a bosonic 
character. The proof of Eq.\ (\ref{tom}), and its extension to 
shorter wavelength fluctuations which are approximately 
bosonic, suggests that the demarcation 
at $k = \alpha^{-1}$ will not be a sharp divide, but will mark 
a gradual crossover. However, as the discussion of the 
$|k| \approx 2k_{F}$ fluctuations makes clear, there is always a 
short-wavelength regime in which density fluctuations are 
demonstrably non-bosonic, and $\alpha$ is a measure of the 
wavelength at which the crossover occurs. 
It has been shown that most generally $\alpha$ satisfies 
\beqa
2(k_{F} - k_{0})/3 \leq \alpha^{-1} \lesssim 2k_{F} \, .
\label{alpha}
\eeqa
The further specification of $\alpha$ depends on the particular 
system and interactions considered, and in general requires 
methods beyond those of bosonization: Any straightforward 
extension of the method of proof of Eq.\ (\ref{tom}), for 
example, quickly becomes intractable. Chapter \ref{ch3} contains 
discussions of how $\alpha$ is 
determined for several one-component systems by comparison 
with results by other methods. In chapter \ref{ch6}, $\alpha$ 
is determined for the Kondo lattice model 
by using the results of numerical simulations.

\subsection{\label{old2.2.3}Completeness of the Bose generated states} 

The second element required to give bosonization a 
rigorous foundation is to prove that the states generated 
by the bosons $b^{\dg}_{k\sig}$ span the entire subspace 
${\cal H}_{k_{0}}$. The completeness of the Bose generated 
states was first noted by Overhauser (1965), and 
was later proved by Schick (1968). 
Completeness relies on the degeneracy introduced by 
the linear dispersion approximation Eq.\ (\ref{2.1.6}), 
and as for Tomonaga's result above   
may be formulated as a theorem.   \\  \\
{\bf Theorem}: For non-negative integers 
$m_{k\sig}$, and `generalized' non-interacting ground-states 
$|\{N_{r\sig'}\}\rangle$, the orthonormal states
\beqa
|\Psi\{m_{k\sig},N_{r\sig'}\}\rangle =  
\prod_{0<|k|< \alpha^{-1}} \prod_{\sig}
\frac{1}{\sqrt{m_{k\sig}!}}
\left(b^{\dg}_{k\sig}\right)^{m_{k\sig}}|\{N_{r\sig'}\}\rangle
\label{2.2.7}
\eeqa
span ${\cal H}_{k_{0}}$, provided $k_{0} \ll k_{F}$. \\ \\
Before proceeding to a proof of this theorem, a note is in order
on the generation of the states 
$|\Psi\{m_{k\sig},N_{r\sig'}\}\rangle$.  
It is straightforward to verify that the 
operators $b^{\dg}_{k\sig}$ may 
map states in ${\cal H}_{k_{0}}$ to states outside, depending 
in a complicated fashion on the particular state in 
${\cal H}_{k_{0}}$ which is acted 
on. If this occurs, then the next operator $b^{\dg}_{k'\sig''}$ 
used in the construction of a particular state 
$|\Psi\{m_{k\sig},N_{r\sig'}\}\rangle$,
as in Eq.\ (\ref{2.2.7}), will be acting on a state outside 
${\cal H}_{k_{0}}$. In this case, there is no guarantee that 
$b^{\dg}_{k'\sig''}$ is bosonic: Tomonaga's result 
Eq.\ (\ref{tom}) holds only in ${\cal H}_{k_{0}}$. To prevent 
this, the operators $b^{\dg}_{k\sig}$ used to construct a state 
$|\Psi\{m_{k\sig},N_{r\sig'}\}\rangle$
are to be understood to be projected onto the subspace 
${\cal H}_{k_{0}}$. (See Prugove\v{c}ki (1981), for 
example, for details on the projection of linear operators 
onto Hilbert subspaces.) 

There are at least two distinct ways to prove the completeness 
theorem. The first is a direct counting of states argument. This 
method calculates the degeneracy of each energy level. The 
number of states within each energy level as generated by the 
Bose states Eq.\ (\ref{2.2.7}) is then shown to equal the number 
of states generated by the manifestly complete set of 
non-interacting electron eigenstates. 
This was the argument originally 
given by Schick (1968). (A more accessible version  
is given by Sch\"{o}nhammer and Meden (1996).)
The second method of proof follows Haldane's (1981)  
treatment for the Luttinger model, and uses the grand partition 
function to count the states. The second method of proof is used 
below as it is more elegant, and accounts automatically for 
systems in which the number of electrons may vary. The latter 
property is useful for the derivation of Bose representations 
in section \ref{old2.3}. Note, however, that the 
first method may better aid an intuitive understanding of the 
equivalence of the Bose and Fermi states in 
1D many-electron systems (Sch\"{o}nhammer and Meden 1996). 

To proceed with the proof, first note that on 
${\cal H}_{k_{0}}$, 
\beqa
[H_{0},b^{\dg}_{k\sig}] = v_{F}|k|b^{\dg}_{k\sig}
\label{2.2.8}
\eeqa
where $|k| < \alpha^{-1}$. The verification of Eq.\     
(\ref{2.2.8}) is straightforward, and is similar to the 
proof of Eq.\ (\ref{tom}). The appearance of the  
Fermi velocity $v_{F}$ in Eq.\ (\ref{2.2.8}) signals its 
dependence on the linear 
dispersion approximation of Eq.\ (\ref{2.1.6}); the 
commutator contains further terms as higher orders 
in $|k| - k_{F}$ are included in the dispersion 
relation. 

It follows from Eq.\ (\ref{2.2.8}) that the 
boson states $|\Psi\{m_{k\sig},N_{r\sig'}\}\rangle$ are 
eigenstates of $H_{0}$:
\beqa 
H_{0}|\Psi\{m_{k\sig},N_{r\sig'}\}\rangle = 
v_{F}\left\{\sum_{k, \sig}|k|m_{k\sig} + \sum_{r, \sig'} 
\pi N^{2}_{r\sig'}/L \right\} 
|\Psi\{m_{k\sig},N_{r\sig'}\}\rangle\, , 
\label{2.2.9}
\eeqa
with the second contribution being the non-interacting 
energy of the `generalized' ground-states 
$|\{N_{r\sig'}\}\rangle$. 
The grand partition function for the boson generated 
states $|\Psi\{m_{k\sig},N_{r\sig'}\}\rangle$ in the 
non-interacting system is then 
\beqa
Z_{b} &=& \sum_{m_{1\uparrow}=0}^{\infty}
      \sum_{m_{1\downarrow}=0}^{\infty}
      \sum_{m_{-1\uparrow}=0}^{\infty} \cdots 
      \sum_{m_{-n_{\rm max}\downarrow}=0}^{\infty}
      \sum_{N_{+\uparrow}=-\infty}^{\infty} \cdots
      \sum_{N_{-\downarrow}=-\infty}^{\infty}
\nonumber \\
  && \quad \times \exp \left( 
2u\sum_{n \neq 0}\sum_{\sig = \uparrow, \downarrow} 
|n|\, m_{n\sig} + u\sum_{r = \pm} 
\sum_{\sig' = \uparrow, \downarrow}N^{2}_{r\sig'} \right)
\label{2.2.10}
\eeqa
where $u = -\beta\pi v_{F}/L$, $\beta = 1/k_{B}T$, and where 
integers $n$ have been used to label the momentum states $k$ 
according to $k = 2\pi n/L$. $n_{\rm max}$ is the greatest 
integer $< L/2\pi \alpha$. $Z_{b}$ factorizes into separate 
components due to the bosons and to $N_{r\sig'}$. The first 
component further factorizes  
to give
\beqa
\prod_{n\neq 0}^{\pm n_{\rm max}}\prod_{\sig}\left( 
\sum_{m_{n\sig}=0}^{\infty}w^{2|n|m_{n\sig}}\right) 
&=& \prod_{n\neq 0}^{\pm n_{\rm max}}\prod_{\sig}\left( 
 1 - w^{2|n|}\right)^{-1}
\nonumber \\
&=& \left(\prod_{n=1}^{n_{\rm max}}
\left(1 - w^{2n}\right)^{-1}\right)^{4}\, ,
\label{2.2.11}
\eeqa
where $w = e^{u} = e^{-\beta\pi v_{F}/L}$ 
and where the first equation involves summing a 
geometric progression. The component due to $N_{r\sig'}$ is 
easily evaluated, and the total result is 
\beqa 
Z_{b} = \left(\prod_{n=1}^{n_{\rm max}}
\left(1 - w^{2n}\right)^{-1}\right)^{4}
\left(\sum_{n=-\infty}^{\infty}w^{(n^{2})}\right)^{4}\, . 
\label{2.2.12}
\eeqa
The calculation of the grand partition function using 
non-interacting electron states is standard (Haldane 1981, 
Huang 1987), and gives 
\beqa
Z_{f} = \left(\prod_{n=1}^{\infty}\left(1+w^{2n-1}\right)^{2}
 \right)^{4} 
\label{2.2.13}
\eeqa
at temperatures $\beta^{-1} \ll v_{F}k_{F}$. Noting the product 
and sum forms for the theta function of the third kind 
(Gradshteyn and Ryzhik 1965, Haldane 1981), 
\beqa
\vartheta_{3}(0) = \sum_{n = -\infty}^{\infty} w^{(n^{2})} 
= \prod_{n=1}^{\infty}\left(1+w^{2n-1}\right)^{2}
\left(1-w^{2n}\right) \, ,
\label{2.2.14} 
\eeqa
it is clear that $Z_{b} = Z_{f}$ provided the product in 
Eq.\ (\ref{2.2.12}) may be extended from $n_{\rm max}$ to 
$\infty$. This is valid provided the temperature satisfies 
$\beta^{-1} \ll v_{F}\alpha^{-1}$. Since both the Bose and the 
Fermi states generate the same energy levels (the eigenstates 
of $H_{0}$), and since the partition functions are 
sums of positive quantities, if any states were missing from the 
Bose generated states, then $Z_{b} < Z_{f}$ at certain 
temperatures. The equality $Z_{b} = Z_{f}$ for $\beta^{-1} 
\ll v_{F}\alpha^{-1}$ thus
establishes the completeness of the Bose generated states 
of Eq.\ (\ref{2.2.7}) for excitation energies 
$\ll v_{F}\alpha^{-1}$. 

In order to prove the completeness of the boson states within 
the subspace ${\cal H}_{k_{0}}$, note that it is necessary to 
establish 
that $Z_{b} = Z_{f}$ for excitation energies up to the maximum 
$\approx v_{F}k_{0}$ for states in ${\cal H}_{k_{0}}$.  
Completeness in ${\cal H}_{k_{0}}$ thus requires that 
$k_{0} \ll \alpha^{-1}$. Indeed a weaker version of this 
condition may be established by simple physical arguments: 
On ${\cal H}_{k_{0}}$ an electron constructed out 
of states near $rk_{F}$, $r = \pm$, is spread over a spatial 
range $\approx 1/k_{0}$. To describe this electron completely 
using bosonic density fluctuations, it is required that the 
fluctuations remain bosonic down to similar wavelengths. 
This gives $k_{0} \lesssim \alpha^{-1}$. 
The extra condition $k_{0} \ll \alpha^{-1}$ coming from the proof 
above is to render the Boltzmann weight $e^{-1/k_{0}\alpha}$ 
statistically negligible. This does not imply a vanishingly small 
$k_{0}$, as is clearly seen in the derivation of Sch\"{o}nhammer 
and Meden (1996). Thus bosonization contrasts 
with Landau's Fermi liquid theory, for example, in which the 
quasiparticle excitations rigorously exist only at the Fermi 
surface. To derive the condition $k_{0} \ll k_{F}$ given in the 
statement of the completeness theorem, 
note that $\alpha^{-1} \geq 2(k_{F} - k_{0})/3$ using Tomonaga's 
rigorous result Eq.\ (\ref{tom}) for $\alpha^{-1}$. Thus 
$k_{0} \ll \alpha^{-1}$ is assured provided $k_{0} \ll k_{F}$. 

For systems with interactions strong enough that 
$k_{0} \approx \alpha^{-1}$, the boson states Eq.\ (\ref{2.2.7}) 
may fail to generate all the states in ${\cal H}_{k_{0}}$. In 
this case bosonization by itself is insufficient to determine 
all the low-energy properties of the interacting system. 
However, provided that $\alpha^{-1}$  remains 
non-zero, the boson states 
Eq.\ (\ref{2.2.7}) will still generate the long-wavelength 
behaviour correctly, although they will fail to reproduce the 
short-wavelength properties. In this case,  
bosonization provides a {\it partial} description 
of the low-energy properties of 
the interacting system, and must in principle be supplemented 
with a description of the short-wavelength behaviour due to 
non-bosonic density fluctuations.

\section{\label{old2.3}Bose Representations}

The completeness of the Bose generated states in 
${\cal H}_{k_{0}}$ (for small enough $k_{0}$) 
guarantees the existence of Bose 
representations for Fermi operators within ${\cal H}_{k_{0}}$. 
The representations are 
very useful. They provide an alternative formulation for 
problems involving 1D electrons, which often 
makes the solution easier, and they simplify the 
calculation of physical quantities such as correlation 
functions. This section derives  
the Bose representations for 
Fermi operators which appear in the standard 
Luttinger model bosonization. The final forms are  
familiar, although their derivation 
here from a realistic system with a finite Fermi sea 
is new. The derivation makes 
explicit the limitations on the validity of the Bose 
representations, which is entirely obscured in the Luttinger 
model, and provides proof by construction that a careful 
treatment has no need of limiting procedures, or non-trivial 
normal-ordering conventions, which are essential elements in 
the Luttinger model bosonization. This provides additional
physical insight by avoiding manipulations which are 
meaningless in a realistic condensed matter system.  
The derivation makes manifest all of the elements of the 
Luttinger model representation which need reinterpretation in 
applications to condensed matter. These are discussed in  
detail in chapter \ref{ch3}. Some elements, in particular the Bose field 
commutators, have not previously been emphasized. 

The completeness theorem Eq.\ (\ref{2.2.7}) leads to a simple, 
though rigorous prescription for deriving Bose representations. 
By virtue of Tomonaga's result Eq.\ (\ref{tom}), the prescription 
is also relatively easy to implement. \\ \\
{\bf Prescription for Bose representations}: 
The Bose representation for an operator $O$ is an 
operator $O_{B}(\rho_{r\sig}(k))$ which satisfies 

(i) $[\rho_{r\sig}(k),O_{B}] =[\rho_{r\sig}(k),O]$ for 
all $\rho_{r\sig}(k)$ with $0 < |k| < \alpha^{-1}$, and 

(ii) $\langle \{N_{r\sig}\}|O_{B}|\{N_{r'\sig'}\}\rangle = 
\langle \{N_{r\sig}\}|O|\{N_{r'\sig'}\}\rangle$.   \\ \\
If these conditions are satisfied, then $O_{B}$ will 
reproduce the matrix elements of $O$ between all states 
in ${\cal H}_{k_{0}}$, for small enough $k_{0}$, and 
$O_{B} = O$ as an operator identity within the subspace 
${\cal H}_{k_{0}}$. 
This follows immediately from the completeness 
of the Bose generated states Eq.\ (\ref{2.2.7}) in 
${\cal H}_{k_{0}}$ for $k_{0} \ll k_{F}$.
The following sections derive Bose representations for 
$H_{0}$, for the Fermi fields $\psi_{\sig}(x)$, and 
for operators bilinear in the Fermi fields. 

\subsection{\label{old2.3.1}Non-interacting hamiltonian}

$H_{0}$ may be given a Bose representation beginning from 
the commutation relations Eq.\ (\ref{2.2.8}), and the 
non-interacting 
energy of the states $|\{N_{r\sig}\}\rangle$ as 
given in Eq.\ (\ref{2.2.9}).  
A form which reproduces these commutation relations, and  
which satisfies condition (ii) above, is  
\beqa
H_{0B} = \frac{v_{F}\pi}{L}\sum_{r, \sig}N^{2}_{r\sig} + 
v_{F}\sum_{0 < |k| < \alpha^{-1}}\sum_{\sig}
|k|b^{\dg}_{k\sig}b^{}_{k\sig} \, ,
\label{2.3.1}
\eeqa
as is straightforwardly verified.
As discussed in section \ref{old2.2.2}, the summation restriction 
$|k|< \alpha^{-1}$ is a necessary element in 
bosonization for realistic condensed matter systems, and acts to 
exclude shorter wavelength non-bosonic fluctuations.
It is convenient, however, to formally allow 
$k$ to range over all non-zero values. This both 
simplifies manipulations, and permits direct comparisons to be 
made with Luttinger model results. An unrestricted 
summation with identical physical content is achieved by 
attaching a weight $\Lambda_{\alpha}(k)$ to each density 
fluctuation component $\rho_{r\sig}(k)$, with 
$\Lambda_{\alpha}(k)$ an 
even function of $k$ and satisfying
\beqa
\Lambda_{\alpha}(k) \approx \left\{
\begin{array}{lc} 1 & {\rm when}\,\, |k| < \alpha^{-1}, \\
              0 & {\rm otherwise.} 
\end{array} \right. 
\label{2.3.2}
\eeqa
Summations over $\rho_{r\sig}(k)$ for $|k| < \alpha^{-1}$ are 
equivalent to unlimited summations with a step-function 
weight $\Lambda_{\alpha}(k) = \theta(\alpha^{-1} - |k|)$. 
In this case the Bose representation for $H_{0}$ reads 
\beqa
H_{0B} =  \frac{v_{F}\pi}{L}\sum_{r, \sig}N^{2}_{r\sig} +         
v_{F}\sum_{k \neq 0}\sum_{\sig}
|k|b^{\dg}_{k\sig}b^{}_{k\sig} \Lambda^{2}_{\alpha}(k)
\label{2.3.3} 
\eeqa
and $\Lambda_{\alpha}(k)$ acts as a cut-off function 
on bosonic density fluctuations. 

Eq.\ (\ref{2.3.3}) is rigorously correct when 
$\Lambda_{\alpha}(k)$ is the step-function cut-off; with 
this choice of cut-off function, 
$H_{0B}$ reproduces exactly the commutation relations 
Eq.\ (\ref{2.2.8}) required by condition (i) above. However, 
the discontinuity in the step-function cut-off implies a sharp 
limit on bosonic fluctuations. 
As discussed in section \ref{old2.2.2}, a sharp cut-off is 
unlikely. It is probable that $\alpha$ marks instead a 
gradual crossover in the properties of the density fluctuations. 
It is therefore useful to consider also some smooth 
(differentiable) cut-off functions. Smooth cut-off 
functions may be better suited to describing the expected gradual 
crossover to non-bosonic behaviour in the density fluctuations. 
As long as Eq.\ (\ref{2.3.2}) is satisfied, 
the choice of smoother cut-off functions 
will not affect the description, except for minor modifications 
over length scales near the bosonic limit $\alpha$. 
To investigate 
in detail the effects of using different cut-off functions, three 
representative choices will be considered at various stages:
\beqa
{\rm Step-function:}\,\quad \Lambda_{\alpha}(k) 
   & = & \theta(\alpha^{-1} - |k|) \, ,\nonumber \\
{\rm Gaussian:} \,\quad \Lambda_{\alpha}(k) 
   & = &  e^{-\alpha^{2} k^{2}/2}\, , \nonumber \\
{\rm Exponential:} \,\quad \Lambda_{\alpha}(k) 
   & = & e^{-\alpha |k|/2} \, . \label{2.3.4} 
\eeqa
These choices all satisfy Eq.\ (\ref{2.3.2}); 
the step-function cut-off exactly, and the Gaussian and 
exponential cut-offs in the approximate sense. Choices other 
than those in Eq.\ (\ref{2.3.4}) are obviously possible.

{\bf Representation in terms of Bose fields}: 
It is conventional to write Bose representations in 
terms of Bose fields relating to density and current 
excitations (Haldane 1981, Voit 1994). Define charge and spin density 
fluctuation components 
\beqa
\rho_{r}(k)  =  \sum_{\sig}\rho_{r\sig}(k) \, ,
\quad \quad \quad
\sig_{r}(k)  =  \sum_{\sig}\sig\rho_{r\sig}(k)\, , 
\label{2.3.5}
\eeqa   
where $\sig = +1, -1$ corresponding to $\sig = \uparrow, 
\downarrow$, respectively.
These are the right- and left-moving components of the 
operators which appear in a Fourier 
expansion (cf.\ Eq.\ (\ref{2.1.2})) of the charge and spin 
density at $x$. Similarly define charge and spin components 
for the number operators: 
$N^{\rho}_{r} = \sum_{\sig}N_{r\sig}$ and $N^{\sig}_{r} = 
\sum_{\sig}\sig N_{r\sig}$.
Now define hermitian Bose fields  
\beqa
\phi_{\nu}(x) & = &  \frac{\pi x}{L}(N^{\nu}_{+} + N^{\nu}_{-}) 
- i\sum_{k \neq 0} \frac{\pi}{kL}
[\nu_{+}(k) + \nu_{-}(k)]\Lambda_{\alpha}(k)e^{ikx} \, ,
\nonumber \\ 
\theta_{\nu}(x) & = & \frac{\pi x}{L}(N^{\nu}_{+} - N^{\nu}_{-}) 
- i\sum_{k \neq 0} \frac{\pi}{kL}
[\nu_{+}(k) - \nu_{-}(k)]\Lambda_{\alpha}(k)e^{ikx} \, ,
\label{2.3.6}
\eeqa
where $\nu = \rho, \sig$ labels charge and spin. The field 
$\phi$ relates to number excitations, and the field 
$\theta$ to current excitations. 
The physical significance of the Bose fields is as potentials. 
$(1/\pi)\partial_{x}\phi_{\rho}(x)$ gives the 
excitation charge density at $x$, $\sum_{r\sig} 
\rho_{r\sig}(x) - N_{e}/L$. Similarly, 
$(1/\pi)\partial_{x}\phi_{\sig}(x)$ gives 
the spin density at $x$. For the current fields, 
$(1/\pi)\partial_{x}\theta_{\rho}(x)$ gives the average 
charge current density at $x$, $\sum_{r\sig}
r\rho_{r\sig}(x)$; while  
$(1/\pi)\partial_{x}\phi_{\theta}(x)$ gives the average 
spin current density at $x$. 
A third Bose field relating 
to momentum is defined on the basis of the currents:
$\Pi_{\nu}(x) =  \partial_{x}\theta_{\nu}(x)$.
Substitution Eqs.\ (\ref{2.3.6}) in Eq.\ (\ref{2.3.3}) gives 
\beqa
H_{0B} = \frac{v_{F}}{4\pi}\sum_{\nu}\int_{L}dx \left\{
\Pi^{2}_{\nu}(x) + [\partial_{x}\phi_{\nu}(x)]^{2}
\right\} \, .
\label{2.3.7}
\eeqa
For a non-linear dispersion, recall that the commutator 
Eq.\ (\ref{2.2.8}) has corrections. In this case $H_{0B}$  
contains further terms representing interactions between the 
bosonic fluctuations (Haldane 1981). The further terms are 
weak and relatively harmless; they disappear in a 
renormalization group transformation, although they may 
alter the velocities of the elementary excitations in 
interacting systems. See section \ref{old3.2.3} for an example in 
the 1D Hubbard model. 

\subsection{\label{old2.3.2}Fermi field operators}

On ${\cal H}_{k_{0}}$, the real space electron 
destruction operator may be 
decomposed into right- and left-moving components 
$\psi_{\sig}(x) = \sum_{r}\psi_{r\sig}(x)$ where, following 
Eq.\ (\ref{1.22}),
\beqa
\psi_{r\sig}(x) = L^{-1/2}\sum_{k_{F}-k_{0}  
< rk < k_{F}+k_{0}} c^{}_{k\sig}\, 
e^{ikx}  \, .
\label{2.3.8} 
\eeqa
Due to the finite Fermi sea, the components 
$\psi_{r\sig}(x)$ do not satisfy simple commutation 
relations with the bosonic density fluctuations:
\beqa
[\rho_{r\sig}(k), \psi_{r'\sig'}(x)] = 
-\delta_{r,r'}\delta_{\sig,\sig'}e^{-ikx}L^{-1/2}
\sum_{k_{F}-k_{0}+rk < rk' < 
k_{F}+k_{0} +rk}c^{}_{k'\sig}\,e^{ik'x}.
\label{2.3.9}
\eeqa
The commutation relations simplify (to Luttinger 
model form, cf.\ Eq.\ (\ref{A.5})) 
only in the asymptotic limit of 
very long-wavelength fluctuations with respect to $k_{0}$:
\beqa
[\rho_{r\sig}(k), \psi_{r'\sig'}(x)] = 
-\delta_{r,r'}\delta_{\sig,\sig'}e^{-ikx}
\psi_{r\sig}(x) \,\,\,\, {\rm for }\,\,  k/k_{0} \rightarrow 0.
\label{2.3.10}
\eeqa
Any Bose representation derived on the basis 
of this commutation relation will be asymptotically 
exact, but will not have the same status as, for 
example, $H_{0B}$, which holds exactly within the 
subspace ${\cal H}_{k_{0}}$.  An asymptotic Bose 
representation for $\psi_{r\sig}(x)$ can be derived 
from Eq.\ (\ref{2.3.10}) in the same manner as in the 
Luttinger model (Haldane 1981). 
Define the ladder operator 
\beqa
U^{\dg}_{r\sig} = L^{-1/2} \int_{L}dx\,\, e^{-irk_{F}x}
e^{-i\Phi^{\dg}_{r\sig}(x)}\psi_{r\sig}(x)
e^{-i\Phi_{r\sig}(x)}\, ,
\nonumber \\
\Phi_{r\sig}(x) = r\frac{\pi x}{L} N_{r\sig} 
- ir\sum_{rk > 0}\frac{2\pi}{kL}\rho_{r\sig}(k) 
\Lambda_{\alpha}(k)e^{ikx}\, . 
\label{2.3.11}
\eeqa
Using Eq.\ (\ref{2.2.6}), the action of the ladder 
operator on `generalized' ground-states is given by
\beqa
U^{\dg}_{r\sig}|\{N_{r'\sig'}\}\rangle 
= L^{-1/2} \int_{L}dx\,\, e^{-irk_{F}x}
e^{-i\Phi^{\dg}_{r\sig}(x)}\psi_{r\sig}(x)
e^{-ir\pi N_{r\sig}x/L}|\{N_{r'\sig'}\}\rangle\, .
\label{2.3.11a}
\eeqa
Using Eq. (\ref{2.3.8}) for the 
electron field operator $\psi_{r\sig}(x)$, 
and using 
\beqa
e^{-iyN_{r\sig}} \psi_{r\sig}(x) = \psi_{r\sig}(x) 
e^{-iy(N_{r\sig}-1)} 
\label{2.3.11aa}
\eeqa
for real numbers $y$, Eq.\ (\ref{2.3.11a}) 
may be written
\beqa
U^{\dg}_{r\sig}|\{N_{r'\sig'}\}\rangle &=& 
L^{-1}\int_{L}dx
\left\{\prod_{rk < 0}\left(1 + \sum_{n = 1}^{\infty}
(-r2\pi\rho_{r\sig}(k)\Lambda_{\alpha}(k)/kL)^{n}e^{inkx}/n!
\right)\right\}
\nonumber \\
& \times & \sum_{k_{F} - k_{0} < rk < k_{F} + k_{0}}
c^{}_{k\sig}\, e^{i[k - r(k_{F} + \pi(2N_{r\sig}-1)/L)]x}
|\{N_{r'\sig'}\}\rangle
\label{2.3.11b}
\eeqa
where the product comes from expanding an  exponential. 
It straightforward 
to verify that the unit term is the only element from the 
product which gives a non-zero contribution. 
Using Eq.\ (\ref{1.23}), Eq.\ (\ref{2.3.11b}) 
reduces to 
\beqa
U^{\dg}_{r\sig}
|N_{r\sig},\{N_{{\overline r} {\overline {\sig}}} \} \rangle  &=& 
\sum_{k_{F}-k_{0} < rk < k_{F} + k_{0}} c^{}_{k\sig}\, 
\delta_{k, r(k_{F} + \pi (2N_{r\sig} - 1)/L)}
|N_{r\sig},\{N_{{\overline r} {\overline {\sig}}} \} \rangle 
\nonumber \\
&=& |N_{r\sig}-1,\{N_{{\overline r} 
{\overline {\sig}}} \} \rangle  
\label{2.3.12} 
\eeqa
where ${\overline r}{\overline {\sig}}$ label
the indices other than $r\sig$. (Note that the derivation of 
the Luttinger model analogue of Eq.\ (\ref{2.3.12}) in 
Haldane (1981) is erroneous, although the conclusion 
is correct.) Using the asymptotic commutation 
relation Eq.\ (\ref{2.3.10}), it is 
straightforward to verify that the ladder operators 
commute with asymptotically long-wavelength 
density fluctuations: $[\rho_{r\sig}(k), 
U^{\dg}_{r'\sig'}] = 0$ for $k/k_{0} \rightarrow 0$ 
but $k \neq 0$. If Eq.\ (\ref{2.3.11}) is now inverted 
to solve for $\psi_{r\sig}(x)$, there results 
the asymptotic Bose representation   
\beqa
\psi_{r\sig B}(x) & = & {\cal N}(\alpha)e^{ir(k_{F} + \pi/L)x} 
e^{i\Psi_{r\sig}(x)}U^{\dg}_{r\sig} \, ,
\nonumber \\
\Psi_{r\sig}(x) & = & r\frac{2\pi xN_{r\sig}}{L} 
-ir\sum_{k \neq 0} \frac{2\pi}{kL}\rho_{r\sig}(k)
\Lambda_{\alpha}(k)e^{ikx} \nonumber \\
& = & \{\theta_{\rho}(x) + r\phi_{\rho}(x)
+ \sig[\theta_{\sig}(x) + r\phi_{\sig}(x)]\}/2  \, ,
\label{2.3.13}
\eeqa 
where the asymptotic normalization factor 
\beqa
{\cal N}(\alpha) = L^{-1/2}\exp\left\{\sum_{k > 0} 
\frac{\pi}{kL}\Lambda^{2}_{\alpha}(k)\right\} \, .
\label{2.3.14}
\eeqa
The summation in Eq.\ (\ref{2.3.14}) diverges to $+\infty$ 
in the thermodynamic limit ($L \rightarrow \infty$ with 
$N_{e}/L$ constant) for all choices of 
$\Lambda_{\alpha}(k)$ satisfying Eq.\ (\ref{2.3.2}). 
However, since ${\cal N}(\alpha)$ contains a prefactor which   
goes to zero in this limit, it is necessary to carry 
out the summation {\it before} taking the thermodynamic 
limit in order to treat the entire term consistently. 
For the step-function cut-off the summation may be evaluated 
analytically (Gradshteyn and Ryzhik 1965), and for 
finite $L$ gives 
\beqa
{\cal N}(\alpha) = \sqrt{\gamma/2\pi\alpha} + {\cal O}(1/L)
\label{2.3.15}
\eeqa
where $\gamma = 1.781\cdots$ is Euler's constant. Since the 
main contribution to the summation in Eq.\ (\ref{2.3.14}) 
comes from the region of small $k$, where 
$\Lambda_{\alpha}(k)$ is always unity for cut-offs satisfying 
Eq.\ (\ref{2.3.2}) (cf.\ also Eqs.\ (\ref{2.3.4})), 
it is clear that ${\cal N}(\alpha)$ 
will be bounded as $L \rightarrow \infty$ 
for all permitted choices of cut-off function. In fact from 
the closed form result for the step-function cut-off, together 
with corrections introduced as smoother cut-offs satisfy 
Eq.\ (\ref{2.3.2}) only approximately, it is easily  
established that in the thermodynamic limit, and for all 
permitted cut-off functions, the asymptotic normalization 
\beqa
{\cal N}(\alpha) \propto 1/\sqrt{\alpha}
\label{2.3.16}
\eeqa
with the constant of proportionality  
depending on the particular form of $\Lambda_{\alpha}(k)$. 

The Bose representation for $\psi_{r\sig}(x)$ has been derived 
by inverting Eq.\ (\ref{2.3.11}) for the ladder operator 
$U^{\dg}_{r\sig}$, and follows a similar derivation in the 
Luttinger model (Haldane 1981). With respect to the two 
conditions given at the beginning of this 
section on the properties to be satisfied by a Bose 
representation, it is easily verified that 
$\psi_{\sig B}(x) = \sum_{r}\psi_{r\sig B}(x)$ 
reproduces correctly the matrix elements between the 
states $|\{N_{r\sig}\}\rangle$, and so condition 
(ii) is satisfied. However, $\psi_{r\sig B}(x)$ 
only satisfies the correct commutation 
relations with asymptotically long-wavelength density 
fluctuations $\rho_{r\sig}(k)$ with $k/k_{0} 
\rightarrow 0$, and so condition (i) is only partially 
satisfied: $\psi_{\sig B}(x)$ will reproduce correctly the 
effects of destroying an electron at $x$ only over separations 
which are asymptotically large with respect to $1/k_{0}$. This 
contrasts with the Luttinger model result, Eq.\ 
(\ref{A.7}), in which 
the analogous Bose representation is exact. 

Since Bose representations preserve hermitian conjugation, the 
asymptotic Bose representation for the creation operator 
$\psi^{\dg}_{\sig}(x)$ is just the sum over $r$ 
of components conjugate to $\psi_{r\sig B}(x)$:
\beqa
\psi^{\dg}_{r\sig B}(x) = {\cal N}(\alpha)e^{-ir(k_{F}-\pi/L)x} 
e^{-i\Psi_{r\sig}(x)}U_{r\sig}\, ,
\label{2.3.17}
\eeqa 
where Eq.\ (\ref{2.3.11aa}) has been used, and where 
$U_{r\sig}$ adds an electron of spin $\sig$ to a 
state near $rk_{F}$, analogous to Eq.\ (\ref{2.3.12}).  

The effects of a non-linear dispersion on the Bose 
representation for Fermi field operators is to add higher 
harmonics in $k_{F}$; with a non-linear 
dispersion, the representations Eqs.\ (\ref{2.3.13}) and 
(\ref{2.3.17}) contain weaker higher harmonics at $3k_{F}$, 
$5k_{F}$, and so on.

\subsection{\label{old2.3.3}Operators bilinear in the Fermi fields}

In the Luttinger model the Bose representations for the 
Fermi fields is exact, and the representations for 
operators bilinear in the Fermi fields follows directly 
from the single Fermi field Bose 
representation (cf.\ Appendix \ref{appa}).
For the realistic system the Bose representation for the 
single Fermi fields is only asymptotically valid, and the 
representations for Fermi bilinears must be derived on a 
case by case basis, according to the prescription given at 
the beginning of this section. 

On the subspace ${\cal H}_{k_{0}}$, 
real space Fermi operator pairs may be written
\beqa
\psi^{\dg}_{\sig}(x)\psi_{\sig'}(x') = 
\sum_{r,r'}\psi^{\dg}_{r\sig}(x)\psi^{}_{r'\sig'}(x') 
+ \delta_{\sig,\sig'}{\rm const.}
\label{2.3.18}
\eeqa
where the right- and left-moving components are defined 
in Eq.\ (\ref{2.3.8}), and where const. depends on 
$x-x'$ and on $k_{0}$.
For asymptotically long-wavelength density fluctuations, 
the component pairs 
$\psi^{\dg}_{r\sig}(x)\psi^{}_{r'\sig'}(x')$ satisfy 
simple commutation relations which are the product form 
of Eq.\ (\ref{2.3.10}):
\beqa
[\rho_{r\sig}(k), 
\psi^{\dg}_{r'\sig'}(x)\psi^{}_{r''\sig''}(x')] = 
(\delta_{r,r'}\delta_{\sig,\sig'}e^{-ikx} 
-\delta_{r,r''}\delta_{\sig,\sig''}e^{-ikx'}) 
\psi^{\dg}_{r'\sig'}(x)\psi^{}_{r''\sig''}(x') \nonumber \\
\label{2.3.19}
\eeqa
provided $k/k_{0} \rightarrow 0$. It follows immediately 
that the {\it asymptotic} Bose representation for 
$\psi^{\dg}_{r\sig}(x)\psi^{}_{r'\sig'}(x')$ will have 
the same functional form in the density fluctuations 
as the product 
$\psi^{\dg}_{r\sig B}(x)\psi^{}_{r'\sig' B}(x')$ of the 
single Fermi field representations derived in section \ref{old2.3.2}. 
The simple product form will reproduce the 
commutation relations Eq.\ (\ref{2.3.19}), and condition 
(i) is asymptotically satisfied. It remains to satisfy 
condition (ii); to check that the correct matrix elements 
between the states $|\{N_{r\sig}\}\rangle$ are 
generated by the product form with the normalization 
factor ${\cal N}^{2}(\alpha)$ coming from the single Fermi 
field Bose representations. 

{\bf Off-diagonal bilinears}: For off-diagonal bilinears, 
$\psi^{\dg}_{r\sig B}(x)\psi^{}_{r'\sig' B}(x')$ in which 
$r \neq r'$ and/or $\sig \neq \sig'$, 
it is easily verified that the correct matrix 
elements are generated with the same normalization 
as in the single Fermi field case. Thus, 
\beqa
\psi^{\dg}_{r\sig}(x)\psi^{}_{r'\sig'}(x') = 
\psi^{\dg}_{r\sig B}(x)\psi^{}_{r'\sig' B}(x')
\label{2.3.20}
\eeqa
provided $r \neq r'$ and/or $\sig \neq \sig'$. The Bose 
representation Eq.\ (\ref{2.3.20}) will reproduce 
correctly the effects of a backscattering interaction, 
or a spin-flip interaction, but only over separations 
which are asymptotically large with respect to $1/k_{0}$.

{\bf Asymptotic diagonal bilinears}: 
For diagonal bilinears $\psi^{\dg}_{r\sig}(x)\psi^{}_{r\sig}(0)$, 
in which $r = r'$ and $\sig = \sig'$, the situation 
is complicated by the existence of a finite core of 
occupied states, and the normalization taken 
over from the single Fermi field representations is not 
sufficient. To determine the correct normalization in this 
case, note first that for diagonal bilinears the ladder 
operators cancel out. It is then convenient to write the  
diagonal bilinear representation in the form
\beqa
\psi^{\dg}_{r\sig}(x)\psi^{}_{r\sig}(0)_{B} = 
{\overline {\cal N}}^{2}_{rx}(\alpha)
e^{-irk_{F}x}e^{-i\Psi_{r\sig}(x)}e^{i\Psi_{r\sig}(0)}\, .
\label{2.3.21}
\eeqa
(It is assumed that $x \neq 0$ for reasons to be made clear
shortly.)
Eq.\ (\ref{2.3.21}) preserves the functional form of the bosonic 
fluctuations in 
$\psi^{\dg}_{r\sig B}(x)\psi^{}_{r\sig B}(0)$, and thus 
satisfies the correct asymptotic commutation relations. 
Additional functional dependences, which are 
required in the diagonal case, are included in a 
generalized normalization factor 
${\overline {\cal N}}^{2}_{rx}(\alpha)$. In accordance    
with condition (ii), this factor must be chosen so that 
$\sum_{r}\psi^{\dg}_{r\sig}(x)\psi^{}_{r\sig}(0)_{B}$ 
reproduces correctly the expectation values
\beqa
\langle\{N_{r\sig}\}|\psi^{\dg}_{\sig}(x) 
\psi^{}_{\sig}(0)|\{N_{r\sig}\}\rangle = 
\frac{i}{2\pi x}\left(e^{-i(k_{F}+2\pi N_{+\sig}/L)x} 
- e^{i(k_{F}+2\pi N_{-\sig}/L)x}\right).
\label{2.3.22}
\eeqa
Some computation gives 
\beqa
\langle\{N_{r\sig}\}|\psi^{\dg}_{r\sig}(x) 
\psi^{}_{r\sig}(0)_{B}|\{N_{r\sig}\}\rangle =& 
\nonumber \\ 
\quad {\overline {\cal N}}^{2}_{rx}(\alpha)&
e^{-ir(k_{F} + 2\pi N_{r\sig}/L)x}\, \,   
\exp\left\{\int_{0}^{\infty}\frac{e^{irkx}-1}{k}
\Lambda_{\alpha}^{2}(k)\, dk\right\}
\nonumber \\
\label{2.3.23}
\eeqa
and the expectation values Eq.\ (\ref{2.3.22}) will 
be correctly reproduced with the choice 
\beqa
{\overline {\cal N}}^{2}_{rx}(\alpha) = 
\frac{ir}{2\pi x}
\exp\left\{\int_{0}^{\infty}\frac{1-e^{irkx}}{k}
\Lambda_{\alpha}^{2}(k)\, dk \right\}.
\label{2.3.24}
\eeqa
The integral may be evaluated in closed form for the 
exponential and step-function cut-offs. For the step-function 
cut-off the result is (Lebedev 1965)
\beqa
{\overline {\cal N}}^{2}_{rx}(\alpha) 
= \frac{\gamma}{2\pi\alpha} 
e^{-{\rm Ei}(irx/\alpha)}  \, ,
\label{2.3.25}
\eeqa
where the exponential integral 
\beqa
{\rm Ei}(z) = \int_{-\infty}^{z}\frac{e^{t}}{t}\, dt \,  . 
\label{2.3.26} 
\eeqa 
For the exponential cut-off function, 
\beqa
{\overline {\cal N}}^{2}_{rx}(\alpha) = 
\frac{ir\,  {\rm sign}(x)}{2\pi\alpha} 
e^{-ir\tan^{-1}(x/\alpha)}\sqrt{1 + (\alpha/x)^{2}} \, ,
\label{2.3.27}
\eeqa
using Dwight (1961), p.\ 235. 

In the Luttinger model, the diagonal bilinear Bose 
representations take the same normalization factor as the 
off-diagonal product representations. The same is true here, 
but only in a limiting sense. To see this, note that 
the bilinear Fermi fields 
$\psi^{\dg}_{r\sig}(x)\psi^{}_{r\sig}(0)$ which are 
diagonal in $r$ and $\sig$ are related directly to density 
fluctuation components $\rho_{r\sig}(k)$ in a manner 
distinct from the off-diagonal bilinears: The diagonal 
bilinear products destroy an $r$-moving electron of 
spin $\sig$ at $0$, and create it again at $x$. It follows 
immediately that an accurate description of this process will 
require density fluctuations $\rho_{r\sig}(k)$ with wavelengths 
$1/k \approx x$. The Bose representation Eq.\ (\ref{2.3.21}) 
is valid only for asymptotically long-wavelength density 
fluctuations $\rho_{r\sig}(k)$ with $k/k_{0} \rightarrow 0$, 
and will thus provide an accurate description only if 
$k_{0}x \rightarrow \infty$. In other words, the separation
$x$ appearing in the diagonal Fermi bilinear 
$\psi^{\dg}_{r\sig}(x)\psi^{}_{r\sig}(0)$ 
is identical to the asymptotic separations  
over which the Bose representation Eq.\ (\ref{2.3.21}) 
is valid. It is important to emphasize that no 
such identification is possible for the off-diagonal Fermi 
bilinears. Indeed the backscattering 
and spin-flip interactions described by the off-diagonal 
Fermi bilinears are generally short-range and often taken to 
be on-site: $x \rightarrow 0$. This has no relation to 
the asymptotic separations which are 
well-described by the off-diagonal Bose representation Eq.\ 
(\ref{2.3.20}). The asymptotic separation in this case is 
instead that of standard scattering theory, in which the 
scattered particle is treated as free of the scattering 
potential at large distances:  
The off-diagonal Bose representation describes well the 
properties of a backscattered or spin-flipped electron, but 
only at large distances from the scattering centre. 

Since $\psi^{\dg}_{r\sig}(x)\psi^{}_{r\sig}(0)_{B}$ 
of Eq.\ (\ref{2.3.21}) is accurate only for $k_{0}x 
\rightarrow \infty$, it is useful to perform an 
asymptotic expansion for the 
normalization constants ${\overline {\cal N}}^{2}_{rx}(\alpha)$ 
using the small parameter $1/k_{0}x$. For weak interactions, 
for which $\alpha < 1/k_{0}$, $\alpha/x \ll 1$ is the 
appropriate expansion parameter. To second order this gives 
\beqa
{\overline {\cal N}}^{2}_{rx}(\alpha) & = & 
\frac{\gamma}{2\pi\alpha}\left(1 + 
\frac{ir\alpha e^{irx/\alpha}}{x}\right)\,\,\,\, \quad
{\rm Step-function}, 
\nonumber  \\
{\overline {\cal N}}^{2}_{rx}(\alpha) & = & 
\frac{1}{2\pi\alpha}\left(1 + 
\frac{ir\alpha}{x}\right)\,\,\,\,  \quad \quad
{\rm Exponential}. 
\label{2.3.28}  
\eeqa
The general form for the expansions, valid for all 
permitted choices of cut-off function, is 
\beqa
{\overline {\cal N}}^{2}_{rx}(\alpha) \propto \frac{1}{\alpha}
[ 1 + {\cal O}(\alpha/x)].
\label{2.3.29}
\eeqa
This follows from the closed form result for the step-function 
cut-off, together with small corrections as other 
choices of cut-off function satisfy Eq.\ (\ref{2.3.2}) in the 
approximate
sense. The constant of proportionality depends on the 
particular form for $\Lambda_{\alpha}(k)$; it is $\gamma/
2\pi$ for the step-function cut-off, and $1/2\pi$ for the 
exponential cut-off function. These constants coincide with 
those for the off-diagonal Fermi product representations 
provided corrections of ${\cal O}(\alpha/x)$ are negligible. 
The Luttinger model achieves this for all non-zero $x$ with 
the unphysical 
choice (cf.\ Appendix \ref{appa}) of $\alpha \rightarrow 0$. 
Of course the Luttinger model then has to contend with a 
divergent leading term, which is also unphysical.

{\bf On-site diagonal bilinears}: 
The derivation of the required Bose representations for Fermi 
bilinears is almost complete. There remains one case not 
covered above. That is, the representation of diagonal 
Fermi bilinears over zero separation, or density 
operators $\rho_{r\sig}(x) = \psi^{\dg}_{r\sig}(x)
\psi^{}_{r\sig}(x)$. These already have a Bose representation 
in their Fourier expansion (cf.\ Eq.\ (\ref{2.1.2})):
\beqa
\rho_{r\sig}(x)  =  L^{-1}\sum_{k} \rho_{r\sig}(k)
\Lambda_{\alpha}(k)\, e^{ikx}\, .
\label{2.3.30}
\eeqa
As for $H_{0B}$, this Bose representation applies 
as an operator identity on the entire  
subspace ${\cal H}_{k_{0}}$, and is distinguished from the 
other representations that apply only asymptotically.
Note that the commutation relations of the density operators 
with the density fluctuations are nothing like the 
operator-valued asymptotic commutation relations 
Eq.\ (\ref{2.3.19}). 
Instead, c-number commutation relations are satisfied:
\beqa
[\rho_{r\sig}(k),\rho_{r\sig}(x)] = \frac{rk}{2\pi}e^{-ikx}
\label{2.3.31}
\eeqa
for $|k| < \alpha^{-1}$ and using the step-function cut-off.
The difference between the commutation relations for the 
asymptotic and on-site diagonal bilinears is the formal 
manifestation of the physical arguments given above. There is no 
analogous difference in the commutation relations for  
off-diagonal products as the separation $x$ is varied. 
Note that the Luttinger 
model generates a similar difference between the Bose 
representation for $\psi^{\dg}_{r\sig}(x) \psi^{}_{r\sig}(0)$ 
when $x \neq 0$, to that when $x = 0$. This is achieved 
through a non-trivial normal ordering convention, and a 
prescription for the correct taking of limits (Haldane 1981)
(see also Appendix \ref{appa}). 
These manipulations, which appear bizarre from a condensed 
matter perspective, are necessary in the Luttinger model, 
since the representation must interpolate smoothly between 
$x = 0$ density operators, which satisfy c-number commutation 
relations, and $x \neq 0$ diagonal bilinears which satisfy 
the operator-valued commutation relations Eq.\ (\ref{2.3.19}) 
{\it exactly} for all non-zero $x$. The Luttinger model 
manipulations do not have a condensed matter analogue, but are 
clearly not a necessary prerequisite to derive Bose 
representations.

To summarize results for the 
diagonal Fermi bilinears, the Bose representation for 
$\sum_{r}\psi^{\dg}_{r\sig}(x)\psi^{}_{r\sig}(x') + 
{\rm const.}$, which is the diagonal portion of Eq.\ 
(\ref{2.3.18}), is given by 
\beqa
\sum_{r}{\overline {\cal N}}^{2}_{rx-x'}(\alpha)
e^{-irk_{F}(x-x')}e^{-i\Psi_{r\sig}(x)}e^{i\Psi_{r\sig}(x')}
\label{2.3.32} 
\eeqa
for $k_{0}(x-x') \rightarrow \infty$, while for $x=x'$ it is 
given by 
\beqa
\frac{N_{e}}{2L} + \frac{1}{2\pi}\partial_{x}[
\phi_{\rho}(x) + \sig \phi_{\sig}(x)],
\label{2.3.33} 
\eeqa
where Eq.\ (\ref{2.3.30}) has been rewritten in terms of 
Bose fields. The complete Bose representation for 
$\psi^{\dg}_{\sig}(x)\psi^{}_{\sig}(x')$ is then obtained 
by adding elements (cf. Eq.\ (\ref{2.3.20})) which are 
off-diagonal in $r$ and $r'$. 

For a non-linear dispersion, the Bose representations for  
the Fermi bilinears contain weak contributions at higher 
multiples of $2k_{F}$; at $4k_{F}$, $6k_{F}$, and so on
(Voit 1994). This is similar to the appearance of higher 
odd multiples of $k_{F}$ in the Bose representation for the  
single Fermi fields discussed in section \ref{old2.3.2}.

\section{\label{old2.4}Bosonizing Lattice Systems}

In applications of bosonization to lattice systems, 
it is standard to first take the continuum limit, and then 
to use results taken over from the Luttinger model. 
This procedure is unnecessary, 
and has caused confusion regarding the interpretation of 
Luttinger model quantities in realistic lattice systems.   
(See, for example, Emery (1979), 
where the continuum limit leads to the identification of 
$\alpha$ with the lattice 
spacing $a$. This is discussed further in section \ref{old3.1.3}.) 
It is not necessary to take a continuum limit in order to use 
bosonization: Bose representations follow from the two 
theorems of section \ref{old2.2}, and these rely only on the 
{\it structure} of the 1D state space, as opposed to any 
peculiarities of a continuum electron system. 

Bosonization for lattice systems begins from fluctuation 
components $\rho_{r\sig}(k)$ for the lattice:
\beqa
\rho_{+ \sig}(k) &=& \sum_{0 < \overline{k} \leq \pi/a} 
c^{\dg}_{\overline{k}-\frac{k}{2}\sig}
c^{}_{\overline{k}+\frac{k}{2}\sig} \, , \nonumber \\
\rho_{- \sig}(k) &=& \sum_{-\pi/a < \overline{k} \leq 0} 
c^{\dg}_{\overline{k}-\frac{k}{2}\sig}
c^{}_{\overline{k}+\frac{k}{2}\sig} \, ,  
\label{2.4.1}
\eeqa
which are the lattice equivalent of Eq.\ (\ref{2.1.4}),  
with $a$ the lattice spacing. (It suffices here to 
consider only one band, and so the band index is suppressed.) 
The right- and left-moving density fluctuation 
components add to give the Fourier components in an expansion 
of the site number operator
\beqa
n_{j\sig} &=& c^{\dg}_{j\sig}c^{}_{j\sig} \nonumber \\
&=& N^{-1}\sum_{k \in {\rm FBZ}} \rho_{\sig}(k)\, e^{ikja} \, ,
\nonumber \\
\rho_{\sig}(k) &=& \sum_{r = \pm} \rho_{r\sig}(k) \, ,
\label{2.4.1a}
\eeqa
where $N$ is the number of sites and FBZ denotes the first Brillouin zone. 

For a band at or less than half-filling, 
$n = N_{e}/N \leq 1$, it is straightforward to verify that 
the theorems of section \ref{old2.2},  
Eqs.\ (\ref{tom}) and (\ref{2.2.7}), go through for the 
lattice case exactly as for the continuum. 
For $n > 1$ there are trivial modifications to 
the conditions of the theorems due to the Brillouin zone 
boundaries at $\pm \pi/a$. 

The derivation of Bose representations for lattice 
electrons now proceeds exactly as in section \ref{old2.3} for a 
continuum system. Bose fields are defined by 
\beqa
\phi_{\nu}(j) &=& \frac{\pi ja}{L}(N^{\nu}_{+} 
+ N^{\nu}_{-}) 
-i\sum_{k \neq 0}\frac{\pi}{kL}[\nu_{+}(k) 
 + \nu_{-}(k)] \Lambda_{\alpha}(k)e^{ikja}, 
\nonumber \\ 
\theta_{\nu}(j) &=& \frac{\pi ja}{L}(N^{\nu}_{+} 
- N^{\nu}_{-}) 
-i\sum_{k \neq 0}\frac{\pi}{kL}[\nu_{+}(k) 
 - \nu_{-}(k)] \Lambda_{\alpha}(k)e^{ikja},
\nonumber \\
\Pi_{\nu}(j) &=& \frac{\pi}{L}(N^{\nu}_{+} 
- N^{\nu}_{-}) 
+ \frac{\pi}{L}\sum_{k \neq 0}[\nu_{+}(k) 
 - \nu_{-}(k)] \Lambda_{\alpha}(k)e^{ikja} 
\nonumber \\
 &=& \partial_{x}\theta_{\nu}(j),
\label{bosefields}
\eeqa
where $\nu = \rho, \sig$ labels charge and spin, and 
where the charge and spin density fluctuation components are 
defined as in Eqs.\ (\ref{2.3.5}) for the continuum, 
but using Eqs.\ (\ref{2.4.1}) for $\rho_{r\sig}(k)$.
($\partial_{x}\psi_{\nu}(j)$, $\psi = \phi, \theta$, 
is shorthand for $\partial_{x}\psi_{\nu}(x/a)$ 
evaluated at $x = ja$.) The number of electrons at 
$j$ is given by $n + (a/\pi)\partial_{x} 
\phi_{\rho}(j)$, and the spin at $j$ is given by 
$(a/\pi)\partial_{x} \phi_{\sig}(j)$. Similarly, 
the average $\nu$ currents at $j$ are given 
by $(a/\pi)\Pi_{\nu}(j)$. 

Analogous to the derivation of Eq.\ (\ref{2.3.7}), 
the non-interacting hamiltonian for the lattice 
system with nearest neighbour hopping, 
\beqa  
H_{0} = -t\sum_{j, \sig}
\left(c^{\dg}_{j\sig}c^{}_{j+1\sig} + {\rm h.c.}
\right) \, ,
\label{2.4.3}
\eeqa
has the Bose representation
\beqa
H_{0B} = \frac{v_{F}a}{4\pi}\sum_{\nu,j}
\left\{ \Pi_{\nu}^{2}(j) 
+ [\partial_{x}\phi_{\nu}(j)]^{2}\right\}\, ,  
\label{Ho}
\eeqa
to an additive constant depending on $n$, 
and where the nearest neighbour dispersion 
Eq.\ (\ref{1.2.5}) has been linearized about the 
Fermi points as in Eq.\ (\ref{2.1.6}). 
The Fermi velocity is given by
\beqa
v_{F} = -2t\, \left. 
\frac{d \cos (ka)}{dk}\right|_{k = k_{F}} 
= 2at\sin(\pi n/2)\, .
\label{Vf}
\eeqa

Bose representations for the Fermi 
bilinears $c^{\dg}_{j\sig}c^{}_{l\sig'} = 
\sum_{r,r'}c^{\dg}_{rj\sig}c^{}_{r'l\sig'} 
+ \delta_{\sig, \sig'}{\rm const.}$ 
may be constructed from the single site operator 
Bose representation
\beqa
c_{rj\sig B} &=& \sqrt{\frac{Aa}{2\alpha}}\,\,
e^{ir(k_{F}+\pi/L)ja} \,
e^{i\Psi_{r\sig}(j)}\, U^{\dg}_{r\sig}\, , 
\nonumber \\
\Psi_{r\sig}(j) &=& \{ \theta_{\rho}(j) 
+ r\phi_{\rho}(j) 
+ \sig[\theta_{\sig}(j) + r\phi_{\sig}(j)]\}/2\, ,
\label{Crjsigbose}
\eeqa
where $U^{\dg}_{r\sig}$ removes an electron of spin $\sig$ 
in a state near $rk_{F}$, and commutes with asymptotically 
long-wavelength density fluctuations $\rho_{r\sig}(k)$,  
exactly analogous to the continuum case 
(cf.\ Eq.\ (\ref{2.3.11}) and sequel). 
It is convenient for later purposes to write the 
normalization constant for the lattice system in the 
form shown, where $A$ is a dimensionless constant 
depending on the cut-off function 
$\Lambda_{\alpha}(k)$. For example, 
$A = \gamma/\pi \approx 0.5$ 
for the step-function cut-off.
The representation Eq.\ (\ref{Crjsigbose}) for $c_{rj\sig}$ 
reproduces the correct commutation relations only with 
asymptotically long-wavelength fluctuations 
$\rho_{r\sig}(k)$, and the normalization constant is 
correct only to leading order in $\alpha/(j - l)a$ for 
asymptotic diagonal Fermi bilinears, as 
in the continuum case (cf.\ Eqs.\ (\ref{2.3.28})).  

The Bose representation for the diagonal on-site 
bilinears, i.e.\ the number operators, may be 
obtained directly from their Fourier expansion 
Eq.\ (\ref{2.4.1a}): 
\beqa
\sum_{r} c^{\dg}_{rj\sig}c^{}_{rj\sig B} = 
\frac{a}{2\pi}\partial_{x}
[\phi_{\rho}(j) + \sig \phi_{\sig}(j)] 
\label{diag}
\eeqa
to an additive constant depending on $n$.
As for $H_{0B}$, and in contrast to $c_{rj\sig B}$, 
this representation is exact. 

%%%%%%%%%%%%%%%%%%%%%%%%%%%%%%%%%%%%%%%%%%%
%% chapter 4
%%%%%%%%%%%%%%%%%%%%%%%%%%%%%%%%%%%%%%%%%%%%

\cleardoublepage
\chapter{\label{ch3}Features of the Bosonization}

The bosonization formalism derived in the previous chapter 
is self-contained, and within clearly stated 
limits it provides a rigorous alternative description for 1D 
electrons. It may be applied immediately in a 
variety of interacting 1D many-electron systems. The formalism 
is applied to the 1D Hubbard model in section \ref{old3.2} 
for later use in \ref{old5.4}. 
In chapter \ref{ch5}, the 1D Kondo lattice model is solved
using (this Abelian) bosonization. The non-Abelian extension
of this bosonization is given in Appendix \ref{appc}. 

Before proceeding to applications, it is useful to compare  
the formalism of chapter \ref{ch2} with the standard Luttinger model 
bosonization. The standard derivation of Bose representations,  
given in Appendix \ref{appa}, begins from a Luttinger 
model approximation, which replaces the finite Fermi sea of the 
system of interest with two infinite Dirac seas. In chapter \ref{ch2} 
this approximation was avoided, and 
Bose representations were derived beginning from the original 
realistic system with a finite Fermi sea. 
The resulting Bose representations, although familiar, are in 
several respects different from the corresponding Luttinger 
model representations. The new features are:

(i) An interpretation for $\alpha$ in realistic systems. 
The length $\alpha$ measures the minimum wavelength for 
density fluctuations which satisfy bosonic 
commutation relations. $\alpha$ depends in the general case on 
the number of electrons through $k_{F}$, and on the interactions 
through the bandwidth cut-off $k_{0}$. While 
$\alpha \rightarrow 0$ in the Luttinger model, 
$\alpha \gtrsim {\cal O}(k_{F})^{-1}$ in a realistic system. 
Common interpretations for $\alpha$ are that it is `something 
like' the lattice spacing (Emery 1979), or that 
$\alpha^{-1}$ is the effective bandwidth (in momentum 
units) (Luther and Peschel 1974). $\alpha$ as the minimum 
bosonic wavelength includes these interpretations 
as special cases in models where they are correct. 
In particular, there is some justification for these 
interpretations in one-component systems with standard 
electron-electron interactions. This may be established by a 
direct comparison with  
results obtained by other methods (S\'{o}lyom 1979).
However, the interpretation of $\alpha$ given here is 
more general, and applies for more 
complex systems and interactions. In particular, it applies for 
two-component systems such as the Kondo lattice model, 
in which compelling evidence for rejecting the standard 
interpretations is presented in chapters \ref{ch5} and 
\ref{ch6}. 

(ii) $\alpha$ is kept finite in a consistent manner 
throughout the bosonization formalism. As well as requiring 
a finite $\alpha$ in the normalization factors, it is also 
required that $\alpha$ be kept finite in 
the Bose field commutators. The first of these requirements 
has long been known, for otherwise the normalization factors 
diverge. This divergence is  responsible, for example, for the 
infinite spin gap in the attractive backscattering 1D 
electron gas (Luther and Emery 1974). 
The second requirement, of keeping $\alpha$ finite in 
the Bose field commutators, has previously been neglected. It 
is a missing element in the {\it ad hoc} mappings from 
the Luttinger model back to realistic systems. A correct  
treatment of the Bose field commutators is an 
essential formal aspect in the bosonization description of the 
ground-state phases of the 1D Kondo lattice. 

(iii) Bose representations for various Fermi operators do 
not have the same ranges of validity. In particular, the 
representations for the non-interacting hamiltonian $H_{0}$ 
and the density operators $\rho_{r\sig}(x)$ are operator 
identities in the subspace ${\cal H}_{k_{0}}$ 
defined by the bandwidth 
cut-off $k_{0}$. Representations for Fermi field operators 
$\psi_{r\sig}(x)$ are only valid for asymptotically 
long-wavelength fluctuations with respect to $1/k_{0}$. 
This holds also for the representations of 
the off-diagonal Fermi bilinears 
$\psi^{\dg}_{r\sig}(x)\psi^{}_{r'\sig'}(x')$, in which 
$r \neq r'$ and/or $\sig \neq \sig'$. The Bose 
representation for the diagonal bilinears 
$\psi^{\dg}_{r\sig}(x)\psi^{}_{r\sig}(0)$ when $x \neq 0$ 
is similarly only asymptotically valid, and will 
provide an accurate description only when 
$k_{0}x \rightarrow \infty$. These differences are not 
apparent in the Luttinger model, in which all 
representations are exact.

Section \ref{old3.1} considers these differences in more detail, and on 
a more formal level. Section \ref{old3.1.1} discusses normalization 
constants, and compares the standard Luttinger model constants 
with those of chapter \ref{ch2}. Section \ref{old3.1.2} calculates the Bose 
field commutators for a realistic system, and shows how the 
Luttinger model commutators may be reobtained as a limiting 
case. Section \ref{old3.1.3} considers the interpretation of $\alpha$, 
and gives a critique of previous interpretations, together with 
their relationship to the interpretation of chapter \ref{ch2}.

Anticipating the qualitatively different results which are 
obtained for the Kondo lattice using the bosonization formalism 
of chapter \ref{ch2}, as opposed to the Luttinger model bosonization, 
it is useful to consider an application to a simple 
one-component system. Section \ref{old3.2} considers the bosonization 
solution of the 1D Hubbard model, for later use in \ref{old5.4}. 
This is a standard application 
of bosonization, and there are only minor differences between 
the results obtained using the different Bose representations. 
%% The formalism of chapter \ref{ch2} is still to be preferred; for example, 
%% a non-zero coefficient is automatically obtained for the leading 
%% order term in the momentum distribution expansion near $k_{F}$  
%% (cf.\ Eq. (\ref{3.2.7})). 
However, no qualitatively new 
behaviour is obtained by using the formalism of chapter \ref{ch2}. 
The reason for this is because the Hubbard model is a simple
single-component system; being one-component, the minimum 
physically meaningful length scale in the system is the average 
inter-component (i.e.\ inter-electron) 
spacing.\footnote{The standard Hubbard model, as 
considered here, does not contain impurities or other 
lattice imperfections that may act to localize the electrons below 
this scale.} Further, the Hubbard model is simple in the sense that 
the density fluctuations are qualitatively bosonic down to the order 
of the 
inter-electron spacing: Tomonaga's theorem Eq.\ (\ref{tom}) holds 
in a qualitative sense well beyond its rigorous bound. This is 
established by comparison with the exact Bethe ansatz solution 
at strong-coupling (cf.\ section \ref{old3.2.3}). Thus, in the Hubbard 
model, $\alpha$ acts as a short-distance cut-off, much as in field  
theory, and delimits the minimum length scale in the system. 
In this case $\alpha$ may be made arbitrarily small, as in  
the Luttinger model, without overlooking any important physical 
processes. 

The bosonization solution of the 1D Hubbard model is considered 
in section \ref{old3.2}. The Hubbard model is diagonalized in 
section \ref{old3.2.1}, and the elementary charge and spin excitations 
are exhibited. These results will also be useful in chapter \ref{ch5},  
when repulsive interactions are introduced between the 
conduction electrons in the Kondo lattice. 
%%The zero temperature momentum distribution is 
%%calculated in section \ref{old3.2.2}. This calculation, which proves that 
%%the momentum distribution remains continuous at the Fermi 
%%momentum $k_{F}$, constitutes a proof of the failure of Landau 
%%Fermi liquid theory for 1D systems. 
In section \ref{old3.2.3}, the 
bosonization solution of the 1D Hubbard model is compared with 
the exact Bethe ansatz solution. Perhaps surprisingly, it is 
found that bosonization gives a qualitatively complete 
description of the low-energy properties of the Hubbard model 
even for strong couplings $U/t \gg 1$.

\section{\label{old3.1}Comparisons with Luttinger Model Results}

\subsection{\label{old3.1.1}Normalization constants}

In the Luttinger model there is a single (divergent) 
normalization factor ${\cal N}_{L}$ attached to the 
Bose representation for each Fermi field (cf.\ 
Eq.\ (\ref{A.7})):  
\beqa
{\cal N}_{L} = \lim_{\alpha \rightarrow 0} 
\frac{1}{\sqrt{2\pi \alpha}}. 
\label{2.5.1}
\eeqa
In the Bose representation for realistic  
systems, in which $\alpha$ measures the minimum wavelength 
for bosonic density fluctuations, there are two 
independent normalization factors. For the single Fermi 
fields the normalization is  
\beqa
{\cal N}(\alpha) = L^{-1/2}\exp\left\{
\sum_{k>0} \frac{\pi}{kL}\Lambda_{\alpha}^{2}(k)\right\} 
\propto \frac {1}{\sqrt{\alpha}} \, ,
\label{2.5.2}
\eeqa
(cf.\ Eqs.\ (\ref{2.3.14}) and (\ref{2.3.16})), where the 
constant of proportionality depends on the choice of cut-off 
function $\Lambda_{\alpha}(k)$. For example, the constant 
is $1/\sqrt{2\pi}$ for the 
exponential cut-off function $\Lambda_{\alpha}(k) = 
\exp(-\alpha |k|/2)$, and is $\sqrt{\gamma/2\pi}$, 
$\gamma = 1.78\cdots$ for the step-function cut-off 
$\Lambda_{\alpha}(k) = \theta(\alpha^{-1} - |k|)$.  
The square of Eq.\ (\ref{2.5.2}) is also the normalization 
factor for off-diagonal Fermi bilinears 
$\psi^{\dg}_{r\sig}(x)\psi^{}_{r'\sig'}(x')$ in which 
$r \neq r'$ and/or $\sig \neq \sig'$ (cf.\ Eq.\ 
(\ref{2.3.20})). 
In Bose representations for realistic systems there is a 
second normalization for the asymptotic diagonal bilinears 
$\psi^{\dg}_{r\sig}(x)\psi^{}_{r\sig}(0)$  where $x$ is 
large compared with $\alpha$;
\beqa
{\overline {\cal N}}^{2}_{rx}(\alpha) = 
\frac{ir}{2\pi x}
\exp\left\{\int_{0}^{\infty}\frac{1-e^{irkx}}{k}
\Lambda_{\alpha}^{2}(k)\, dk \right\} \propto 
\frac{1}{\alpha}\left[ 1 + {\cal O}\left(
\frac{\alpha}{x}\right) \right] \, ,
\label{2.5.3}
\eeqa
(cf.\ Eqs.\ (\ref{2.3.24}) and (\ref{2.3.29})). The constant
of proportionality in Eq.\ (\ref{2.5.3}) again depends on the 
cut-off function, and coincides with the square of the constant 
in Eq.\ (\ref{2.5.2}) (cf.\ Eq.\ (\ref{2.3.28})). 
Three points are worth noting: 

(i) The two normalizations for 
the realistic system are consistent with the 
single Luttinger model normalization only when $\alpha
\rightarrow 0$. The unphysical choice for $\alpha$ 
decouples the normalization constant for the 
asymptotic diagonal Fermi 
bilinears by quenching terms of order $1/x$ for 
all $x \neq 0$. The diagonal Fermi bilinears may then 
be expressed as a straightforward product of single Fermi 
field Bose representations, as for the off-diagonal products.
  
(ii) The constants of proportionality in the 
normalizations for the realistic system depend 
on the form of the cut-off function $\Lambda_{\alpha}(k)$ near 
the bosonic limit $\alpha$. 
For example, for the step-function cut-off the 
constant is $\sqrt{\gamma/2\pi}$, $\gamma = 1.78 \cdots $, 
independent of the value of $\alpha$. The factor 
$1/\sqrt{2\pi}$ in the Luttinger model factor ${\cal N}_{L}$ 
comes from the conventional choice of an exponential 
cut-off function, whose status
in  that model is as a formal device introduced by hand to 
remove divergences (Haldane 1981). 

(iii) The normalizations relate to the description of the 
electrons in terms of bosons; they normalize the 
Fermi fields only over length scales $> \alpha$ which are 
described by bosonic fluctuations. The  
Luttinger model obscures this somewhat by having bosonic 
fluctuations over all length scales. Here it is manifest, in 
Eq.\ (\ref{2.5.2}) for example, in which the true 
Fermi field normalization 
$L^{-1/2}$ is increased by a factor depending on the  
cut-off function $\Lambda_{\alpha}(k)$.

\subsection{\label{old3.1.2}Bose field commutators}

Since the Luttinger model normalization diverges, 
it has long been realised that $\alpha$ needs 
reinterpreting when $1/\alpha$ appears in the final 
result of a calculation for a realistic system.  
The reinterpretation of $\alpha$ is not carried out in a  
consistent way. Specifically, $\alpha$ is taken to 
be vanishingly small throughout a calculation, which 
is necessary in the Luttinger model in order to 
obtain the correct form for the density operators 
$\rho_{r\sig}(x)$ (cf.\ Eq.\ (\ref{A.9})) for example, 
but then $\alpha$ is kept finite 
in certain ($\alpha \rightarrow 0$ divergent) terms 
because the limiting result is manifestly non-physical. 
The derivation given in chapter \ref{ch2} of Bose representations 
for realistic systems permits $\alpha$ to be be kept finite 
in a consistent manner throughout the bosonization formalism.
This reveals that there are missing elements in the mapping 
from the Luttinger model back to the realistic system; 
it is not sufficient solely to keep $\alpha$ finite in the 
$\alpha \rightarrow 0$ divergent terms. 

The missing elements are conveniently summarized by 
considering the Bose field commutators. 
The Bose fields $\phi_{\nu}(x)$ and  
$\theta_{\nu}(x)$, where $\nu = \rho,\sig$ 
labels charge and spin, have been introduced 
for convenience, and follow common 
usage in the Luttinger model (Voit 1994). 
The central difference of the Bose fields used here to 
their Luttinger model counterparts is the (finite) cut-off 
function $\Lambda_{\alpha}(k)$ on bosonic fluctuations. 
This has a significant effect on the 
short-distance behaviour of the Bose field commutators. 

\begin{table}
\centering
\begin{tabular}{|c||c|c|} 
\hline 
Cut-off function & 
$[\phi_{\nu}(x),\theta_{\nu}(0)]/2i$ & 
$i[\phi_{\nu}(x),\Pi_{\nu}(0)]/2$ \\   \hline
Step-function: $\theta(\alpha^{-1} - |k|)$ & 
${\rm sign}(x){\rm Si}(|x|/\alpha)$ & 
$\frac{\sin(x/\alpha)}{x}$ \\ 
Gaussian: $e^{-\alpha^{2}k^{2}/2}$ &
$\frac{\pi}{2}{\rm erf}(x/2\alpha)$ & 
$\frac{{\sqrt \pi}}{2\alpha}e^{-(x/2\alpha)^{2}}$ \\
Exponential:  $e^{-\alpha |k|/2}$ & 
$\tan^{-1}(x/\alpha)$ & 
$\frac{\alpha}{\alpha^{2} + x^{2}}$ \\        \hline
Luttinger model & 
${\rm sign}(x) \pi/2$ & $\pi \delta(x)$ \\    \hline
\end{tabular}
\caption{ Bose field commutators for different choices 
of the cut-off function $\Lambda_{\alpha}(k)$. 
Si is the sine integral (Lebedev 1965) 
and erf is the error function (Dwight 1961).  
Luttinger model results (corresponding to the exponential 
cut-off with $\alpha \rightarrow 0$) are given for 
comparison. Commutators are plotted in Figs.\ \ref{oldfig3.1} 
and \ref{old3.2}.}
\end{table}

\begin{figure}[htb]
\centering
\includegraphics[scale=0.7,angle=-90]{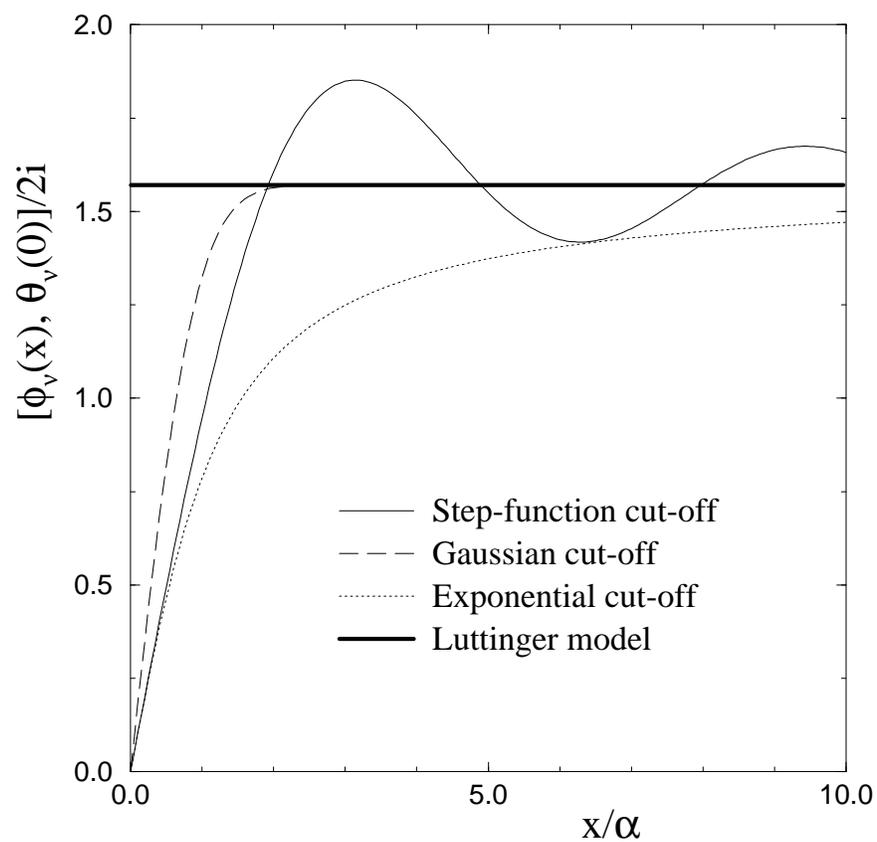}
\vspace{3.0mm}
\caption{\label{oldfig3.1}The Bose field commutator 
$[\phi_{\nu}(x), \theta_{\nu}(0)]/2i$, from 
Table 3.1, for different choices of the cut-off 
function $\Lambda_{\alpha}(k)$.}
\end{figure}

To evaluate the Bose field commutators, note that on 
${\cal H}_{k_{0}}$, and for $|k|, |k'| < \alpha^{-1}$, 
Eqs. (\ref{tom}) and (\ref{2.3.5}) 
give
\beqa
[\nu_{r}(k), \nu'_{r'}(k')] 
= \delta_{\nu, \nu'}\, \delta_{r, r'}\, 
\delta_{k, -k'}\, \frac{rkL}{\pi}\, .
\label{2.5.4}
\eeqa
Hence 
\beqa
[\nu_{+}(k) \pm \nu_{-}(k), 
\nu'_{+}(k') \pm \nu'_{-}(k')] = 0  \, ,
\label{2.5.5} 
\eeqa
and
\beqa
[\nu_{+}(k) \pm \nu_{-}(k), 
\nu'_{+}(k') \mp \nu'_{-}(k')]  
= \delta_{\nu, \nu'}\, \delta_{k, -k'}\, 
\frac{2kL}{\pi}\, .
\label{2.5.6}
\eeqa
Since $[N_{r\sig}, \rho_{r'\sig'}(k)] = 0$ 
for all $|k| < \alpha^{-1}$
(cf.\ Eq.\ (\ref{tom})), it follows immediately from 
Eqs.\ (\ref{2.5.5}) and (\ref{2.5.6})
that all the commutators between the Bose 
fields $\phi_{\nu}(x), \theta_{\nu}(x)$ and 
$\Pi_{\nu}(x)$ will vanish except for 
\beqa
[\phi_{\nu}(x), \theta_{\nu}(x')] = 
\sum_{k \neq 0}\frac{2\pi}{kL}
\Lambda^{2}_{\alpha}(k)\, e^{ik(x - x')} \, ,
\label{2.5.7}
\eeqa
and
\beqa
[\phi_{\nu}(x), \Pi_{\nu}(x')] = 
-i\frac{2\pi}{L}\sum_{k \neq 0}
\Lambda^{2}_{\alpha}(k)\, e^{ik(x - x')} \, . 
\label{2.5.8}
\eeqa
For later reference, note the relation
\beqa
[\partial_{x}\phi_{\nu}(x), \theta_{\nu'}(x')] =
-[\phi_{\nu}(x), \Pi_{\nu'}(x')] \, .
\label{2.5.9}
\eeqa
In the thermodynamic limit, Eq.\ (\ref{1.3}) may be 
used to convert the sums in Eqs.\ (\ref{2.5.7}) and 
(\ref{2.5.8}) to integrals, giving
\beqa
[ \phi_{\nu} (x) , \theta_{\nu '} (0) ]     =   
2 i \, \delta_{\nu , \nu '} \int_{0}^{\infty}dk\, 
\frac{\sin (k x)}{k} \Lambda^{2}_{\alpha} (k) \,  ,
\label{2.5.10}  
\eeqa
and
\beqa
[ \phi _{\nu} (x) , \Pi_{\nu '} (0) ]   & = & 
-2 i \delta_{\nu , \nu '}  \int_{0}^{\infty}dk\, 
\cos ( k x) \Lambda^{2}_{\alpha} (k) \, .  
\label{2.5.11}
\eeqa
Using Dwight (1961), the integrals in 
Eqs.\ (\ref{2.5.10}) and (\ref{2.5.11})
may be evaluated in closed form for the three cut-off 
functions of Eq.\ (\ref{2.3.4}). 
The results are listed in Table 3.1, and are plotted 
in Fig.\ \ref{oldfig3.1}  for 
$[\phi_{\nu}(x), \theta_{\nu}(0)]$ and in Fig.\ \ref{oldfig3.2} for 
$[\phi_{\nu}(x), \Pi_{\nu}(0)]$.  
It is clear that the commutators coincide with 
the Luttinger model forms for each choice of 
(continuous) cut-off function for $x \gtrsim \alpha$, but are 
smoothed over shorter distances. With the realistic 
bosonization, the system is described 
by bosonic density fluctuations over lengths 
beyond ${\cal O}(\alpha)$. The smoothing of the commutators 
below ${\cal O}(\alpha)$ is governed by the form of the cut-off 
function, and reflects the minimum length which can be 
described by bosonic density fluctuations: 
$[\phi_{\nu}(x), \Pi_{\nu}(0)]$, which has canonical 
$\delta$-function form in the Luttinger model,  
remains non-zero for a finite range of $x$, and signifies that 
the Bose fields do not distinguish separations below this 
range. 
 
A note is in order on the commutators for the step-function 
cut-off. At short-distances $x \lesssim \alpha$, the 
commutators show the same behaviour as the smoother 
cut-off functions, but have a slowly damped 
oscillatory behaviour for $x \gtrsim \alpha$, which 
is not shared by the smoother cut-offs. 
The oscillations are due to the discontinuity of the 
step-function cut-off at $\alpha^{-1}$. For example, 
the commutator $[\phi_{\nu}(x), \Pi_{\nu}(0)]$  has the form 
of the amplitude function for Fraunhofer diffraction from a 
single slit of `width' $\alpha^{-1}$. 
The diffraction pattern is due to the
hardness of the slit edges, which generate  
coherent interference between `wavelets' $\rho_{r\sig}(k)$. 
As discussed in section \ref{old2.2.2}, in particular in the extension 
of Tomonaga's result Eq.\ (\ref{tom}) to slightly 
shorter wavelength fluctuations, an abrupt cut-off 
on bosonic density fluctuations is unlikely. 
The monotonic commutators due to continuous cut-off 
functions, which share the same features as  
the Luttinger model commutators for $x \gg \alpha$,  
will be present in real systems in general. 

The lattice Bose fields $\phi_{\nu}(j), \theta_{\nu}(j)$, 
and $\Pi_{\nu}(j)$ were defined in Eqs.\ (\ref{bosefields})  
analogous to their continuum counterparts. The results 
above on the continuum Bose field commutators are easily 
verified to carry over directly to the lattice 
case, with the trivial modification $x \rightarrow ja$.

\subsection{\label{old3.1.3}Common interpretations of $\alpha$}

There are two interpretations of $\alpha$ in common use: 
(a) $\alpha^{-1}$ is interpreted as the bandwidth 
(in momentum units) (Luther and Peschel 1974), and (ii) $\alpha$ 
is interpreted as `something like' the lattice spacing 
$a$ of an underlying lattice model (Emery 1979). 
These interpretations seem to be at least partially 
correct in some one-component systems, which can be 
established by direct comparison 
with results obtained by other methods, such as the 
renormalization group, direct diagram summation, and 
the exact Bethe ansatz solution.  

The interpretation of $\alpha^{-1}$ as the bandwidth in 
momentum units (i.e. $\alpha^{-1} = k_{0}$) 
was proposed by Luther and Peschel (1974).
They noticed that the non-interacting 
expectation value $\langle 0| 
\psi^{\dg}_{\sig}(x) \psi^{}_{\sig}(0) |0\rangle$, $x \neq 0$,  
is reproduced correctly using the Luttinger model Bose 
representation with a finite $\alpha$, provided only that the 
(constant) density of states is altered to give  $\sum_{k} 
\rightarrow (L/2\pi)\int dk\, e^{-\alpha|k - k_{F}|}$. In 
other words, they focus on $\alpha$ as it appears in the 
non-interacting Hamiltonian, and where it acts as a type of
bandwidth cut-off. Luther and Peschel's interpretation is 
partially supported by other calculations. In the attractive
1D backscattering electron gas, a comparison between the 
charge-density response function obtained by 
bosonization, and that obtained by renormalization group 
methods, gives $\alpha^{-1} = k_{0}$ to leading logarithmic 
order (S\'{o}lyom 1979). This is plausible in 
terms of the interpretation of chapter \ref{ch2}, since 
on ${\cal H}_{k_{0}}$ the electrons are spread over a spatial 
range $\approx 1/k_{0}$, and thus $\alpha^{-1} \gtrsim 
1/k_{0}$ for a complete description by bosonization 
(cf.\ section \ref{old2.2.3}). However, the  
comparison with the renormalization group results 
indicates problems with Luther and Peschel's interpretation 
already in the next-to-leading logarithmic contributions. 

Following this early attempt, and the problems with it,  
a series of papers appeared around the late 1970's which 
attempted to clarify the status of $\alpha$
(S\'{o}lyom 1979, Apostol 1983).  
The results of this work were inconclusive, due in 
the main to a focus on relating $\alpha$ to traditional 
cut-offs in the 1D electron liquid. The problem  
is that $\alpha$ is not a cut-off in any of the usual 
senses; $\alpha^{-1}$ acts as a bandwidth cut-off only in the 
non-interacting hamiltonian. In the Bose representation of 
the interaction term in the Tomonaga-Luttinger model, 
by contrast, $\alpha^{-1}$ acts as a momentum transfer cut-off. 
(An example of this is given in section \ref{old3.2.1}, when the 
Hubbard model interaction is bosonized.) The only 
conclusion to be drawn is that $\alpha$ cannot 
unambiguously be identified with any of the conventional 
cut-offs, but acts in place of both of them (S\'{o}lyom 1979). 
This is understood immediately when $\alpha$ is 
identified as the minimum bosonic wavelength, and if it is  
assumed that bosonization provides a complete description. 

A different approach to interpreting $\alpha$ is due to 
Emery (1979). Emery's view, which now  
has wide currency, is that $\alpha$ is `something like' the 
lattice spacing $a$ of an underlying lattice model. 
Emery arrives at this conclusion by taking the continuum 
limit of a lattice model, and then bosonizing. 
He then identifies the length $\alpha$, which comes from the  
bosonization, with the lattice spacing $a$. 
It is useful  first to comment on Emery's argument, and 
then to assess the validity of the interpretation. 

Emery's argument (Emery 1979) using the continuum limit 
of a lattice model is spurious. It is not necessary to take the 
continuum limit of a lattice model 
in order to apply bosonization. 
Rather, bosonization follows from 
the structure of the many-particle 1D state space, and 
Bose representations may be derived, as in section \ref{old2.4}, 
beginning from a basis of Bloch states. In this case, 
$\alpha$ is again identical to the minimum wavelength for 
bosonic density fluctuations. Notwithstanding this, it has 
become commonplace to simply write $\alpha = a$ in 
bosonization formulae for continuum systems. This confuses the 
limit $\alpha \rightarrow 0$, which comes from the Luttinger 
model, with the continuum limit $a \rightarrow 0$. The limits 
are physically distinct, and both are unnecessary from the 
point of view of bosonization. A clearer approach would be 
to bosonize the lattice system directly, in which case the 
normalization constants are $\propto a/\alpha$ 
(cf.\ Eq.\ (\ref{Crjsigbose})). 

Emery's interpretation makes sense if it assumed that 
(i) density fluctuations remain bosonic down to 
wavelengths of the order of the average inter-particle 
spacing, i.e.\ well beyond 
the rigorous bound prescribed by Eq.\ (\ref{tom}), and 
(ii) the system is one-component, so that the average 
inter-particle spacing is the smallest length scale in 
the system. With these assumptions, Emery's interpretation 
is essentially correct; $\alpha$ in this case acts as 
a harmless short-distance cut-off, and it is reasonable 
to take this cut-off at the lattice spacing of an underlying 
lattice model. Assumption (i) is qualitatively correct
at least in some systems. This is shown for the 1D  
Hubbard model in section \ref{old3.2.3} by comparison with  
results from the exact Bethe ansatz solution.  
However, Emery's interpretation fails if assumption (ii) 
is not met; while density fluctuations may remain at 
least qualitatively bosonic down to wavelengths of the 
order of the inter-particle spacing, this may not be 
the minimum length scale in a two-component system. 
An example of this is the 1D Kondo lattice at partial 
conduction band filling.

\section{\label{old3.2}The 1D Hubbard Model via Bosonization}

The Hubbard model is a single-band model of electrons 
on a lattice, and in 1D has the hamiltonian 
\beqa
H_{\Hub} = -t\sum_{j, \sig}
\left(c^{\dg}_{j\sig}c^{}_{j+1\sig} + {\rm h.c.}\right) 
+ U\sum_{j} n_{j\uparrow}n_{j\downarrow}\, ,
\label{3.1}
\eeqa
where the nearest neighbour hopping $t>0$, and  
the Hubbard interaction $U>0$ is repulsive.
The Hubbard model was introduced
in 1963 to describe $3d$-electrons in transition  
metals (Hubbard 1963, Gutzwiller 1963), and has since been  
studied intensively as it is one of the simplest models for strongly 
correlated electrons on a lattice
\footnote{A brief overview of the strongly correlated
electron systems in given in Appendix \ref{appd}.}. 
$H_{\Hub}$ is derived in section \ref{old1.2.2} 
(cf.\ Eq.\ (\ref{1.2.11})) for tightly bound electrons, 
whose Wannier wavefunctions are strongly localized at 
their lattice sites. As discussed earlier, the 1D Hubbard model 
is a simple single-component system, and as such it allows a
straight forward bosonization solution (see, eg, Emery (1979), 
Haldane (1981), Fradkin (1991), Schulz (1991) and Shankar (1995),
for a finite temperature bosonization see, Bowen and Gul\'{a}csi (2001)).
Here a bosonization based on chapter \ref{ch2} is implemented, 
for further use in section \ref{old5.4}.

\subsection{\label{old3.2.1}Bosonization solution}

To use bosonization to solve the Hubbard model, it is 
convenient first to rewrite the Hubbard interaction 
in terms of density fluctuation operators. Using Eqs.\ 
(\ref{1.25}) and (\ref{1.26}), 
\beqa
V_{\Hub} &=& U\sum_{j}n_{j\uparrow}n_{j\downarrow} 
\nonumber \\
&=& \frac{U}{2N}\sum_{q, q', k, k' \in {\rm FBZ}}
\sum_{G, \sig} c^{\dg}_{k\sig}c^{}_{k+q\sig}
c^{\dg}_{k'-\sig}c^{}_{k'+q'-\sig}\, 
\delta_{-q', q+G}
\nonumber \\
&=& \frac{U}{2N}\sum_{q, q'}\sum_{G, \sig}
\rho_{\sig}(q)\rho_{-\sig}(q') \, \delta_{-q', q+G} \, .  
\label{3.2}
\eeqa
The second form here is the lattice equivalent of Eq.\ 
(\ref{1.2.7}), but includes a summation over reciprocal 
lattice vectors $G = 2\pi m/a$, $m$ an integer, to 
account for Umklapp scattering. (Crystal momenta which 
differ by a reciprocal lattice vector are equivalent.) 
The third form in Eqs.\ (\ref{3.2}) uses the definition 
Eq.\ (\ref{2.4.1a}) of density fluctuation operators for 
the lattice. 

To bosonize the Hubbard interaction, that is to  
write $V_{\Hub}$ in terms of density fluctuation components which 
satisfy the bosonic algebra Eq.\ (\ref{tom}), it is necessary to 
assume that for small enough $U$ the ground-state lies within 
a subspace ${\cal H}_{k_{0}}$ defined with a bandwidth cut-off 
$k_{0}$ as described in section \ref{old2.1}. This is expected quite 
generally since having a macroscopic population of electrons in 
excited states far from the Fermi surface is energetically 
expensive due to the dispersion $\varepsilon(k)$. Thus, when 
the kinetic energy is large compared with the interaction energy, 
only those states close to the Fermi surface are expected to 
be affected. The particle-hole excitations represented by the 
density fluctuations in $V_{\Hub}$ then fall into three 
categories. (i) Small momentum transfer or forward scattering 
bosonic excitations $\rho_{\sig}(q)$ with $q < \alpha^{-1}$ (and 
$k_{0} < \alpha^{-1}$). 
These are the only density fluctuations which 
will be considered in $V_{\Hub}$ in this section. (In this case 
the continuum limit of the Hubbard model coincides with the 
Tomonaga-Luttinger model, or equivalently with the $g$-ology 
model with only $g_{2}$ and $g_{4}$ processes included 
(S\'{o}lyom 1979).) 
There are two remaining classes of density fluctuation processes 
in $V_{\Hub}$ which are consistent with existence in the subspace 
${\cal H}_{k_{0}}$; (ii) $2k_{F}$ momentum transfer or 
backscattering particle-hole excitations $\rho_{\sig}(q)$ with 
$|q| \approx 2k_{F}$. These correspond to the excitation of an 
electron from one Fermi point to the opposite Fermi point. 
Backscattering processes 
may be written in terms of bosonic fluctuations indirectly, by 
using the Fermi site operator Bose representation Eq.\ 
(\ref{Crjsigbose}) for each Fermi operator in $\rho_{\sig}(k)$ 
separately. 
Backscattering is neglected in the bosonization solution 
of the Hubbard model because for $U > 0$ it 
is irrelevant in the description of the long-wavelength 
properties. This is established by scaling the  
effective bandwidth $2v_{F}k_{0}$   
to smaller values, which corresponds to an equivalent system at 
long-wavelengths. Under the scaling, 
the effective backscattering interaction 
coupling scales to zero for all positive initial couplings $U$. 
(The details of the scaling procedure are reviewed by S\'{o}lyom 
(1979).) (iii) The third category of particle-hole excitations is 
Umklapp scattering. When 
the electron density or filling $n = N_{e}/N$ satisfies 
$n  = 1$ (called 
half-filling), Umklapp processes involving a momentum 
transfer of $4k_{F}$ are permitted within ${\cal H}_{k_{0}}$,  
because $4k_{F} = 2\pi /a$ is a reciprocal lattice vector, 
and falls in the equivalence class of 
zero crystal momentum. By ignoring Umklapp scattering, it is 
thus assumed that the filling is away from half. The effects 
of Umklapp scattering at half-filling 
are formally similar to the effects of attractive backscattering.   
However, instead of a spin gap, the Umklapp processes open a 
charge gap, and the half-filled Hubbard model is insulating
(Emery, Luther and Peschel 1976).  

The Hubbard interaction Eq.\ (\ref{3.2}), including only 
bosonic density fluctuations representing forward scattering, 
is given by 
\beqa
V_{\Hub} &=& \frac{U}{2N}\sum_{|q| < \alpha^{-1}}\sum_{\sig}
\rho_{\sig}(q)\rho_{-\sig}(-q) 
\nonumber \\
&=& \frac{U}{4N}\sum_{|q| < \alpha^{-1}}
\left[ \rho(q)\rho(-q) - \sig(q)\sig(-q)\right]\, , 
\label{3.3}
\eeqa
where the charge and spin density fluctuation operators are 
defined by 
\beqa
\rho(k) = \sum_{\sig}\rho_{\sig}(k)\, , \quad \quad 
\sig(k) = \sum_{\sig}\sig\rho_{\sig}(k) \, .
\label{3.4}
\eeqa
analogous to Eqs.\ (\ref{2.3.5}) for the continuum. 
This way of writing $V_{\Hub}$ 
corresponds to choosing the hard cut-off $\Lambda_{\alpha}(q) 
= \theta(\alpha^{-1} - |q|)$ on the density fluctuations, 
similar to Eq.\ (\ref{2.3.1}). As in chapter \ref{ch2}, it is 
useful to consider more general $\Lambda_{\alpha}(q)$.  

To solve the Hubbard model it is convenient and instructive 
to cast the hamiltonian in the form of true bosonic operators, 
instead of using the density fluctuations $\rho_{r\sig}(q)$ 
either directly, or through the Bose fields introduced in 
chapter \ref{ch2}. (The formulation in terms of Bose fields is used 
extensively in the Kondo lattice in chapter \ref{ch5}.) 
The charge and spin bosons corresponding to the 
right- and left-moving components of 
$\rho(q)$ and $\sig(q)$ are defined as 
\beqa
C_{q} &=& \left\{ 
\begin{array}{lc}
\sqrt{\frac{\pi}{|q|L}}\, \rho_{+}(q)\, ,
& 0 < q < \alpha^{-1}\, ,  \\
\sqrt{\frac{\pi}{|q|L}}\, \rho_{-}(q)\, ,
& -\alpha^{-1} < q < 0 \, ,  
\end{array} \right.
\nonumber \\
S_{q} &=& \left\{ 
\begin{array}{lc}
\sqrt{\frac{\pi}{|q|L}}\, \sig_{+}(q)\, ,
& 0 < q < \alpha^{-1} \, , \\
\sqrt{\frac{\pi}{|q|L}}\, \sig_{-}(q)\, ,
& -\alpha^{-1} < q < 0 \, , 
\end{array} \right.
\label{3.5}
\eeqa
which is similar to Eq.\ (\ref{2.2.4}) for the bosons 
$b_{k\sig}$. $C_{q}$ and $S_{q}$ satisfy canonical Bose 
commutation relations as in Eq.\ (\ref{2.2.5}), and the 
mixed commutators containing both a $C_{q}$ and an $S_{q'}$ 
operator (or their conjugates) all vanish. In terms of these 
operators the Hubbard hamiltonian takes the form 
\beqa
&& H_{\Hub} =
\sum_{q > 0} q\Lambda_{\alpha}^{2}(q)\left\{ 
v_{F}\left[ C^{\dg}_{q}C^{}_{q} +  C^{\dg}_{-q}C^{}_{-q}
+S^{\dg}_{q}S^{}_{q} + S^{\dg}_{-q}S^{}_{-q}\right] 
+ (Ua/2\pi) \right. \nonumber \\
&\times& \left. \left[ C^{\dg}_{q}C^{}_{q} 
+\, C^{\dg}_{-q}C^{}_{-q}+ 
C^{\dg}_{q}C^{\dg}_{-q} + C^{}_{q}C^{}_{-q}
- S^{\dg}_{q}S^{}_{q} - S^{\dg}_{-q}S^{}_{-q}
- S^{\dg}_{q}S^{\dg}_{-q} - S^{}_{q}S^{}_{-q}\right] \right\}\, , 
\nonumber \\
\label{3.6}
\eeqa
to an additive constant depending on the total number of 
electrons. Eq.\ (\ref{3.3}) has been used to write the  
interaction term in Eq.\ (\ref{3.6}).  
The bosonized hopping term in Eq.\ 
(\ref{3.6}) has been written using Eq.\ (\ref{Ho}) (cf.\ 
also the lattice equivalent of Eq.\ (\ref{2.3.3})), 
and the dispersion has been linearized about 
the Fermi points with the Fermi velocity $v_{F}$ given 
by Eq.\ (\ref{Vf}). The number operators $N_{r\sig}$ have 
been neglected in bosonizing the hopping term, which is  
valid in the thermodynamic limit: The number operators come   
with prefactors $1/L$ (cf.\ Eq.\ (\ref{2.3.3})), and 
their effect vanishes in the absence of an applied magnetic 
field. (A similar conclusion holds for the fermionic ladder  
operators $U_{r\sig}$ (Voit 1994)).

The bosonized 1D Hubbard hamiltonian  
is a bilinear form in bosons, and 
may be diagonalized using standard methods. 
A straightforward coordinate transformation may be 
used, as in reference (Mahan 1990). Equivalently,  
and in a more useful form for the calculation of 
physical quantities, a Bogoliubov unitary transformation can 
be used. Consider applying the transformation $\exp(W)$ where 
\beqa
W &=& W_{\rho} + W_{\sig} \, , \nonumber \\
W_{\rho} &=& \lambda_{\rho}\sum_{0<q<\alpha^{-1}}
\left(C^{}_{q}C^{}_{-q} - C^{\dg}_{-q}C^{\dg}_{q}\right)\, , 
\nonumber \\
W_{\sig} &=& \lambda_{\sig}\sum_{0<q<\alpha^{-1}}
\left(S^{}_{q}S^{}_{-q} - S^{\dg}_{-q}S^{\dg}_{q}\right)\, ,
\label{3.7}
\eeqa
where the real numbers $\lambda_{\nu}$, $\nu = \rho, \sig$, are 
chosen so that the transformed hamiltonian is 
diagonal. For real $\lambda_{\nu}$, $W^{\dg} = -W$ is 
anti-hermitian, and the transformation $\exp(W)$ is 
unitary. Transformed operators $\tilde{O} = e^{-W}Oe^{W}$ 
may be calculated using the standard commutator expansion 
\footnote{For more details on the canonical transformations
in general, see section \ref{sectionSW}, including Appendix 
\ref{ch4sw}, and section \ref{old5.1.2}.} 
\beqa
\tilde{O} = O + \frac{1}{1!}[O, W] 
+ \frac{1}{2!}[[O, W], W] 
+ \frac{1}{3!}[[[O, W], W], W] + \cdots \; . 
\label{comseries} 
\eeqa 
Using the bosonic commutation relations for $C_{q}$ and $S_{q}$ 
when $|q| < \alpha^{-1}$, it follows that 
\beqa
\tilde{C}_{q} &=& C_{q}\left(1 + \frac{\lambda_{\rho}^{2}}{2!}
+ \frac{\lambda_{\rho}^{4}}{4!} + \cdots\right)
- C^{\dg}_{-q}\left(\lambda_{\rho} + 
\frac{\lambda_{\rho}^{3}}{3!}
+ \frac{\lambda_{\rho}^{5}}{5!} + \cdots\right)
\nonumber \\
&=& \cosh (\lambda_{\rho})\, C^{}_{q} - 
 \sinh (\lambda_{\rho}) \, C^{\dg}_{-q}\, , 
\label{3.9}
\eeqa
and similarly for $\tilde{S}_{q}$, with $\lambda_{\rho}$ 
replaced by $\lambda_{\sig}$. Substituting the transformed 
operators into the hamiltonian Eq.\ (\ref{3.6}) yields the 
transformed bosonized Hubbard model. This contains diagonal 
terms $C^{\dg}_{q}C^{}_{q}$ and $S^{\dg}_{q}S^{}_{q}$, 
together with off-diagonal terms $C^{\dg}_{q}C^{\dg}_{-q}$, 
$S^{\dg}_{q}S^{\dg}_{-q}$, and their conjugates. The 
off-diagonal terms cancel identically provided 
$\lambda_{\rho}$ and $\lambda_{\sig}$ satisfy 
\beqa
\tanh (2\lambda_{\rho}) &=& \left[ 1+ (2\pi v_{F}/Ua) 
\right]^{-1} \, , \nonumber \\ 
\tanh (2\lambda_{\sig}) &=& \left[ 1- (2\pi v_{F}/Ua) 
\right]^{-1} \, . 
\label{3.10}
\eeqa
With these values, the transformed Hubbard hamiltonian reads
\beqa
\tilde{H}_{\Hub} = \sum_{q \neq 0} |q| 
\Lambda_{\alpha}^{2}(q) 
\left( v_{\rho}C^{\dg}_{q}C^{}_{q} 
+ v_{\sig}S^{\dg}_{q}S^{}_{q}\right) \, , 
\label{3.11}
\eeqa
to an additive constant, where the charge and spin velocities 
are 
\beqa
v_{\rho} = v_{F} \sqrt{1 + \frac{Ua}{\pi v_{F}}} \, ,
\quad \quad  
v_{\sig} = v_{F} \sqrt{1 - \frac{Ua}{\pi v_{F}}} \, .
\label{3.12}
\eeqa
From the diagonal form of the transformed hamiltonian  
Eq.\ (\ref{3.11}), it is clear that the low-energy excitations 
of the Hubbard model are long-wavelength oscillations of the 
charge and spin density. The excitations satisfy Bose 
statistics, and the different oscillations move with different 
velocities. This is an example of spin-charge separation, though   
note that corrections due to a non-linear dispersion introduce 
a coupling between the bosonic modes (Matveenko and Brazovskii 1994).

\subsection{\label{old3.2.3}Comparison with the exact Bethe ansatz solution}

The 1D Hubbard model was solved exactly by 
Lieb and Wu (1968) using the Bethe  ansatz.
The ansatz, which was used initially by Bethe (1931)
for a Heisenberg chain of interacting spins,  
postulates an asymptotic form for the 
two-particle scattering wavefunction, which turns out to   
be exact at all separations in 1D systems. The 1D 
Hubbard model thus provides an important testing ground for 
the results of bosonization, as the results by 
bosonization may be compared with the exact solution. 

The fact that the 1D Hubbard model has been solved exactly 
does not reduce the bosonization solution to an academic 
exercise. The complexity of the Bethe ansatz solution limits 
its practical use in calculating low-energy properties. 
It can be used to calculate the ground-state 
energy, excitation spectrum, and some thermodynamic 
properties. However, the complexity of the ground-state 
wavefunction has so far prevented any calculation of  
correlation functions, at least at finite parameter values
\footnote{Within the framework of exact solutions, an extension of the
Bethe ansatz to the quantum inverse scattering method 
(Korepin, Bogoliubov and Izergin 1993) allows correlations
functions to be determined.}.
Bosonization is useful in filling this gap; 
%%as illustrated in section \ref{old3.2.2}, 
a tractable method for calculating correlation functions 
is to use Bose representations. 

To compare the bosonization solution with the exact solution, 
it is convenient to focus on the low-energy excitations, for  
which both solutions make clear predictions. (In the following  
the filling $n < 1$ is assumed. For a comparison of the bosonization
results with the Bethe ansatz solution for the half filled band, see 
Gul\'{a}csi and Bedell (1994).)
At low-energy the Bethe ansatz solution gives two elementary 
excitations for the 1D Hubbard model; holons and spinons. 
The holons carry charge but not spin, while the spinons 
possess spin but do not carry charge. In limiting cases 
the elementary excitations have the dispersion 
\beqa
\varepsilon^{h}(k) = 
\left\{ \begin{array}{lcc}
4t\cos[ka/2] - 2t\cos[k_{F}a] & 
\quad \, & {\rm when}\, \,  U/t \ll 1 \, , \\
2t\cos[ka] & \quad \, & {\rm when}\, \, U/t \gg 1 \, ,
\end{array} \right. 
\label{3.3.1a}
\eeqa 
for the holons, and 
\beqa
\varepsilon^{s}(k) = 
\left\{ \begin{array}{lc}
2t(\cos[ka] - \cos[k_{F}a]) & {\rm when}\, \,  
U/t \ll 1 \, , \\
(2\pi t^{2}/U)\left(n - \frac{\sin[2\pi n]}{2\pi}\right)
\cos[ka/n] & {\rm when}\, \, U/t \gg 1 \, ,
\end{array} \right. 
\label{3.3.1b}
\eeqa 
for the spinons.

The holons and spinons are related to the charge and spin 
boson excitations, respectively, in the bosonization 
solution Eq.\ (\ref{3.11}). The spin velocity $v_{\sig}$ 
is obtained directly from the exact solution as the velocity 
of long-wavelength spin waves (Coll 1974, Schulz 1990). 
The identification of the charge velocity is more subtle. 
The first thing to note is that the $4k_{F}$ part of the 
density-density correlation function from the bosonization 
solution involves only 
charge degrees of freedom, and shows a power-law decay 
indicative of gapless excitations. Now, from the exact 
solution there are known to be gapless 
excitations at $4k_{F}$ which, somewhat misleadingly, are 
called `particle-hole' excitations (Coll 1974).   
The charge velocity in bosonization is 
then identified with the velocity of these  
`particle-hole' excitations (Schulz 1990, 1991). 
At weak-coupling 
(i.e. to leading order in $U/t$), these identifications recover 
the bosonization results Eqs.\ (\ref{3.12}) for the velocities. 

When the coupling strength $U$ is increased, it is expected 
that the bandwidth cut-off $k_{0}$ (cf.\ section \ref{old2.1}) will 
increase, until eventually $k_{0} = k_{F}$ and no core of 
fully occupied states remains. This is exhibited explicitly 
by Ogata and Shiba (1990). In this case  
Tomonaga's theorem gives a rigorous upper bound on bosonic 
density fluctuations at the wave vector 
$2(k_{F} - k_{0})/3 = 0$; in a rigorous sense no bosonic 
fluctuations remain. Schulz (1990) showed that the 
long-wavelength density fluctuations remain qualitatively 
bosonic even in the strong-coupling regime, and thus that 
bosonization continues to provide a qualitatively complete 
description of the low-energy properties of the 1D Hubbard 
model as $U$ is increased to large values. To achieve this,  
Schulz (1990) used results by methods beyond those of bosonization, 
in particular renormalization group and Bethe ansatz 
results. The argument is as follows: (i) In the small $U$ 
perturbative regime, 
interactions renormalize to the weak-coupling fixed-point 
described by the bosonization. 
(ii) The Bethe ansatz solution 
does not show any singular behaviour as $U$ is increased; 
the small $U$ and large $U$ regimes are in the same phase. 
It follows that the low-energy excitations of the model 
are still described by the charge and spin bosons 
of the bosonization solution, even at strong-coupling. 
What changes are the values of the charge and spin 
velocities. Schulz (1990) was able to determine 
these at arbitrary coupling strengths by solving an integral 
equation in the Bethe ansatz. The velocities take the 
values as in Eq.\ (\ref{3.12}) at weak-coupling, but are 
altered as the coupling is increased. For later purposes, 
it is worth noting that as the coupling strength is 
increased, the spin velocity does not go complex, as 
naively suggested by the bosonization result, but goes 
smoothly to zero as $U \rightarrow \infty$. 

%%%%%%%%%%%%%%%%%%%%%%%%%%%%%%%%%%%%%%%%%%%
%% chapter 5
%%%%%%%%%%%%%%%%%%%%%%%%%%%%%%%%%%%%%%%%%%%%

\cleardoublepage
\chapter{\label{ch5}Bosonization Solution of the 1D Kondo Lattice}

Since the early numerical approaches by Troyer and W\"{u}rtz (1993)
and Tsunetsugu, Sigrist and Ueda (1993), the 
unsolved problem in the Kondo lattice has been the nature of 
the ferromagnetic-paramagnetic phase transition. In the 1D 
Kondo lattice with $J > 0$ there has been convincing evidence 
(cf.\ section \ref{old4.4.2})
that the unquenched localized spins order ferromagnetically 
at stronger coupling, and undergo a more or less sharp 
transition to an RKKY-type paramagnetic phase as the coupling  
is decreased. The transition occurs at a filling-dependent 
critical coupling $J_{c}$ in the weak to intermediate 
range; $0 < J_{c}/t \lesssim 4$. There also has been some  
evidence that a similar transition occurs for 
ferromagnetic coupling $J < 0$, both 
in 1D and in higher dimensions. The 
ferromagnetic-paramagnetic transition during the
early 90's has been 
accessed only by numerical simulations. The numerical work 
helps to identify where the transition occurs, and determines 
in rough outline some of 
the properties of the system on either side 
of the transition. Beyond this nothing was known of the nature 
of the transition, and the numerics were silent on the 
underlying physical processes which lead to the transition.  
The aim of this chapter is to derive the analytic description 
of the ferromagnetic-paramagnetic phase transition 
in the 1D Kondo lattice. The results presented in this 
chapter are mostly based on Honner and Gul\'{a}csi (1997a,1998b).
More details of the derivations can be found also in Honner 
and Gul\'{a}csi (1997b,1997c,1998a,1999). 

In the early 90's the existence of a large ferromagnetic phase in the 
$J > 0$ Kondo lattice (cf.\ Fig.\ \ref{oldfig4.2}) came as something 
of a surprise. Based on the interactions known to operate in 
single- and dilute-impurity Kondo 
models, and following an early and influential treatment by 
Doniach (1977), the Kondo lattice with an 
antiferromagnetic coupling has usually been discussed in 
terms of a competition between Kondo singlet formation, and the 
RKKY interaction. It is possible that this characterization 
is sufficient when the conduction band in the 
Kondo lattice is half-filled. (Note, however, 
the criticism of Varma (1984)). 
When the conduction band filling is below half, Doniach's 
characterization is certainly incomplete.
Neither Kondo singlet formation nor the RKKY interaction are 
sufficient to account for known properties of the lattice. 
The extensive region of ferromagnetism at stronger 
coupling cannot be explained in terms of RKKY, which 
operates at most at weak-coupling, nor can it be explained 
in terms of Kondo singlet formation, since Kondo singlets are 
magnetically inert; the localized spin is quenched by the 
conduction electrons and loses its magnetic properties. 
For several years after the ferromagnetic phase was first 
identified, the published papers remained silent on the 
physical mechanism responsible for the ordering 
(Sigrist, Tsunetsugu and Ueda 1991, Sigrist, Ueda, 
and Tsunetsugu 1992, Sigrist, {\it et al.} 1992b, 
Troyer and W\"{u}rtz 1993, Tsunetsugu, Sigrist and 
Ueda 1993, Moukouri and Caron 1995).  

For $J < 0$, the ferromagnetic phase was expected.
In this regime the Kondo lattice is an effective model for 
the manganese oxide perovskites (cf.\ section \ref{old4.2.1}), and the 
ferromagnetic metallic properties of these compounds led 
Zener (1951) to introduce the double-exchange interaction. 
For $J < 0$, the ferromagnetism is due to the 
excess of conduction  
electrons over localized spins. The conduction electrons hop 
to neighbouring vacant sites, and tend to preserve their spin as 
they hop (called coherent electron hopping). This 
tends to align the underlying localized spins, as discussed in 
detail in section \ref{old4.3.3}. The effective ferromagnetic interaction 
between the localized spins, which is mediated by coherently 
hopping electrons, was termed double-exchange by Zener (1951)
and is the mechanism responsible for 
ferromagnetism in the $J < 0$ Kondo lattice. 

It is clear from the derivation in section \ref{old4.3.3} that 
the ferromagnetism in the $J > 0$ Kondo lattice is due to 
double-exchange interaction, very similar to the $J < 0$ case. 
The only significant difference is that for $J > 0$ the 
double-exchange is tempered by residual on-site Kondo 
singlets, which remain non-magnetic and do not participate  
in the magnetic ordering. The  
double-exchange was first mentioned in relation to 
ferromagnetism in the partially-filled $J > 0$ Kondo 
lattice by Yanagisawa and Harigaya (1994),  
and then presented more 
systematically by comparison with the $J < 0$ Kondo lattice 
by Yanagisawa and Shimoi (1996). As mentioned earlier, 
independently, Honner and Gul\'{a}csi (1997a,1998b) incorporated 
double-exchange in a theory of the ferromagnetic-paramagnetic
transition  (described in this chapter), and 
this has formed the basis for interpreting later numerical 
simulations. 

To describe the ferromagnetic-paramagnetic transition in the 
1D Kondo lattice it is necessary to describe coherently 
hopping conduction electrons at stronger coupling. These  
electrons 
mediate the double-exchange ferromagnetic ordering of the 
localized spins. It is necessary also to describe nearly-free 
electrons at weak-coupling. Scattering processes at  
weak-coupling are restricted to states close to the 
non-interacting conduction electron Fermi points at 
$\pm k_{F}$ (cf.\ chapter \ref{ch1two}), and can give rise via 
backscattering to the $2k_{F}$ correlations in the localized 
spins. The basic idea behind the theory of this  
chapter is that coherent hopping at strong-coupling,  
together with the more standard 
nearly-free scattering at weak-coupling, can be described using 
the bosonization formalism derived in chapter \ref{ch2}. Before ploughing 
through the mathematics, it is useful to give some physical 
motivation as to why and exactly how this idea is expected to 
work. 

In chapter \ref{ch2} it was proved {\it a priori} that bosonization 
provides a complete description of 1D electron systems for 
which the interactions are not too strong. The measure of  
interaction strength is the deformation of the momentum 
distribution from its 
non-interacting step-function form (cf.\ sections \ref{old2.1} 
and \ref{old2.2} for the detailed specification). This is expected to be 
the case quite generally for small enough interaction 
couplings, and was first verified explicitly  
in the numerical simulation of Moukouri and Caron (1995) for 
the 1D Kondo lattice at small $J$, see also the review of
Shibata and Ueda (1999), for recent results, see McCulloch, 
{\it et al.} (2002). Thus bosonization will provide an 
accurate description of the nearly-free conduction electrons 
at weak-coupling, as is standard. 

Bosonization describes the conduction electrons in terms 
of the collective density fluctuations defined 
in Eq.\ (\ref{2.4.1}). These involve the coherent 
superposition of large numbers of particle-hole pairs. Although 
composed of fermions, the collective fluctuations satisfy 
the bosonic commutation relations of Eq.\ (\ref{tom}). As the 
coupling $|J|$ is increased, density fluctuations of ever 
decreasing wavelengths are required to describe the electrons. 
Based on the rigorous result Eq.\ (\ref{tom}), together 
with comparisons with results by other methods 
(cf.\ section \ref{old3.2.3}), it is expected that the density 
fluctuations will remain qualitatively bosonic down to 
wavelengths $\alpha$, where 
$\alpha = {\cal O}(k_{F}^{-1})$ is of the 
order of the average spacing between the conduction electrons. 
Below $\alpha$ the density fluctuations are single-electron
like, and cease to obey a bosonic algebra (cf.\ section 
\ref{old2.2.2}, especially Eq.\ (\ref{B.3})). 
In the bosonization formalism, the density fluctuations with 
wavelengths shorter than $\alpha$ are excluded by employing 
a cut-off function $\Lambda_{\alpha}(k)$ as in Eq.\ 
(\ref{2.3.4}). This is 
necessary in order for the Bose fields to obey simple c-number 
commutation relations, and is thus necessary for the 
derivation of Bose representations. The effect of using the 
Bose representation for a 
conduction electron site operator 
is then to keep the electron finitely 
delocalized over a length $\approx \alpha$; the non-bosonic 
density fluctuations over shorter wavelengths are 
excluded, and by Fourier analysis any wave packet constructed 
using the bosonic fluctuations will be spread in real space over 
a range $\Delta x \approx \alpha$. 

The basic physical idea (Honner and Gul\'{a}csi 1997a,1998b) 
is that the finite 
delocalization over $\alpha$ can describe coherent conduction 
electron hopping over the same range, and will generate a 
double-exchange ordering between the localized spins at 
stronger coupling. There are several reasons which lend  
credibility to this: (i) The double-exchange 
interaction is mediated by a single conduction 
electron;\footnote{In more complicated cases it is  
conceivable that double-exchange may be mediated by 
two or several conduction electrons. Double-exchange is 
not, however, a collective effect involving a large number of  
conduction electrons.} indeed the ferromagnetic phase becomes 
more robust as the conduction band filling is reduced. 
This sets the groundwork for the idea within the conceptual 
framework of bosonization. Bosonization makes a qualitative 
distinction between collective bosonic density fluctuations 
at wavelengths beyond $\alpha$, and single-electron type 
behaviour over smaller length scales. 
(ii) Each conduction electron has on average $1/n = N/N_{e}$ 
localized spins to screen, and in a simple  
picture of the double-exchange interaction, coherent conduction 
electron hopping will occur over a range $\approx a/n$, 
where $a$ is the lattice spacing. This is observed in 
numerical simulations with small numbers of conduction 
electrons (Sigrist, Ueda and Tsunetsugu 1992a, 
Zang, R\"{o}der, Bishop and Trugman 1997).  
Now $a/n$ is just the average inter-electron 
spacing, and this is fully consistent with its identification 
with the $\alpha$ of bosonization. This gives quantitative 
support for the identification of $\alpha$ with the range for 
coherent conduction electron hopping. 
(iii) A conduction electron represented using bosonization 
is spread over the range $\alpha$ at stronger coupling, and 
preserves its spin over this range. This is the 
the fundamental property required for double-exchange  
(cf.\ section \ref{old4.3.3}). Thus bosonization can reproduce 
coherent hopping (i.e.\ preservation of spin over several 
lattice sites), and in the correct regime of strong coupling.
These reasons make plausible the idea that the
bosonization of chapter \ref{ch2} 
can describe coherent conduction electron hopping, as 
well as the standard behaviour at weak-coupling. 
The final justification comes from a detailed implementation 
of the bosonization, which is contained in this chapter, and 
in chapter \ref{ch6}. 

A final comment is in order before proceeding to a chapter 
outline, and then the calculations themselves. The comment 
relates to the importance of finite $\alpha$ for bosonized 
conduction electrons in the partially-filled Kondo lattice. 
This contrasts with the situation in some of 
the standard systems to 
which bosonization is applied. (An example is given in 
section \ref{old3.2}.) In these systems $\alpha$ is still finite, 
but this does not lead to any qualitatively new behaviour. 
The reason is that the standard systems are single 
component systems. This means that the minimum length scale in 
the system is set by the average inter-particle spacing of the 
bosonized objects. The density fluctuations are at 
least qualitatively bosonic down to this scale (cf.\ section 
\ref{old3.2.3}), and in these cases $\alpha$ acts as a relatively harmless 
short-distance cut-off; it may have to be kept finite to avoid 
unwanted divergences, but may be made arbitrarily small without 
overlooking any important physical processes. 
In the two-component Kondo lattice, by contrast, the minimum 
scale is set by the lattice spacing $a$ between the localized 
spins. This is less than the average inter-conduction electron 
spacing for a partially-filled conduction band. In this 
case $\alpha$ for the bosonized conduction 
electrons is greater than the minimum length scale in the 
system; physical processes between neighbouring localized spins,  
which are mediated by single conduction electrons, 
will be overlooked if 
$\alpha$ is made arbitrarily small. This is the reason why 
double-exchange cannot be described within a Luttinger model 
bosonization. 

The organization of this chapter is as follows: Section 
\ref{old5.1} derives an effective hamiltonian for the localized spins in 
the 1D Kondo lattice in several steps. In section \ref{old5.1.1} 
the formulae of chapter \ref{ch2} are used to bosonize the conduction 
band, and the Kondo lattice hamiltonian is written in terms of 
conduction electron Bose fields. In section \ref{old5.1.2} a 
canonical transformation is applied to the bosonized hamiltonian. 
This rewrites the hamiltonian in terms of a basis of states in 
which the ordering induced on the localized spins is more 
clearly apparent. A crucial term in the transformed hamiltonian 
is a ferromagnetic coupling term between nearby localized spins. 
Section \ref{old5.1.3} discusses this term in detail, and shows that its 
properties are fully consistent with a double-exchange ordering. 
This provides a formal verification of the physical ideas, 
discussed above, which underlie the use of the bosonization. 
In section \ref{old5.1.4} an effective hamiltonian for 
the localized spins is obtained by taking expectation values 
for the conduction electron Bose fields. 
Section \ref{old5.2} analyses the phase transition region of the 
effective hamiltonian, in which it reduces to a transverse-field 
Ising chain.\footnote{The similarities between the transition in  
the Kondo lattice model, and that in the transverse-field Ising 
chain, have been investigated by 
Juozapavicius, Caprara and Rosengren (1997)  
on the basis of this mapping.} 
Section \ref{old5.2.1} derives the generic 
ferromagnetic-paramagnetic transition which is  
determined by the effective 
hamiltonian. This fixes the ground-state phases of the 
1D Kondo lattice at partial band filling. 
To determine the properties of the localized 
spins near the transition, the effects of the particular form 
of the transverse-field are discussed in section \ref{old5.2.2}. These 
are shown to be reproduced by taking the transverse-field as a
random variable with an appropriate distribution. 
The properties of the localized spins near the transition are 
then listed in section \ref{old5.2.3} on the basis of known results for 
the random transverse-field Ising chain.

\section{\label{old5.1}Effective Hamiltonian for the Localized Spins}

\subsection{\label{old5.1.1}Bosonized Kondo lattice hamiltonian}

As discussed in chapters \ref{ch2} and \ref{ch3}, a large class of 1D 
many-electron systems may be described using bosonization 
techniques: The electron fields 
may be represented in terms of collective 
density operators which satisfy  
bosonic commutation relations. Bose  
representations provides a non-perturbative
description which, in general, is far easier to manipulate than
a formulation in terms of fermionic operators. In
the 1D Kondo lattice, the conduction band may be  
bosonized, but not the localized spins. This is because the 
spins are strictly localized, and their Fermi velocity 
vanishes. Moreover, since there is no direct interaction between 
the localized spins in the Kondo lattice, it is not possible to use 
bosonization via a direct Jordan-Wigner transformation. 
Only the conduction band is therefore bosonized.

The bosonized Kondo lattice is obtained by substituting the
Bose representations for lattice fermions (cf.\ section \ref{old2.4}) 
into $H_{\KL}$. Using Eqs.\ (\ref{Ho}), (\ref{Crjsigbose}) and 
(\ref{diag}) in Eq.\ (\ref{4.11}) and collecting terms gives  
bosonized 1D Kondo lattice hamiltonian 
(Honner and Gul\'{a}csi 1997a,1998b): 
\beqa
H_{\KL} &=& \frac{v_{F}a}{4\pi}\sum_{j,\nu}
\left\{ \Pi_{\nu}^{2}(j) + [\partial_{x}\phi_{\nu}(j)]^{2}
\right\}
 + \frac{Ja}{2\pi}\sum_{j}
[\partial_{x}\phi_{\sig}(j)] S_{j}^{z}
\nonumber \\
&+& A\frac{Ja}{2\alpha}\sum_{j}\left\{
\cos [\phi_{\sig}(j)] + \cos[2k_{F}ja + \phi_{\rho}(j)]\right\}
\left(e^{-i\theta_{\sig}(j)}S_{j}^{+} + {\rm h.c.}\right) 
\nonumber \\
&-& A\frac{Ja}{\alpha} \sum_{j}\sin[\phi_{\sig}(j)]
\sin[2k_{F}ja + \phi_{\rho}(j)]S_{j}^{z}\, , 
\label{bklm}
\eeqa
where the Bose fields are defined in Eqs.\ (\ref{bosefields}), 
and $A$ is a dimensionless constant depending on the cut-off 
function $\Lambda_{\alpha}(k)$.
Since the aim is to describe phase transitions, the thermodynamic 
limit $N \rightarrow \infty$, $n = N_{e}/N$ constant, has been 
assumed. The effects of fermionic ladder operators vanish in this 
limit (cf.\ the discussion following the bosonization 
of the Hubbard model in Eq.\ (\ref{3.6}), and Voit (1994), 
and they have been dropped from Eq.\ (\ref{bklm}). 

With the same ease, any Kondo lattice type hamiltonian 
or extended Kondo lattice models, see sections 
\ref{old5.4} - \ref{7.2} for examples, can be bosonized. 
For example, the bosonized hamiltonian of the 
spin-anisotropic case can be written as 
\beqa
H_{\AKL} &=& \frac{v_{F}a}{4\pi}\sum_{j,\nu}
\left\{ \Pi_{\nu}^{2}(j) + [\partial_{x}\phi_{\nu}(j)]^{2}
\right\}
 + \frac{J_{z} a}{2 \pi}\sum_{j}
[\partial_{x}\phi_{\sig}(j)] S_{j}^{z}
\nonumber \\
&+& A\frac{J_{\perp} a}{2 \alpha}\sum_{j} 
\cos [\phi_{\sig}(j)] 
\left(e^{-i\theta_{\sig}(j)}S_{j}^{+} + {\rm h.c.}\right) 
\nonumber \\
&+& A\frac{J_{\perp} a}{2 \alpha}\sum_{j}
\cos[2k_{F}ja + \phi_{\rho}(j)] 
\left(e^{-i\theta_{\sig}(j)}S_{j}^{+} + {\rm h.c.}\right) 
\nonumber \\
&-& A\frac{J_{z} a}{\alpha} \sum_{j}\sin[\phi_{\sig}(j)]
\sin[2k_{F}ja + \phi_{\rho}(j)]S_{j}^{z}\, , 
\label{bklm-anisotropic}
\eeqa
This is the well-known form used by, eg, Zachar, Kivelson 
and Emery (1996), and Novais, {\it et al.} (2002b), except 
that in their usage the bosonic fields 
are renormalized with $\sqrt{2 \pi}$, and for the 
numerical constant $A$ the value $1 / \pi$ was used. 
(For an accurate determination of $A$ close to the critical 
line, see section \ref{old6.2}.) 

The Kondo lattice bosonized hamiltonian Eq.\ (\ref{bklm}), 
generates the same behaviour as the 
original hamiltonian provided the conduction electrons are not 
strongly localized. In particular, the bosonized hamiltonian 
does not directly describe the formation of localized Kondo 
singlets. At strong antiferromagnetic coupling the electrons 
localize, with each forming a Kondo singlet with the localized 
spin at the same site (cf.\ section \ref{old4.3.2}).  
The localized singlet formation is governed by the 
short-range properties of the spin-flip bilinears 
$c^{\dg}_{cj\sig}c^{}_{cj-\sig}$. However, the Bose 
representation for these terms, derived from Eq.\  
(\ref{Crjsigbose}), is reliable only at long-wavelengths
and describes the properties of a spin-flipped conduction 
electron only at large distances from the site of the scattering 
localized spin. 
This provides a good description at weaker couplings, as  
usual in the bosonization of 1D Fermi systems, but may be  
insufficient when the coupling is strong enough that the 
electron becomes trapped on-site by the localized spin. 

A second point to note about the bosonized hamiltonian concerns 
spin-rotation symmetry. $SU(2)$ symmetry is manifest in the  
original hamiltonian, Eq.\ (\ref{4.11}), for both the 
conduction electrons and the localized spins, but is obscured
in the bosonized version. This is due to 
the use of Abelian bosonization, which treats the 
conduction electron spin $z$ direction on a special footing, 
and breaks the $SU(2)$ electron spin-rotation 
symmetry down to $U(1)$. To see the effect of this,
note that the original Kondo lattice hamiltonian preserves both 
the total spin $S_{\mbox{\scriptsize{tot}}}$ as well as its 
$z$ component $S^{z}_{\mbox{\scriptsize{tot}}}$, and at
stronger coupling in the ferromagnetic phase may be
decoupled into subspaces with different values of 
$S^{z}_{\mbox{\scriptsize{tot}}}$. (See, for example, 
section \ref{old4.3.3}.)
Abelian bosonization effectively singles out the subspace
with maximal $S^{z}_{\mbox{\scriptsize{tot}}}$ in the
ferromagnetic phase (cf.\ Eq.\ (\ref{tbklm}) below). 
However, a non-Abelian bosonization (cf.\ Appendix \ref{appc})
will only confirm the results of this chapter 
and in particular Eq.\ (\ref{tbklm})
to a full ${\bf S}_{j} {\bf \cdot} {\bf S}_{j'}$  
interaction, for details see Eq.\ (\ref{appc8.2})
of Appendix \ref{appc}.

\subsection{\label{old5.1.2}Canonical transformation}

A simple and rigorous method for determining the ordering 
induced on the localized spins by the electrons is to choose a 
basis of states in which competing effects become more 
transparent. This is achieved by applying a canonical 
transformation which changes to a basis of states in which 
the conduction electron spin degrees of freedom are coupled 
directly to the localized spins. Consider the transformation
\beqa
\exp({\bf S})\, , \quad {\rm S}=i \frac{Ja}{2\pi v_{F}}
\sum_{j}\theta_{\sig}(j) \,S_{j}^{z}\, .
\label{5.1.1} 
\eeqa
Since $\theta_{\sig}(j)$ and $S^{z}_{j}$ are hermitian, 
S satisfies ${\rm S}^{\dg} = -{\rm S}$, and $\exp({\rm S})$ is 
unitary:
\beqa
\exp\left({\rm S}^{}\right)\, \exp\left({\rm S}^{\dg}\right)
= 1 = 
\exp\left({\rm S}^{\dg}\right)\, \exp\left({\rm S}^{}\right)\, .  
\label{5.1.2}
\eeqa
Transformed operators 
$\tilde{O} = e^{-{\rm S}}Oe^{{\rm S}}$ may be 
calculated using the standard commutator expansion  
Eq.\ (\ref{comseries}). For more details on the infinite 
order canonical transformations, see section \ref{sectionSW} 
and Appendix \ref{ch4sw}.

The localized spin operators commute with all the Bose  
fields, since the former are bilinear in $f$-electron operators
(cf.\ Eqs.\ (\ref{4.4}) and (\ref{4.4a})), 
while the latter are bilinear forms in conduction electron 
operators since they are based on the density fluctuations 
Eq.\ (\ref{2.4.1}).
It follows immediately that the transformation $e^{\rm S}$ 
does not affect the $S^{z}_{j}$ configuration:
\beqa
\tilde{S}^{z}_{j} = e^{-{\rm S}}S^{z}_{j}e^{\rm S} 
= S^{z}_{j} \, . 
\label{5.1.3}
\eeqa
This is most important, since it means that a ferromagnetic 
ordering along $z$ in the transformed basis corresponds to 
the same ordering in the original (physical) basis. 
Using the commutation relations Eq.\ (\ref{4.5}) for the 
localized spin operators, together with the first two of 
Eqs.\ (\ref{4.4}), it is straightforward to show that
\beqa
[S^{\pm}_{j}, S^{z}_{l}] = \mp\delta_{j, l}\, S^{\pm}_{j}\, .
\label{5.1.4}
\eeqa
Using Eq.\ (\ref{comseries}), it follows that 
$S^{\pm}_{j}$ transforms as
\beqa
\tilde{S}^{\pm}_{j} &=&  S^{\pm}_{j}\left\{ 
1 + \left(\mp i\frac{Ja}{2\pi v_{F}}\theta_{\sig}(j)\right) + 
\frac{1}{2!}\left(
\mp i\frac{Ja}{2\pi v_{F}}\theta_{\sig}(j)\right)^{2}
+ \cdots \right\}
\nonumber \\
&=& S^{\pm}_{j}
\exp\left\{\mp i \frac{Ja}{2\pi v_{F}} 
\theta_{\sig}(j)\right\}\, .
\label{5.1.5}
\eeqa
The transformation thus rotates the localized spins in the 
$xy$-plane depending on the on-site conduction electron spin 
current field.  

The spin current Bose field commutes with all the Bose fields, 
except for the spin number field $\phi_{\sig}(j)$ (cf.\ 
section \ref{old3.1.2}). From the lattice equivalent of Eq.\ (\ref{2.5.7}),
$[\phi_{\sig}(j),\theta_{\sig}(l)]$ is a c-number, and only the 
first two terms of the commutator 
series of Eq.\ (\ref{comseries}) are non-zero. Thus 
\beqa
\tilde{\phi}_{\sig}(j) &=& \phi_{\sig}(j) + K(j) \, , 
\nonumber \\
K(j) &=& i \frac{Ja}{2\pi v_{F}} \sum_{l}
[\phi_{\sig}(j), \theta_{\sig}(l)]\, S_{l}^{z} \, .
\label{Kj}
\eeqa
The commutator $[\phi_{\sig}(j),\theta_{\sig}(l)]$ is finite  
for $j - l$ arbitrarily large (cf.\ Fig.\ \ref{oldfig3.1}). It follows that 
the transformed spin number field $\tilde{\phi}_{\sig}(j)$ 
is highly non-local:
Since $[\phi_{\sig}(j),\theta_{\sig}(0)] \rightarrow 
{\rm sign}(j)\, i\pi$ for $ja \gg \alpha$ (cf.\ Table 3.1), 
$K(j)$ is a long-range object which essentially 
adds all the localized spins to the right of site $j$, 
and subtracts from that sum of all the localized spins to the 
left of $j$. This term is discussed further in section \ref{old5.2}.

Related to the transformation of the spin number Bose field is 
the transformation of the actual conduction electron spin. 
The electron spin at $j$ is proportional to 
$\partial_{x}\phi_{\sig}(j)$ (cf.\ section \ref{old2.4}). 
Using the lattice equivalents of Eqs.\ (\ref{2.5.9}) and 
(\ref{2.5.11}), only the first two terms of the commutator 
series Eq.\ (\ref{comseries}) are non-zero, and so 
\beqa
\widetilde{\partial_{x}\phi}_{\sig}(j) = 
\partial_{x}\phi_{\sig}(j) - 
i\frac{Ja}{2\pi v_{F}}\sum_{l}[\phi_{\sig}(j), \Pi_{\sig}(l)]\, 
S^{z}_{l}\, .
\label{tphi} 
\eeqa
The Bose field commutator $[\phi_{\sig}(j), \Pi_{\sig}(l)]$ is 
non-zero for $\vert j - l \vert \lesssim \alpha/a$ 
(cf.\ Fig \ref{oldfig3.2}), and 
the conduction electron spin at $j$ is transformed depending on 
the nearby $S^{z}$ configuration.

After some algebraic manipulation (Honner and Gul\'{a}csi 1997a,1998b), 
the above results give the 
transformed Kondo lattice hamiltonian
\beqa
\tilde{H} & = &  \frac{v_{F}a}{4\pi}\sum_{\nu,j}
\left\{ \Pi_{\nu}^{2}(j)
+ [\partial_{x}\phi_{\nu}(j)]^{2} \right\}
\nonumber \\
& & -  \frac{J^{2}a^{2}}{4\pi^{2}v_{F}}
\sum_{j,l}\left\{\int_{0}^{\infty}dk\,\cos[k(j-l)a]
\Lambda_{\alpha}^{2}(k)\right\}S_{j}^{z}S_{l}^{z}
\nonumber \\
& &  + A\frac{Ja}{2\alpha}\sum_{j}
\left\{ \cos [K(j)+\phi_{\sig}(j)]
+ \cos [2k_{F}ja+\phi_{\rho}(j)] \right\}
\nonumber \\
& & \quad \quad \quad \times 
\left( e^{-i (1+Ja/2\pi v_{F})
\theta_{\sig}(j)}S_{j}^{+} + {\rm{h.c.}}\right)
\nonumber \\
& &  - A\frac{Ja}{\alpha}  \sum_{j}
\sin [K(j)+\phi_{\sig}(j)]
\sin [2k_{F}ja + \phi_{\rho}(j)] S_{j}^{z}
\label{tbklm}
\eeqa
provided that the cut-off function is not too `soft': 
$\Lambda_{\alpha}^{m}(k) \approx \Lambda_{\alpha}(k), m=2,3,4.$
(Discrepancies near $|k| \approx \alpha^{-1}$ introduce 
negligible corrections.) Note that the canonical transformation 
has been carried out exactly up to infinite order and not perturbatively, 
i.e.\ there has been no artificial truncation of the commutator 
series of Eq.\ (\ref{comseries}). (The c-number Bose field commutators 
are essential for this. Other infinite order canonical transformations, 
have been presented in section \ref{sectionSW} 
and Appendix \ref{ch4sw}.) It follows that the transformed 
hamiltonian of Eq.\ (\ref{tbklm}) is identical to the bosonized 
hamiltonian, and is an exact rewriting of Eq.\ (\ref{bklm}) in 
terms of a new basis of states in which the conduction band 
and localized spins are interwoven. 

A transformation closely related to $\exp({\rm S})$ of 
Eq.\ (\ref{5.1.1}) was first used by Emery
and Kivelson (1992) in the single-impurity Kondo model  
This was later generalized by Zachar, 
Kivelson and Emery (1996) to the 1D Kondo lattice.  
The usage here is very different. The aim  
of Zachar, Kivelson and Emery (1996)
was to describe the conduction electrons, 
and the transformation is used to remove the spin current 
field $\theta_{\sig}(j)$ from the hamiltonian (see also
the comment at the end of section). 

Here the aim is to describe interactions between the localized 
spins in which the conduction electrons are the mediators. 
The form of the transformation is then chosen so as to make 
explicit a ferromagnetic  ordering of the localized spins. 
This effect was entirely missed in the previous work due to a 
Luttinger model bosonization (Zachar, Kivelson and Emery 1996).  
The further factor $Ja/2\pi v_{F}$ in Eq.\ (\ref{5.1.1}) 
is chosen so that terms of the form 
$[\partial_{x}\phi_{\sig}(j)]S_{j}^{z}$ exactly cancel 
in the transformed hamiltonian. This permits the ground-state 
$S_{j}^{z}$ configuration to be chosen independent of the 
on-site conduction electron spin density in the transformed 
basis. 

It should be mentioned that not all the
results derived in Zachar, Kivelson and Emery 
(1996) are correct (Honner and Gul\'{a}csi 1997a,1998b). 
The problem is that the canonical transformation is 
carried out incorrectly; instead of the 
term $\sum_{l}{\rm sign}(j-l)S^{z}_{l}$, which is Zachar, 
Kivelson, and Emery's (1996) analogue of the long-range 
object $K(j)$ (cf.\ Eq.\ (\ref{Kj})), they 
obtain the integer $j$. This gives $(-1)^{j}$ factors in 
their transformed hamiltonian, and these factors are central to 
their later results. What appears to have happened is that they 
confused the long-range object analogous to $K(j)$ 
with the disorder term in the 
Jordan-Wigner transformation (cf.\ Appendix \ref{appb}), 
and treated it as a soliton-type 
term. However, $K(j)$ counts the spins on {\it both} 
sides of $j$, not just on one side, and the factors 
$(-1)^{j}$ are then spurious. The correct treatment of 
$K(j)$ is given in section \ref{old5.2}. 

\subsection{\label{old5.1.3}Double-exchange ordering}

The important new term in the transformed hamiltonian 
Eq.\ (\ref{tbklm}) is the second:
\beqa
-  \frac{J^{2}a^{2}}{4\pi^{2}v_{F}}
\sum_{j,l}\left\{\int_{0}^{\infty}dk\,\cos[k(j-l)a]
\Lambda_{\alpha}^{2}(k)\right\}S_{j}^{z}S_{l}^{z}\, .
\label{FMterm}
\eeqa
It represents a non-perturbative effective interaction 
between the localized spins, and is the only non-perturbative
interaction to be derived for the Kondo lattice; other 
effective interactions, namely the RKKY interaction 
at weak-coupling (Ruderman and Kittel 1954,
Kasuya 1956, Yosida 1957), and the strong-coupling effective 
interaction of Sigrist {\it et al.} (1992b), are both 
perturbative. Ferromagnetism in the Kondo lattice is 
considered in some detail in this section. First the properties 
of the interaction Eq.\ (\ref{FMterm}) are 
analysed, and a description is given of 
how it arises from the bosonization of the 
conduction band. Second, the properties of the double-exchange 
interaction in the Kondo lattice are briefly presented, 
mainly on the basis of Anderson and Hasegawa's (1955) semiclassical 
analysis. The interaction described by Eq.\ (\ref{FMterm}) shares 
these properties, and is identified  as the double-exchange 
interaction in the Kondo lattice. 

\begin{figure}[htb]
\centering
\includegraphics[scale=0.7,angle=-90]{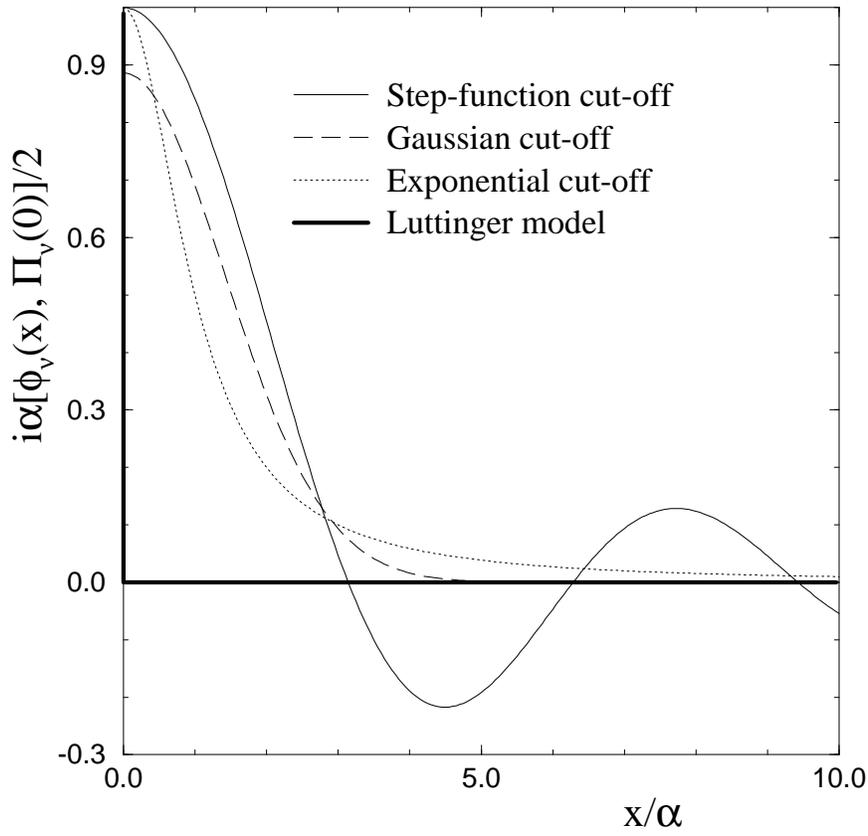}
\vspace{3.0mm}
\caption{\label{oldfig3.2}The Bose field commutator 
$i\alpha [\phi_{\nu}(x), \Pi_{\nu}(0)]/2$ 
for different choices of the cut-off function 
$\Lambda_{\alpha}(k)$, from Table 3.1. 
The commutator is scaled with 
$\alpha$ for the 
three choices of cut-off function. The Luttinger model 
$\delta$-function 
for the commutator is shown for comparison, and is unscaled.}
\end{figure}

\begin{figure}[ht]
\centering
\includegraphics[scale=0.7,angle=-90]{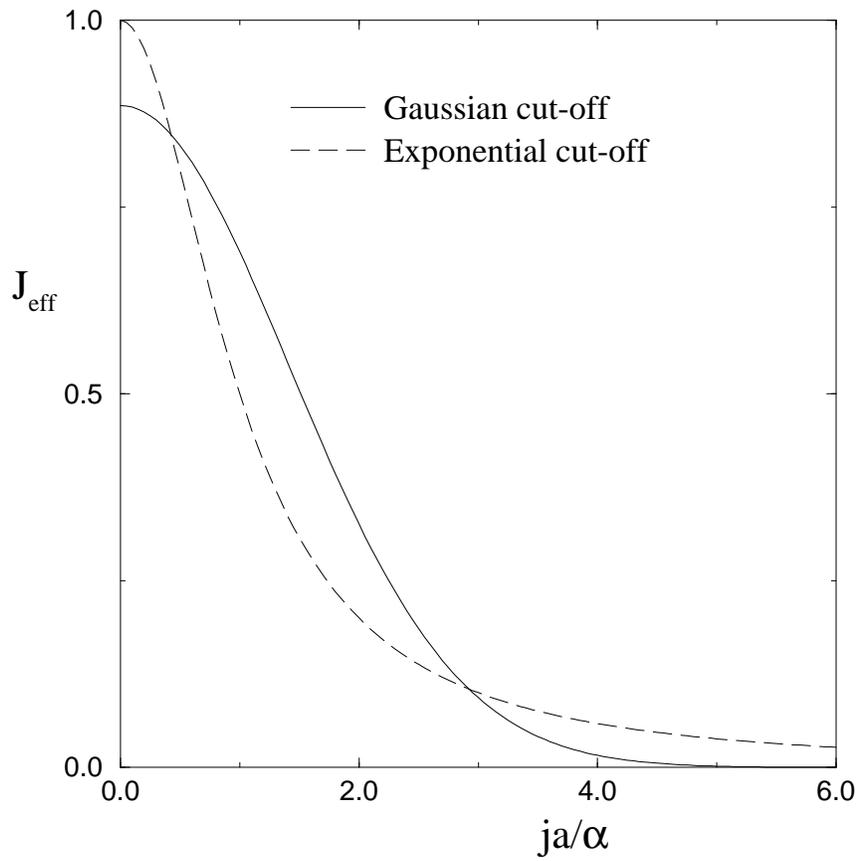}
\vspace{3.0mm}
\caption{\label{oldfig5.1}The range in real space of the ferromagnetic 
interaction Eq.\ (\ref{FMterm}) for exponential 
$\exp -(\alpha|k|/2)$ and Gaussian 
$\exp -(\alpha^{2} k^{2}/2)$ cut-off 
functions $\Lambda_{\alpha}(k)$. 
$J_{\eff}$ is the interaction strength 
in units of $\alpha J^{2}a^{2}/4\pi^{2}v_{F}$.}
\end{figure}

Eq.\ (\ref{FMterm}) possesses the following properties: 
(i) the term originates, via bosonization and 
then the canonical transformation, from the terms $H_{0}$ 
and the forward scattering part of 
$(J/2) \sum_{j} (n_{j\uparrow} - n_{j\downarrow}) 
S^{z}_{j}$ ($n_{j\sig} = c^{\dg}_{j\sig}c^{}_{j\sig}$) in 
the Kondo lattice hamiltonian. (Note that the Bose 
representations for the electrons in these terms are exact.) 
(ii) Eq.\ (\ref{FMterm}) is 
independent of the sign of $J$, and takes the same form  
for any magnitude $S_{j}^{2}$ of the localized spins. 
(iii) Since Eq.\ (\ref{FMterm}) is of order $J^{2}$, 
whereas the remaining terms in the transformed hamiltonian 
Eq.\ (\ref{tbklm}) are of order $J$, the interaction    
Eq.\ (\ref{FMterm}) dominates the ordering of the localized 
spins as $J$ increases.  
(iv) Eq.\ (\ref{FMterm}) is ferromagnetic for all 
(differentiable) choices of the cut-off function 
$\Lambda_{\alpha}(k)$.\footnote{For the step function cut-off 
$\Lambda_{\alpha}(k) = \theta(\alpha^{-1} - |k|)$, Eq.\ 
(\ref{FMterm}) gives a ferromagnetic interaction at 
short-range, similar to the interaction determined by 
the continuous cut-off functions. However, due to interference 
set up by the discontinuity, the step-function cut-off 
determines a weak oscillating interaction at longer range 
(cf.\ Fig.\ \ref{oldfig3.2}). This is not shared by the smoother 
cut-off functions. As discussed in detail in sections \ref{old2.2.2} 
and \ref{old3.1.2}, a discontinuously sharp cut-off on bosonic density 
fluctuations is unlikely in a real system, and the long 
range oscillations are artefacts resulting  
from a singular cut-off function. 
Because of this, the step-function cut-off is 
not considered in this chapter, and attention is restricted 
to continuous cut-off functions.}
To give examples of the form of the ferromagnetic interaction 
determined by Eq.\ (\ref{FMterm}) in real space, consider 
Gaussian and exponential cut-off  
functions defined by 
\beqa
\Lambda_{\alpha}(k) = \left\{ 
\begin{array}{lc}
\exp(-\alpha^{2}k^{2}/2) & {\rm Gaussian,} \\
\exp(-\alpha|k|/2) & {\rm Exponential.} 
\end{array} \right. 
\label{cutoffs}
\eeqa
For these cut-off functions, the integral in 
Eq.\ (\ref{FMterm}) reduces to
\beqa
\int_{0}^{\infty} dk \,\cos(kja) \Lambda^{2}_{\alpha}(k) 
 = \left\{ 
\begin{array}{lc}
(\sqrt{\pi}/2\alpha) \exp-(ja/2\alpha)^{2} & {\rm Gaussian,} \\
\alpha/(\alpha^{2} + (ja)^{2}) & {\rm Exponential.} 
\end{array} \right. 
\label{integral}
\eeqa
The integral is positive
and non-negligible for $ja \lesssim \alpha$. 
The form of the ferromagnetic interaction for Gaussian and 
exponential cut-off functions is shown in Fig.\ \ref{oldfig5.1}. 
It is clear that the length $\alpha$ 
characterizes the effective range of the ferromagnetic 
interaction  determined by Eq.\ (\ref{FMterm}).

The interaction Eq.\ (\ref{FMterm}) 
originates from the bosonization of the  
conduction band as follows: 
At wavelengths beyond $\alpha$, the  
electrons are involved in collective density
fluctuations Eq.\ (\ref{2.4.1}). These fluctuations 
involve large numbers of electrons, and satisfy bosonic 
commutation relations Eq.\ (\ref{tom}). 
At wavelengths below $\alpha$, the density fluctuations are not  
collective, and do not satisfy bosonic 
commutation relations. Since bosonization describes 
fluctuations only over separations beyond $\alpha$, 
the bosonization description is equivalent to keeping the 
electrons 
finitely delocalized over $\alpha$, with the electrons 
preserving their spin over this range.
Eq.\ (\ref{FMterm}) is the ordering consequently 
induced on the localized spins by the finitely 
delocalized electrons, and arises formally from  
the Bose field commutator 
$[\phi_{\sig}(j), \Pi_{\sig}(l)]$. This commutator 
takes canonical $\delta$-function form in a Luttinger 
model bosonization (cf.\ Fig.\ \ref{oldfig3.2}), but is 
smeared over a range $\alpha \gtrsim {\cal O}(k_{F})^{-1}$ 
for a realistic conduction band, which has a finite Fermi 
sea (cf.\ section \ref{old2.2.2}). The smearing of the 
Bose field commutator reflects the 
inability of the Bose fields to distinguish separations 
below $\alpha$.

Turning now to briefly review previously known properties of 
the double-exchange interaction, recall from section 
\ref{old4.3.3} that double-exchange was proposed many years ago by 
Zener (1951) to explain ferromagnetic 
ordering in manganese oxide perovskites, 
and is of much current interest in relation to 
colossal magnetoresistance materials (see section 
\ref{old4.2.1}). The essential 
characteristic required for a system to exhibit 
double-exchange ordering is that the number of  
electrons $N_{e}$ be less than the number $N$ of localized 
spins. In this case, when the conduction electron 
hopping is turned on, the electrons gain energy by hopping 
to unoccupied sites, since they gain energy both by 
screening the unpaired localized spin, together with a 
gain in kinetic energy. Since electrons tend to 
preserve their spin as they hop, called coherent hopping 
(Zang, {\it et al.} 1997), this tends to align the 
underlying localized spins. This was demonstrated 
microscopically in section \ref{old4.3.3} for both ferromagnetic 
and antiferromagnetic coupling. The ferromagnetic 
coupling induced on the localized spins by coherently 
hopping conduction electrons is the 
double-exchange mechanism, and is generated by the 
conduction electron kinetic energy and 
the diagonal part of the on-site interaction between electrons 
and localized spins. Double-exchange is always 
ferromagnetic, and dominates 
at stronger couplings (Zener 1951). It 
is the physical basis of 
the ferromagnetism rigorously established in the 1D 
Kondo lattice  by Sigrist {\it et al.} (1991,1992b).  
Since double-exchange ordering 
requires only $N > N_{e}$ and a non-vanishing 
hopping, its existence (as opposed to other properties 
such as its effective range)
does not depend on the sign of $J$, 
nor on the magnitude of the localized spins. 
This is clear from the analysis of Anderson and 
Hasegawa (1955); their result does not
depend on the sign of $J$, except for some numerical 
prefactors, and is semiclassical with trivial 
quantum modifications when $J < 0$. In the full quantum 
treatment of section \ref{old4.3.3}, the only modification when 
$J > 0$ is a residual Kondo singlet formation. Since 
Kondo singlets are non-magnetic, they can be projected out 
without qualitatively affecting the double-exchange magnetic 
ordering (cf.\ Eq.\ (\ref{afmgsdeform})).

Properties (i)-(iv) above for the interaction described by 
Eq.\ (\ref{FMterm}) are identical to those summarized above 
for a double-exchange interaction. This  
leads to the identification of Eq.\ (\ref{FMterm}) with the 
double-exchange interaction in the partially-filled 
1D Kondo lattice (Honner and Gul\'{a}csi 1997a,1998b,1999). 
Coherent conduction electron hopping, 
which generates double-exchange ordering, 
is described in the bosonization by electrons 
finitely delocalized over lengths $\alpha$,  
and $\alpha$ measures the effective 
range of the double-exchange interaction, 
as in Fig.\ \ref{oldfig5.1}. 
Note from chapter \ref{ch2}, Eq.\ (\ref{alpha}), 
that $\alpha$ enters the bosonization description 
as an undetermined but finite length; from bosonization, we 
know only that $\alpha \gtrsim {\cal O}(k_{F})^{-1}$, and that 
in general $\alpha$ will be a function both of filling 
$n$ and coupling $J$. 
In chapter \ref{ch6}, $\alpha$ is determined 
in a special case by using the results of this chapter 
together with available numerical data. 

Since the interaction described by Eq.\ (\ref{FMterm}) 
is short-range for all finite $\alpha$ 
(correct for all finite $n$, cf.\ Fig.\ \ref{oldfig6.5}), 
it is approximated in the usual way by its nearest-neighbour form
$-{\cal J}\sum_{j} S^{z}_{j}S^{z}_{j+1}$, where 
\beqa
{\cal J} = \frac{J^{2}a^{2}}{2\pi^{2}v_{F}} 
\int_{0}^{\infty}dk\, \cos(ka) \Lambda^{2}_{\alpha}(k)\, .
\label{calJ}
\eeqa
The critical properties described in section \ref{old5.2} 
are not affected by this approximation.

\subsection{\label{old5.1.4}Effective hamiltonian}

The aim in this chapter is 
to use the transformed hamiltonian Eq.\ (\ref{tbklm}) 
to determine the ground-state properties of the  
localized spins in the partially-filled 1D Kondo lattice. 
The concentration on the localized spins is motivated 
by the results of numerical simulations. 
Simulations on large chains have been carried out 
on the partially-filled 1D Kondo lattice model using
quantum Monte Carlo (Troyer and W\"{u}rtz 1993), and using the 
density-matrix renormalization group 
(Moukouri and Caron 1995, Caprara and Rosengren 1997),
and the results have been 
summarized in section \ref{old4.4.2}. The results uniformly show that the
correlations between the localized spins are
much stronger than the correlations between the conduction 
electrons, and the ferromagnetic-paramagnetic 
transition is signalled 
by the crossover from ferromagnetic to incommensurate 
(generally $2k_{F}$) correlations in the structure factor 
Eq.\ (\ref{lsss}) of the 
localized spins. The corresponding electron correlations 
are observed to weakly track those of the 
localized spins, and the electron 
momentum distribution shows no dramatic change as the 
ferromagnetic-paramagnetic transition line is crossed 
(Moukouri and Caron 1995). The freezing 
of the electron spin degrees of freedom  
occurs only at very strong coupling deep in the 
ferromagnetic phase. 

An effective hamiltonian for the localized spins is  
obtained from Eq.\ (\ref{tbklm}) by taking appropriately 
chosen expectation values for the conduction electron Bose 
fields.  
Since the Bose fields enter only in the weak-coupling 
terms of order $J$ in Eq.\ (\ref{tbklm}), the 
Bose fields are approximated by their non-interacting $J=0$ 
expectation values:
\beqa
\langle \phi_{\nu}(j) \rangle_{0} = 
\langle \theta_{\sig}(j) \rangle_{0} = 0\, .   
\label{approx}
\eeqa
This holds for the charge 
density field $\phi_{\rho}(j)$, since at weak-coupling 
the charge structure factor is free 
electron-like (Troyer and W\"{u}rtz 1993). For the spin fields 
Eq.\ (\ref{approx}) follows from 
real-space renormalization group studies 
(Jullien, Fields and Doniach 1977), which 
show that the spin degrees of freedom of the 1D Kondo 
lattice flow to the non-interacting fixed point at 
weak-coupling.\footnote{This property is 
specific to the Kondo lattice (Jullien, Fields and Doniach 1977); 
for the single-impurity Kondo hamiltonian at zero temperature, 
the spin coupling renormalizes to infinity for all 
$J > 0$, as discussed in section \ref{old4.3.2}.} 
Note that Eq.\ (\ref{approx})  is 
further supported by a study of the 
1D Kondo lattice with $t$-$J$ interacting conduction electrons 
(Moukouri, Chen and Caron 1996).
Using a combination of exact diagonalization 
and the density-matrix renormalization group, 
the same ordering was observed to be induced 
on the localized spins as in the pure Kondo lattice
 of Eq.\ (\ref{4.11}). 
This confirms the insensitivity of the ordering 
to the details of the conduction electron behaviour. 
The transformed hamiltonian Eq.\ (\ref{tbklm}) now 
reduces to an effective hamiltonian for the localized spins: 
\beqa
H_{\eff} &=& -{\cal J}\sum_{j}S^{z}_{j}S^{z}_{j+1} 
\nonumber \\
&+& A\frac{Ja}{\alpha}\sum_{j} 
\{ \cos[K(j)] + \cos[2k_{F}ja]\}S^{x}_{j}
\nonumber \\
&-& A\frac{Ja}{\alpha}\sum_{j}
\sin[K(j)] \sin[2k_{F}ja]S^{z}_{j}\, . 
\label{Heff}
\eeqa

In sections \ref{old5.2} and \ref{old5.3}, 
the effective hamiltonian $H_{\eff}$ of 
Eq.\ (\ref{Heff}) 
is analysed to determine the ground-state properties 
of the localized spins as a function of conduction band 
filling $n$ and antiferromagnetic 
Kondo coupling strength $J > 0$.\footnote{There are minor 
changes in the derivation for ferromagnetic couplings $J < 0$. 
For clearness in the exposition, it is however convenient to fix 
the sign of $J$ as positive for the remainder of this chapter. 
It is straightforward to verify 
that all the results go through 
for $J < 0$ with trivial modifications. The phase diagram 
and ground-state properties for $J < 0$ are discussed in 
section \ref{old6.3}.} Since the 
ferromagnetic 
 double-exchange coupling ${\cal J}$ is of order $J^{2}$ 
(cf.\ Eq.\ (\ref{calJ})), it is immediate from 
Eq.\ (\ref{Heff}) that $H_{\eff}$ 
determines a ferromagnetic
 ordering for the localized spins at stronger 
couplings $J/t \gg 1$ for all fillings $n < 1$. 
It will become clear below that 
the ferromagnetic ordering is gradually destroyed as the 
coupling $J$ is 
lowered. The destruction of the ferromagnetic order is  
determined by the second term of Eq.\ (\ref{Heff}); 
the effective 
hamiltonian takes the form of a transverse-field Ising chain 
in the phase transition region, and the Kondo lattice undergoes a 
quantum ferromagnetic to paramagnetic transition at a 
filling dependent critical 
coupling $J_{c}$. The critical coupling is of the order of the 
hopping at 
most conduction band fillings. The determination of the 
ferromagnetic-paramagnetic transition, and a discussion of the 
properties of the localized spins near 
the phase boundary, are contained in section \ref{old5.2}. In 
section \ref{old5.3}, the weak-coupling regime $J/t \ll 1$ is discussed. 
At weak coupling the double-exchange ordering 
is ineffective, and $H_{\eff}$ reduces to a system of 
free localized spins in fields determined by 
conduction electron scattering (the last two terms of 
Eq.\ (\ref{Heff})). It is shown that $H_{\eff}$ 
determines dominant $2k_{F}$ (RKKY-like) correlations in the 
localized spins at weak-coupling.

\section{\label{old5.2}Ferromagnetic-Paramagnetic Transition}

The analysis of $H_{\eff}$ begins 
by evaluating the long-range object 
$K(j)$ in the strong-coupling ferromagnetic phase.
$K(j)$, given by Eq.\ (\ref{Kj}), originates 
with the canonical transformation S. It has some similarity 
to the disorder term in the Jordan-Wigner transformation 
(cf.\ Eq.\ (\ref{B.9}) in Appendix \ref{appb}), but 
instead of counting the $S^{z}_{j'}$ only over sites to the 
left of $j$, it counts also the $S^{z}_{j'}$ to the 
right.\footnote{A failure to realise this point may be 
the reason for the erroneous evaluation of the canonical 
transformation by Zachar, Kivelson and Emery (1996), 
see section \ref{old5.1.2}.} 
In the thermodynamic limit $N \rightarrow \infty$, and 
using the large $j - j'$ form 
$[\phi_{\sig}(j), \theta_{\sig}(j')] = 
{\rm sign}(j - j')\, i\pi$ for the Bose field commutator 
(cf.\ Table 3.1),  
Eq.\ (\ref{Kj}) may be written in the form 
\beqa
K(j) = \frac{Ja}{2v_{F}}\sum_{l=1}^{\infty} \epsilon_{j}(l)
\, ,
\label{Kj2}
\eeqa
where $\epsilon_{j}(l) = S^{z}_{j+l} - S^{z}_{j-l}$. 
$\epsilon_{j}(l)$ has  
possible values $0, \pm 1$ for large $l$. 
For small $j - j'$, the commutator
$[\phi_{\sig}(j), \theta_{\sig}(j')]$ grows smoothly from 
zero at $j = j'$, to  ${\rm sign}(j - j')\, i\pi$ 
at $(j - j) = {\cal O}(\alpha/a)$, as illustrated in 
Fig.\ \ref{oldfig3.1}. The exact form of the 
commutator at short-range depends on the choice of cut-off 
function $\Lambda_{\alpha}(k)$, but all that is required here  
is the general form: The effect of short-range corrections
to the Bose field commutator is just to allow  
$\epsilon_{j}(l)$ to take values between $-1$ and $1$ for 
$l \lesssim \alpha/a$. $K(j)$ is then similarly 
smoothed, and takes values between integral multiples 
of $Ja/2v_{F}$. 

By writing $K(j)$ in the form of 
Eq.\ (\ref{Kj2}), it is clear that $K(j)$ vanishes in the 
ferromagnetic  
phase in a thermodynamically large system. 
Indeed $K(j)$ will not be appreciable until the system is 
strongly disordered. It follows that any transition out of the 
ferromagnetic 
phase will be governed by the first two terms of the 
effective hamiltonian Eq.\ (\ref{Heff}). For convenience,
these terms are collected in the hamiltonian $H_{\crit}$ (with 
`crit' for critical):
\beqa
H_{\crit} = -{\cal J}\sum_{j}S^{z}_{j}S^{z}_{j+1} 
+ A\frac{Ja}{\alpha}\sum_{j} 
\{1 + \cos(2k_{F}ja)\}S^{x}_{j}\, .
\label{Hcrit}
\eeqa
$H_{\crit}$ is a quantum transverse-field Ising chain, and 
a discussion of its properties occupies the remainder of this 
section \ref{old5.2}.  A great deal is already known about the 
transverse-field Ising chain, and the following discussion is 
essentially a summary of known results as they relate to 
the transition in the Kondo lattice. Of particular importance, 
it is known that 
$H_{\crit}$ undergoes a quantum phase transition from a 
ferromagnetic phase, to a disordered paramagnetic phase. 
The transition is formally determined in section \ref{old5.2.1} 
below, but before 
proceeding it is perhaps useful to consider the physics of 
the ferromagnetic-paramagnetic transition in the Kondo lattice 
model. 

$H_{\crit}$   
describes the double-exchange ferromagnetic ordering being 
gradually destroyed as the coupling $J$ is lowered. 
The first term of $H_{\crit}$, which describes double-exchange, 
has been discussed extensively in section \ref{old5.1.3} above. The 
destruction of the double-exchange ordering may be 
understood physically as 
follows:  As the coupling $J$ is decreased the 
conduction electrons become less strongly bound to the localized 
spins, and tend to extend over spatial ranges beyond 
the effective range $\alpha$ for double-exchange ordering. 
Double-exchange becomes less effective, 
and regions of ordered localized spins begin to 
interfere as the conduction electrons extend. 
The interference leads to spin-flip processes, and are 
embodied in $H_{\crit}$ in the transverse-field (the second 
term of Eq.\ (\ref{Hcrit})). 
The transverse-field in 
$H_{\crit}$ includes two low-energy spin-flip processes by 
which the conduction electrons disorder the localized spins. 
One spin-flip process is backscattering, and is accompanied 
by a momentum transfer of $2k_{F}$ from the conduction 
electrons to the 
localized spins. Since the chain of localized spins 
will tend to order so as to reflect this transfer, the 
transverse-field corresponding to backscattering spin-flips 
is sinusoidal with modulation $2k_{F}$. The other low-energy 
spin-flip process in $H_{\crit}$ is forward scattering. 
This involves zero momentum 
transfer to the localized spins, and the 
corresponding transverse-field is a constant (i.e.\ 
has modulation zero). 

It will become clear in section \ref{old5.2.1}
that either forward or backscattering 
spin-flip processes separately 
are sufficient to destroy the ferromagnetic order, 
and bring on a ferromagnetic-paramagnetic 
phase transition in the Kondo lattice. However,  
for incommensurate conduction band filling, 
the backscattering spin-flip processes introduce a 
competing periodicity in the chain of localized spins. 
It turns out that this has non-trivial consequences for certain 
properties of the localized spins near the transition. 
In section \ref{old5.2.1} the ferromagnetic-paramagnetic transition 
is determined for arbitrary transverse-fields. Section \ref{old5.2.2} 
then compares the properties of the localized spins which are 
disordered 
due to forward scattering (constant transverse-field), with 
the properties in which the spin disorder is  due to 
backscattering (incommensurately modulated transverse-field). 
It is  pointed out that the special properties resulting from 
incommensurate backscattering are 
at least qualitatively reproduced 
by treating the full transverse-field in $H_{\crit}$ as 
a random variable with the appropriate (displaced cosine) 
distribution. A summary of the 
properties of the random transverse-field Ising chain, as they 
relate to the transition region in the Kondo lattice, is then 
given in section \ref{old5.2.3}.

\subsection{\label{old5.2.1}Critical line for the phase transition}

Using a Jordan-Wigner transformation to spinless 
fermions $c^{}_{j}, c^{\dg}_{j}$ as detailed in Appendix \ref{appb}, 
and in the thermodynamic limit, $H_{\crit}$ may be 
written 
\beqa
H_{\crit} = \sum_{j,l}\left\{ c^{\dg}_{j}A_{jl}c^{}_{l} 
+ \frac{1}{2}\left( c^{\dg}_{j}B_{jl}c^{\dg}_{l} + 
{\rm h.c.}\right) \right\} 
\label{matrixform}
\eeqa
to an additive constant, where $(A_{jl})$ and 
$(B_{jl})$ are real 
symmetric and antisymmetric matrices, respectively, 
with non-zero entries 
\beqa
A_{jj} &=& h_{j} \equiv A\frac{Ja}{\alpha}
\{1 + \cos(2k_{F}ja)\} 
\nonumber \\
A_{jj+1} &=& A_{j+1j} = B_{jj+1} = -B_{j+1j} 
= -{\cal J}/4\, .
\nonumber 
\eeqa
The quadratic form Eq.\ (\ref{matrixform}) may be 
diagonalized for any transverse-field $h_{j}$ by 
using the method of Lieb, Schultz and Mattis (1961). 
This gives 
\beqa
H_{\crit} = \sum_{k} \omega_{k} \eta^{\dg}_{k}
\eta^{}_{k} 
\label{Hcritdiag}
\eeqa
to an additive constant, where $\eta^{\dg}_{k}, 
\eta^{}_{k}$ are creation and annihilation operators for 
free spinless fermions, and 
where the energies $\omega^{2}_{k}$ are eigenvalues 
of the symmetric matrix $(A + B)(A - B)$. As the 
coupling $J$ is decreased, $H_{\crit}$ undergoes 
a quantum order-disorder transition from a 
ferromagnetic phase, to a quantum disordered paramagnetic phase, 
signalled by the breakdown of long-range 
correlations between the localized spins, 
and a continuously vanishing spontaneous 
magnetization. The critical line for the transition
is determined by the critical coupling $J_{c}$ which 
solves (Pfeuty 1979)
\beqa
{\cal J}^{N} - 2^{N}\prod_{j=1}^{N}h_{j} = 0 
\label{implicit}
\eeqa
as $N \rightarrow \infty$. The free energy of the localized 
spins becomes non-analytic 
at points satisfying Eq.\ (\ref{implicit}). 

The ferromagnetic-paramagnetic
transition at the coupling $J_{c}$ is generic to 
Ising chains with a transverse-field, 
and does not assume a 
particular form for the transverse-field $h_{j}$ 
(Pfeuty 1979). For example, 
considering only forward scattering spin-flip processes 
in the Kondo lattice, 
the transverse-field $h_{j} = AJa/\alpha$ is a 
constant. Solving Eq.\ (\ref{implicit}) with 
$h_{j} = AJa/\alpha$ gives the quantum 
critical line for the ferromagnetic-paramagnetic transition at
\beqa
\frac{J_{c}}{t} = 
\frac{8\pi^{2}A\sin(\pi n /2)}{\alpha \int_{0}^{\infty}
dk\, \cos(ka)\Lambda^{2}_{\alpha}(k)}\, .
\label{Jc}
\eeqa
A detailed discussion of the properties of the Ising chain with 
a constant transverse-field, which describes the 
ferromagnetic-paramagnetic
transition in the Kondo lattice 
 with backscattering neglected, is given 
by Pfeuty (1970). 
As a second example, considering only backscattering 
spin-flip processes in the Kondo lattice, the transverse-field 
$h_{j} = AJa \cos(2k_{F}ja)/\alpha$ is sinusoidal. In real 
heavy fermion materials, the number of available conduction 
electrons per localized spin will in general be irrational 
(Strong and Millis 1994). 
In this case the transverse-field due to 
backscattering has an incommensurate modulation $2k_{F}$ 
with respect to the underlying lattice of localized spins. 
Nonetheless a ferromagnetic-paramagnetic transition still occurs. 
As shown by Satija and Doria (1989), the solution of 
Eq.\ (\ref{implicit}) for incommensurately modulated 
transverse-fields yields a coupling $J_{c}$ as in 
Eq.\ (\ref{Jc}) for the constant transverse-field. Thus the 
critical line for the ferromagnetic-paramagnetic 
transition in the Kondo lattice with only
backscattering spin-flip processes coincides with the critical 
line for the transition 
in the Kondo lattice with only forward scattering.

\subsection{\label{old5.2.2}Effects of the form of the transverse-field} 

While the ferromagnetic-to-paramagnetic 
transition itself is largely independent 
of the details of the transverse-field,  
there are significant differences in the properties of the 
localized spins on either side of the transition depending on 
the particular form of the transverse-field. 
The differences are most clearly apparent in the wavefunctions 
corresponding to the free fermions $\eta_{k}$ of the 
diagonalized hamiltonian Eq.\ (\ref{Hcritdiag}). 
The wavefunctions are 
always extended, or Bloch-like, for a constant transverse-field
(Pfeuty 1970).
For an incommensurately modulated transverse-field, the behaviour 
of the wavefunctions is far more complex. The Ising chain 
with an incommensurate transverse-field has been studied 
extensively by Satija and Doria (1989), and Satija (1990,1994). 
The model is important as it has localized states in 1D, and thus 
provides a link to random systems (Satija 1994).
Numerical studies (Satija and Doria 1989) show 
that the wavefunctions corresponding to the free fermions 
$\eta_{k}$ are localized in the disordered paramagnetic 
phase, and undergo 
a spectral transition at the ferromagnetic-paramagnetic 
phase boundary.
In the ferromagnetic phase, the wavefunctions are 
self-similar and the eigenvalue spectrum forms a 
Cantor set. Since the correlation functions for the localized 
spins are determined by the wavefunctions, the Kondo lattice with 
backscattering possesses far different properties 
to the Kondo lattice with only forward scattering 
spin-flip processes, 
even though both undergo a ferromagnetic-paramagnetic 
transition. Differences in 
the eigenvalue spectrum $\omega_{k}$ of the diagonalized 
hamiltonian Eq.\ (\ref{Hcritdiag}) lead to a similar 
conclusion regarding thermodynamic properties; since the 
Kondo lattice model 
with incommensurate backscattering has a fractal eigenvalue 
spectrum, its thermodynamics are far different to the 
Kondo lattice 
with forward scattering, which has the more standard 
(cosine-type) eigenvalue spectrum (Pfeuty 1970). 

The situation becomes yet more complex when considering all 
possible low-energy spin-flip processes available to 
the conduction electrons. $H_{\crit}$ includes forward 
scattering with zero momentum transfer, and is represented by a 
constant transverse-field. $H_{\crit}$ also includes 
backscattering with an incommensurate momentum transfer 
$2k_{F}$, and is represented by a $2k_{F}$ 
sinusoidal transverse-field. $H_{\crit}$ does not include 
spin-flip interactions with momentum transfers at higher 
harmonics of $2k_{F}$: at $4k_{F}$, $6k_{F}$, and so on. The 
higher harmonics will arise in a bosonization treatment which 
includes non-linear corrections to the conduction electron 
dispersion relation (see section \ref{old2.3.3}). These 
corrections are  very weak compared with the forward 
and backscattering spin-flip processes, and it is usual to 
neglect them. However, the addition 
of (even weak) higher harmonics in $2k_{F}$ to the  
transverse-field $h_{j}$ will greatly alter the solution of Eq.\ 
(\ref{implicit}). Instead of one solution, there now occur 
an infinite number of solutions to Eq.\ (\ref{implicit}), 
and these occupy a finite region of the parameter 
space (Satija and Doria 1989). The series of solutions is 
reflected in numerical studies 
on the Ising chain with a transverse-field containing 
more than one incommensurate harmonic 
(Satija and Doria 1989, Satija 1990). 
The region of spectral transitions becomes broadened, and 
the wavefunctions 
corresponding to the free fermions $\eta_{k}$ of Eq.\ 
(\ref{Hcritdiag}) are observed to undergo a cascade of 
transitions between extended, critical, and 
localized behaviour. The transitions in the wavefunctions occupy 
a finite region of the parameter space, 
and coincide with the solutions of Eq.\ (\ref{implicit}) 
at which the free energy becomes  
non-analytic. While the region of spectral 
transitions becomes broadened, this is not the case for the 
magnetic transition. The ferromagnetic-paramagnetic phase
transition, signalled 
by the vanishing of long-range correlations between the 
localized spins, is observed to remain sharp (Satija 1990).  

The behaviour observed by Satija {\it et al.} in the 
numerical studies 
discussed above is qualitatively identical to the 
behaviour of the Ising chain with a random transverse-field. 
(Properties of the random transverse-field Ising chain are 
discussed extensively by Fisher (1992,1995), and are 
summarized in section \ref{old5.2.3}.) To see this identification, 
note that the central 
feature of Fisher's treatment of the random transverse-field 
Ising chain is that 
dilute regions of ferromagnetic order may survive into the 
paramagnetic phase, and 
similarly that dilute regions of disorder may continue into 
the ferromagnetic phase. This feature is at the heart of 
Fisher's (1995) results, and is shared by $H_{\crit}$ for 
incommensurate $k_{F}$: As discussed above, a broadened region 
of spectral transitions about the true 
ferromagnetic-paramagnetic transition occurs 
in the Kondo lattice with incommensurate conduction band 
filling. Thus 
there are small regions in the paramagnetic phase in which 
the localized spins exhibit behaviour normally associated with 
the ferromagnetic phase, and vice versa. To further pursue the 
identification between a random transverse-field, and 
that present in $H_{\crit}$ for the Kondo lattice, 
recall that the 
spectral transitions occur at points satisfying Eq.\ 
(\ref{implicit}) at which the free energy becomes non-analytic. 
There is an immediate identification between these non-analytic 
points, and the Griffiths (1969) singularities 
present in random models, in which 
thermodynamic quantities such as the magnetization become 
singular in a range of parameter space about the non-random 
transition. (See Fisher (1992,1995) for the Griffiths 
regions in the random transverse-field Ising chain.) 

The behaviour of $H_{\crit}$ for incommensurate conduction band 
filling admits of a natural physical interpretation. The 
conduction band does not share the periodicity of the lattice 
of localized spins, 
and is unable to either totally order or totally disorder the 
lattice as the ferromagnetic-paramagnetic transition is 
crossed. There remain dilute 
regions of double-exchange ordered 
localized spins into the paramagnetic phase, as 
only a quasi-commensurate fraction of the conduction electrons 
become weakly-bound, and become free to scatter along the chain, 
at the ferromagnetic-paramagnetic transition. 
The remaining ordered regions are 
dilute enough that no long range correlations remain, but 
their existence dominates the low-energy properties of the 
localized spins near the transition.   

These considerations above suggest (Honner and Gul\'{a}csi 
1997a,1998b) that the full transverse-field 
$h_{j} = AJa\{1 + \cos(2k_{F}ja)\}/\alpha$ of $H_{\crit}$ 
be taken as a random variable, so that $h_{j}$ is chosen from the 
displaced cosine distribution $\rho(h)dh$ where
\beqa
\rho(h) = \frac{\alpha}{\pi AJa}
\frac{1}{\sqrt{1- (\alpha h/AJa - 1)^{2}}}\, . 
\label{rhoh}
\eeqa
As discussed above, this treatment of $h_{j}$ does not 
alter the basic ferromagnetic-paramagnetic transition 
described by $H_{\crit}$, 
and thus is not needed in order to plot the phase diagrams 
of the Kondo lattice in chapter \ref{ch6}. However, it does 
account for the properties of the localized spins near the 
ferromagnetic-paramagnetic transition as observed by 
Satija and Doria (1989) in their numerical simulations. 
Note finally that at low conduction band filling 
the treatment of 
$h_{j}$ in Eq.\ (\ref{rhoh}) follows an analogous treatment 
in spin glass systems (cf.\  Honner and Gul\'{a}csi (1997a,1998b)). 

\subsection{\label{old5.2.3}Properties of the localized spins near 
criticality}

Results on the random transverse-field Ising chain 
may be obtained from Fisher (1992,1995), who uses 
an approximate real-space renormalization group 
analysis, which nonetheless yields 
asymptotically exact results at low-temperatures
near criticality. Following Fisher (1992,1995), the 
critical coupling for the Kondo lattice model is given by
\beqa
\frac{J_{c}}{t} = 
\frac{4\pi^{2}A\sin(\pi n /2)}{\alpha \int_{0}^{\infty}
dk\, \cos(ka)\Lambda^{2}_{\alpha}(k)}\, .
\label{Jctot}
\eeqa
The critical line thus retains the form 
of Eq.\ (\ref{Jc}) for forward or 
backscattering separately, but is down by a factor of 2 as 
both spin-flip disorder processes are included. 
It is convenient to measure deviations from criticality using  
Fisher's (1995) criterion , which for the distribution 
Eq.\ (\ref{rhoh}) gives 
\beqa
\delta = \left[ {\rm var}(\log h) \right]^{-1} 
\log \left\{ 
\frac{4\pi^{2} A t
\sin(\pi n/2)}{J\alpha\int_{0}^{\infty}dk\, 
\cos(ka)\Lambda_{\alpha}^{2}(k)} \right\} ,
\label{delta}
\eeqa
where the measure of randomness for the displaced cosine 
distribution is 
\beqa
{\rm var}(\log h) = \sum_{n=1}^{\infty} 
\left\{\frac{1}{n}
\frac{1.3. \cdots .(2n-1)}{2.4. \cdots .(2n)}
\sum_{m=1}^{2n-1}\frac{1}{m} \right\} 
- \log^{2}2\, . 
\label{5.29} 
\eeqa
$\delta = 0$ on the critical line, is positive 
in the disordered paramagnetic phase, and negative in the 
ferromagnetic phase. 

The distinctive feature of Fisher's renormalization group  
analysis is that it focuses on anomalous clusters of 
double-exchange ordered localized spins  
which survive for small $\delta$ into the paramagnetic 
phase, and similarly, rare disordered regions 
close to criticality in the ferromagnetic phase. These are due 
to the incommensurability of the conduction band 
filling with respect to the lattice of localized spins, 
and the consequent inability of the conduction 
band, as a single many-body entity, to either 
totally order or totally disorder the localized spins 
as the transition is crossed. It is the anomalous 
ordered (disordered) regions of localized spins in 
the paramagnetic (ferromagnetic) phase which are responsible 
for the
Griffiths singularities. Although these anomalous 
regions are very dilute, they dominate the 
low-energy properties of the spin chain. Thus, while 
typical correlations are much as in the 
constant transverse-field Ising chain, the 
measurable mean correlations are dominated by 
the anomalous regions, and consequently greatly alter the  
low-energy behaviour of the localized spins.

An important prediction of theory presented here 
of the phase transition in the Kondo lattice 
is that the spontaneous magnetization grows 
continuously from criticality into the ferromagnetic phase: 
For the random transverse-field (Fisher 1992,1995) 
\beqa
M_{0}(\delta) \sim (-\delta)^{\beta}\, ,
\quad \quad \delta < 0\, ,
\label{Mo}
\eeqa
where $\beta = (3 - \sqrt{5})/2 
\approx 0.38$.\footnote{For 
certain quasi-commensurate fillings, for 
example $n = 1/2$, the spontaneous magnetization still 
grows continuously, but the critical exponent may be 
reduced to its value $1/8$ as in the constant transverse-field 
Ising chain, see also Pfeuty (1970).} 
This prediction disagrees with numerical 
diagonalization results on small systems 
(Tsunetsugu, Sigrist and Ueda 1993), 
which see a discontinuous jump 
in $M_{0}$ at least at larger fillings, 
but note that regions 
of intermediate $M_{0}$ have been observed in  
related studies 
(Moukouri, Chen and Caron 1996), and in  
small systems at lower fillings 
(Tsunetsugu, Sigrist and Ueda 1993).  
Indeed a discontinuous jump in 
$M_{0}$ immediately above the transition 
seems difficult 
to understand in a thermodynamically large system, 
given that the ordering is due to 
double-exchange, and that the electron 
spin degrees of freedom are not frozen until deep 
into the ferromagnetic phase (Moukouri and Caron 1995).\footnote{It is 
interesting to note that some of the features of the transition 
in the random transverse-field Ising chain are loosely 
similar to those of first order transitions in random 
classical systems. See Fisher (1995) for more details.} 

Using Fisher's (1995) results, it useful to 
summarize the properties 
of the 1D Kondo lattice which are relevant to the transition  
region of small $\delta$. The 
mean spin-spin correlation function is defined by 
\beqa
{\overline C}(x) = 
{\overline{\langle S^{z}_{j}S^{z}_{j+x}\rangle}}\, ,
\label{5.30}
\eeqa
where the average is over $\rho(h)$, and where for 
convenience $x$ denotes a continuous and 
positive variable. ${\overline C}(x)$ is dominated  
by atypically large correlations and for small 
$|\delta|$ decays as   
\beqa
{\overline C}(x) \sim \left\{ 
\begin{array}{llc}
M_{0}^{2}(\delta) + {\rm const.}|\delta|^{2\beta}
(\xi/x)^{5/6}e^{-3(\pi x/\xi)^{1/3}}e^{-x/\xi} 
& x \gg \xi  & {\rm Ferromagnetic,} \\
x^{-\beta} & 
{\rm as}\,\, {x \rightarrow \infty} 
& {\rm Critical,} \\
\delta^{2\beta}
(\xi/x)^{5/6}e^{-(3/2)(2\pi^{2} x/\xi)^{1/3}}e^{-x/\xi} 
& x \gg \xi & {\rm Paramagnetic,} 
\end{array} \right. 
\nonumber 
\eeqa
where the 
correlation length $\xi \approx 1/\delta^{2}$. (For 
typical pairs of spins the correlation length 
$\xi \approx \delta^{-1}$; the exponent is the same as in 
the Ising chain with a constant transverse-field.) Note that 
${\overline C}(x)$ decays more rapidly to 
$M_{0}^{2}(\delta)$ in the ferromagnetic phase, than it decays to 
zero in the paramagnetic phase. 

It is important to remark, that the form of ${\overline C}(x)$
obtained above suggests that a Luttinger liquid like behaviour 
only appears on the critical line. Here, 
the flipping of the spin polarons (for details, see section \ref{old6.1}), 
which determines the quantum dynamics of the model, 
is exponentially slow at long length scales
and completely vanishes. This suggests that the
quantum fluctuations are asymptotically absent on the critical line,
and the model becomes classical (Gul\'{a}csi 1997b). No fluctuations
on the critical line means that each polaron will act as an either
fully ordered or disordered chain. This observation explains why
a standard bosonization {\it \'{a} la} Luttinger model (cf.\ Appendix
\ref{appa}) will not work for the Kondo lattice model. 

At low temperatures $T$, 
${\overline C}(x,T)$ decays exponentially at large 
distances with a correlation length 
\beqa
\xi_{T} \approx \left\{
\begin{array}{lc}
e^{2\Gamma_{T}|\delta|}/4\delta^{2} & 
\Gamma_{T}|\delta| \rightarrow \infty \quad 
{\rm Ferromagnetic,} \\ 
4\Gamma^{2}_{T}/\pi^{2} & {\rm Critical,} \\
(\delta^{2} + \pi/\Gamma_{T}^{2})^{-1} & 
\Gamma_{T}\delta \gg 1 \quad {\rm Paramagnetic,}
\end{array} \right.
\label{5.31} 
\eeqa
where $\Gamma_{T} = 
\log( {\rm max}\{{\cal J}, h_{j}\}/T)$ 
at fixed $J$ and $n$ close to the transition. In the 
ferromagnetic phase $\xi_{T}$ diverges as a continuously variable 
power law of $T$. The 
correlation lengths $\xi_{H}$ for the long-range 
exponential decay of the correlations 
${\overline C}(x, H)$ in small 
applied fields $H$ along $z$ have 
identical functional forms to those of $\xi_{T}$ above.  
In the ferromagnetic phase, $\xi_{H} \sim 
H^{-2|\delta|}$ as $H \rightarrow 0$. This reflects the 
development of long-range order, and shows a power 
law dependence on $H$. (See Fisher (1995) for more details).

The magnetization in small positive applied fields 
$H$ along the $z$ direction is obtained (Fisher 1995) 
using an exact critical scaling function. Close to 
the critical line this gives
\beqa
M(\delta ,H) \sim \left\{
\begin{array}{lc}
M_{0}(\delta)[1 + {\cal O}(\delta H^{2|\delta |}\log H)] 
& {\rm Ferromagnetic,} \\
|\log H |^{-\beta} & {\rm Critical,} \\
\delta^{1+\beta} H^{2\delta} |\log H | & {\rm Paramagnetic,} 
\end{array} \right. 
\label{M} 
\eeqa
at $T = 0$. Close to the transition in both phases the 
magnetization 
is highly singular. In the paramagnetic phase the 
magnetization has a power 
law singularity with a continuously variable exponent 
$2\delta$, and the linear susceptibility is 
infinite for a range of $\delta$ into the paramagnetic 
phase. The susceptibility remains infinite 
(with a continuously variable exponent) close to the 
transition into the ferromagnetic phase. The low temperature 
linear susceptibility $\chi(T)$ takes the form 
\beqa
\chi (T) \sim \left\{ 
\begin{array}{lc}
T^{2\delta -1}(-\delta)^{-2(1-\beta)} & {\rm Ferromagnetic,} \\
T^{-1} |\log T|^{2(1-\beta)} & {\rm Critical,} \\
\delta^{-4(1-\beta)}T^{2\delta - 1}(\log T)^{2} 
& {\rm Paramagnetic.} 
\end{array} \right.
\label{chiT}
\eeqa
$T \chi (T)$ vanishes as $T \rightarrow 0$ in the paramagnetic 
phase, and diverges as $T \rightarrow 0$ in the ferromagnetic 
phase. Note that the latter property was conjectured 
by Troyer and W\"{u}rtz (1993) on the basis of their quantum 
Monte Carlo results. The zero-field 
specific heat of the localized spins 
at low temperatures close to the 
transition is given by 
\beqa
C_{v}(T) \sim \left\{
\begin{array}{lc}
|\log T|^{-3} & {\rm Critical,} \\
|\delta|^{3} T^{2|\delta|}
[1 + {\cal O}(T)^{2|\delta|}] & 
{\rm Ferromagnetic, Paramagnetic.}
\end{array}
\right.
\nonumber
\eeqa
It is interesting to remark that the above equation,
similarly to Eqs.\ (\ref{M}) and (\ref{chiT}) give a 
paramagnetic to ferromagnetic phase transition with 
continuously variable exponent. The singularity at
the phase transition is of a power law,
see Eq.\ (\ref{M}), but with variable exponent. 
This is a rather different
behaviour than for conventional field theory models in
one dimension. The only other example where variable exponents
is known to exist is the two dimensional statistical
mechanical eight-vertex model, or the equivalent 
one dimensional quantum XYZ model (Baxter 1982). 

The exponent obtained in Eq.\ (\ref{M}) is $2 \delta$, where
$\delta$ is given in Eq.\ (\ref{delta}). From this equation 
it can be seen that
$\delta$ will depend on $\alpha$, $n$ and $J$. However, 
on the phase transition
curve, $\delta$ can be approximated as $\ln 2.5 \sin (\pi n) / J$
(Honner and Gul\'{a}csi 1997a), for details see also section \ref{old6.2}. 
Hence for low electron concentration, i.e., $n \ll 1$ the dominant
contribution comes from $J \rightarrow 0$, which means that 
$\delta \rightarrow \infty$, an infinite order transition, i.e.,  
a Kosterlitz-Thouless type. As we increase $n$ the order of the
transition decreases, and close to half-filling
both $n$ and $J$ are constants of the
same order, see e.g., Fig.\ \ref{Ian-kondo-PRB-fig1}, 
thus the exponent of the power law singularity is 
$2 \delta \approx 2$. Thus the phase transition will look
more or less as a second order transition. Indeed, for $n > 0.5$ the 
phase transition obtained via density-matrix renormalization group
is of second order, for details see section \ref{KLM-DMRG}.

%%%%%%%%%%%%%%%%%%%%%%%%%%%%%%%%%%%%%%%%%%%
%% chapter 6
%%%%%%%%%%%%%%%%%%%%%%%%%%%%%%%%%%%%%%%%%%%%

\cleardoublepage
\chapter{\label{ch6}Ground-State Phase Diagrams for the 1D Kondo Lattice}

In the previous chapter, the ground-state phases of the localized 
spins in the partially-filled 1D Kondo lattice were determined 
using a description of the conduction electrons  
based on the bosonization formalism of chapter \ref{ch2}. 
It was found that the system 
undergoes an order-disorder transition from a 
ferromagnetic phase at stronger 
coupling, to a quantum disordered paramagnetic phase at weaker 
coupling. The critical line for 
the transition is given by Eq. (\ref{Jctot}) for an
antiferromagnetic Kondo coupling $J > 0$. Using the 
methods of chapter \ref{ch5}, it is straightforward 
to verify that a very similar derivation goes through for 
a ferromagnetic coupling $J < 0$, and the same critical line of 
Eq.\ (\ref{Jctot}) is obtained for $|J_{c}|$, together with 
the same critical properties. 
That the ferromagnetic-paramagnetic transition is the same for 
either sign of the coupling is not surprising 
given the origin of the transition: Double-exchange 
ferromagnetism can be described semiclassically, as 
in Anderson and Hasegawa's (1955) original 
treatment, and its character is the same for 
either sign of the coupling. The disordered paramagnetic phase 
is likewise expected to be largely independent of the sign of the 
coupling, since in this phase the 
conduction electrons are weakly bound.  

The differences due to the sign of the coupling occur with 
the effective range $\alpha$ of the double-exchange 
interaction. Thus far it has only been assumed that 
$\alpha$ is large enough for the ferromagnetic interaction 
Eq.\ (\ref{FMterm}) 
to be non-negligible at a distance of one lattice 
spacing.\footnote{At partial conduction band filling  
this is assured by bosonization. In section \ref{old2.2.2}  
it was shown that $\alpha$ is limited by the inter-particle 
spacing of the conduction electrons. The inter-particle 
spacing is greater than a lattice spacing 
for a partially-filled conduction band.} Beyond this, 
$\alpha$ is not determined. Bosonization gives only the 
limits for $\alpha$, as in Eq.\ (\ref{alpha}), and as in 
chapter \ref{ch3} it is necessary to use methods beyond those of 
bosonization in order to determine $\alpha$ more precisely.
In this chapter, $\alpha$ will be determined for the 
Kondo lattice by using 
the critical line equation derived in chapter \ref{ch5}, 
together with 
numerically determined phase transition points. 
This will give information 
on the effective range of the double-exchange interaction 
in a thermodynamically large system, 
and will allow the phase diagrams to be plotted. 

This chapter is organized as follows: the first part 
contains the details of the Kondo lattice ground-state 
phase diagram, following the results of the previous 
chapter, but supplemented with a rigorous determination 
$\alpha$. The results presented in this  part 
are mostly  based on Honner and Gul\'{a}csi (1997a,1998b).
More details can be found also in Honner 
and Gul\'{a}csi (1997c,1999). 
Accordingly, in section \ref{old6.1} the 
rigorous result for the Kondo lattice with one conduction 
electron is recalled, in parallel to a simplified description of the 
Kondo lattice at small conduction band fillings. 
These arguments suggest a form $\alpha/a \propto \sqrt{t/J}$ 
at the ferromagnetic-paramagnetic transition. The 
constant of proportionality is evaluated using
numerically determined points, cf.\ 
Fig.\ \ref{oldfig6.1} for $J > 0$, and Fig.\ \ref{oldfig6.3} for $J < 0$.
Ground-state phase diagrams are then given in 
section in section \ref{old6.2} for $J > 0$ (cf.\ 
Fig.\ \ref{oldfig6.2}), and in section \ref{old6.3} for $J < 0$ 
(cf.\ Fig.\ \ref{oldfig6.4}). In section \ref{old6.4}, 
$\alpha$ is given as a function of conduction 
band filling, and the physical origin of the 
differences in $\alpha$ depending on the sign of the 
coupling is briefly discussed.

In the second part of this chapter, the numerical results
of a non-Abelian density-matrix renormalization group method
are presented. The numerical method is described in detail
by McCulloch and Gul\'{a}csi (2000,2001,2002), and as such,
just a brief description is enclosed. In contrast,
the obtained results (McCulloch, {\it et al.} 1999, 2001,2002)
concerning the Kondo lattice phase diagram are discussed in
full.

\section{\label{old6.1}Low Density Form for $\alpha$}

In principle, $\alpha$ is a function of both filling and 
coupling (cf.\ section \ref{old2.2.2}). At the ferromagnetic-paramagnetic 
phase transition, the critical line equation (\ref{Jctot}) 
renders either the filling or the coupling redundant, and 
$\alpha$ may be considered to be 
a function of only one of the Kondo lattice parameters, at 
least on the phase transition line. In the following, 
a simplified characterization of double-exchange is given 
at low conduction band filling Honner and Gul\'{a}csi 1998b). 
This determines a form 
for $\alpha$ as a function of the coupling $J$. 

In a simplified picture, double-exchange may be described 
by the Kondo lattice hamiltonian 
Eq.\ (\ref{4.11}) with spin-flip interactions ignored:
\beqa
H_{\doex} = -t\sum_{j}\left(c^{\dg}_{j\sig}c^{}_{j+1\sig}
 + {\rm h.c.} \right) 
 + J/2\sum_{j} (n_{j\uparrow} - n_{j\downarrow})S^{z}_{j}\, .
\label{doex}
\eeqa
The occupation of a site by an 
electron with the same spin as the localized 
spin costs an energy $J/2$. 
These states as excluded as a first approximation valid 
at stronger couplings. 
At low conduction band filling, consider a finitely 
delocalized electron of spin $\sig$ spread over 
sites $j$ for which the localized spins 
$S^{z}_{j}$ have spin $-\sig$ for $J > 0$ (or  
similarly spin $\sig$ if $J < 0$). From Eq.\ (\ref{doex}), 
the wavefunction $\psi_{\sig}(x)$ for the electron, in the 
continuum limit, satisfies the nonlinear Scr\"{o}dinger equation
\beqa
\partial^{2}_{x}\psi_{\sig}(x) 
+ (Jm_{e}/2)|\psi_{\sig}(x)|^{2} \psi_{\sig}(x) 
= 2 m_{e} E\psi_{\sig}(x)
\label{nls}
\eeqa
with $m_{e}$ the bare electron mass. 
The electron gains energy 
due to its occupation $|\psi_{\sig}(x)|^{2}$ of a point with 
localized spin $S^{z}_{x}$ of spin $-\sig$, and 
this generates the non-linearity. At low conduction band 
filling, the electrons form separate double-exchange 
ordered regions as shown by Sigrist, Ueda and Tsunetsugu (1992a)
and Zang, {\it et al.} (1997) for the Kondo lattice with two conduction 
electrons. It is appropriate therefore to look for  
finitely delocalized solutions of Eq.\ (\ref{nls}). 
These are the well-known soliton solutions (Makhankov 1989) 
and have wavefunctions 
\beqa
\psi_{\sig}(x) = B\, e^{ix}\,
{\rm sech}\left(B\sqrt{Jm_{e}/4}\,(x-x_{0})
\right)\, . 
\label{soliton}
\eeqa
where B and $x_{0}$ are constants. The simplified picture 
of the Kondo lattice at low conduction 
band filling is then of a gas of solitons
(Honner and Gul\'{a}csi 1998b,1999). The solitons may  
be pictured as spin polarons (Holstein 1961, Sigrist, Tsunetsugu 
and Ueda 1991), and describe the dressing of each 
electron by a cloud of localized spins which tend to 
align opposite 
to the conduction electron spin for $J > 0$ (and tend to align 
parallel to the electron spin for $J < 0$). 
The spatial extension of the polarization  
cloud, ie, the width of the spin polaron,  
characterizes the range of the indirect ferromagnetic ordering 
induced on the localized spins by the electron, 
and is equivalent to the effective range $\alpha$ of the 
double-exchange interaction as described in 
section \ref{old5.1.3}.\footnote{The spin polaron picture for the 
ferromagnetic ordering induced on the localized spins was used 
by Sigrist {\it et al.} (1991). It is 
equivalent to the double-exchange picture, and is perhaps 
more illustrative at low conduction band filling.} 

This spins polarons correspond to a dressing of the electron 
by a cloud of antiparallel local spins for $J > 0$, or by 
a cloud of parallel local spins for $J < 0$. This represents
a bound state of kink and antikink domain walls, eg, 
in the case of antiferromagnetic coupling ($J > 0$), of
a typical form  
$$\ldots \: \Uparrow \Uparrow \Uparrow \Uparrow
\: \Downarrow \Downarrow \Downarrow \:
\Uparrow \Uparrow \: \Downarrow \Downarrow
\Downarrow \Downarrow \: \ldots$$ 
Here, 
$\Uparrow \Uparrow \Uparrow \Uparrow$, 
$\Downarrow \Downarrow \Downarrow$, 
$\Uparrow \Uparrow$, $\Downarrow \Downarrow \Downarrow \Downarrow$, 
are the spin polarons, with average polaronic width $\alpha$. 
The short hand notation, eg, $\Uparrow \Uparrow \Uparrow \Uparrow$, 
represents a superposition of states. That is, 
$\Uparrow \Uparrow \Uparrow \Uparrow$ is a superpositions of the
following states
\begin{eqnarray}
&& \vert \Uparrow \downarrow, \: \Uparrow 0, \: \Uparrow 0, \: 
\Uparrow 0 \rangle \, ,
\nonumber \\
&& \vert \Uparrow 0, \: \Uparrow \downarrow, \: \Uparrow 0, \: 
\Uparrow 0 \rangle \, , 
\nonumber \\
&& \vert \Uparrow 0, \: \Uparrow 0, \: \Uparrow \downarrow, \: 
\Uparrow 0 \rangle \, , 
\nonumber \\ 
&& \vert \Uparrow 0, \: \Uparrow 0, \: \Uparrow 0, \: 
\Uparrow \downarrow \rangle \, , 
\label{marha}
\end{eqnarray}
where $\Uparrow$ and $\uparrow$ referrers to the impurity and conduction 
electron spins, respectively. 
The polaronic length scale competes with the length scale set by
the free conduction electron mean free path and introduces
simultaneously competing time scales: slow motion of the
polarons with low energy dynamics, and fast motion of the
free electrons (within the length scale $\alpha$, see Eq.\
(\ref{marha})) with high energies.

From Eq.\ (\ref{soliton}), the polarization cloud decays 
exponentially at large distances with a characteristic 
length scale proportional to $\sqrt{t/J}$. 
This gives a low density 
form $\alpha/a \propto \sqrt{t/J}$ which will be assumed 
in order to plot the ferromagnetic-paramagnetic 
transition lines in sections \ref{old6.2} and \ref{old6.3}. 

Note that at vanishingly small conduction band 
filling $n \rightarrow 0$, it is possible to determine 
$\alpha$ using the exact 
solution of Sigrist {\it et al.} (1991) for the 
Kondo lattice with one conduction electron. 
As summarized in section \ref{old4.4.2}, the polarization cloud  
decays exponentially for small $J$ in the ferromagnetic 
phase, with a characteristic length $\alpha/a = \sqrt{2t/J}$.
This gives rigorous support, at least as $n \rightarrow 0$, 
to the form identified for $\alpha$ above, which is 
based on the simplified treatment of double-exchange 
through $H_{\doex}$.

\section{\label{old6.2}Phase Diagram for the Antiferromagnetic Kondo Lattice}

The behaviour identified for the 1D Kondo lattice in 
chapter \ref{ch5}   
is in complete qualitative agreement with the results 
of numerical simulations on larger systems 
(Troyer and W\"{u}rtz 1993, Moukouri and Caron 1995, 
Caprara and Rosengren 1997, McCulloch, {\it et al.} 1999,2001,2002). 
To establish quantitative agreement,   
i.e.\ to plot the critical line,
two obstacles are presented. 
The critical line equation with both forward and 
backscattering spin-flip interactions included may be  
written
\beqa
J_{c}a = 
\frac{2\pi^{2}Av_{F}}{\alpha \int_{0}^{\infty}
dk\, \cos(ka)\Lambda^{2}_{\alpha}(k)}\, .
\label{fullJc}
\eeqa
The first obstacle in using Eq.\ (\ref{fullJc}) is
somewhat trivial, and relates 
to the global scaling of the critical line: the number $A$ 
comes from the normalization of the Bose representations 
for spin-flip and backscattering electron interactions. 
It depends significantly on the cut-off function 
$\Lambda_{\alpha}(k)$, and moreover relates to the 
normalization of Bose representations only in the limit 
of long wavelengths (cf. section \ref{old2.3} and 
Eq.\ (\ref{Crjsigbose})). 
From the bosonization developed in chapter \ref{ch2} 
(cf.\ Eq.\ (\ref{alpha})),   
$\alpha \gtrsim {\cal O}(k_{F}^{-1})$, and so  
$\alpha$ will diverge as the filling $n \rightarrow 0$.
From Eq.\ (\ref{fullJc}) it follows that 
\beqa
\frac{J_{c}}{t} \rightarrow 2\pi^{3}A\, n\, , \quad \,
{\rm as}\,\,\,  n \rightarrow 0\, .
\label{smalln}
\eeqa
Note the agreement with the 
exact solution of Sigrist {\it et al.} (1991) 
for the Kondo lattice with one conduction electron; 
the system is ferromagnetic for all 
finite $J$. Recall also that the exact solution gives 
$\alpha/a = \sqrt{2t/J}$ for small $J$. $\alpha$ 
diverges at criticality in agreement with the result above 
from bosonization as $n \rightarrow 0$.  
Using numerical results for $J_{c}$ 
at the smallest available filling ($J_{c}/t = 0.455$ 
at $n = 0.2$ from the infinite-size density-matrix 
renormalization group simulation 
of Caprara and Rosengren (1997)), it is concluded 
from Eq.\ (\ref{smalln}) that $2\pi^{2}A \approx 0.7$. This 
fixes the constant $A$ for $J > 0$.

\begin{figure}[ht]
\centering
\includegraphics[scale=0.7,angle=-90]{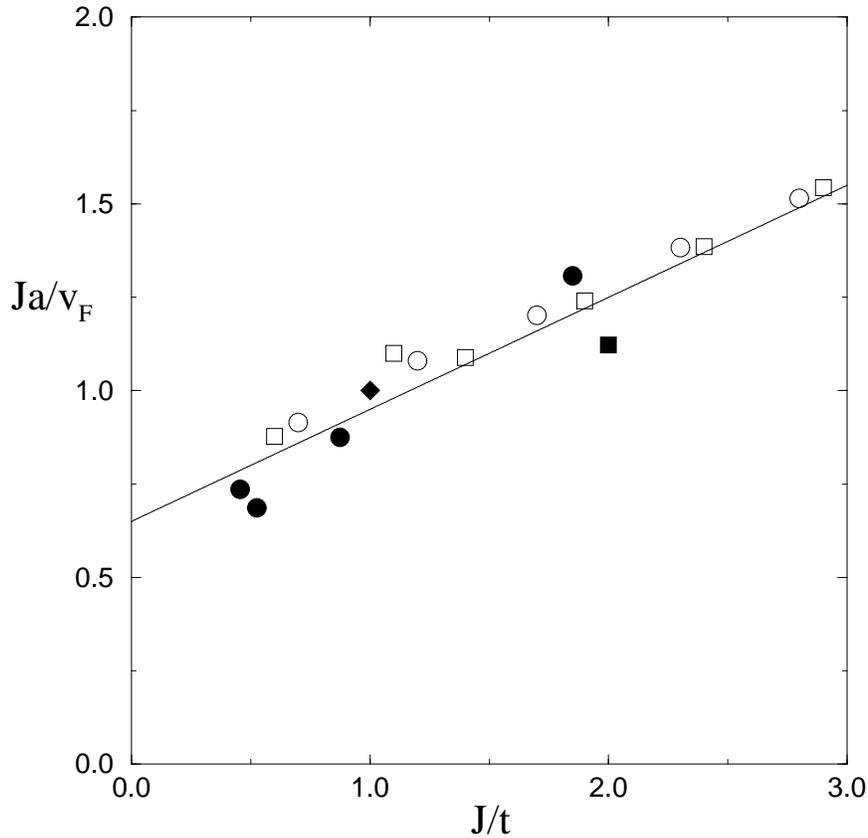}
\vspace{3.0mm}
\caption{\label{oldfig6.1}Plot of the dimensionless parameter $Ja/v_{F}$, 
which characterizes double-exchange ferromagnetism, against 
coupling $J > 0$ for numerically determined 
ferromagnetic-paramagnetic transition points $J_{c}$: 
the filled diamond is the quantum Monte Carlo
result for systems up to 24 sites from 
reference (Troyer and W\"{u}rtz 1993); open 
circles and squares are exact numerical diagonalization 
results for the 8 and 9 site chain, respectively, 
from reference (Tsunetsugu, Sigrist and Ueda 1993); 
the filled square is the density-matrix renormalization 
group result for systems up to 75 sites from 
reference (Moukouri and Caron 1995);  
the filled circles are infinite-size density-matrix 
renormalization group  results 
from reference (Caprara and Rosengren 1997). 
$J_{c}a/v_{F} \approx 0.7$ for vanishing $J$. 
The straight 
line of best fit is given, and shows good agreement 
with the spread of numerical results, together with 
the expected result as $J \rightarrow 0$.} 
\end{figure}

The second obstacle to using Eq.\ (\ref{fullJc}) to plot 
the critical line is less trivial, and relates to the 
dependence of $\alpha$ on $J$ and $n$. $\alpha$ measures 
the effective range of the double-exchange interaction 
(cf.\ section \ref{old5.1.3} and Fig.\ \ref{oldfig5.1}), and is a non-trivial 
quantity in a thermodynamically large system.
In section \ref{old6.1} the form $\alpha/a \propto \sqrt{t/J}$ 
was obtained at low conduction band filling on the 
critical line. $\alpha$
enters the critical line equation in the  
denominator of the right hand side of 
Eq.\ (\ref{fullJc}), and the proportionality constant 
may be determined by using numerically determined 
ferromagnetic-paramagnetic transition points. For Honner and
Gul\'{a}csi (1998b) the available numerical data 
(Troyer and W\"{u}rtz 1993, Tsunetsugu, Sigrist and
Ueda 1993, Moukouri and Caron 1995, Caprara and Rosengren 1997) 
allowed the fit of the dimensionless parameter $Ja/v_{F}$ against
$J$, see Fig.\ \ref{oldfig6.1}. 
$Ja/v_{F}$ characterizes double-exchange in the theory of 
chapter \ref{ch5}, and gives 
the denominator of Eq.\ (\ref{fullJc}) at criticality. 
The functional dependence is linear, and supports the form 
$\propto \sqrt{t/J}$ for $\alpha/a$ even at larger 
fillings. Fig.\ \ref{oldfig6.1} shows the straight line of best fit. 
This line 
gives $2\pi^{2}A = 0.65$ as $n \rightarrow 0$, in 
agreement with the estimate $\approx 0.7$ from $n = 0.2$
given in the previous paragraph, and gives 
$\alpha/a = \sqrt{2.1 t/J}$, in very good agreement with the 
exact result $\sqrt{2 t/J}$ obtained at vanishing filling with 
the exact solution (Sigrist, {\it et al.} (1991)). 
It is reasonable to conclude that the small
deviations in the numerically determined points for 
$Ja/v_{F}$ from the straight line are in Fig.\ \ref{oldfig6.1} are 
reflections of the different critical values determined in 
different simulations (cf.\ Fig.\ \ref{oldfig4.2}). 
Having determined $\alpha$ on the critical line, 
Eq.\ (\ref{fullJc}) determines the critical line at
\beqa
\frac{J_{c}}{t} 
= \frac{1.3 \sin(\pi n/2)}{1-0.6\sin(\pi n/2)}\, , 
\quad  \quad J > 0\, .
\label{Jcafm} 
\eeqa
The resulting phase diagram is given in Fig.\ \ref{oldfig6.2}. 

A comparison of Eq.\ (\ref{Jcafm}) with the new non-Abelian
density-matrix renormalization group results is given in
Fig.\ \ref{Ian-kondo-MOS-fig4}. The results of McCulloch,
{\it et al.} (1999) does not allow a more accurate determination
of $2\pi^{2}A$ as the points given in Fig.\ \ref{Ian-kondo-MOS-fig4}
are the boundaries of the fully polarized ferromagnetic state,
rather than the the phase transition line. However, the `state of the
art' results of McCulloch, {\it et al.} (2001,2002) give a curve 
\beqa
\frac{J_{c}}{t} 
= \frac{1.358 \sin(\pi n/2)}{1 - 0.583\sin(\pi n/2)} \, , 
\label{Jcafm_new} 
\eeqa
which implies $2\pi^{2}A = 0.679$ at the critical line, in 
perfect agreement with the previously obtained values.  

\section{\label{old6.3}Phase Diagram for the Ferromagnetic Kondo Lattice}

To plot the critical line for a ferromagnetic coupling 
$J < 0$, the analysis of 
section \ref{old6.2} is followed, and available numerically determined 
transition points for the $J < 0$ Kondo lattice are used to 
determine the constant of proportionality in 
$\alpha/a \propto \sqrt{t/J}$.  
Yunoki {\it et al.} (1998)
determine the ferromagnetic-paramagnetic transition 
for classical spins 
via Monte Carlo, and for quantum spins 3/2 via the 
density-matrix renormalization group. 
The resulting transition lines, with coupling 
$J$ correspondingly scaled, are very close, and their 
points may be used within the spin 1/2 Kondo lattice 
(cf.\  chapter \ref{ch4} and Zang, {\it et al.} (1997)).
In Fig.\ \ref{oldfig6.3} the dimensionless parameter 
$Ja/v_{F}$ is plotted against $J$ for numerically determined 
points. The straight line of best fit gives very good 
agreement with the points, and gives strong support for the 
form $\alpha/a \propto \sqrt{t/J}$ even for large fillings. 
The line of best fit determines the constant of proportionality 
as $\sqrt{0.7}$ (cf.\ $\sqrt{2.1}$ for $J > 0$). 
As in section \ref{old6.2}, this determines the critical line at 
\beqa
-\frac{J_{c}}{t} = \frac{0.7 \sin(\pi n/2)}{1 - \sin(\pi n/2)}\, .
\label{Jcfm}
\eeqa

The resulting phase diagram is given in Fig.\ \ref{oldfig6.4}. 
The critical line diverges for $J < 0$ close to half-filling, 
and differs from the $J > 0$ Kondo lattice for which the line 
remains finite. The phase separated region 
seen by Yunoki {\it et al.} (1998) has not 
been included in 
Fig.\ \ref{oldfig6.4}. Phase separation is observed in the classical 
spin simulation in the paramagnetic region from $J_{c}/t = 4$. 
It is not observed in the quantum simulation until 
$J/t = 6$, 
and then occurs away from the ferromagnetic-paramagnetic 
transition closer to half filling. Any phase 
separation involves strongly localized  
electrons, and strong on-site localization is not described 
well by the bosonization of the conduction 
electrons, as discussed in section \ref{old5.1.1}. 

However, recent numerical simulations have questioned the existance
of a phase separated region. Several new simulations
(Horsch, Jaklic and Mack 1999, Batista, {\it et al.} 1998, 2000,
Garcia, {\it et al.} 2002, Koller, {\it et al.} 2003) suggests 
that the phase separated region close to half filling is rather 
a phase dominated by ferromagnetic polarons with one single trapped 
charge carrier. This is in perfect agreement with the polaronic
picture first proposed by Honner and Gul\'{a}csi (1998b), and derved
in details in section \ref{old6.1}. It may even be that the previously
attributed phase separation regime is actually a polaronic liquid,
as it appears (see Fig.\ \ref{Ian-kondo-PRB-fig1}) for $J>0$.

\section{\label{old6.4}Range of Double-Exchange at Criticality}

The lines of best fit  Figs.\ \ref{oldfig6.1} and \ref{oldfig6.3} 
determine the effective range 
$\alpha$ of the double-exchange interaction on the 
transition line. Choosing the exponential cut-off 
function $\Lambda_{\alpha}(k) = e^{-\alpha|k|/2}$ for 
simplicity, it follows from Eq.\ (\ref{fullJc}) that
\beqa
\frac{1}{\alpha\int_{0}^{\infty}dk\, 
\cos(ka) \Lambda_{\alpha}^{2}(k)} = 
1 + (a/\alpha)^{2}\, .
\label{alphaexp}
\eeqa
For this choice of cut-off function, the line of 
Fig.\ \ref{oldfig6.1} for $J > 0$ gives $\alpha/a =
\sqrt{2.1/J}$ at the ferromagnetic-paramagnetic 
transition. This compares with 
the result $\sqrt{2/J}$ obtained in the exact solution 
of the Kondo lattice with one conduction electron 
(Sigrist, {\it et al.} (1991) 
just above the critical point at vanishing $J$. 
The filling dependence of  
$\alpha$ may be determined by using Eq.\ (\ref{Jcafm}) to 
write $\alpha/a = \sqrt{2.1/J}$ in terms of $n$. The result 
is plotted in Fig.\ \ref{oldfig6.5}.
The effective range $\alpha$ of the double-exchange 
interaction on the transition line for $J < 0$ 
may be determined in a similar fashion.
For an exponential cut-off function, 
it follows that $\alpha/a = \sqrt{0.7/J}$. As 
a function of filling, this relation may be used  
together with the critical line Eq.\ (\ref{Jcfm}) 
to plot $\alpha/a$ against $n$ and is plotted 
also in Fig.\ \ref{oldfig6.5}.
The vanishing of the effective range close to half 
filling is the reason the critical line diverges for $J < 0$. 

The different filling dependence of $\alpha$ for $J > 0$ and 
$J < 0$ is shown in Fig.\ \ref{oldfig6.5}. Different effective ranges 
$\alpha$ for the double-exchange interaction for different signs 
of the coupling is due to the different 
infinite $|J|$ symmetries of the sites containing 
localized conduction electrons. For a ferromagnetic coupling,  
this is a triplet with energy  $-|J|/4$, while for 
Kondo couplings $J > 0$ the on-site symmetry is singlet with the 
lower energy $-3J/4$ (cf.\ Table 4.1). The gain in 
energy for double-exchange per site and per conduction 
electron in either case is $-|J|/4$. 
For $J < 0$, the system thus gains just as 
much energy from double-exchange as it does by 
forming localized triplets, and the triplets have 
minimal effect for a non-vanishing hopping. This is clear 
from the ground-state Eq.\ (\ref{fmgs}) for the two-site 
Kondo lattice (see also Fig.\ \ref{oldfig4.1}).
The effective range $\alpha$ vanishes smoothly as 
the number of excess localized spins declines: $\alpha 
\rightarrow 0$ as $n \rightarrow 1$. For $J > 0$ the 
situation is far more complex. The on-site singlet 
energy is lower than the gain for double-exchange, 
and there is a complicated co-existence between the 
two effects. (This is why the exact solution 
for the Kondo lattice with one conduction electron is difficult 
for $J > 0$, whereas it is trivial for $J < 0$ 
(Sigrist {\it et al.} (1991).)
Fig.\ \ref{oldfig6.5} indicates a saturation 
$\alpha \approx a$ as $n \rightarrow 1$ for $J > 0$, 
in contrast to the $J < 0$ behaviour. 
It is beyond the methods presented here to understand the 
reason for this behaviour close to half-filling. 
What is required for an accurate determination  
is a detailed description 
of localized Kondo singlet formation, on a par with the 
double-exchange ordering and weak-coupling spin-flip 
disorder scattering that are described by the effective 
hamiltonian Eq.\ (\ref{Heff}). 

\section{\label{6.5}Numerical Results}

As presented in section \ref{old4.4.2} earlier 
failures to detect the ferromagnetic phase 
boundaries for any doping level of the Kondo lattice using 
density-matrix renormalization group can be attributed to 
the fact that the existing numerical packages do not keep
track of the total spin or magnetisation,  
which is a crucial criteria in determining the 
ferromagnetism. Neglecting these parameters, the
previous density-matrix renormalization group methods,
for details see section \ref{old4.4.2}, could not reveal 
the full complexity of the model. Thus a modification of the
density-matrix renormalization group algorithm 
to explicitly include
the total spin quantum number was necessary. This has been
achieved by McCulloch and Gul\'{a}csi (2000,2001,2002). In
the following the results obtained by this `state of the art'
density-matrix renormalization group method will be presented. 
After a short introduction, in section \ref{DMRG}, 
an overview of the new non-Abelian 
density-matrix renormalization group method will be presented. 
Following in section \ref{KLM-DMRG} by the details of the numerically
obtained phase diagram.

\subsection{\label{intro-DMRG}Why the Density-Matrix Renormalization Group?}

The original density-matrix renormalization group 
algorithm (White 1992, 1993) was developed to
overcome the problems that arise in one dimensional interacting
systems when standard renormalization group procedures are
applied. The starting point for all 
density-matrix renormalization group 
and renormalization
group calculations is a block Hamiltonian $H_A$, with some given
set of boundary conditions.
The traditional renormalization group method follows closely the
Kadanoff real space scaling ideas (Gul\'{a}csi and Gul\'{a}csi 1998), whereby
two or more of such blocks are joined to create a superblock
$H_{A \: A^{\prime} A^{\prime \prime} \ldots}$ with a
basis that is a direct product of the basis states of each individual
block. This new Hamiltonian is diagonalized and only the lowest,
e.g. $m$ eigenstates are kept. With these $m$ eigenstates we form
a new block, $H_{B}$ and the iteration is continued by forming
a new superblock, $H_{B \: B^{\prime} B^{\prime \prime} \ldots}$
whose lowest $m$ eigenvalues are kept, etc.

This method was extremely successful in certain cases 
(Wilson 1975, Krishna-murthy, Wilkins and Wilson 1980). 
However, it proved to be a very poor
approximation to one dimensional interacting electron systems
(Bray and Chui 1987). In these systems, long range correlations
and collective effects are paramount in describing the properties
of the system.  In the traditional renormalization group technique
these collective effects are suppressed because of the boundary
conditions that apply to each block. If open boundary
conditions are used, then eigenstates of $H_A$
will have the wavefunction going to zero at the edges of the block.
Thus when the blocks are joined and the old block edges become adjacent
and in the middle of the lattice, the resulting basis is not good
at representing collective effects that span the whole system.
A similar argument applies in the case of periodic boundary conditions.
With the current understanding of 
one dimensional systems and knowing how different these systems
are from their three dimensional counter parts is
seems almost obvious that such an `old fashioned' scheme will
not work.

An electron sea in one dimension has particular characteristics
which place it outside the Fermi liquid framework. The
important properties are the following (Voit 1995, Gul\'{a}csi 1997a):
(1) The low-energy
excitations of the one dimensional electron sea are collective,
and exhibit a phenomena called spin and charge separation.
This contradicts the three dimensional picture
in which the excitations resemble individual
electron excitations; (2) The collective excitations do not
interact at low energies, and one dimensional systems with
simple forward scattering interactions have a universal
structure composed of independent harmonic oscillators; and, 
as a consequence of this, (3) the momentum distribution
is continuous: in the one dimensional electron sea the spectral
weight $Z$ vanishes, and the system does not support low-energy
quasi-particle excitations\footnote{For more details see also chapter
\ref{ch2} and section \ref{old3.2.3}.}. 

The theory of one the dimensional electron sea has been in
detailed presented in chapter \ref{ch2}.
Here a brief overview is presented, for the benefit of readers 
who are not experts in the field. The theory was
was pioneered by Haldane (1981) and has 
been developed since by numerous
theoretical physicists over the years and is known as the
`Luttinger liquid' theory.  In this theory the
destruction of quasi-particles and the fact that $Z = 0$ is due
to the non-zero momentum transfer processes. It is possible
to go beyond perturbation theory and, consistent with
$Z = 0$, one indeed finds that the elementary excitations at
low energy are collective bosonic charge and spin fluctuations.
The origin of this so-called `spin and charge separation',
can be described in a number of ways all of which refer to the
nature of the one-dimensional phase space. The one-dimensional
Fermi surface consists of only two (traditionally called `left'
and `right') points. Scattering processes which take place on
the same Fermi point give rise to spin and charge separation.
This phenomenon is seen in all one-dimensional electron systems
and it seems that it is not related to other characteristics of
one dimensional systems, namely the renormalized correlation
function exponents (the $Z = 0$ effect mentioned earlier).

Accordingly, the main property of one dimensional conductors
is that their low lying excited states are charge and spin
collective density modes. Due to these density modes the
traditional numerical renormalization group methods break down.
This problem was recognized first by White (1992,1993)  
who introduced the central idea of 
density-matrix renormalization group, namely to keep the
`most probable' states of a smaller section of the lattice
in such a way that additional lattice sites can be added
without undesirable effects arising from the
external boundary conditions.
This idea follows from the requirement that
for a collective density mode to be operational it must have the
largest possible overlap between the blocks i.e., must
have the largest eigenvalues of the corresponding block density
matrix.

Because of these reasons, 
in past years, the density-matrix renormalization group method
has been extensively used to study one and two
dimensional strongly correlated electron systems (White 1998)
\footnote{A brief overview of the strongly correlated
electron systems in given in Appendix \ref{appd}.}.
This method became very popular when it was realized that it enabled
a level of numerical accuracy for one dimensional systems
that was not possible using other methods 
(Nishino, {\it et al.} 1999,Peschel, {\it et al.} 1999).

\subsection{\label{DMRG}Non-Abelian Density-Matrix Renormalization Group}

One major drawback of density-matrix renormalization group
is that calculations are performed in a
subspace of purely Abelian symmetries, such as the $U(1)$ symmetries 
of total particle number and the z component of the total spin.  Thus 
one can only obtain a few states in different total particle number 
and z component of total spin sectors (Noack and White 1993). 
For models where 
ferromagnetism emerges the situation worsens, that is, to determine 
magnetization, a combination of methods must be employed which will 
artificially raise the energy of the higher spin state 
(Daul and Noack 1998) 
within the chosen z component total spin sector.

In recognizing the imperative need to introduce a 
density-matrix renormalization group method which
has a total spin quantum number naturally implemented, a number of
unsuccessful attempts were previously made 
({\"O}stlund and Rommer 1995, Sakamoto and Kubo 1996, 
Sierra and Nishino 1997, Daul 2000). McCulloch and Gul\'{a}csi
(2001,2002) were the first to solve completely this problem
by showing that non-Abelian symmetries
can be naturally accommodated into density-matrix renormalization group.
In their form, the starting point of a calculation are
the matrix elements of the single site operators which are relatively
simple to calculate; the number of such elements varies
inversely with the dimension of the irreducible representations of the
global symmetry group, and thus is {\it reduced} for a larger
global symmetry group. For example (McCulloch and Gul\'{a}csi 2002), 
all single
site operators of a spin chain are represented as $1 \times 1$ matrices, 
independent of the magnitude of the actual spins. 

It is not the purpose of this section to give a complete description
of the density-matrix renormalization group, only to present 
the essential elements of the new algorithm. For more
details please refer to McCulloch and Gul\'{a}csi (2002). 
In order to take into account all $SU(2)$ spin components 
{\bf the first step} is to calculate the eigenstates by using 
Clebsch-Gordan coefficients. In principle, it is
not difficult to calculate eigenstates of $SU(2)$ for a finite
system by using the Clebsch-Gordan transformation 
(Biedenharn and Louck 1981)
especially in density-matrix renormalization group where 
the system is built one or two lattice sites at a time. 

Generally, the tensor product of two basis vectors, labelled
here by subscripts 1 and 2, can be written as 
\begin{equation}
\vert j m^z (j_1 j_2 \epsilon_1 \epsilon_2) \rangle \:
= \: \sum_{m^z_1,m^z_2} C^{j_1 j_2 j}_{m^z_1 m^z_2 m^z} \:
\vert j_1 m^z_1 (\epsilon_1) \rangle \: \vert j_2 m^z_2 (\epsilon_2) \rangle \; ,
\label{eq:ClebschGordan}
\end{equation}
where $C^{j_1 j_2 j}_{m^z_1 m^z_2 m^z}$ is the Clebsch-Gordan coefficient, 
$j$ is the total spin quantum number, such as $S^2 \vert j \rangle
= j(j+1) \vert j \rangle$,
$m^z$ is the projection of the spin onto the $z-$axis
and $\epsilon$ is an index that encapsulates the additional 
labels used in density-matrix renormalization group
(ie, to label the $\epsilon$'th basis state of the given quantum numbers).
Bracketed labels are not associated with a quantum number. Constructing
basis states in this way in 
density-matrix renormalization group
suffers from two problems (McCulloch and Gul\'{a}csi 2002). Applying this
transformation involves two summations for each operator matrix element.
This impacts severely on the computational effort required to construct
the block, and especially the superblock, operators. Secondly, the direct 
application of the usual 
density-matrix renormalization group
reduced density matrix to a wavefunction 
constructed from some $(jm^z)$ subspace of Eq.\ (\ref{eq:ClebschGordan}) 
does not commute with the $SU(2)$ generators. 

Despite the additional overhead of the Clebsch-Gordan transformation, this
construction of $SU(2)$ invariant 
density-matrix renormalization group
works well for small values of $j$
(McCulloch and Gul\'{a}csi 2000,2001). 
However, further improvements are possible as pointed out by 
McCulloch and Gul\'{a}csi (2002). In {\bf the second step}, 
the projection quantum number $m^z$ is completely
eliminated using the Wigner-Eckart theorem,
\begin{equation}
\langle j^{\prime} m^{z \prime} (\epsilon^{\prime})
\vert T^J_M \vert j^{} m^{z} (\epsilon^{}) \rangle \: = \:
C^{j^{} J j^{\prime}}_{m^{z} M m^{z \prime}}
\langle j^{\prime} (\epsilon^{\prime}) \vert \vert \vektor{T}^J 
\vert \vert j^{} (\epsilon^{}) \rangle \; ,
\label{eq:WignerEckartTheorem}
\end{equation}
for the $M$'th component of an operator $\vektor{T}^J$ transforming as
a rank $2J+1$ tensor. The quantity $\langle j^{\prime} (\epsilon^{\prime})
\vert \vert \vektor{T}^J \vert \vert j^{} (\epsilon^{}) \rangle$ is the 
{\it reduced matrix element} (Biedenharn and Louck 1981)
and is independent of 
the projection quantum numbers.  This operator can be considered to 
act on a reduced basis, given by the complete set of basis vectors 
$\vert \vert j (\epsilon) \rangle$.  
With this construction, all steps of the 
density-matrix renormalization group algorithm
can be performed using only the reduced basis. The importance of this
is that, unlike equation (\ref{eq:ClebschGordan}), there is no
summation involved. The only essential difference from the standard 
density-matrix renormalization group 
formulation is the quantum number-dependent $9j$ factor multiplying
each subspace.  Thus, there is no significant computation penalty for
using the $SU(2)$ formulation, as long as the $9j$
coefficients can be calculated efficiently. 

Thus, a new formulation of 
density-matrix renormalization group has been achieved 
by McCulloch and Gul\'{a}csi (2002), 
in which the states transform as $2j+1$
dimensional irreducible representations of $SU(2)$. However, it is clear
that the general formulation is essentially independent of the details
of the $SU(2)$ algebra -- given an arbitrary compact global symmetry
group the only modifications to the formulation is a different series
expansion corresponding to Eq.\ (\ref{eq:ClebschGordan}). 
It is worth noting that in the $SU(2)$ formulation, the basis vectors
are exact eigenstates of total spin even after the truncation.  This
is not true, for example, if other attempts to force the ground-state
to be in a particular total spin state by adding some suitably
chosen multiple of $S^2$ to the hamiltonian. Mixing of total spin
states due to numerically near-degenerate states will still occur.
Calculations involving long range
interactions are also affected by the lack of explicit symmetries.
Using a $U(1)$ symmetric basis labelled by the $z$-component of 
spin only, interaction terms no longer transform as exact
representations of $SU(2)$ after a truncation.
This can lead to situations where, even
for a large number of kept states, the ground-state is a broken
symmetry N\'eel type state (White and Scalapino 1998) and only converges
slowly to an eigenstate of $S^2$. This
is purely an artefact of the 
density-matrix renormalization group
algorithm when appropriate
symmetries are not explicitly preserved.

The computational advantage of the non-Abelian construction is two 
fold: (1) each reduced basis element corresponds to $2j+1$ basis 
states of the old representation, thus the storage requirement for 
the block operators is reduced for an equivalent number of block 
states. This allows, for virtually no increase in
CPU time, to {\it i}) investigate long chains, with 
several hundreds of  lattice sites, using hundreds (in most cases
infinity) number of kept states, and 
gain {\it ii}) an improvement of at least four orders of
magnitude in the cumulative truncation error and the fractional
error in the ground-state energy; 
(2) the superblock basis can be projected onto an exact 
subspace of arbitrary total spin. As well as reducing the size of 
the target Hilbert space, this greatly simplifies the calculation 
of excited states that have total spin less than the total spin of 
the ground-state. This is very useful for investigating magnetic 
phase transitions (McCulloch, {\it et al.} 1999,2001,2002)
or other type of topological ordering (Kruis, {\it et al.} 2002).
The method is easily adaptable to 2D systems, and it was tested
already on the 2D t-J model (McCulloch, Bishop and Gul\'{a}csi 2001). 
For ferromagnetic target states 
(or more generally, target representations with a dimension greater 
than one), it is possible to calculate to first order the splitting 
of the degenerate states due to a symmetry breaking field, trivially 
in the case of a uniform magnetic field $h$ (where the splitting is 
just $h m^z$, for $m^z = -j,-j+1,\ldots,j$), or in other cases by 
calculating the projection of the wavefunction and the symmetry 
breaking operator onto each $z$-component of spin using the 
Wigner-Eckart theorem Eq.\ (\ref{eq:WignerEckartTheorem}).

\subsection{\label{KLM-DMRG}Phase Diagram}

The problem of
earlier density-matrix renormalization group studies, as presented in
section \ref{old4.4.2}, was that due to the low accuracy, doping 
levels beyond quarter filling, $n = 0.5$,
were impossible to analyse
\footnote{This limitation is still valid: even 
recent results (Xavier, Novais and Miranda 2002,
Xavier, {\it et al.} 2003) do not cover the region 
of doping close to half-filling.}. 
This was addressed by McCulloch, {\it et al.}
(1999), which represented the first application of the  
non-Abelian density-matrix renormalization group, 
see Fig.\ \ref{Ian-kondo-MOS-fig4}. 

Here the total impurity spin of the system for $n = 0.5$, 
$0.6$, $0.7$, $0.8$, $0.9$ and $0.95$ fillings have been calculated. 
As the phase transition line is approached (see dashed curve in
Fig.\ \ref{Ian-kondo-MOS-fig4}) the total spin of the impurities 
will be different from zero. In every analysed filling 
factor case the total impurity spin gradually increases with $J$ and finally 
the system reaches a fully polarized state. Hence, in  
Fig.\ \ref{Ian-kondo-MOS-fig4} the stars (McCulloch, {\it et al.} 1999)
represent the fully polarized impurity spins states. 
Accordingly, these points are situated above the true
thermodynamic phase transition.  As a comparison 
we also plotted the previous numerical results (presented earlier
in Fig.\ \ref{oldfig4.2}).   

The most significant contribution of McCulloch, {\it et al.} (1999) 
is to establish the
existence of a ferromagnetic regime extending up to half filling, see 
Fig.\ \ref{Ian-kondo-MOS-fig4} for $n = 0.9$ and $n = 0.95$ filling 
factors. For $n = 0.95$ the impurity spins are 
already fully polarized above $J \approx 3.5$. This proves that,
contrary to the previous suggestions
\footnote{As mentioned in section \ref{old4.4.2}, earlier results based
on simple mean-field, slave-boson or Gutzwiller approximations (Tsunetsugu, 
Sigrist and Ueda 1997) could not predict ferromagnetism close
to half filling.}, ferromagnetism will exist for any band filling. 

More recent non-Abelian density-matrix renormalization group results
(McCulloch, {\it et al.}  2001, 2002; Juozapavicius, {\it et al.} 2002)
confirmed this picture. The obtained results (McCulloch, {\it et al.}
2001,2002) can be summarized with the phase diagram of Fig.\
\ref{Ian-kondo-PRB-fig1}, which will be analysed hereafter. 

As it can be seen from Fig.\ \ref{Ian-kondo-PRB-fig1}, the 
main feature dominating the Kondo lattice is the impurity spin
ferromagnetic ordering, as such confirming the results of
chapters \ref{ch5}. 
This element was missing in the early approaches,
which concentrated on the competition between
Kondo singlet formation at large $J$ and the RKKY interaction
in the weak coupling limit (Jullien, Fields and Doniach 1977,
Tsunetsugu, Sigrist and Ueda 1997, Shibata and Ueda 1999). This picture
is borrowed from the single impurity Kondo model and is
inadequate for the lattice case. 

Starting the analysis of the phase diagram for large $J$,
it can be seen that all the conduction electrons form 
singlets with the localized spins (Sigrist, {\it et al.} 1992b). 
The uncoupled $f$ spins order ferromagnetically 
in a mechanism similar to the $J < 0$ case. 
Here, there is no competition between Kondo singlet
formation and double-exchange. The fully polarized state 
[with $S = (L-N)/2$] appears for any value of $n < 1$ 
as proven in chapter \ref{ch5}. As $J$ is lowered, the
Kondo lattice can be rigorously mapped into a random transverse
field Ising model, see section \ref{old5.2}, hence the phase transition (the
solid curve in Fig.\ \ref{Ian-kondo-PRB-fig1}) is identical to a  
quantum order - disorder transition. It should be emphasized 
that this is also true for
the second ferromagnetic phase, as will be shown later on.

The phase transition obtained via density-matrix renormalization group
fits exceptionally well this
picture, confirming the bosonization result of chapter \ref{ch5}. 
The open circles on Fig.\ \ref{Ian-kondo-PRB-fig1} 
correspond to points at which the energy of the ferromagnetic 
state crosses the energy of the singlet state. Since the phase
transition is second order, this is only an upper bound on the
true transition line. However the partially polarized region is
very small, of the order of $J/t \sim 0.01$, which is why
this phase transition has not previously been observed to be
continuous. Examples for the energy versus the
magnetization curves are given by McCulloch, {\it et al.} (2002). 

Below the solid curve, Fig.\ \ref{Ian-kondo-PRB-fig1}, 
the Kondo singlets are not
inert anymore and they greatly contribute to the properties of
the Kondo lattice. Excluding the Kondo triplet states, the 
conduction electron wave function in the
continuum limit satisfies a nonlinear Schr\"{o}dinger equation,
see Eq.\ (\ref{nls}), 
which has finitely delocalized solitonic solutions of the form
given in Eq.\ (\ref{soliton}). 
This corresponds to a dressing of the conduction electrons by a cloud
of antiparallel local spins, i.e., spin polarons are formed. The
polaronic length scale competes with the length scale set by the 
conduction electron 
mean free path and introduces competing time scales: slow motion of
the polarons with low energy dynamics and fast motion of the free 
conduction electrons
with high energies. This scenario resembles a two-fluid picture with
intrinsic inhomogeneities which involves spin fluctuations and
short-range spin correlations, which was called
(McCulloch, {\it et al.} 2002) a {\sl polaronic liquid}.

Finite temperature density-matrix renormalization group 
results (Shibata and Tsunetsugu 1999)
confirmed the presence
of short-range $f$ spin correlations in the van-Hove singularities.
Consequently the structure factor peaks at $2 k_F - \pi$,
where $k_F$ is the Fermi point determined by the filling of the 
conduction band.
This means that the localized $f$ spins, even though they
are completely immobile, contribute to the volume of the Fermi sea.
This conventionally is called a {\sl large} Fermi surface, the
effect of which is also seen in the momentum distribution
function, see Fig.\ \ref{Ian-kondo-PRB-fig3}. 
As the polarons are formed the peak
of $S(k)$ shifts from the small $J/t$ value of $2 k_F$:
the slow motion of the spin polarons will dominate the
low energy dynamics of the quasiparticles. This shift
can be seen clearly in Fig.\ \ref{Ian-kondo-PML-fig2}. 
This proves
that the appearance of the large Fermi surface is a
dynamical effect since it involves local inhomogeneities,
impurity spin fluctuation, and short-range correlations
of the $f$ spins. This is the Griffiths phase described
in section \ref{old5.2}. 

The large Fermi surface is conventionally explained by
reference to the periodic Anderson model ancestry 
(Yamanaka, Oshikawa and Affleck 1997, 
Shibata and Ueda 1999, Oshikawa 2000, Sinjukow and Nolting 2002). 
However, in the view of section \ref{old5.2} the Griffiths 
phase with its local polaronic inhomogeneities gives rise 
to the large Fermi surface. 
For $n < 0.5$ the width of the polarons is over several lattice spacings
(diverging for $n \rightarrow 0$, see also Fig.\ \ref{oldfig6.5}) hence the
energy needed to excite these polarons is too large for this
effect to happen. The polarons will not contribute to the low
energy dynamics and as such a small Fermi surface is active,
see Figs.\ \ref{Ian-kondo-PRB-fig3} and \ref{Ian-kondo-PML-fig2},
and the discussion in section \ref{old5.3}.
This has been confirmed in earlier density-matrix renormalization group
calculations (Troyer and W\"{u}rtz 1993, Moukouri 
and Caron 1995, Caprara and Rosengren 1997), for a review see
Shibata and Ueda (1999), and more recently 
(McCulloch, {\it et al.} 2001,2002, Xavier, Novais and Miranda 2002). 
So it seems that there has to be a small - large Fermi
surface crossover at quarter filling, see the dotted 
line in Fig.\ \ref{Ian-kondo-PRB-fig1}. Density-matrix
renormalization group results of Xavier, {\it et al.} (2003)
suggest that this is a dimerized phase, but this has yet to be
proven rigoirously.

An interesting phenomenon appears as $J$ is further lowered.
The residual weight attached to the Kondo singlets vanishes,
hence all conduction electrons which participated in the formation of these
singlets, become delocalized. The distance between these conduction electrons
is much larger than the lattice spacing, and below
$J \le 2 {\sqrt{n}} \sin ( \pi n )$ their continuum
limit takes the regular quantum sine-Gordon form
(Zachar, Kivelson and Emery 1997). 
In the bosonization language of section \ref{old5.1.4}, this
means that the spin Bose fields, $\Phi_{\sigma}$ cannot be
approximated by their noninteracting expectation values,
rather by their expectation value corresponding to a
sine-Gordon (sG) model, $\Phi_{\sigma} \approx \langle
\Phi_{\sigma} \rangle_{\rm{sG}}$. However, the charge degrees
of freedom not being affected by the sine-Gordon spin gap,
their corresponding Bose fields, $\Phi_{\rho}$ may be still
approximated by their noninteracting values. Extending the
bosonized results of section \ref{old5.1} to a finite
$\langle \Phi_{\sigma} \rangle_{\rm{sG}}$, it is obtained
(McCulloch, {\it et al.} 2002) 
the critical hamiltonian governing this second phase
transition
at intermediate $J$ values in the form:
\begin{eqnarray}
H_{\rm crit.} &=& - J^2 {\cal{A}} / (2 \pi^2 v_F) 
\sum_{j} {\bf S}^{z}_{j} {\bf S}^{z}_{j + 1} 
\nonumber \\
&+& 2 J {\cal{B}} 
\sum_{j} \{ 1 - ( \langle \Phi_{\sigma} \rangle^{2}_{\rm{sG}} / 2 ) 
[1 + J / (2 \pi v_F) ]^2 + \cos ( 2 k_F j ) \} {\bf S}^{x}_{j} \, , 
\end{eqnarray}
where ${\cal{A}}$ and ${\cal{B}}$ are functions which depend only
on the cutoffs introduced by the bosonization
scheme, see section \ref{old6.2}. Following closely 
chapter \ref{ch5} it can be proven (McCulloch, {\it et al.} 2002)
that the critical behaviour of this ferromagnetic transition for 
the intermediate
this $J$ case is of a random transverse-field Ising
model type, where the transverse field
\begin{equation}
h_j = 2 J {\cal{B}} \{ 1 - ( \langle \Phi_{\sigma} \rangle^{2}_{\rm{sG}} / 2 ) 
[1 + J / (2 \pi v_F) ]^2 + \cos ( 2 k_F j ) \} \, , 
\end{equation}
is driven by a displaced cosine distribution
of the form: 
\begin{equation}
\rho (h) = [1 / (2 \pi J {\cal{B}} ) ] \{ 1 - [ h / (2 J {\cal{B}} ) 
+ ( \langle \Phi_{\sigma} \rangle^{2}_{\rm{sG}} / 2 ) 
[1 + J / (2 \pi v_F) ]^2 - 1 ]^2 \}^{-1/2} \, .
\end{equation}
Accordingly, the these transitions emerging at
intermediate values of $J$ are of a quantum order -
disorder type. These transitions are driven by spin polarons,
contrary to the ferromagnetic phase emerging at high $J$ values, which
is given by the uncoupled $f$ spins. 
The new critical line is (McCulloch, {\it et al.} 2002):
\begin{equation}
J_c = \alpha ({\cal {A}}, {\cal {B}}) 
\sin (\pi n /2) / [1 - \beta ({\cal {A}}, {\cal {B}})] 
- \gamma ({\cal {A}}, {\cal {B}}, 
\langle \Phi_{\sigma} \rangle^{2}_{\rm{sG}}) \, .
\end{equation}
The bosonization (conformal field-theory) arguments
do not determine the magnitude of $\alpha$, $\beta$
and $\gamma$, accordingly these constants are used as
fitting parameters to the numerically obtained points.
The best fits are the dashed curves in Fig.\ \ref{Ian-kondo-PRB-fig1}. 

Below the second ferromagnetic region the Kondo lattice 
reduces to a system of free
localized spins in fields determined by conduction electron 
scattering: dominant $2k_{F}$ modulations are manifest, 
see Figs.\ \ref{Ian-kondo-PRB-fig3} and \ref{Ian-kondo-PML-fig2}, 
superimposed on
an incoherent background. This reflects the momentum transferred
from the conduction electrons band to the spin chain 
in backscattering interactions,
together with incoherent forward scattering. This case is referred
to as an RKKY liquid in Fig.\ \ref{Ian-kondo-PRB-fig1} as 
the scattering processes give an RKKY-like
correlation for the $f$ spins, see section \ref{old5.3} for details,
even though the RKKY interaction strictly diverges in one dimension. 

In closing this section, it has to be mentioned that the Kondo lattice 
still holds surprises. As it was presented in Juozapavicius, 
{\it et al.} (2002), just below the second ferromagnetic phase of
Fig.\ \ref{Ian-kondo-PRB-fig1} two other smaller ferromagnetic 
regions have been found, see Fig.\ \ref{Ian-kondo-PMB-fig4}. 
At around $n=0.8$ a third ferromagnetic phase develops 
up to $n\ge 0.8125$. More detailed calculations show that 
other smaller ferromagnetic regions appear as $J$ is decreased. 
All the regions which were successfully mapped out are presented
in Fig.\ \ref{Ian-kondo-PMB-fig4}. The size of these ferromagnetic
phases are becoming vanishingly small with lower $J$. 
The overall gap size (between the $S = 0$ and fully polarized states)
rapidly decreases with the filling and the accuracy is insufficient to
obtain credible data above $n=0.9$\footnote{This is the reason why the
second ferromagnetic region above $n=0.9$ is plotted as a striped area
in Fig.\ \ref{Ian-kondo-PMB-fig4}, where the accuracy was insufficient 
to determine the phase boundaries}. Otherwise, in the calculation
of Juozapavicius, {\it et al.} (2002) the typical truncation 
error $< 10^{-7}$ allowed energy gaps as small as one millionth part of the
ground-state energy to be calculated, 
within $J=\pm 0.1$ accuracy for all the points below $n=0.9$. 
This behaviour of the Kondo lattice is very intriguing and has not been
explained.

%%%%%%%%%%%%%%%%%%%%%%%%%%%%%%%%%%%%%%%%%%%
%% chapter 7
%%%%%%%%%%%%%%%%%%%%%%%%%%%%%%%%%%%%%%%%%%%%

\cleardoublepage
\chapter{\label{ch8}Other Properties of the 1D Kondo Lattice}

The previous chapters have proven that the 
main feature dominating the Kondo lattice is the impurity spin 
ferromagnetic ordering. It was shown, see section \ref{old5.1.3}, 
that this ferromagnetism 
is due to the double-exchange interaction which appears as a consequence
of an excess of localized spins over conduction electrons: 
each conduction electron has to screen more than one localized spin, and
since hopping is energetically most favourable for conduction electrons
which preserve their spin, this tends to align the localized
spins. This element was missing in the early approaches,
which concentrated on the competition between
Kondo singlet formation at large $J$ and the RKKY interaction
in the weak coupling limit (Jullien, Fields and Doniach 1977). 
This picture is borrowed from the single impurity Kondo model 
and is inadequate for the lattice case (Le Hur 1998) 
\footnote{Here a full non-Abelian bosonization approach is used to study
the fixed points of the 1D Kondo lattice, and it is shown that
the fixed points differ from the ones of the single impurity Kondo  
model.}. 

However, it is necessary also to describe nearly-free 
electrons at weak-coupling. Scattering processes at  
weak-coupling are restricted to states close to the 
non-interacting conduction electron Fermi points at 
$\pm k_{F}$ (cf.\ chapter \ref{ch1two}), and can give rise via 
backscattering to the $2k_{F}$ correlations in the localized 
spins. The basic idea behind the presented theory 
is that coherent hopping at strong-coupling,  
together with the more standard 
nearly-free scattering at weak-coupling, can be described using 
the bosonization formalism derived in chapter \ref{ch2}. 
As such, the weak-coupling regime of the effective hamiltonian
will be considered in the following, in section \ref{old5.3}. 
The ground-state of the localized 
spins is obtained, as is the ground-state 
correlation function between the localized spins. 
The correlations oscillate with a dominant wave vector at 
$2k_{F}$ of the conduction electrons. 

The following sections consider the effects of {\it i}) repulsive 
interactions between the conduction electrons, {\it ii}) presence
of electron-phonon couplings, and {\it iii}) the effect of 
diluting the array of impurity spins. 
With all these extensions, the ferromagnetic-paramagnetic 
transition still occurs, but the ferromagnetic phase in the 
presence of the Hubbard term becomes larger and the 
critical coupling is pushed to 
lower values of $J$. For infinite $U$, the 
critical coupling $J_{c} = 0$, and ferromagnetism occupies the 
entire phase diagram. The electron-phonon couplings, on the other
hand, makes the spin polarons more robust, diminishing the ferromagnetic 
region. Diluting the array of impurity spins again reduces the
effect of the double-exchange, and as such decreases the ferromagnetic
region: it also induces antiferromagnetism. This 
distinguishes the dilute Kondo lattice model from
the conventional Kondo lattice.

\section{\label{old5.3}Weak-Coupling: RKKY-like Behaviour}

Well below the transition, where $Ja/v_{F}$ is small 
and the ferromagnetic interaction Eq.\ (\ref{FMterm}) is 
negligible, the ordering of the localized spins is 
governed by the last two terms of 
the effective hamiltonian Eq.\ (\ref{Heff}). 
To determine the dominant correlations in this 
strongly disordered phase, 
it suffices to take eigenvalues for 
$\epsilon_{l}(j)$ in the 
long-range object $K(j)$ (cf.\ Eq.\ (\ref{Kj2})). 
$K(j)$ then fluctuates about zero incoherently, 
depending on the global $S^{z}_{j}$ configuration.
The effective hamiltonian corresponds to free 
localized spins in  
$x$ and $z$ fields determined by 
conduction electron scattering. 
The free localized spin problem is straightforwardly 
diagonalized by standard methods (Wagner 1986), and 
yields the ground-state $S^{z}_{j}$ configuration 
\beqa 
|\psi_{0} \rangle = \exp\left\{ 
i\sum_{j} \tan^{-1}\left(
\frac{\cos[K(j)] 
+ \cos(2k_{F}ja)}{\sin[K(j)]\sin(2k_{F}ja)} \right)
S^{y}_{j}\right\} |\downarrow\rangle \, ,
\label{5.3.1}
\eeqa
where $|\downarrow \rangle$ is the state with 
$S^{z}_{j} = -1/2$ for all $j$. The 
dominant $2k_{F}$ modulations in 
$|\psi_{0} \rangle$ are manifest, 
and are superimposed on an incoherent background: 
\beqa
\langle\psi_{0}| S^{z}_{j} S^{z}_{j+x} 
|\psi_{0}\rangle 
\approx \sin[2k_{F}ja] \sin[2k_{F}(ja+x)] 
\label{5.3.2}
\eeqa
to an incoherent normalization. Eq.\ (\ref{5.3.1}) is in agreement
with the elementary Raleigh-Schr{\"o}dinger perturbation theory
performed by Aristov (1997) and earlier by Yafet (1987), see 
section \ref{old4.3.1} for details, and 
reflects the momentum transferred
from the conduction electrons band to the spin chain 
in backscattering interactions,
together with incoherent forward scattering.

This weak-coupling behaviour has been observed already 
in the early numerical simulations (Troyer and W\"{u}rtz 1993, Moukouri 
and Caron 1995, Caprara and Rosengren 1997), and is 
called the RKKY liquid phase,\footnote{Recall from section \ref{old4.3.1} 
that the RKKY interaction strictly diverges in 1D, and there  
is no lower bound on the ground-state energy for 
the RKKY hamiltonian, even for arbitrarily small  
$J$ (Sigrist, {\it et al.} 1992b). The divergence is  
typical of perturbation expansions in 1D, and does 
not occur in higher dimensions.} see section \ref{KLM-DMRG}
and Fig.\ \ref{Ian-kondo-PRB-fig1}. 
The review of Shibata and Ueda (1999) lists all the earlier 
numerical results, which all support a dominant $2k_{F}$ 
modulation in the weak-coupling limit. The recent numerical
results (McCulloch, {\it et al.} 2001,2002, Xavier, Novais and 
Miranda 2002, Xavier, {\it et al.} 2003) also confirm this. 
The results of McCulloch, {\it et al.} (2002,2003) are 
presented in Figs.\ \ref{Ian-kondo-PRB-fig3} and 
\ref{Ian-kondo-PML-fig2}, which clearly indicates a 
$2k_{F}$ modulation superimposed on an incoherent background. 

In spite off the overwhelming proof that  
$2k_{F}$ modulations are dominant in the weak-coupling 
limit of the spin-isotropic Kondo lattice, they are nevertheless 
neglected {\it a priori} in recent approaches
to the spin-anisotropic Kondo lattice (Novais, 
{\it et al.} 2002a). These neglected terms correspond to 
(Honner and Gul\'{a}csi 2002)
\beqa
H_{2k_{F}} &=&  \frac{J_{\perp}a}{2\pi\alpha}\sum_{j}
\left\{ e^{-i\sqrt{2\pi}\theta_{\sig}(j)}
\cos\left[2k_{F}ja + \sqrt{2\pi}\phi_{\rho}(j)\right]
S_{j}^{-} + {\rm H.c.} \right\}
\nonumber \\
&+& \frac{J_{z}a}{\pi\alpha} \sum_{j}
\sin\left[\sqrt{2\pi}\phi_{\sig}(j)\right]
\sin\left[2k_{F}ja + \sqrt{2\pi}\phi_{\rho}(j)\right]S_{j}^{z} \, ,
\label{marhasag}
\eeqa
which includes {\it every} interaction term involving
charge degrees of freedom\footnote{In Eq.\ (\ref{marhasag}) 
the notations of Novais, {\it et al.} (2002a,2002b)
were used, instead of Eq.\ (\ref{bklm-anisotropic}).}.
Neglecting these terms led Novais, {\it et al.} 
(2002a,2002b) to incorrectly predict an ordered 
phase for the spin-anisotropic Kondo lattice at arbitrarily
weak-coupling. The correct phase diagram of the 
spin-anisotropic Kondo lattice has been determiend earlier,
see Chen, {\it et al.} (1999). 

To determine the effect of backscattering within the 
context of the calculation performed by Novais {\it et al.} 
(2002a,2002b), Honner and Gul\'{a}csi (2002) noticed their 
claim (Novais, {\it et al.} 2002b) 
that their calculation is equivalent (Novais, {\it et al.} 2002a) 
to performing a unitary transformation $\exp(U)$
with $U = -i a J_{z} / (\sqrt{2\pi} v_{F})
\sum_{j}\theta_{s}(j) \,S_{j}^{z}$ (cf.\ Eq.\ (\ref{5.1.1}) of 
section \ref{old5.1.2}). However, this 
transformation has already been
carried out {\it exactly}, ie., up to infinite order, on the 
hamiltonian with all terms included 
(Honner and Gul\'{a}csi 1997a,1998b), for details 
see also section \ref{old5.1.2}.
As $|J_{z}| \rightarrow 0$, where for ferromagnetic coupling 
Novais, {\it et al.} (2002a,2002b) 
claim the Kondo lattice is ordered, the bosonic fields
$\phi_{\rho}, \theta_{\rho}$ and $\phi_{\sig}, \theta_{\sig}$
take their noninteracting values to a very good approximation, 
and the transformed hamiltonian reduces to free spins in magnetic 
fields determined in principle by backscattering conduction electrons,
and the ground state spin configuration reduces to 
Eq.\ (\ref{5.3.1}). Thus, the spins form a paramagnetic ground state
with dominant correlations at $2k_{F}$, independent of the sign of
$J_{z}$ for $|J_{z}| \rightarrow 0$.

\section{\label{old5.4}Effects of Coulomb Repulsion}

To determine the effects on the ordering of the 
localized spins due to 
interactions between the electrons, consider adding 
to the standard Kondo lattice hamiltonian of Eq.\ (\ref{4.11}) 
the Hubbard interaction term (cf.\ Eq.\ (\ref{3.3}))
\beqa
V_{\Hub} &=& U\sum_{j} n_{j\uparrow}n_{j\downarrow} 
\nonumber \\
  &=& \frac{U}{4N}\sum_{k} 
  [(\rho_{+}(k) + \rho_{-}(k))(\rho_{+}(-k) + \rho_{-}(-k))
\nonumber \\ 
  & & \quad
  -(\sig_{+}(k) + \sig_{-}(k))(\sig_{+}(-k) + \sig_{-}(-k))],
\label{5.4.1}
\eeqa
where the charge and spin density fluctuation components are 
defined in Eq.\ (\ref{3.4}). 
It will be sufficient for the purposes here to consider only 
forward scattering contributions to $V_{\Hub}$. (For weak 
repulsive interactions, $U$ small and positive,  
backscattering interactions renormalize to zero 
(S\'{o}lyom 1979).) Within the Bose description, this is equivalent 
to attaching the weight $\Lambda_{\alpha}(k)$ to the 
density fluctuations in $V_{\Hub}$. The interaction 
then reduces to standard Tomonaga-Luttinger--type, with 
forward scattering interactions described by bosonic 
density fluctuations, exactly as in the bosonization of the 
Hubbard model in section \ref{old3.2}. 
The pure conduction band part $H_{0} 
+ V_{\Hub}$ of the interacting Kondo lattice 
may now be straightforwardly 
diagonalized via a Bogoliubov transformation
$\exp(W)$, where $W = \sum_{\nu = \rho,\sig}
W_{\nu}$ and 
\beqa
W_{\nu} = 
\frac{\pi}{2L}\log\left(\frac{v_{\nu}}{v_{F}}\right) 
\sum_{k>0}\frac{1}{k}[\nu_{+}(k)\nu_{-}(-k) - 
\nu{-}(k)\nu_{+}(-k)]\Lambda_{\alpha}^{2}(k)\, ,
\label{5.4.2}
\eeqa
exactly as in section \ref{old3.2}. The charge and spin velocities are 
\beqa
v_{\rho} = v_{F} \sqrt{1 + a U /\pi v_{F}}\,\, ,
\quad \quad 
v_{\sig} = v_{F} \sqrt{1 - a U/\pi v_{F}}\, \, .      
\label{5.4.3} 
\eeqa
By comparison with the exact Bethe ansatz solution, 
these velocities are correct to leading order in 
$U$. Corrections to the velocities at stronger 
couplings are given by Schulz (1991): The 
spin velocity $v_{\sig}$ does not go complex as $U$ 
increases, but smoothly goes to zero. This is discussed 
in detail in section \ref{old3.2.3}. 
Under $W$ the Bose fields transform as 
\beqa
\tilde{\phi}_{\nu}(j) = \sqrt{v_{F}/v_{\nu}}\, 
\phi_{\nu}(j)\, ,
\quad \quad 
\tilde{\theta}_{\nu}(j) &=& \sqrt{v_{\nu}/v_{F}}\, 
\theta_{\nu}(j)\, ,
\label{5.4.4} 
\eeqa
while
\beqa
\tilde{H}_{0} &+& \tilde{V_{\Hub}} 
= \sum_{\nu = \rho ,\sig}H_{\nu}, 
\nonumber \\
H_{\nu} &=& \frac{v_{\nu}a}{4\pi}\sum_{j} 
\left\{ \Pi_{\nu}^{2}(j) + [\partial_{x} 
\phi_{\nu}(j)]^{2}\right\} 
\label{5.4.5}
\eeqa
to an additive constant. The details of the calculations 
leading to these results are given in section \ref{old3.2.1}. 

Under the transformation $W$, 
the bosonized Kondo lattice with interactions between 
the electrons takes the same basic 
form as the original bosonized hamiltonian of 
Eq.\ (\ref{bklm}). The only differences are that 
the first term in Eq.\ (\ref{bklm}) is replaced 
by $\sum_{\nu}H_{\nu}$, and that the Bose fields 
in the remaining terms are replaced by their scaled 
forms as in Eq.\ (\ref{5.4.4}). Proceeding much 
as in the original $U = 0$ problem, choose the 
transformation $\exp({\rm S})$ where
\beqa
{\rm S} = i\frac{Ja}{2\pi}\sqrt{\frac{v_{F}}{v_{\sig}^{3}}} 
\sum_{j} \theta_{\sig}(j)\,S^{z}_{fj}, 
\label{5.4.6}
\eeqa
and obtain a transformed hamiltonian similar to 
Eq.\ (\ref{tbklm}). The important difference is 
that the prefactor of the double-exchange ferromagnetic term of 
Eq.\ (\ref{FMterm}) is increased: 
\beqa
\frac{J^{2}a^{2}}{4\pi^{2}v_{F}} \rightarrow
\frac{J^{2}a^{2}}{4\pi^{2}v_{F}}\,
\frac{1}{1 - a U/\pi v_{F}}\,.
\label{5.4.7}
\eeqa
Following exactly the analysis of section \ref{old5.1},   
an effective hamiltonian for the localized spins is 
obtained. This 
determines a quantum order-disorder transition between 
ferromagnetic and quantum disordered paramagnetic phases, with a 
critical coupling $J_{c}(U)$ which is down from the 
$U = 0$ critical coupling by a factor 
$1 - Ua/\pi v_{F}$. For stronger interactions between 
the conduction electrons, the spin velocity of 
Schulz (1991) should be used in the 
transformation Eq.\ (\ref{5.4.6}). 

The effect of a repulsive Hubbard interaction between 
the electrons is then as follows. Double-exchange 
is characterized by the enhanced dimensionless 
constant $Ja\sqrt{v_{F}/v_{\sig}^{3}}$. The ferromagnetic phase 
becomes more robust, and the ferromagnetic-paramagnetic 
phase boundary is 
pushed to lower values of $J$. This is  expected 
on physical grounds; the repulsion between the 
electrons tends to keep the regions of double-exchange 
ordered localized spins from overlapping. Spin-flip disorder 
processes, which occur when the regions  
intermingle, are thereby 
reduced. The result here is consistent 
with the numerical work of Moukouri {\it et al.} (1996)
on the Kondo lattice with $t$-$J$ 
interacting electrons; reduced critical 
couplings $J_{c}/t \approx 0.8, 1, 1.2$ are 
determined for fillings $n = 0.5, 0.7, 0.9$, 
respectively. Moreover, since 
$v_{\sig} \rightarrow 0$ 
as $U \rightarrow \infty$ (Schulz 1991), so  
that $J_{c}/t \rightarrow 0$, the result here coincides 
with the rigorous result of Yanagisawa and Harigaya (1994) 
for infinite repulsive electron interactions. 

\section{\label{7.4}Effect of Electron-Phonon Interaction}

As presented in section \ref{old4.2} the 
initial understanding
of the properties of manganites was based on the double-exchange
mechanism within the Kondo lattice (Searle and Wang 1970, 
Kubo and Ohata 1972). However, 
Millis {\it et al.} (1995) argued 
that the predictions of the Kondo lattice disagree with the 
experimental data by an 
order of magnitude or more. They propose that 
the discrepancy arises from the neglect of strong 
electron-phonon coupling.  Accordingly, the effect of the
electron-phonon coupling could be of importance in understanding 
the real materials. This effect has been taken into account via the
bosonization formalism of chapters \ref{ch2} and \ref{ch5} by 
Gul\'{a}csi, Bussmann-Holder and Bishop (2003,2004), details of which
calculation are presented in this section. 

In 1D, the electron-phonon coupling could be of either 
inter-site (Su-Schrieffer-Heeger 1980a,1980b) or on-site
(Holstein 1980b) character. The model that have been studied
by Gul\'{a}csi, Bussmann-Holder and Bishop (2003,2004) assumes a 
dispersionless phonon mode with frequency $\omega$.
The neglect of the dispersion of bare phonons is not essential since 
it is absent in the Holstein model and the acoustic phonons are decoupled
from the low energy electronic spectrum in the continuum limit of the 
Su-Schrieffer-Heeger model (Hirsch and Fradkin 1982, 1983, Fradkin
and Hirsch 1983). In this approximation the Holstein 
coupling to dispersionless phonons will have the form: 
\begin{equation}
H_{\Hol} = \sum_{i} \Big( \beta q^{}_{i} n^{}_{ci} + \frac{K}{2} q^{2}_{i} + 
\frac{1}{2 M} p^{2}_{i} \Big) \; , 
\label{ph-egy}
\end{equation}
with the conduction electron density at site $i$ of $n^{}_{c i}$, the 
lattice displacement $q^{}_{i}$, its conjugate momentum $p^{}_{i}$, the 
coupling strength $\beta$, the spring constant $K$ and the ionic mass
$M$. While the Su-Schrieffer-Heeger coupling to phonons is 
\begin{equation}
H_{\SSH} = 
\sum_{i} \Big[ \sum_{\sigma} \alpha_{\sigma} (q^{}_{i + a} - q^{}_{i}) 
(c^{\dagger}_{c i \sigma} c^{}_{c i + a \sigma} + 
c^{\dagger}_{c i + a \sigma} c^{}_{c i \sigma}) 
+ \frac{K}{2} (q^{}_{i + a} - q^{}_{i})^2 
+ \frac{1}{2 M} p^{2}_{i} \Big] \; , 
\label{ph-ot}
\end{equation}
where $\alpha_{\sigma}$ denotes the coupling strength and the rest 
of the notations are as the same as in chapters \ref{ch4} and \ref{ch5}.
Thus, the starting hamiltonian is 
\begin{equation}
H_{\KLph} \: = \: H_{\KL} \: + \: H_{\Hol} \: + \: H_{\SSH} \, , 
\label{ph-new-egy}
\end{equation}
where $H_{\KL}$ is defined in Eq.\ (\ref{4.11}). 

These phononic contributions may not describe the full 
complexity of the phononic couplings observed in real materials 
because of the phase space constraint of any one dimensional 
calculation. Still the results capture the essence of the Kondo
lattice coexisting with phonons, and being an exact solution, it 
represents a vital source of information due to the lack of similar 
solutions applicable to colossal magnetoresistance materials. 

In standard bosonization language, Eq.\ \ref{ph-egy} 
simplifies to $H^{\rm ph} + H^{\rm el-ph}_{1} + H^{\rm el-ph}_{2}$,
where 
\begin{eqnarray}
H^{\rm ph} \: &=& \: \frac{1}{2 N} \sum_{p} \left[ \Pi^{2}_{0} (p) + 
\omega^{2}_{0} \Phi^{2}_{0} (p) \right]
\nonumber \\
&+& \: \frac{1}{2} \int dx \left[ \Pi^{2}_{\pi} (x) + 
\omega^{2}_{\pi} \Phi^{2}_{\pi} (x) \right] \, ,
\label{ph-ketto}
\end{eqnarray}
and $\omega_{0} = \omega_{\pi} = \sqrt{ K / M}$ are their 
respective phonon frequencies. 
The electron-phonon forward scattering term is simply
\begin{equation}
H^{\rm el-ph}_{2} \: = \: \gamma_{2} \frac{{\sqrt2}}{N} \sum_{p} 
[ \rho_{+} (-p) + \rho_{-} (-p) ] \Phi_{0} (p) \; . 
\label{ph-harom}
\end{equation}
On the other hand, 
the rapidly oscillating phonon assisted backward scattering
term will pick up an extra factor $\exp [ \pm i \pi \delta n x]  
\equiv \exp [ \pm i (2 k_F - \pi) x ]$, in the form:  
\begin{equation}
H^{\rm el-ph}_{1} \: = \: \gamma_{1} \sum_{\nu = \pm, \sigma} \int dx 
[ \Psi^{\dagger}_{\nu, \sigma} \Psi^{}_{-\nu, \sigma} 
e^{i \pi \nu \delta n x} ] \Phi_{\pi} (x) \; , 
\label{ph-negy}
\end{equation}
with $\gamma_{1} = \gamma_{2} = \beta / \sqrt{M}$, where we used the same
subscripts for backward and forward scattering  as in g-ology (Voit 1994). 

In the continuum limit, the 
Su-Schrieffer-Heeger term in contrast to the Holstein coupling, 
gives only two terms $H^{\rm ph} + H^{\rm el-ph}_{- 1}$,
the standard phononic component and a rapidly oscillating  back 
scattering term. The fact that the forward scattering term is missing 
will have serious consequences on its applicability. $H^{\rm ph}$ 
is given in Eq.\ \ref{ph-ketto}, while the back scattering term 
$H^{\rm el-ph}_{- 1}$ differs from Eq.\ \ref{ph-negy} only
in a form factor:
\begin{equation}
H^{\rm el-ph}_{- 1} \: = \: \gamma_{- 1} \sum_{\nu = \pm, \sigma} 
\int dx [ i \nu \Psi^{\dagger}_{\nu, \sigma} \Psi^{}_{-\nu, \sigma} 
e^{i \pi \nu \delta n x} ] \Phi_{\pi} (x) \; , 
\label{ph-negy-new}
\end{equation}
with $\gamma_{- 1} = 4 \alpha_{\sigma} / \sqrt{M}$. 

The Su-Schrieffer-Heeger coupling was found 
(Gul\'{a}csi, Bussmann-Holder and Bishop 2003, 2004)
to be irrelevant to forward scattering processes and
to the Kondo lattice model away from half filling. Thus 
Eq.\ (\ref{ph-negy-new}) can be neglected in the present case,
similarly the term from Eq.\ (\ref{ph-negy}). With the remaining
on-site coupling Holstein terms, 
the calculations proceed similarly as in  
chapters \ref{ch2} and \ref{ch5} 
(see also Gul\'{a}csi, Bussmann-Holder 
and Bishop (2003,2004)) and, as such, the bosonization form of
Eq.\ (\ref{ph-new-egy}) is 
\beqa
H_{\KLph} &=& \frac{v_{F}a}{4\pi}\sum_{j,\nu}
\left[ \Pi_{\nu}^{2}(j) + [\partial_{x}\phi_{\nu}(j)]^{2}
\right] + \frac{Ja}{2\pi}\sum_{j}
[\partial_{x}\phi_{\sig}(j)] S_{j}^{z} 
\nonumber \\
&+& \frac{1}{2 N} \sum_{p} \left[ \Pi^{2}_{0} (p) + 
\omega^{2}_{0} \Phi^{2}_{0} (p) \right]
\nonumber \\
&+& \frac{1}{2} \int dx \left[ \Pi^{2}_{\pi} (x) + 
\omega^{2}_{\pi} \Phi^{2}_{\pi} (x) \right] 
\nonumber \\
&+& \gamma_{2} \frac{{\sqrt2}}{N} \sum_{p} 
\left[ \rho_{+} (-p) + \rho_{-} (-p) \right] \Phi_{0} (p) 
\nonumber \\
&+&  \frac{Ja}{4 \pi \alpha}\sum_{j}\left\{
\cos [\phi_{\sig}(j)] + \cos[2k_{F}ja + \phi_{\rho}(j)]\right\}
\left(e^{-i\theta_{\sig}(j)}S_{j}^{+} + {\rm h.c.}\right) 
\nonumber \\
&-& \frac{Ja}{4 \pi \alpha} \sum_{j}\sin[\phi_{\sig}(j)]
\sin[2k_{F}ja + \phi_{\rho}(j)]S_{j}^{z} \, .
\label{ph-new-ketto}
\eeqa
Except the lattice contributions, the electronic part is identical to
Eq.\ (\ref{bklm}), where for the $A$ constant the calculated value 
from chapter \ref{ch5} has been used. 

The canonical transformation follows section \ref{old5.1.2} 
and the final result takes the same form as the original
bosonized hamiltonian from Eq.\ (\ref{tbklm}). The
only difference is that the prefactor of the double-exchange 
ferromagnetic term of 
Eq.\ (\ref{FMterm}) is decreased: 
\beqa
\frac{J^{2}a^{2}}{4\pi^{2}v_{F}} \rightarrow
\frac{J^{2}a^{2}}{4\pi^{2}v_{F}} \, 
\frac{1}{1 + a \beta^2 /\pi K v_{F}} \, . 
\label{ph-ot-new}
\eeqa

Following the same analysis of section \ref{old5.1},   
an effective hamiltonian for the localized spins is 
obtained. This 
determines a quantum order-disorder transition between 
ferromagnetic and quantum disordered paramagnetic phases, with a 
critical coupling $J_{c}(U)$ which is higher than the  
$U = 0$ critical coupling by a factor 
$1 + a \beta^2  / \pi K v_{F}$. 

The renormalization of the 
spinon-holon velocities appears here due to the 
phonon terms which act oppositely to the Hubbard 
contribution from Eq.\ (\ref{5.4.7}), obtained in 
the previous section.   
While the Hubbard term leads to a localization 
of the spinons and an increased hopping of the holons, thus 
supporting a magnetic ground state, the phonons delocalize 
the spins, but localize the charges and act destructively 
on the magnetic properties. 
This means that the Hubbard term acts destructively on the 
spin polarons, as discussed in detail in section \ref{old6.1}.
On the other hand the phonons, as expected, are favouring polarons. 
As a consequence of which, the average width of the spin polarons
will increase compared to Fig.\ \ref{oldfig6.5}, the conventional 
Kondo lattice case. 

It is worth mentioning that the 
Hubbard term alone already suffices to establish two time 
scales for the holon-spinon dynamics, but an important 
renormalization of the critical properties of the system 
is achieved through the variable phonon coupling. The 
competition between the Hubbard and 
the phonon term obviously vanishes for $U = \beta^2 / K$.  

It is interesting to note the discrepancy between 
infinite dimensional calculations and the present one dimensional 
result. Many calculations to model colossal magnetoresistance
materials have been made in dynamical mean-field theory, 
which is an infinite dimensional 
approximation and therefore incapable of capturing spatial 
inhomogeneities. Gul\'{a}csi, Bussmann-Holder and Bishop 
(2003,2004) approached the same problem via 
a one dimensional approximation, but with 
techniques able to describe fluctuations of short-range order. 
Their results show that strong intrinsic spatial inhomogeneities of 
Griffiths type dominate the behaviour of the 
Kondo lattice. Consequently the inhomogeneities exhibit clear 
statistical scaling properties as a function of the proximity
to a {\sl quantum} (order-disorder) {\sl critical point}. 
The phonons enhance the inhomogeneities, which in a good 
approximation behave as a supercritical 
(metastable) phase of a two fluid model.

\section{\label{7.2}The dilute Kondo lattice model}

In the last two sections different extensions of the Kondo lattice 
model have been presented, which, based on the conclusions of section
\ref{old4.2}, may shed some light on the complexity of the real materials. 
In the present section a third extension will be discussed, again due
its relevance to real materials. In using the Kondo lattice to model
real materials, one of the most frustrating problems is the 
lack of antiferromagnetism from its ground state properties. 
Most manganites exhibit antiferromagnetism and the periodic
Anderson model even in 1D has an antiferromagnetic solutions
(Yanagisawa and Shimoi 1996). 

The solution to this problem is to bring the Kondo 
lattice even more closer to model real materials,
where as a function of doping levels the number of
impurity atoms is not necessarily equal to the lattice 
sites. That is, the Kondo lattice becomes {\it diluted}.
This problem has been studied by Gul\'{a}csi, {\it et al.}
(2003,2004) for the 1D case and as expected it has been shown
that for an incommensurate case of dilution the model
exhibits antiferromagnetism. 

In order to analyse this {\it dilution} effect, the 
Kondo lattice hamiltonian has to be re-written in order 
to take into account the `missing' impurity sites. Thus, the hamiltonian
of the 1D Kondo lattice from Eq.\ (\ref{4.11}) is written
\begin{equation}
H_{\DKL} = 
-t \sum^{L-1}_{i = 1, \sigma} 
( c^{\dagger}_{c i \sigma}
c^{}_{c i + a \sigma} + {\rm h.c.} ) + 
J \sum^{L}_{i = 1} {\cal P} \: {\bf S}_{c i} {\bf \cdot}
{\bf S}^{}_{i} \: {\cal P} \, ,
\label{dklm.1}
\end{equation}
where $L$ is the number of sites and $t > 0$ is the conduction
electron hopping. We measure the Kondo coupling $J$ in units
of the hopping $t$. We denote by $N_{f}$ ($n_{f} = N_{f} / L$)
the number (concentration)  of impurities and $N_{c}$ ($n_{c} =
N_{c} / L$) the number (concentration) of conduction electrons.
The constraint $N_{f} \le L$ is imposed by ${\cal P}$, which
is an operator that projects out a pre-determined set of $f$-spins.
The rest of the notations are the same as in chapter \ref{ch4}. 

The hamiltonian of Eq.\ (\ref{dklm.1}) has been named the
dilute Kondo lattice (DKL) model by Gul\'{a}csi, {\it et al.}
(2003,2004), where the behaviour of the DKL has been studied both by an
analytical approach, based on a standard bosonization scheme,
and by numerical calculations. The latter were performed using
the newly developed non-Abelian density-matrix renormalization-group
(DMRG) algorithm (McCulloch and Gul\'{a}csi 2000, 2001, 2002),
for details see cf.\ section \ref{6.5}. 

The bosonization follows chapters \ref{ch2} and \ref{ch5}, and 
the final result (Gul\'{a}csi, {\it et al.} 2003,2004) 
\begin{eqnarray}
H_{\DKL}  \: &=& \: \frac{v_{F}a} {4 \pi} \sum_{j,\nu}
\left\{ \Pi_{\nu}^{2}(j) + [\partial_{x}\phi_{\nu}(j)]^{2} \right\}
\: + \: \frac{Ja}{2 \pi}\sum_{j} [\partial_{x}\phi_{\sigma}(j)] S_{j}^{z}
\nonumber \\
&+& \: \frac{Ja}{4 \pi \alpha} \sum_{j} \left\{
\cos [\phi_{\sigma}(j)] + \cos[2k_{F}j + \phi_{\rho}(j)] \right\}
\left(e^{-i \theta_{\sigma}(j)} S_{j}^{+} + {\rm h.c.} \right)
\nonumber \\
&-& \: \frac{Ja}{2 \pi \alpha} \sum_{j} \: \sin[\phi_{\sigma}(j)]
\sin[2k_{F}j + \phi_{\rho}(j)] S_{j}^{z} \; .
\label{dklm.1_new}
\end{eqnarray}
has the same basic form as the original bosonized 
hamiltonian of Eq.\ (\ref{bklm})
\footnote{As in the previous section, here also the calculated
value, see chapter \ref{ch5}, of the constant $A$ has been used. },
except that the impurity spin, i.e. terms containing $S_{j}^{z}$,
$S_{j}^{+}$ and $S_{j}^{-}$, only contribute if there is an
$f$ spin at site $j$.

The most straightforward method to determine the ordering of the
localized spins is by applying a unitary transformation 
as in section \ref{old5.1.2}. 
In the new transformed basis the double
exchange interaction leading to ferromagnetism is clearly
exhibited and we obtain the effective Hamiltonian, similarly
to  Eq.\ (\ref{tbklm}), for the localized spins:
\begin{eqnarray}
\tilde{H}_{\DKL} &=&  - {\frac{J^2 a^2}{2 \pi^2 v_{F}}}
\sum_{i, j} {\frac{\alpha}{\alpha^2 + [(i - j)a]^2}} \:
S^{z}_{i} S^{z}_{j}
\nonumber \\
&+& {\frac{J a}{2 \pi \alpha}} \sum_{i}
\{ \cos[K(i)] + \cos[2 k_{F} i] \} S^{x}_{i}
\nonumber \\
&-& {\frac{J a}{2 \pi \alpha}} \sum_{i}
\sin[K(i)] \sin[2 k_{F} i] S^{z}_{i} \; .
\label{dklm.2}
\end{eqnarray}
The term $K(i)$, defined in Eq.\ (\ref{Kj}) and
introduced by the unitary transformation
(for details, see section \ref{old5.1.2}), 
counts all the $S^{z}_{i}$'s to the right of the site $i$
and subtracts from those to the left of $i$: $K(i) =
(J a/ 2 v_{F}) \sum_{l = 1}^{\infty} ( S^{z}_{i + l} -
S^{z}_{i - l} )$ (see also section \ref{old5.2} and
Eq.\ (\ref{Kj2})). This term gives the crucial difference between
the Kondo lattice and the dilute Kondo lattice, 
as will be explained later on.
The most important term in Eq.\ (\ref{dklm.2}) is the first one,
which clearly shows that a ferromagnetic coupling is present 
also for the dilute Kondo lattice model (Gul\'{a}csi, 
{\it et al.} 2003,2004). This coupling is non-negligible 
for $N_{c} < N_{f}$ and
$i - j \le \alpha / a$, ie, when $i - j$ is less then the width
of the spin polarons\footnote{$\alpha /a$ measures the effective range of
the double-exchange, see Fig.\ \ref{oldfig6.5}, which was shown
to be equivalent to the width of the spin polarons, for details
see sections \ref{old5.1.3} and \ref{old6.1}.},   
and its strength will decrease with increasing 
distance between impurity spins. 

The reader may notice that in Eq.\ (\ref{dklm.2}) an exponential
cutoff function has been used (Gul\'{a}csi, {\it et al.} 2003,2004). 
The bosonization can be formulated independently of the form
of the cutoff, see section \ref{old5.1.3} for details. Thus
Eq.\ (\ref{dklm.2}) is valid with other form of cutoffs,
see Eqs.\ (\ref{cutoffs}) and (\ref{integral}), 
e.g., for Gaussian cutoff the first term of Eq.\ (\ref{dklm.2}) becomes
\begin{equation}
- {\frac{J^2 a^2}{4 \alpha \pi^{3/2} v_{F}}}
\sum_{i, j} e^{ - \left[ \frac{(i - j) a}{2 \alpha} \right]^{2} } \:
S^{z}_{i} S^{z}_{j} \, . 
\label{dklm.2_new}
\end{equation}

For $N_{c} < N_{f}$, the physical picture given by
Eq.\ (\ref{dklm.2}) will be crucially different if the
lattice of impurity spins contains {\sl commensurate}
or {\sl incommensurate} array of holes. If a {\sl commensurate} 
depletion of the impurity spins is present, then the ferromagnetic
term can be approximated in the usual way (see chapter \ref{ch5}) 
by taking $\approx 1 / n_{f}$ for the shortest average
distance between $f$ spins: 
\begin{equation}
- \frac{ \alpha J^2 a ^2 n^2_{f}}{ [2 \pi^2 v_{F}
(a^2 + \alpha^2 n^2_{f}) ] }
\: \sum_{i} S^{z}_{i} S^{z}_{i + a / n_{f}} \, .
\label{dklm.3}
\end{equation}
Lattice sites which are not occupied by $f$ spins are inert and
do not contribute to the ferromagnetic phase. This was verified by
DMRG (Gul\'{a}csi. {\it et al.} 2003,2004): 
the calculated $f-f$ spin correlation functions behave
similarly as those of the normal Kondo lattice. 
The $f$-structure factor has the usual peak at $k/\pi = N_c/N_f$ for
low $J$, hence in the commensurate case the dilute Kondo lattice 
behaves similarly to the standard Kondo lattice model.

To understand the behaviour of the second and third term from Eq.\
(\ref{dklm.2}), we notice that $K (i)$ is vanishingly small for the
commensurate case, as the number of $f$ impurity spins to the left and
to the right of a given site $i$ is the same. So the effective
Hamiltonian will reduce to the random transverse field Ising model, as
presented in section \ref{old5.1.3}. 
However, this ferromagnetic phase
disappears for larger distances between impurities because, as mentioned
earlier, the double exchange interaction vanishes if the average distance
between impurity spins, $1 / n_{f}$, is larger than $\alpha$. This is
very important because it ensures that the single impurity limit,
$n_{f} \rightarrow 0$, is free of ferromagnetism, as it should be.

The {\sl incommensurate} case is more difficult than the commensurate case.
The reason is that in the low concentration limit the properties of
the dilute model 
will be very much dependent on the random distribution of $f$
spins. We may observe phase separation or clusterization processes in
this case. In this limit, where the average distance between impurities
is very large, then the single impurity approximation
seems natural. However, if we look at small doping of $f$ electrons
only, then the main difference compared to the commensurate limit
studied previously is that the $K (i)$ term, in Eq.\ (\ref{dklm.2}), is not
negligible anymore. The impurity $f$ spins are no longer equally
distributed to the left and right of a given site $i$, hence $K(i)
\approx (-1)^{i} (J a/ 2 v_{F})$, which gives rise to a staggered
field. The properties of Eq.\ (\ref{dklm.2}) are then given by the
staggered field Ising model, which gives an antiferromagnetic ordering.
This antiferromagnetic ordering of the impurity spins represents a new
element in the dilute model compared to the Kondo lattice. 
This corresponds to the soliton lattice obtained by Schlottmann (1992) 
in a dynamical mean-field treatment of 
the three dimensional dilute Kondo lattice.

Similar behaviour also occurs above half-filling, i.e. $N_{c} >
N_{f}$, where double exchange (as shown previously) does not
appear. But bosonization still works: the effective Hamiltonian
reduces to the second and third terms of Eq.\ (\ref{dklm.2}), from which
the most dominant term, for low
doping of impurity spins, as in the case described previously,
is a staggered $S^{z}_{i}$ field. As the
first term in Eq.\ (\ref{dklm.2}) is missing in this case, 
the only fluctuation which
can destroy a locked staggered order is $S^{x}_{i}$. For large $J$ ($
4 \lesssim J$) the staggered order wins. While for smaller values of
$J$ the systems will be disordered.

As we approach half filling from both sides, the bosonization approach
breaks down as the strongly oscillating (umklapp) fields start
dominating. The DKLM will undergo a metal-insulator transition as in a
standard quantum sine-Gordon model (Gul\'{a}csi and Bedell 1994) 
by dynamical mass generation. 
A spin gap will also appear. This can be understood easily,
because the half filled dilute Kondo lattice model is equivalent 
to the quarter filled periodic Anderson model, which has an  
antiferromagnetic order (Yanagisawa and Shimoi 1996). 
The only difference from the Kondo lattice is that the
massive solitons obtained for the dilute Kondo lattice 
are of Su, Schrieffer and Heeger (1980a,1980b) type.

The above bosonization results have been confirmed by a 
`state of the art' non-Abelian DMRG analysis by Gul\'{a}csi,
{\it et al.} (2003,2004) in the commensurate
case, for both the $N_{c} < N_{f}$ and $N_{c} > N_{f}$,
where ferro- and antiferro-magnetism, respectively exist. 
For $N_{c} < N_{f}$ ferromagnetism appears at large $J$ values,
see Fig.\ \ref{Fig1-dilute}. 
In the opposite limit, i.e., $N_{c} > N_{f}$, it was confirmed
(Gul\'{a}csi, {\it et al.} 2003,2004) 
the existence of antiferromagnetism by calculating the spin
structure factor $S (k)$ (for details see Fig.\ 2 of Gul\'{a}csi,
{\it et al.} (2003,2004)). 

%%%%%%%%%%%%%%%%%%%%%%%%%%%%%%%%%%%%%%%%%%%
%% conclusions
%%%%%%%%%%%%%%%%%%%%%%%%%%%%%%%%%%%%%%%%%%%%

\cleardoublepage
\chapter{Summary}
\label{summary}  

Low-dimensional interacting many-electron systems are
an important current topic in condensed matter physics. 
The interest is due to two factors. Firstly, many 
materials are sufficiently anisotropic that the motion 
of the electrons is effectively confined to 1D or 2D,
and these materials often exhibit intriguing and complex 
properties that are difficult to understand on the basis 
of 3D systems. For example in some organic conductors,   
such as the Bechgaard salts, the conduction electrons are 
confined to move along a chain of ions, and exhibit 
quasi-1D behaviour. Similarly, in the cuprate high-$T_{c}$ 
superconductors, the electrons responsible for the 
superconductivity move in 2D planes containing copper and 
oxygen. 

The second reason for studying low-dimensional systems is 
as a test bed for ideas in higher dimensions. Strongly-correlated 
many-electron systems, in which interaction effects are all 
important, are notoriously difficult to describe. The standard 
description of interacting 3D many-electron systems is in 
terms of Landau Fermi liquid theory, and is based essentially 
on similarities to the non-interacting case. This 
description fails for many strongly-correlated systems. 
The correct starting point in these cases is uncertain. 
Some simplification, and the possibility for progress, 
may be obtained by considering a reduced state space for 
the system, in particular a 1D state space. The hope is 
then that the properties identified in the reduced state 
space carry over, at least qualitatively, to higher 
dimensions.

These are the reasons why the 1D models in general, and
the 1D Kondo lattice model in particular, has been
studied intensively for the past decade. The Kondo
lattice model is one of the most important canonical
models in the study of strongly correlated systems
\footnote{A brief overview of the strongly correlated
electron systems in given in Appendix \ref{appd}.}. It is 
particularly suited to study a class of rare earth
compounds, called heavy fermion systems, where the main
challenge is to understand the interaction between an 
array of localised moments (generally $f$-electrons in 
lanthanide or actinide ions) and conduction electrons 
(generally $p$- or $d$-band). This review focused on the
theoretical description, as given by the bosonized 
solution, of the one dimensional convetional Kondo 
lattice model at partial band filling. 
All the presented results were
mainly based on Honner and Gul\'{a}csi (1997a,1998b).
More details can be found also in Honner 
and Gul\'{a}csi (1997b,1997c,1998a,1999). 
The review divides naturally into two parts, relating 
to the description of the formalism, and then to the 
application.
 
Bosonization is a well-known method in 1D 
systems, and the presentation in chapter \ref{ch2} 
is a variation on a well-worn theme.  
In the standard Luttinger model approach, 
bosonization is derived beginning from a 
field-theoretic--type approximation to the condensed matter 
system of interest (cf.\ section \ref{old1.2.3}), 
and replaces the finite Fermi sea with 
two infinite Dirac seas. In chapter \ref{ch2} this approximation 
was not made, and bosonization was developed beginning from the 
realistic system with a finite Fermi sea. The resulting   
Bose representations (cf.\ section \ref{old2.3}) are similar to 
the standard Luttinger model results, but reveal several 
new features which clarify the status of bosonization 
in applications to realistic condensed matter systems.
The new features are discussed in chapter \ref{ch3}, 
section \ref{old3.1}, and may be summarized as follows: 

{\it i}) The standard Luttinger model bosonization involves a 
length $\alpha$ which of necessity is infinitesimal, and 
is to be taken to zero at the end of a calculation. This  
sometimes generates manifestly non-physical results in 
condensed matter systems, such as divergent excitation 
spectra, and $\alpha$ is then reinterpreted on an 
{\it ad hoc} basis as a finite physical quantity. The 
reinterpretation has proved problematic, as no consensus 
has been reached on the general meaning of $\alpha$, and 
a finite $\alpha$ is strictly inconsistent with the 
Luttinger model bosonization. In the development of 
bosonization given in chapter \ref{ch2}, $\alpha$ emerges 
naturally from the formalism as the minimum wavelength 
for bosonic density fluctuations. $\alpha$ separates 
longer wavelength density fluctuations, which are 
collective and satisfy Bose statistics, from shorter 
wavelength density fluctuations, which show 
single-electron--type behaviour, and most 
generally $\alpha \gtrsim {\cal O}(k_{F})^{-1}$. 
This interpretation 
of $\alpha$ was compared with previous interpretations 
in section \ref{old3.1.3}, and was shown to include them as 
special cases.  

{\it ii}) In the bosonization developed in chapter \ref{ch2}, 
$\alpha$ is kept finite in a consistent manner 
throughout the bosonization formalism. 
This contrasts with the reinterpretation of $\alpha$ 
in applications of the standard Luttinger model 
bosonization to condensed matter systems, in which 
$\alpha$ is not treated consistently. A consistent  
treatment reveals  
a change in the behaviour of the Bose field commutators 
over lengths below ${\cal O}(\alpha)$ 
(cf.\ section \ref{old3.1.2}, Figs.\ \ref{oldfig3.1} 
and \ref{oldfig3.2}). This has not 
previously been emphasized when $\alpha$ is kept finite 
in the Luttinger model bosonization, and is an important 
formal element in the application of bosonization to 
the Kondo lattice.

{\it iii}) Bose representations for fermionic operators do 
not have the same ranges of validity. This is obscured 
in the Luttinger model, in which all representations are 
exact. In chapter \ref{ch2} it was found that the Bose 
representations for the non-interacting hamiltonian 
and the density operators is exact, but that other 
representations are only asymptotically valid.\footnote{The 
rule of thumb is that if the Luttinger model representation 
(cf.\ Appendix \ref{appa}) involves the normal-ordering convention 
non-trivially, then the corresponding Bose representation 
in the condensed matter system is exact. If the Luttinger 
model representation does not require a non-trivial  
application of the normal-ordering convention, then the  
corresponding Bose representation in the condensed matter 
system is valid only over asymptotically large separations.}

These new properties have intrinsic value, and help to 
clarify the results of bosonization in condensed matter 
systems. However, in applications to simpler one-component systems, 
the use of the bosonization of chapter \ref{ch2} does not reveal 
any qualitatively new behaviour beyond that obtained  
with a Luttinger model bosonization. This is because in 
the simpler systems, such as the 
Hubbard model (cf.\ section \ref{old3.2}), $\alpha$ is of the order 
of the average inter-electron spacing,  and in 
homogeneous one-component 
systems this is the smallest physically meaningful length 
scale. In these cases $\alpha$ acts as a harmless short-distance 
cut-off, much as in field theory, and may be made 
arbitrarily small, as in the Luttinger model, without 
overlooking any important physical processes. This is not 
the case in the Kondo lattice model, 
which is a two-component system. In the partially-filled 
Kondo lattice the inter-conduction electron spacing is 
larger than the lattice spacing between the localized spins, 
and if $\alpha$ for the bosonized conduction band 
is made arbitrarily small, then the double-exchange 
ferromagnetic ordering between neighbouring localized 
spins is overlooked. 

The bosonized solution of the Kondo lattice is 
contained in  
chapters \ref{ch5} and \ref{ch6}. Chapter \ref{ch8}
contains bosonization of some extended models of the Kondo
lattice. Chapter \ref{ch6} deals with the  
ground-state phases of the partially-filled 
1D Kondo lattice model, and a description of the  
ferromagnetic-paramagnetic quantum phase transition.
The bosonization of chapter \ref{ch2}, as opposed to the  
standard Luttinger model bosonization, 
was an essential aspect in this work. The underlying  
physical idea in this work  
was that bosonization could provide a description 
of coherent conduction electron hopping, which is  
a single-electron property, over lengths below 
$\alpha$, together with 
collective density fluctuations,  
as is usual in bosonization, over lengths beyond 
$\alpha$. The coherent 
conduction electron hopping should mediate a 
ferromagnetic double-exchange ordering of the 
localized spins. This was confirmed in section \ref{old5.1.3}, when a 
ferromagnetic interaction Eq.\ (\ref{FMterm}), 
with features characteristic of 
a double-exchange interaction, was exhibited between the 
localized spins after a unitary transformation. A summary 
of the method and main results of chapters \ref{ch5} and 
\ref{ch6} follows.

Chapter \ref{ch5} described double-exchange ferromagnetic ordering 
in the partially-filled $J > 0$ 1D Kondo lattice, and 
the destruction of the ferromagnetic phase by spin-flip 
disorder scattering.  
At weak-coupling deep in the disordered paramagnetic phase, 
the scattering  
determines RKKY-like correlations in the localized spins. 
Kondo singlet formation was been taken into account 
indirectly, via an effective range for the double-exchange 
interaction. The effective range was identified as follows: 
The length $\alpha \gtrsim {\cal O}(k_{F})^{-1}$ originated  
in bosonization of chapter \ref{ch2} 
as the minimum wavelength for density 
fluctuations which satisfy bosonic commutation relations.  
The bosonization describes fluctuations beyond $\alpha$, and 
keeps the conduction electrons finitely delocalized 
over lengths below $\alpha$. The electrons preserve their 
spin over this range. In section \ref{old5.1.3} it was shown that 
this finite delocalization may be identified with the  
length for coherent conduction electron hopping, and  
measures the effective range of the double-exchange 
ferromagnetic ordering induced on the localized spins 
by the electrons
(cf.\ Eq.\ (\ref{FMterm}) and Fig.\ \ref{oldfig5.1}). The reason this 
works is that double-exchange is conceptually a simple 
interaction. It reflects only the tendency for hopping 
electrons to preserve their spin, 
as they move to screen the more numerous localized spins.
Double-exchange is characterized by the dimensionless 
factor $Ja/v_{F}$, with $v_{F}$ the conduction electron 
Fermi velocity and $a$ the lattice spacing. 

A ferromagnetic double-exchange interaction term 
Eq.\ (\ref{FMterm}) was obtained 
between the localized spins. The term was derived
using a unitary transformation and is non-perturbative. 
This contrasts with other interactions derived for the 
localized spins in the Kondo lattice, such as the 
RKKY interaction, which are perturbative. 
The unitary transformation generates an effective 
hamiltonian Eq.\ (\ref{Heff}) for the localized spins. 
The competing affects on the spin ordering are made 
manifest in the effective hamiltonian. The competing effects 
are double-exchange ordering at stronger coupling, and 
spin-flip disorder processes involving nearly free  
electrons at weak-coupling. 
The transition from a double-exchange 
ordered ferromagnetic phase to a quantum disordered 
paramagnetic phase 
was then shown in section \ref{old5.2} to be the quantum order-disorder 
transition of the transverse-field Ising chain 
(cf.\ Eq.\ (\ref{Jc})). 
This describes double-exchange ordered regions 
of localized spins being destroyed as the electrons 
become weakly-bound, and become free to move and scatter 
along the chain. As the coupling $J$ is lowered, the 
transition is signalled by a continuously vanishing 
spontaneous magnetization Eq.\ (\ref{Mo}), and a breakdown 
in long-range correlations between the localized spins. 
It was shown in section \ref{old5.3} that well below the critical line, 
no remnants of the ferromagnetic ordering remain, and 
the effective hamiltonian describes dominant correlations 
in the localized spins at $2k_{F}$ of the conduction band. 

Spin disorder in the Kondo lattice 
occurs through forward and backscattering spin-flip 
processes between the electrons and the localized 
spins. Interesting properties were identified in section \ref{old5.2} 
which result from an incommensurate modulation of the 
backscattering momentum transfer with respect to the 
underlying lattice of localized spins: For 
incommensurate fillings, the conduction band has a 
competing periodicity with respect to the spin chain, 
and the electrons are unable to totally order, or 
totally disorder the spin chain at criticality. This 
leaves anomalous regions of double-exchange ordered 
localized moments close to criticality in the paramagnetic 
phase, as only a quasi-commensurate fraction of the 
electrons become weakly-bound at the transition. 
Similarly, there remain anomalous disordered regions close 
to criticality in the ferromagnetic phase. 
The anomalous regions are 
very dilute, but they dominate the low-energy behaviour of 
the localized spins. The magnetization Eq.\ (\ref{M}) 
is highly singular 
for a finite range of couplings about the critical line:  
The magnetization has a continuously variable 
power law exponent, and the susceptibility is infinite for 
a finite range of couplings even in the paramagnetic 
phase (cf.\ Eq.\ (\ref{chiT})). 

The effect of repulsive conduction electron interactions 
on the ferromagnetic-paramagnetic transition were considered 
in section \ref{old5.4}. It was found that double-exchange is 
characterized in this case by the dimensionless factor 
$Ja\sqrt{v_{F}/v_{\sig}^{3}}$, 
where $v_{\sig}$ is the conduction electron 
spin velocity. This factor 
is enhanced (cf.\ Eq.\ (\ref{5.4.7})) 
over the factor $Ja/v_{F}$ characterizing 
double-exchange with no interaction between the conduction 
electrons. This pushes the critical line to lower values 
of the coupling $J$, 
and for infinitely strong repulsive interactions 
ferromagnetism occupies the entire phase 
diagram for $J \neq 0$. The reason for this behaviour is 
that for infinite repulsive interactions, the double-exchange 
ordered regions are prevented from interfering, and the 
spin-flip disorder processes are ineffective. 

\begin{figure}[ht]
\centering
\includegraphics[scale=0.7,angle=-90]{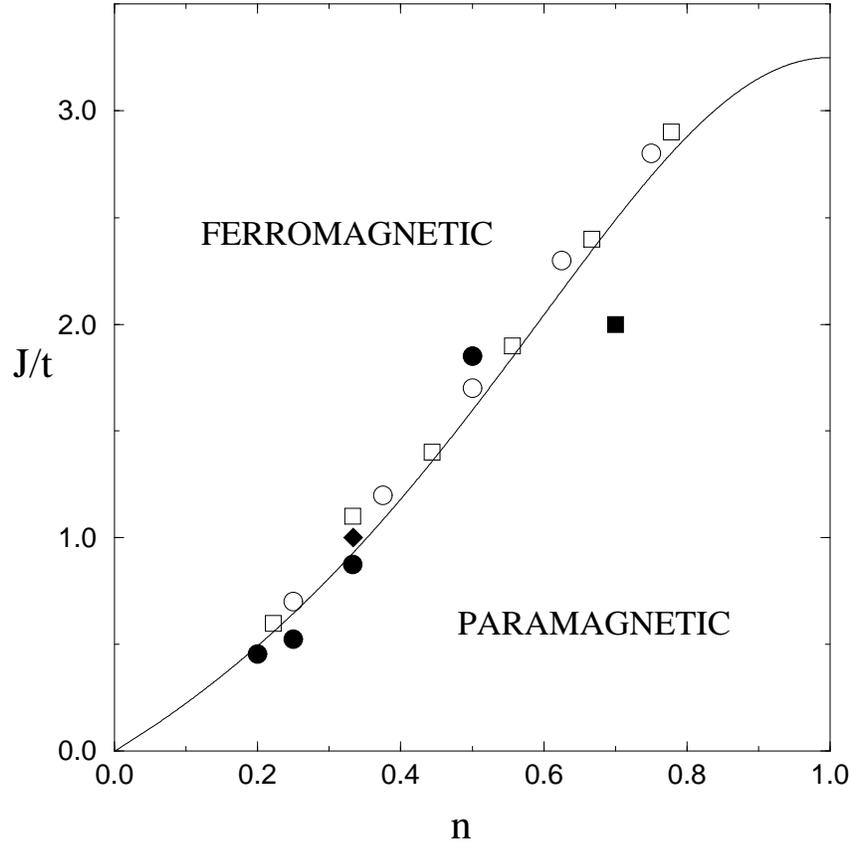}
\vspace{3.0mm}
\caption{\label{oldfig6.2}Ground-state phase diagram 
of the 1D Kondo lattice with 
$J > 0$. The critical line is from 
Eq.\ (\ref{Jcafm}), and uses the line of Fig.\ \ref{oldfig6.1} to 
fix the proportionality constant in $\alpha/a \propto 
\sqrt{t/J}$. 
Numerically determined critical points are as in 
Fig.\ \ref{oldfig6.1}. At incommensurate fillings, 
there are Griffiths singularities in the free 
energy in a finite region of the parameter space 
about the critical line. At small $Ja/v_{F}$ in 
the paramagnetic phase, the 
system presents an RKKY-like behaviour with dominant 
correlations in the localized spins at $2k_{F}$ 
of the conduction band. These properties of 
these phases are discussed at 
length in chapter \ref{ch5}.}
\end{figure}

\begin{figure}[ht]
\centering
\includegraphics[scale=0.7,angle=-90]{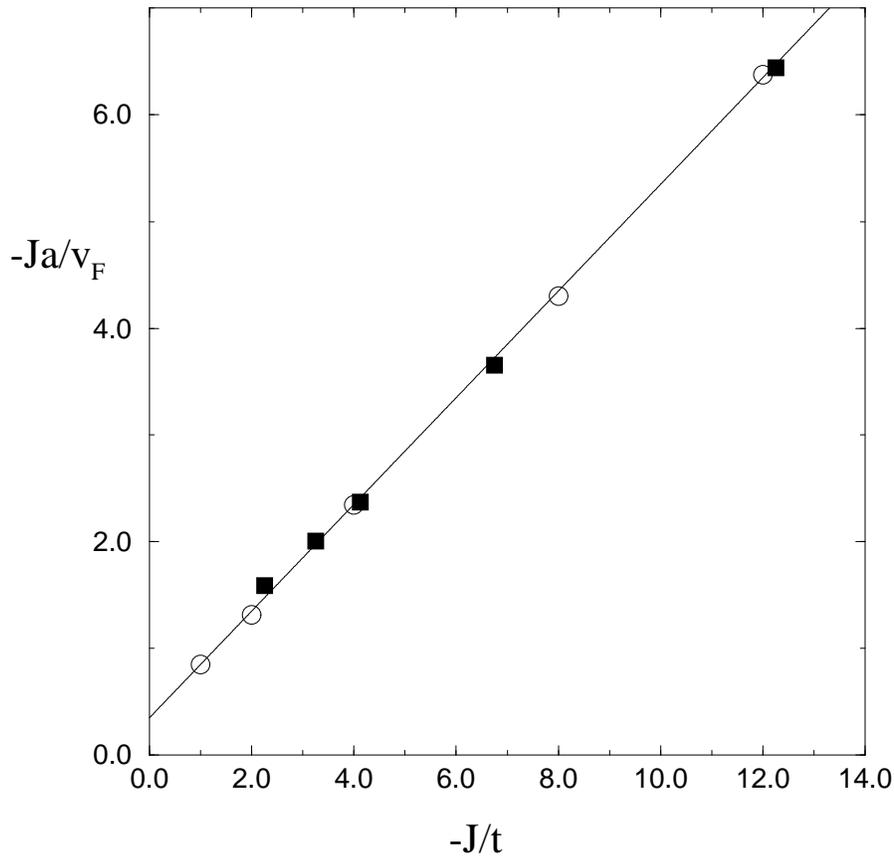}
\vspace{3.0mm}
\caption{\label{oldfig6.3}Plot of the dimensionless parameter $-Ja/v_{F}$, 
which characterizes double-exchange ordering in 
the $J < 0$ Kondo lattice, against coupling $J$ for 
numerically determined ferromagnetic-paramagnetic 
transition points: 
open circles are results on classical 
localized spins using Monte Carlo on systems up 
to 40 sites from Yunoki, {\it et al.} (1998); 
filled squares are density-matrix renormalization group 
 results on a 16 site chain for quantum 
spins 3/2, and a correspondingly normalized 
coupling, of Yunoki,  {\it et al.} (1998). 
The straight line of best fit gives 
very good agreement with all points.}
\end{figure}

\begin{figure}[ht]
\centering
\includegraphics[scale=0.7,angle=-90]{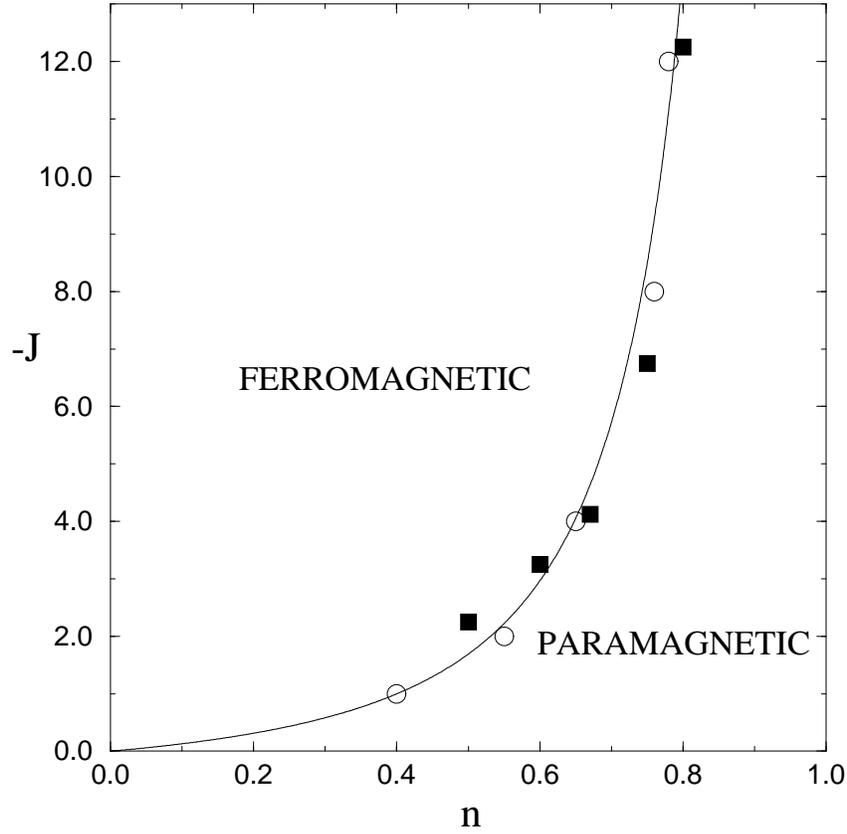}
\vspace{3.0mm}
\caption{\label{oldfig6.4}Phase diagram for the 1D Kondo lattice with a  
ferromagnetic coupling $J < 0$. 
The critical line is from 
Eq.\ (\ref{Jcfm}), and uses the line of Fig.\ \ref{oldfig6.3} 
to determine the constant of proportionality 
in $\alpha/a \propto \sqrt{t/J}$. 
Numerically determined transition points are as in 
Fig.\ \ref{oldfig6.3}. Properties of the localized 
spins close to criticality and at weak-coupling 
are as for the $J > 0$ Kondo lattice. The phase 
separated region identified by Yunoki, {\it et al.} (1998)
for the classical spins from $J_{c}/t = 4$ into the 
paramagnetic phase is not shown. Phase separation 
is observed by Yunoki, {\it et al.} (1998) 
in the quantum simulation 
only at stronger couplings, and away from the 
ferromagnetic-paramagnetic
transition closer to half filling.}
\end{figure}

\begin{figure}[ht]
\centering
\includegraphics[scale=0.7,angle=-90]{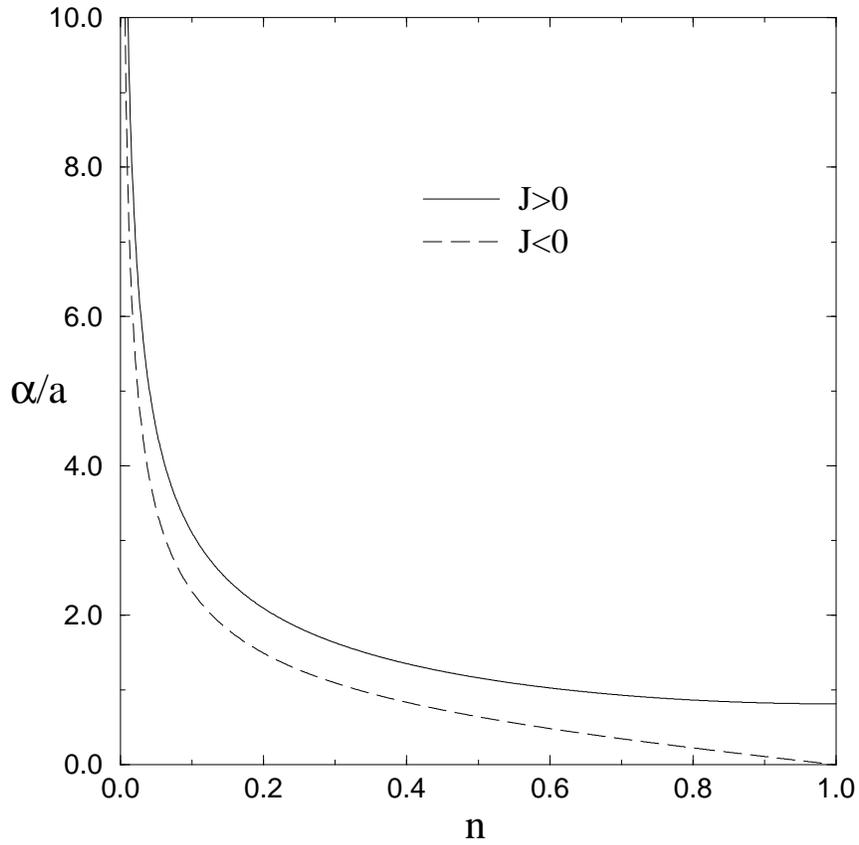}
\vspace{3.0mm}
\caption{\label{oldfig6.5}The effective range $\alpha$ of the double-exchange 
interaction in units of the lattice spacing against 
filling $n$ on the critical line. An 
exponential cut-off function $\Lambda_{\alpha}(k) 
= \exp -(\alpha|k|/2)$ has been chosen. The vanishing 
of the range at half-filling in the $J < 0$ Kondo lattice 
leads to a divergence in the critical line as 
half filling is approached (cf.\ Fig.\ \ref{oldfig6.4}).
$\alpha$ also measures the effective width of the 
corresponding spins polarons, for details see sections 
\ref{old5.1.3} and \ref{old6.1}.}
\end{figure}

\begin{figure}[ht]
\centering
\includegraphics[scale=0.7]{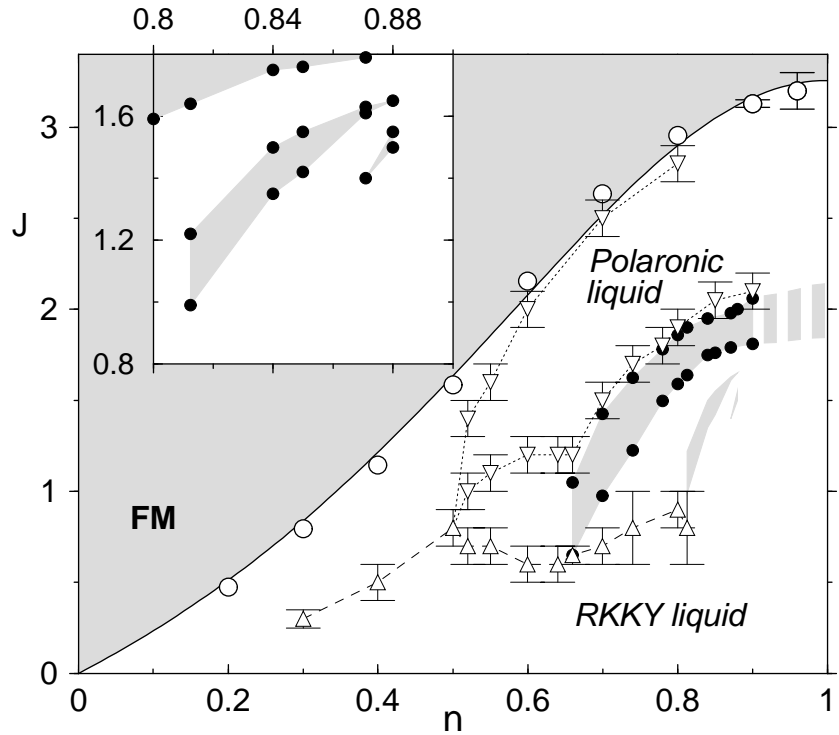}
\vspace{3.0mm}
\caption{\label{Ian-kondo-MOS-fig4}
Similar to Fig.\ \ref{oldfig4.2}. The stars are the 
first non-Abelian density-matrix renormalization data 
(McCulloch, {\it et al.} 1999) compared with the old numerical
results: the filled diamond is a quantum Monte Carlo
point (Troyer and W\"{u}rtz 1993); open circles are the exact 
diagonalization results (Tsunetsugu, Sigrist and Ueda 1993); 
the filled square is a density-matrix renormalization 
group point (Moukouri and Caron 1995); the filled circles 
are the infinite size density-matrix renormalization group 
results (Caprara and Rosengren 1995). The dashed line is
the bosonization result from Fig.\ \ref{oldfig6.2}
(derived in chapters \ref{ch5} and \ref{ch6}) for details
on the analytic form of the curve, see Eq.\ (\ref{Jcafm})
of section \ref{old6.2}.}
\end{figure}

\begin{figure}[ht]
\centering
\includegraphics[scale=0.7]{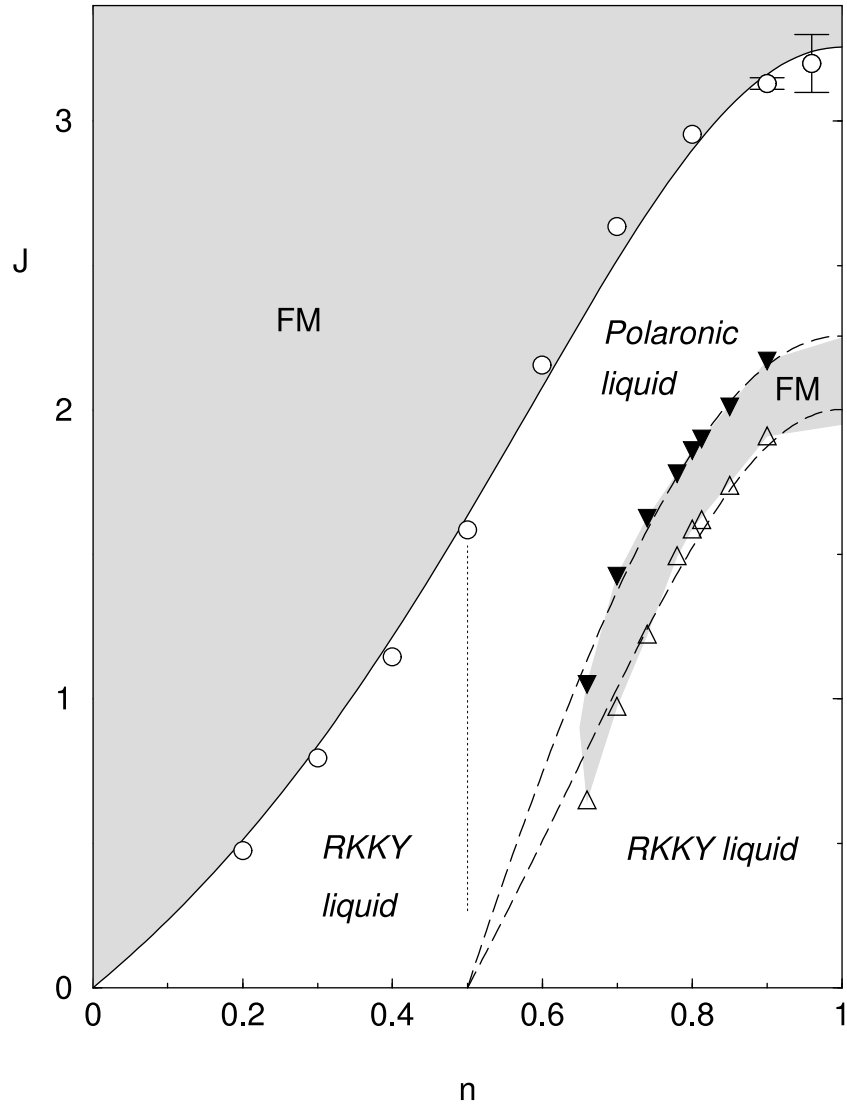}
\vspace{3.0mm}
\caption{\label{Ian-kondo-PRB-fig1} 
The phase diagram of the Kondo lattice as obtained by 
McCulloch, {\it et al.} (2002). The two shaded areas are
the FM phases. The open circles and triangles correspond to
points at which the ferromagnetic energy level crosses the $S=0$ level.
The errors are of the order of the symbol size in this figure.
The dashed curves are the derived phase transition lines, for details
see section \ref{KLM-DMRG}. The solid curve is the bosonization result
similar to Figs.\ \ref{oldfig6.2} and \ref{Ian-kondo-MOS-fig4}.}
\end{figure}

\begin{figure}[ht]
\centering
\includegraphics[scale=0.7]{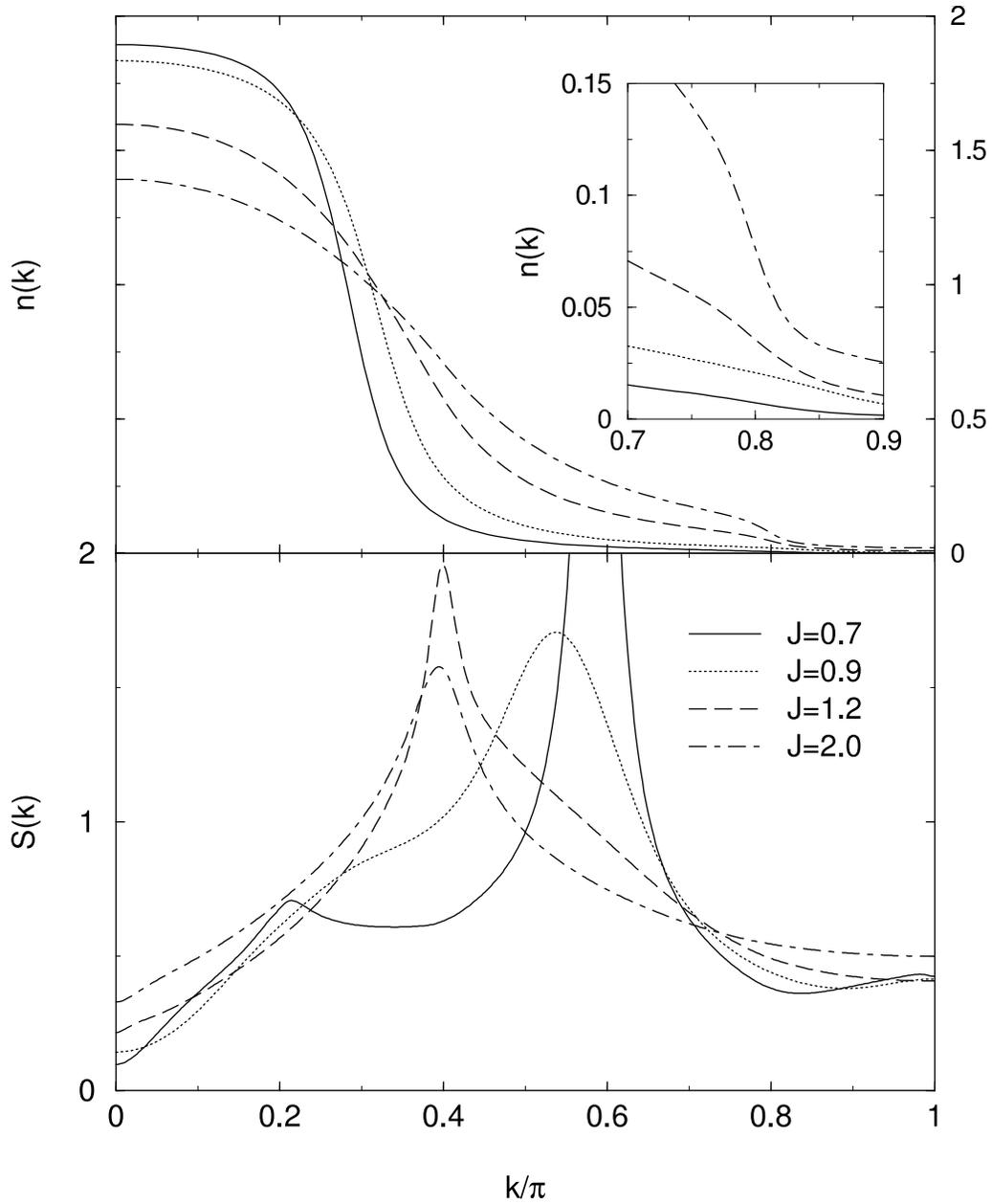}
\vspace{3.0mm}
\caption{\label{Ian-kondo-PRB-fig3} 
Typical $J$ dependence of the spin structure factor, $S(k)$,
and the momentum distribution, $n(k)$ ($n = 0.6$). It can be seen 
that for low $J$ values the small Fermi surface, at $2 k_{F}$ is 
realised, while for large $J$ values the large Fermi surface
emerges at $2 k_{F} - \pi$. Similar plots for other $n$ values
can be found in Juozapavicius, {\it et al.} (2002).}
\end{figure}

\begin{figure}[ht]
\centering
\includegraphics[scale=0.7]{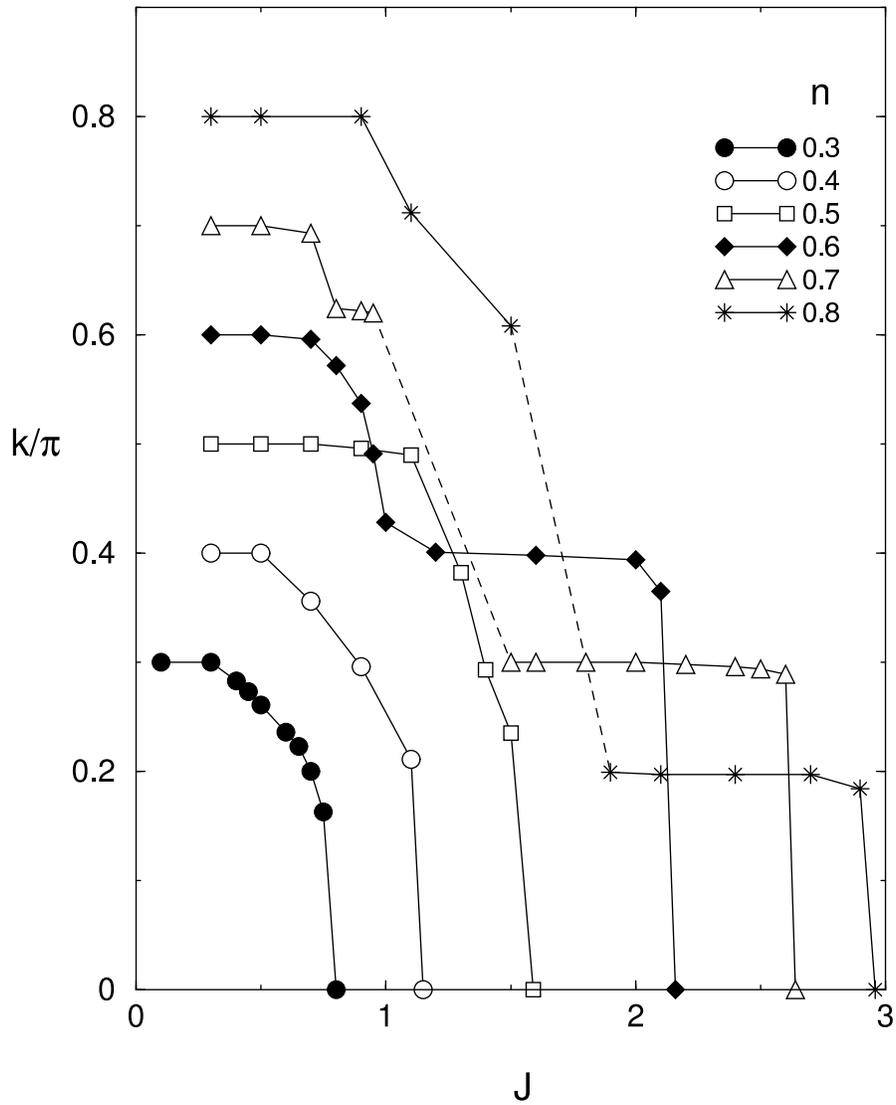}
\vspace{3.0mm}
\caption{\label{Ian-kondo-PML-fig2} 
Several examples of the $f$ spin structure factor, $S(k)$, peak position
as a function of conduction electron doping, $n$. The dotted lines
represent the second ferromagnetic region, where the peak jumps to
$k = 0$ (see Figs.\ \ref{Ian-kondo-PRB-fig1} and \ref{Ian-kondo-PMB-fig4} 
for the second ferromagnetic phase). 
It can bee seen that for $n > 0.5$ the peak moves from
its small Fermi surface value of $2 k_{F}$ to the large Fermi
surface value of $2 k_{F} - \pi$. }
\end{figure}

\begin{figure}[ht]
\centering
\includegraphics[scale=0.7]{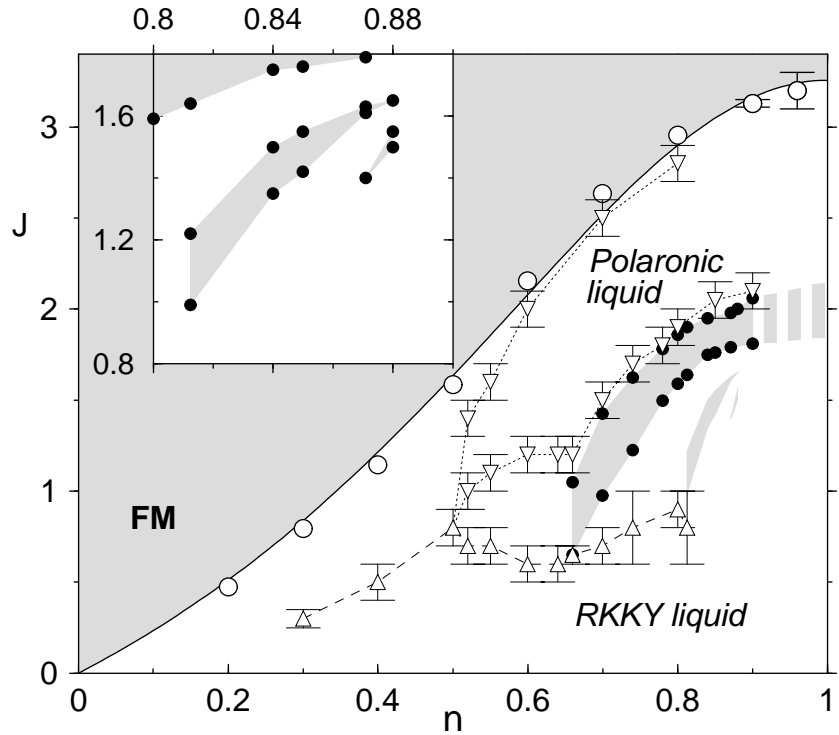}
\vspace{3.0mm}
\caption{\label{Ian-kondo-PMB-fig4} 
The phase diagram of the Kondo lattice as obtained by 
Juozapavicius, {\it et al.} (2002). 
The ferromagnetic phases are
shaded. The polaronic liquid with a large Fermi surface is enclosed by
a dotted curve, while the area below the dashed curve, denoting the
RKKY liquid, has small Fermi surface. The remaining areas have an
intermediate value of the $f$-spin structure factor peak, see Fig.\
\ref{Ian-kondo-PML-fig2}. The circles correspond to
points at which the ferromagnetic energy level crosses the $S=0$ level.
The errors are of the order of the symbol size in this figure.
The solid curve is the bosonization result
similar to Figs.\ \ref{oldfig6.2} and \ref{Ian-kondo-MOS-fig4}.}
\end{figure}

\begin{figure}[ht]
\centering
\includegraphics[scale=0.7]{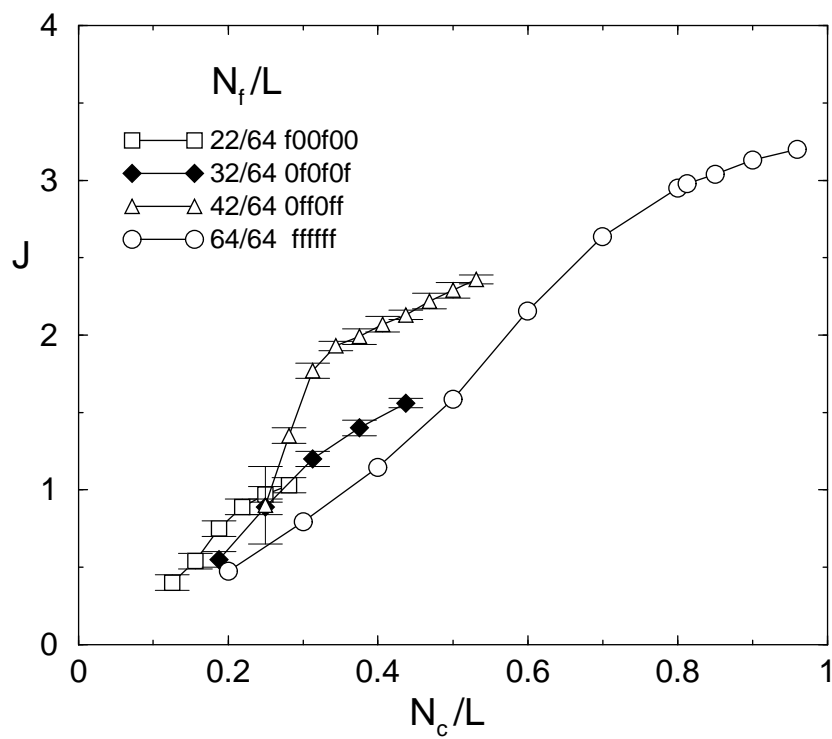}
\vspace{3.0mm}
\caption{\label{Fig1-dilute} The phase diagram of the dilute Kondo model
($J$ is plotted in units of $t$) 
in different commensurate filling cases for $N_{c} < N_{f}$. Legend
shows patterns of dilution. Open circles correspond to the standard
Kondo lattice model. The system of a given dilution pattern is ferromagnetic
above the corresponding solid line and paramagnetic below.}
\end{figure}

Since Kondo singlet formation is taken into account 
in the description of chapter \ref{ch5} only indirectly, the results 
extend also to the Kondo lattice with a ferromagnetic 
$J < 0$ coupling, 
and a ferromagnetic-paramagnetic transition of the same class as 
$J > 0$ was identified. The phase diagrams for the  
Kondo lattice for either sign of the coupling were 
plotted in chapter \ref{ch6}.  Fig.\ \ref{oldfig6.2} gives the phase diagram 
for antiferromagnetic couplings $J > 0$, and 
Fig.\ \ref{oldfig6.4} gives the phase diagram for ferromagnetic 
couplings $J < 0$. The difference between $J > 0$ and   
$J < 0$ is in the effective range $\alpha$ of the 
double-exchange interaction, and is due to the different 
energies for on-site triplets when $J < 0$, to on-site Kondo 
singlets when $J > 0$ (cf.\ section \ref{old6.4}).
The effective range $\alpha$, which enters the 
bosonization of chapter \ref{ch2} as a finite  
but unknown length (with limits given in Eq.\ (\ref{alpha})), 
was determined at the ferromagnetic-paramagnetic transition 
by using the critical line equation derived in chapter \ref{ch5} (cf.\
Eq.\ (\ref{fullJc})), together with numerically determined  
transition points. 
Very good agreement with the numerical results was obtained 
by using a form $\alpha/a \propto t/\sqrt{|J|}$ for 
$\alpha$ at the ferromagnetic-paramagnetic transition 
(cf.\ Figs.\ \ref{oldfig6.1} and \ref{oldfig6.2}). This form for $\alpha$  
was suggested in section \ref{old6.1} on the basis of a simple 
characterization of double-exchange at small conduction 
band fillings $n$, together with an exact result 
at vanishing filling. The 
proportionality constants are different for different 
signs of the coupling, and are fixed by a best fit to the 
numerical data. In Fig.\ \ref{oldfig6.5} the 
corresponding filling dependence of $\alpha$ on the 
transition line was plotted, and shows a form 
$\sim 1/n$ which is expected within a naive view of 
double-exchange ordering. Fig.\ \ref{oldfig6.5} shows that   
$\alpha \rightarrow 0$ as $n \rightarrow 1$ for $J < 0$, 
but remains finite for $J > 0$. This has a significant 
effect on the phase diagrams. For $J > 0$ (Fig.\ \ref{oldfig6.2}) the 
critical line remains finite as $n \rightarrow 1$, while 
for $J < 0$ (Fig.\ \ref{oldfig6.4}) the critical line diverges 
approaching half-filling. (Note that the half-filled Kondo 
lattice is not considered. Double-exchange ferromagnetism 
is absent if the number of conduction electrons 
equals the number of localized spins.) 

The quantum ferromagnetic-paramagnetic phase 
transition identified chapters \ref{ch5} and \ref{ch6} 
is generic to partially-filled 
spin 1/2 Kondo lattices, at least in 1D.
The use of bosonization 
prevents anything more than speculation on 
the ferromagnetic-paramagnetic transition in higher 
dimensional Kondo lattices. 
Note only that {\it i}) double-exchange is not restricted 
to 1D, and should be considered 
in any discussion of partially-filled Kondo lattices 
in higher dimensions. {\it ii}) Numerical work
on the Kondo lattice 
with a ferromagnetic coupling reveals a  
ferromagnetic-paramagnetic
transition in higher dimensions, which is very similar to 
the transition in the 1D case. 

The description of the Kondo lattice given in chapters 
\ref{ch5} and \ref{ch6} is highly successful. It generates 
a ground-state phase diagram in agreement with available 
exact and numerical results for the partially-filled 
antiferromagnetic 1D Kondo lattice, and for the 
partially-filled ferromagnetic 1D Kondo lattice  
Much new information on the 
nature of the ferromagnetic-paramagnetic transition is 
also provided (cf.\ section \ref{old5.2}), and confirmation 
on the correctness of the description has been confirmed
by the recent state of the art density-matrix renormalization
group results of McCulloch, {\it et al.} (1999,2001,2002),
which are presented in detail in section \ref{6.5}. 

It is appropriate to conclude with a simple 
physical picture, suggested by the results of chapters \ref{ch5} 
and \ref{ch6}, which underlies the generic ground-state transition. 
At small fillings in the ferromagnetic phase,  
spin 1/2 Kondo lattices form a 
gas of spin polarons, with each electron dressed 
by a cloud of ordered localized spins. The spatial extent 
of the polarization cloud is the effective range $\alpha$ for 
the double-exchange interaction. For $J > 0$ 
the localized spins tend to align opposite to the spin of the 
conduction electron. For $J < 0$ they tend 
to align parallel to the electron spin. 
As the coupling is lowered, the polarization clouds 
gradually extend and begin to interfere. The interference  
causes spin-flip disorder processes, which eventually destroy 
the ferromagnetic order: The spin-flip  
processes free the electrons from their 
clouds of polarized localized spins, signalling 
the onset of the ferromagnetic-paramagnetic 
phase transition. At couplings just 
below the transition in the paramagnetic phase, 
the electrons are nearly free, and move through 
the system. They scatter from the localized spins as they move, 
and the spin chain is disordered. At weak-coupling, 
the localized spins retain dominant correlations at $2k_{F}$
of the conduction electrons, superimposed on an incoherent 
background. This reflects the momentum transferred from the 
conduction band to the spin chain in 
backscattering interactions, together with 
incoherent forward scattering. 

%%%%%%%%%%%%%%%%%%%%%%%%%%%%%%%%%%%%%%%%%%%
%% acknowledgements
%%%%%%%%%%%%%%%%%%%%%%%%%%%%%%%%%%%%%%%%%%%%

\cleardoublepage
\chapter{Acknowledgments}
\label{acknow}

I wish to thank my past and present collaborators, most notably
Graeme Honner, Ian McCulloch and Raymond Chan, for the research 
undertaken to progress this field of study.  

%%%%%%%%%%%%%%%%%%%%%%%%%%%%%%%%%%%%%%%%%%%
%% appendix 1
%%%%%%%%%%%%%%%%%%%%%%%%%%%%%%%%%%%%%%%%%%%%

\cleardoublepage
\appendix
\chapter{\label{appd}Strongly Correlated Electrons}

As the term `strongly correlated electrons' is often used during this
review, is appropriate to give a brief account on what this actually 
means, which are the models belonging to this category and how can 
these models be tackled from a theoretical point of view.

As the name suggests the term `strongly correlated electrons' represents a 
state of matter where many electrons are strongly interacting with each other. 
For real materials this means that the on-site electron-electron
repulsions are much larger than the energies associated with the
hybridization of atomic orbitals belonging to different atoms. 
This usually happens in systems involving 4$f$ or 5$f$ electrons, 
i.e., rare earth or actinide atoms. But systems with $d$ electrons 
can be strongly correlated, too. A famous example is ${\rm CoO}$, a 
system which treated within the independent electron approximation is 
found to be metallic. In reality, however, ${\rm CoO}$ is an insulator. 
Similarly,  ${\rm La}_{2}{\rm CuO}_{4}$ is an insulator, because of
the strong electron correlations suppress the charge fluctuations
required for a non-vanishing conductivity. 
Thorough discussion of the numerous correlated electron materials
may be found in the book by Fulde (1993), together with references. 

In the following we overview the strongly correlated electron field
by presenting, in section \ref{appd.1}, the most common lattice 
models of strongly correlated electron systems. Is not our purpose
to go through all known models of strongly correlated electrons,
the interested reader is referred to Fradkin (1991) for lattice models
or to Abdalla, Abdalla and Rothe (1991) for continuum models. 
Similarly, in presenting the methods used to study strongly correlated,
see sections \ref{appd.2} and \ref{appd.3}, we limit the discussion 
to the most popular methods specific for 1D, with 
the exception of two of the numerical methods, namely the 
exact diagonalization and quantum Monte Carlo which are 
also used in two dimensions. 

\section{\label{appd.1}Most common lattice models}

From a theoretical point of view, is obvious that the `holy grail'
of the strongly correlated electron system is without any doubt the 
Hubbard model, where the hybridization of atomic orbitals is taken
to be zero and the only interaction taken into account is the 
on-site electron-electron repulsions
\begin{equation}
H_{Hubbard} =  \sum_{i,j,\sigma} t_{ij} c^{\dagger}_{i,\sigma} c_{j,\sigma} +
U \sum_{i} n_{i,\uparrow}n_{i,\downarrow} \, . 
\label{appdhubbard}
\end{equation}
The hopping integral $t_{ij}$ which is usually
acting between nearest-neighbours,
$i.e.$ $t_{ij} = -t$ for $i,j \in$ nearest-neighbours and zero otherwise and an
on-site term of strength $U$, representing the effective screened Coulomb 
interaction. The important component of this interaction term is 
$U \sum_{i} n_{i,\uparrow}n_{i,\downarrow}$,
which gives an energy penalty (in the $U>0$ case) for each double-occupied 
site. A great deal of effort has been devoted to the solution of 
this model, but exact results are still confined to the one-dimensional 
case (for details, see section \ref{old3.2}). 
The Hubbard Model appears in many different forms, but the original 
one was introduced in 1963 in a bid to describe in a simplified
way the effect of correlations for $d$-orbital electrons in transition 
metals (Hubbard 1963, Gutzwiller 1963). 
$H_{Hubbard}$ is derived in section \ref{old1.2.2} 
(cf.\ Eq.\ (\ref{1.2.11})) for tightly bound electrons, 
whose Wannier wavefunctions are strongly localized at 
their lattice sites. As discussed earlier, the 1D Hubbard model 
is a simple single-component system, and as such it allows a
straight forward bosonization solution (see, eg, Emery (1979), 
Haldane (1981), Fradkin (1991), Schulz (1991) and Shankar (1995),
for a finite temperature bosonization see, Bowen and Gul\'{a}csi (2001)).
Here a bosonization based on chapter \ref{ch2} is implemented, 
for further use in section \ref{old5.4}. 

Originated from the idea of Anderson (1987) that the high temperature 
superconductors can be modelled on the basis of the strongly correlated Hubbard
model, the Hubbard model has been intensively investigated at the atomic
limit $U > t_{ij}$. Perturbation and canonical transformation are widely used
to show the antiferromagnetic ground state at half-filling. On the one hand,
the electron hopping can be treated as a perturbation which at the second
order yields an antiferromagnetic coupling due to the virtual 
hopping of electrons.

For sufficiently large $U$, it is well understood that the bare energy band
splits into two subbands. The lower subband is full while the upper 
subband is empty at half-filling, or one electron per atom. The virtual 
electron hopping between the upper and the lower subbands is the main cause
of the antiferromagnetic coupling, which in real space corresponds to virtual 
electron hopping between a pair of singly occupied nearest-neighbour sites.
The spins of the two electrons on this pair of sites are antiparallel. This
implies that the cross subbands hopping is responsible for the 
antiferromagnetic coupling. With respect to this, one can use a canonical
transformation to remove the the cross subbands hopping from the 
Hamiltonian which retains only the motions of electrons in either the lower
or the upper subband. The result of the transformation shows a spin
interaction, denoted by $J$ among the electrons at different sites 
when double occupied sites are ignored in the calculation
\begin{equation}
H_{tJ}= -t \sum_{\langle i,j \rangle,\sigma} c^{\dagger}_{i,\sigma}c_{j,\sigma}
+ J\sum_{\langle i<j \rangle} {\bf S_{i}} \cdot {\bf S_{j}} \, ,
\label{appdtJ}
\end{equation}
where $J = 4 t^2 / U$. 
This model known as the 
$t-J$ model, as it contains the parameters $t$ and $J$, 
shows explicitly the spin interaction 
between electrons. The nature of interaction between electrons
in this model is simpler than that of the Anderson model, but the behaviour
of the model is by no means less complicated. Even thought the $t-J$ 
model has been extensively studied in recent years in connection to
the high temperature superconductors, it was earlier derived 
as the strongly correlated limit of the Hubbard model by Nagaoka (1966), 
Roth (1966), Caron and Pratt (1968), Langer, Plischke and Mattis (1969), 
Visscher (1974) and Ogawa, Kanda and Matsubara (1975). In addition 
to these approaches, several canonical transformations exists
which lead to the same $t-J$ hamiltonian (Kohn 1964, Harris and Lange 
1967, Sokoloff 1970, Chao, Spalek and Oles 1977, Hirsch 1985). 

It is interesting to mention that the $t-J$ model is the only model within the
strongly correlated electron systems which exhibits phase separation. Indeed,
the $t-J$ model belongs
to the class of systems which do not obey the condition of 
Perron-Frobenius (Yosida 1980). This theorem state that if the off diagonal
elements of a matrix are all non-positive and if the matrix is not in a
block diagonal form then the ground state eigenvalue is non-degenerate.
In the case of the t-J hamiltonian the off
diagonal elements are not all non-positive. Thus the above theorem can not
be applied, which implies that the
phenomena of ground state level crossing is present
(Itoyama, McCoy and Perk 1990).
As a direct consequence of this, the thermodynamic system will
be unstable against phase separation (Blatt and Weisskopf 1979). 

At half-filling the exclusion of doubly occupied sites in the $t-J$
model also eliminates the unoccupied site In this case, each site is
a Heisenberg site, with only two states. Hence the $t-J$ model reduces
to a Heisenberg one
\begin{equation}
H_{Heisenberg} = J\sum_{\langle i<j \rangle} {\bf S_{i}} \cdot {\bf S_{j}} \, .
\label{appdHeisenberg}
\end{equation}

Sometimes two orbitals are necessary to describe strongly correlated
electron materials. In the periodic Anderson lattice model
\beqa
H_{PAM} &=& \sum_{k, \sig} 
\varepsilon (k) c^{\dg}_{ck\sig} c^{}_{ck\sig} + 
\varepsilon_{f}\sum_{j, \sig} n_{fj\sig} 
+ U\sum_{j} n_{fj\uparrow}n_{fj\downarrow}  
\nonumber \\
&+& N^{-1/2}\sum_{k,j, \sig}\left( 
V_{k}e^{ikja}c^{\dg}_{fj\sig}c^{}_{ck\sig} 
+ {\rm h.c.}\right) \, ,
\label{appdPAM}
\eeqa
see also Eq.\ (\ref{4.12a}), there are two Hubbard sites per 
Anderson site, one with $U = 0$. The notations in Eq.\ 
(\ref{appdPAM}) are the same as in Eq. (\ref{4.12a}), i.e.,
the conduction electrons are written in terms of Bloch 
states with dispersion $\varepsilon(k)$. 
$\varepsilon_{f}$ is the level of the flat band of 
localized $f$ orbitals. The hybridization $V_{k}$ gives the 
amplitude for a localized $f$-electron to be excited to a 
conduction band Bloch state with crystal momentum 
$k$. In the Kondo lattice model, there is one Hubbard
site (with $U = 0$) and one Heisenberg site per Kondo site: 
\beqa
H_{KLM} = 
-t\sum_{\langle ij \rangle}
\sum_{\sig}c^{\dg}_{ci\sig}c^{}_{cj\sig} 
+ J\sum_{j} {\bf S}_{cj} {\bf \cdot} {\bf S}_{j}  \, , 
\label{appdKLM}
\eeqa 
where the notations are similar to Eq.\ (\ref{4.11}). Detailed 
discussion of the periodic Anderson model are found in section
\ref{old4.2.2}. While the Kondo lattice is analysed thoroughly in
chapters \ref{ch4}, \ref{ch5}, \ref{ch6} and \ref{ch8}. 

These two lattice models are often used to describe heavy fermion 
systems, whereas the Hubbard, $t-J$ and Heisenberg models
are often used for high temperature superconductors. In the
high temperature superconductors values for the parameters of the models
are known fairly well. For example, for the cuprates in the Hubbard
model $U/t \approx 12$ while for the $t-J$ model $J/t \approx 0.3$
It is clear from the value of $U/t$ that any theoretical approach
starting from a non-interacting limit ($U = 0$) is highly 
questionable. In fact, it has been quite difficult to develop 
analytic methods for these models, as high temperature superconductors
are effectively two dimensional. Although progress has been made in
analytical methods, numerical simulations have played a very 
important role. Two of these methods will be discussed in 
section \ref{appd.3}. 

\section{\label{appd.2}Analytic methods}

Bethe Ansatz provides an exact solution for interacting many 
particle models in 1D which satisfy the so-called Yang-Baxter 
conditions (see later). The so-called Bethe Ansatz wave-function,
see Eq.\ (\ref{appdbawf}), was introduced by Bethe (1931) 
to solve the Heisenberg model\footnote{Bethe's highly detailed original 
paper is available in a new English translation by Mahan (1993).}.
Hereafter, we sketch the principal ideas.
The Bethe Ansatz relies on the following
facts. (i) Due to energy and momentum conservation, in 1D
a two-particle collision classically and quantum-mechanically conserves both
momenta individually. The particles then only can be exchanged or phase-shifted,
and the two-particle wave-function asymptotically ($| x_1 - x_2 | \rightarrow
\infty$) obeys
\begin{equation}
\Psi(x_1,x_2) = a e^{i(k_1 x_1 + k_2 x_2)} + b e^{i( k_1 x_2 + k_2 x_1)} \, .
\end{equation}
The Bethe Ansatz postulates this behaviour for all distances between the
particles.
(ii) A three-particle collision does not conserve individual momenta
{\em{except if}} the scattering matrix factorizes. This factorization
implies another conservation law. For $N$ particles, one then
has $N$ conservation laws, expressed by $\{ k'_i \} = \{ k_i \}$. 
(iii) The Hilbert space of the Hamiltonian separates in $N!$ quadrants
each characterized by a permutation $P$ of the $N$ particles, ordered in
one quadrant as $ 1 \leq x_{1} \leq x_{2} \leq \ldots x_{N} \leq L $. 
The $N$-particle wave-function there becomes
\begin{equation}
\Psi ( x_1 , \ldots , x_N ) = \sum_P A[P] e^{i k_{P_i} x_i} \, . 
\label{appdbawf}
\end{equation}
This is what is known as the Bethe Ansatz wave-function. 

Fermi or Bose statistics determines its continuation into the other sectors.
(iv) The amplitude $A[P]$ is determined by the conditions of continuity
of $\Psi$ as $x_i \rightarrow x_{i +1}$ and periodic boundary conditions
$\Psi (x_1, \ldots, x_N ) = \Psi (x_2, \ldots, x_N, x_1 + L)$. The problem
is the computation of $A[P]$. (v) Introducing spin, suppose we have
$N$ electrons, $M$ of which have spin $\downarrow$, on a 
lattice with $L$ sites $x_i$. One must then ensure that the factorization
of the $S$-matrix is not perturbed by the spin indices (Yang-Baxter
conditions). There is then a second permutation $Q$ for the spin labels, and 
the wave function where the $M$ down-spins
occupy the sites $x_1 \ldots x_M$ and the $N-M$ up-spins the sites
$x_{M+1} \ldots x_N$ is denoted by $\Psi(x_1, \ldots , x_M, x_{M+1}, \ldots,
x_N)$. The Bethe Ansatz
postulates that in each quadrant characterized by $Q$, 
i.e. $ 1 \leq x_{Q_1} \leq x_{Q_2} \leq \ldots x_{Q_N} \leq L $, 
the wave function is given by (Lieb and Wu 1968)
\begin{equation}
\label{betwv}
\Psi(x_1, \ldots , x_M, x_{M+1}, \ldots, x_N) = \sum_{P} A[Q,P] \exp 
\left( i \sum_{j=1}^N k_{Pj} x_{Qj} \right) \;\;\;.
\end{equation}
The $N$ numbers $k_i$ are determined from the coupled Lieb-Wu equations
($ u = U / 4 t$)
\begin{eqnarray}
\label{liwu1}
2 \pi I_j & = & L k_j - 2 \sum_{\beta = 1}^{M} \arctan \left(
\frac{\sin k_j - \Lambda_{\beta}}{u} \right) \, , \\
\label{liwu2}
2 \pi J_{\alpha} & = & 2 \sum_{j=1}^{N} \arctan \left( 
\frac{\Lambda_{\alpha} - \sin k_j }{u} \right)
- 2 \sum_{\beta =1}^M \arctan \left( 
\frac{\Lambda_{\alpha} - \Lambda_{\beta}}{2 u} \right) \, , \\
\label{liwubound}
I_j & = & \left\{ \begin{array}{l} \rm integer \\
				   \rm half-odd-integer  \end{array} \right.
\; \rm if \; M = \left\{ \begin{array}{l} \rm even \\
				   \rm odd  \end{array} \right. \, , \\
J_{\alpha} & = & \left\{ \begin{array}{l} \rm integer \\
				   \rm half-odd-integer  \end{array} \right.
\; \rm if \; N-M = \left\{ \begin{array}{l} \rm odd \\
				   \rm even  \end{array} \right. 
\;\;\;. \nonumber
\end{eqnarray}
The total energy and momentum of the system are then
\begin{equation}
\label{bethen}
E = - 2 t \sum_{i=1}^N \cos k_i\;\;\;, \hspace{0.5cm} P = \sum_{i=1}^N k_i
\, .
\end{equation}

Eqs. (\ref{betwv}) -- (\ref{bethen}) give the exact energy and wavefunction
of the 1D Hubbard model. The quantum numbers $k_i$ are the momenta 
of the particles characterizing the orbital
degrees of freedom. Unlike for free particles, they are not equally spaced
but shifted by the presence of the other particles.
The $\Lambda_{\alpha}$ are called rapidities and describe the 
spin state. On the other hand, the integers or half-odd-integers $I_i$ 
and $J_{\alpha}$ are equally spaced. The ground state is obtained by
occupying the levels with minimal $|I_i|$ and $|J_{\alpha}|$. Therefore
the distribution of 
$q_i = 2 \pi I_i / L$ and $p_{\alpha} = 2 \pi J_{\alpha} / L$ is given 
by a Fermi distribution $\Theta(k_{F \uparrow} + k_{F\downarrow} - q_i)$
and $\Theta(k_{F\downarrow} - p_{\alpha})$, respectively. In the absence
of a magnetic field, the ground state has $k_{F\uparrow} + k_{F\downarrow}
= 2k_F$ and $k_{F\downarrow} = k_F$, so that the $q_i$ have a doubled
Fermi wavevector while the $p_{\alpha}$ have the normal $k_F$. 

Unfortunately, very few models in 1D can be solved exactly via 
Bethe Ansatz. It's rather difficult to satisfy the Yang-Baxter conditions. 
The Hubbard model allows an exact solution for any value $U$ (Lieb and
Wu 1968), however, this is not true for the $t-J$ model, which is
solvable only on the supersymmetrical point $J / t = 2$ (Sutherland 1975,
Schlottmann 1987, Bares and Blatter 1990). The single impurity Kondo
model and the Anderson model as discussed in detail in section
\ref{old4.3.2} and Appendix \ref{ch4kondo} 
are solvable via Bethe Ansatz. Their lattice counterpart however is not. 
In this cases the next best option is to obtain an `asymptotic'
exact solution via bosonization. 

The special properties of interacting 1D systems described 
in details in section \ref{intro-DMRG} lead 
naturally to a description in terms of bosonic excitations, 
which are the collective density fluctuations. 
The description of 1D many-electron systems in terms of 
bosonic density fluctuations is called bosonization. 
The central elements in the formalism of bosonization 
are Bose representations; fermionic 
operators, such as the electron 
fields $\psi^{\dg}_{\sig}(x)$ (cf.\ Eq.\ (\ref{1.22})), 
may be written in terms of bosonic density fluctuations. 
This gives a Bose representation for 
$\psi^{\dg}_{\sig}(x)$ which may be substituted back 
into the hamiltonian to give a far simpler 
description of the system. In some cases, the description 
in terms of bosons is so simple that the  
problem may be solved exactly. However, the main power
of the bosonization method lies in the extreme simplicity
in calculating correlation functions. Details of the
method are presented in chapters \ref{ch2} and \ref{ch3},
including the bosonized solution of the Hubbard model.
The Kondo lattice model is solved via bosonization in 
chapter \ref{ch5}. 

In the most complex models, such as the periodic Anderson model,
not even bosonization will work. In these cases the critical
behaviour of the model and the exponents of correlation 
functions can be determined using conformal field theory. 
A `conformal field theory' is a quantum field theory or a statistical 
mechanics model that is covariant under the conformal group. All 
one dimensional systems are conformal invariant and as such can be treated 
with conformal field theory. 

Conformal field theory is a powerful means of characterizing universality
classes in terms of a single dimensionless
number, the central charge $c$ 
(Belavin, Polyakov and Zamolodchikov 1984, Br\'{e}zin and Zinn-Justin
1989, Itzykson and Drouffe 1989).
The critical exponents are the scaling dimensions of 
the various operators in a conformally invariant theory and, generically,
are fully determined by $c$. Both the central charge and
the scaling dimensions can be computed from the finite-size scaling 
properties of the ground state energy and the low-lying excitations
(Belavin, Polyakov and Zamolodchikov 1984, Br\'{e}zin and Zinn-Justin
1989, Itzykson and Drouffe 1989). 
This is important because these quantities can
be computed accurately either by Bethe Ansatz (for models solvable by
the technique) or, in any case, by numerical diagonalization. 

What are the symmetries of systems at a critical point? It is certainly
translationally and rotationally invariant. Quantum field theories,
in addition are Lorentz invariant but in one dimensions, Lorentz invariance
reduces to rotations in the ${\bf x} = (x,t)$-plane. 
A system
at criticality, in addition is characterized by scale invariance,
\begin{equation}
{\bf x} \rightarrow \lambda {\bf x} \, .
\end{equation}
It turns out that the combined rotational and scale invariance implies
that the system is invariant under a wider symmetry group, the global
conformal group, hence the name conformal field theory appeared. 
On a classical level, conformal transformations are general coordinate
transformations which leave the angles between two vectors invariant.

Consider a general coordinate transformation
\begin{equation}
{\bf x} \rightarrow {\bf x} ' = {\bf x} + {\bf \xi} ({\bf x}) \, .
\end{equation}
For this transformation to be conformal, ${\bf \xi}$ must satisfy certain
constraints which can be expressed in a differential
equation (Killing-Cartan equation). In higher
dimensions, this leaves for ${\bf \xi}({\bf x)}$ a polynomial of second degree
in $\bf x$ (with tensor coefficients). 
In one dimensions, however, the Killing-Cartan equation reduces
to the Cauchy-Riemann equation, and therefore all analytic functions
are allowed for conformal transformations.
This group of transformations, 
called local conformal group, is much wider than the global conformal group
encountered before.
It is then natural to switch to complex variables 
$z, \bar{z} = x_1 \pm i x_2$,
so that we have
\begin{equation}
\label{locconft}
z \rightarrow z + \xi^z(z) = f(z)\;\;\;, \hspace{1cm} 
\bar{z} \rightarrow \bar{z} + \bar{\xi}^{\bar{z}} (\bar{z}) = 
\bar{f} (\bar{z}) \;\;\;.
\end{equation}
To determine the algebra corresponding to the local conformal group,
we need the commutation relations of the generators of the transformations.
Since $\xi^z(z)$ and $f(z)$ are analytic, they can be expanded in a
Laurent series
\begin{equation}
\xi^{z}(z) = \sum_{n = -\infty}^{\infty} \xi_n z^{n+1}
\end{equation}
(and a similar equation for $\bar{\xi}(\bar{z})$), and hence 
we find the generators of the local conformal transformations. 
The interested reader is encouraged to consult Br\'{e}zin and Zinn-Justin
(1989) or Itzykson and Drouffe (1989). This infinite dimensional algebra 
is called the classical Virasoro algebra.

We now go to the quantum (or statistical mechanics) case.
How do fields and correlation functions of a quantum field
theory transform under conformal transformations?
Local coordinate transformations are generated by the charges constructed
from the stress-energy tensor $T_{ij}$. 
The stress-energy tensor measures the cost of energy of a change in metric,
i.e., a  change in the action 
\begin{equation}
\delta S = \frac{1}{2 \pi} \: \int T_{ij} \partial_i \xi_j d^2 r \, . 
\end{equation}
Rotational invariance constrains $T_{ij}$ to
be symmetric, and scale invariance requires its trace to vanish; then
conformal invariance does not impose additional constraints
showing that it is implied by rotational and dilatational 
invariance. 

One of the important property of the stress-energy tensor is that 
under a local
conformal transformation to $z' = f(z)$, it transforms as
\begin{eqnarray}
T(z) & \rightarrow & T'(z) = \left( \frac{dz'}{dz} \right)^2 T(z')
+ \frac{c}{12} \{ z' , z \} \, , \\
\{z' , z\} & = & \frac{\partial^3_z z'}{\partial_z z'} - 
\frac{3 (\partial^2_z z')^2}{2 (\partial_z z')^2} \, .
\label{appdset}
\end{eqnarray}

The first term in this equation translates the fact that $T(z)$ is
a field of conformal weight,
while the second term contains the conformal anomaly and the central
charge $c$. The second term is known as the Schwarzian derivative.
Thus, Eq.\ (\ref{appdset}) gives a way to calculate $c$. Knowing $c$ 
the scaling dimensions and correlation functions of the conformal
field theory are determined. 

The second way to obtain $c$ is through finite size scaling,
this is very important in numerical approaches. 
Up to now, we have implicitly assumed that our fields are defined in
the infinite $z$-plane. What happens when we consider finite systems?
Let us use the exponential transformation
\begin{equation}
z = \exp \left( \frac{2 \pi i }{L} u \right) \, , \hspace{1cm}
u = \frac{L}{2 \pi i} \log z
\end{equation}
to map the infinite $z$-plane onto a strip ($u$) of width $L$
with periodic boundary conditions. For the stress-energy tensor 
it is obtained (Belavin, Polyakov and Zamolodchikov 1984, 
Br\'{e}zin and Zinn-Justin 1989, Itzykson and Drouffe 1989)
\begin{equation}
\langle T_{\rm strip} (u) \rangle = \frac{c}{24} \left( \frac{2 \pi}{L}
\right)^2     \, .
\end{equation}
One can now calculate the change in energy associated with another 
(nonconformal) transformation,
a horizontal dilatation of the $u$-strip ($u_1' = (1+ \varepsilon) u_1,
\; u_2' = u_2$) which changes the length of the system, and integrate
to find 
\begin{equation}
E(L) - E(\infty) = \frac{c \pi}{6L} \, ,
\end{equation}
where $E(L)$ is the energy per unit length.
This formula is extremely important because it allows to determine
the value of the central charge from calculations on finite systems. 
Because of these reason, 
in past years, conformal field theory in conjunction with numerical
simulations has been extensively used to study one 
dimensional strongly correlated electron systems. 

\section{\label{appd.3}Numerical methods}

Probably the three most useful numerical methods for studying quantum 
lattice models in the strongly interacting limit are the Quantum Monte 
Carlo (QMC), the Density Matrix Renormalization Group (DMRG) method and 
exact diagonalization. From these methods the DMRG become one of the most 
popular one during the past years to study one dimensional strongly
correlated systems. An extensive description of DMRG can be found in 
sections \ref{intro-DMRG} \ref{DMRG}, where it is shown that the
central idea of DMRG is to keep the `most probable' states at each
step of iteration, instead of the lowest energy states (White 1992,1993).
This idea follows from the requirement that for a collective density 
mode, which are typical to one dimensional conductors, to be operational 
it must have the largest possible overlap between the blocks i.e., must
have the largest eigenvalues of the corresponding block density
matrix. Because of these reasons DMRG enabled
a level of numerical accuracy for one dimensional systems
that was not possible using other methods 
(Nishino, {\it et al.} 1999,Peschel, {\it et al.} 1999). 

The oldest numerical method from the three mentioned, however, is the 
exact diagonalization. In this case, one takes a small system
and finds the ground state exactly (Dagotto 1994). Since the number 
of states in the Hilbert space grows exponentially with the number
of sites it is important to be as efficient as possible. One way to
be more efficient is to use symmetries to reduce the size of the space.
For example, separating states by their total momentum, will work
in most cases. However, the simplest algorithm which can be used 
is the so-called `power method', which is basically a simple
iteration: take an initial guess $\psi$ and multiply by
$(1 - \epsilon H)$, where $H$ is the studied hamiltonian and $\epsilon$
is a small parameter. For $\epsilon$ sufficiently small the maximum
eigenvalue of $(1 - \epsilon H)$ is the ground state of $H$. Repeated
multiplication by a matrix always projects out the eigenvector with maximum
eigenvalue, in this case the ground state. Of course, the starting vector
must not be orthogonal to the ground state. 

In this simple power algorithm, the wave function after $n$th iteration 
looks like a polynomial in terms of $a_{n} \psi H^{n}$, where $\psi$ was the 
initial guess and $a_{n}$ is a parameter dependent on, e.g., $\epsilon$.
The Lanczos algorithm takes this $a_{n}$ parameters to be independent
and determines the minimum energy as a function of $a_{n}$'s. Within
the subspace spanned by $\psi, \psi H, \psi H^2, \ldots$, the Lanczos
algorithm always finds the lowest energy in much shorter time then
the previous power method. 

Another exact diagonalization algorithm, which can sometimes converge
faster than Lanczos (Dagotto 1994) is the Davidson method. This method
uses diagonal elements of the hamiltonian to generate a slightly 
different subspace than that used by the Lanczos method and the
method calculates the lowest energy state within that subspace. The
Davidson method performs much better than Lanczos if the
matrix attached to studied hamiltonian is diagonally dominant
(Dagotto 1994). If the diagonal elements are constant, than it reduces
to the lanczos method. 

Similarly, they are a variety of types of QMC methods, each with
their strengths and weaknesses (Linden 1992). One of the most common
method is the determinant QMC, often used to deal with the Hubbard
model. In this finite temperature method the partition function
is expressed as a path integral
\begin{equation}
Z \: = \: {\rm tr} \: e^{ - \beta H} \, ,
\end {equation}
where $\beta = 1 / k_{B} T$ and $H$ is for example the Hubbard 
hamiltonian from Eq.\ (\ref{appdhubbard}). The first step is the
Suzuki-Trotter breakup to separate the kinetic $H_{kin}$ and 
interaction $H_{int}$ parts of the hamiltonian
\begin{equation}
Z \: \approx {\rm tr} \: [ e^{- \Delta \tau \: H_{kin}} \:
e^{- \Delta \tau \: H_{int}} ]^{L} \, .
\label{appdqmc1}
\end{equation}
The purpose of separating the kinetic energy from the interaction term 
is to facilitate an approximate treatment of $H_{int}$, as we will see
later on. While introducing a `time-slice' for each term in Eq.\
(\ref{appdqmc1}) is borrowed from the standard imaginary time Green's
function method (Fetter and Walecka 1971). 

With the interaction term isolated,a Hubbard-Stratonovitch transformation
can be used to eliminate the two-body terms in $H_{int}$ (Linden 1992).
the price which must be paid for this is the introduction of a sum
over an Ising variable $S_{i, \ell}$ and the interaction term reduces
to a sum of constants, e.g., for the Hubbard model $exp ( - \Delta \tau
\: U \: n_{i, \uparrow} n_{i, \downarrow} ) \approx \sum_{ S_{i, \ell} }
exp ( - \Delta \tau \: S_{i, \ell} \: \lambda (n_{i, \uparrow} - 
n_{i, \downarrow})$, where $\lambda$ is a proportionality constant. 
This transformation is somewhat similar to transforming a $d$
dimensional quantum problem into a $d+1$ dimensional classical one.
With the two-particle interaction terms eliminated, for each Ising
spin variable configuration the non-interacting electron system can
be solved exactly. The solution turns out to involve determinants
of matrices whose dimension is equal to the number of sites. 

The products of the as obtained determinants define a `probability'
function, giving the probability of each Ising spin configuration. 
Accordingly, classical Monte Carlo methods can be applied from 
now on. In practice there are several problems with this 
method (Linden 1992). Firstly, calculating the change in 
probability from flipping one spin requires a substantial calculation 
because of the presence of determinants. If $L$ is the number of sites
in the system, the the best known procedure requires $L^2$ operations
to determine if a spin should be flipped. Secondly, at low 
temperatures this procedure becomes unstable because the matrices 
involved become singular. And thirdly, the famous fermionic 
`sign problem' occurs - the determinants may become negative. 
QMC can still proceed using as the probability of the absolute 
value of the product of the determinants. But the measurements
has to be divided by the average sign of the determinants and as such
their statistical uncertainties blow up when the average sign 
approaches zero. Because of these problems that QMC has  
methods such as DMRG become so popular in one dimension.
However, in two dimension, where DMRG is not yet implemented
properly, the QMC method is more useful. Here in 
recent years several attempts have been made
to `fix' the above mentioned problems, including the sign problem, 
but these are beyond the scope of this brief review.

%%%%%%%%%%%%%%%%%%%%%%%%%%%%%%%%%%%%%%%%%%%
%% appendix 2 
%%%%%%%%%%%%%%%%%%%%%%%%%%%%%%%%%%%%%%%%%%%%

\cleardoublepage
\chapter{\label{ch4kondo}The Single Impurity Kondo Model}

As presented in section \ref{old4.3.2} the Kondo problem was essentially solved
in the 1970's by Anderson's poor man's scaling (Anderson 1970)
and finally by the numerical renormalization group method developed
by Wilson (1975). The results of the renormalization group have 
been confirmed by the celebrated Bethe Ansatz exact solutions (Andrei 1980,
Wiegmann 1980). What actually led to the numerical renormalization
group approach of Wilson (1975) is a rather interesting perturbation 
theory due to Anderson and Yuval (Anderson 1967, 
Anderson and Yuval 1969, Anderson, Yuval and Hammann 1970,
Yuval and Anderson 1970). The main idea of this approach was 
to split the Kondo coupling (see Eq.\ (\ref{4.2.5})) into 
a transverse coupling ($J_{\pm}$ (corresponding to the $S^{x}$ and
$S^{y}$ spin terms of Eq.\ (\ref{4.4})) and a longitudinal one $J_{z}$. 

For $J_{\pm}$ there is no spin flip scattering and as such the ground
state has only two possible states (for 1/2 spins and no external magnetic 
field). This limit can be solved without any problems. If one then applies
a finite temperature perturbation theory in powers of $J_{\pm}$, the
spin flips are generated at a sequence of imaginary times (Fetter and
Walecka 1971). This sequence of imaginary time variables can be 
represented as a one dimensional Ising model, with a coupling which
falls off inversely as the square of the distance as $T \rightarrow 0$,
and hence can be solved exactly. Another representation can map the
sequence of the imaginary time variables into a chain of charged
particles with long range interaction that has a logarithmic form.
This model can also be solved exactly. Even though the resulting 
effective models are exactly solvable, the finite temperature 
perturbation theory leading to them renders these type of solutions 
of the original Kondo model as perturbative\footnote{Only in the 
so-called Toulouse limit (Toulouse 1969), the Anderson and Yuval 
(Anderson 1967, 
Anderson and Yuval 1969, Anderson, Yuval and Hammann 1970,
Yuval and Anderson 1970) solution is exact. The Toulouse limit 
corresponds to a certain value of $J_{z} = J_{\rm T}$, where
$J_{\rm T}$ $\approx$ 0.97 (Wiegmann and Finkel'shtein 1978). 
In this limit, however, the Kondo model reduces to a simple 
quadratic non-interacting model.}. 

This Anderson and Yuval formulation of the Kondo model leads logically
to the approach used later by Wilson (1975) and it originates from 
the (very similar) X-ray absorption problem in which
one electron is excited from a core shell of an atom or ion in 
a metal into the conduction band (see Mahan 1990). This problem is
known as the {\it orthogonality} {\it catastrophe} due to Anderson
(1967), who pointed out that the ground state of the conduction electrons 
with and without the core hole potential are orthogonal in the 
thermodynamic limit. 

After this short historical introduction, a brief summary of the 
single impurity Kondo model results are presented based on 
Andrei, Furuya and Lowenstein (1983), Mahan (1990) and Hewson (1993).
The hamiltonian of the model is that of Eq.\ (\ref{4.2.5}), where 
for simplicity it is usually assumed that the conduction electrons
have rectangular density of states with width 2$D$. When in the 
perturbative scaling approach this band width is reduced by an
amount $\vert \delta D\vert $ the coupling constant ($J > 0$
is considered only) is found to increase by an amount 
$\delta J = 2 J ( \rho(\varepsilon_{F}) J - (\rho(\varepsilon_{F}) J)^2
+ \ldots ) \vert \delta D \vert / D$, where $\rho(\varepsilon_{F})$
is the density of the conduction electrons at the Fermi level, 
given in Eq.\ (\ref{4.2.7}). This process of scaling, by which
decreasing the band width increases the effective interaction,
breaks down at a certain point when the obtained effective coupling
strength becomes too large. This occurs when the scale of the reduce
band width becomes of the order of the Kondo temperature, where
\beqa
T_{K} = {\rm const.} \, D \, \vert 2 \rho(\varepsilon_{F}) J \vert^{1/2} \:  
\exp(-1/ 2 \rho(\varepsilon_{F}) J) \, ,
\label{a1} 
\eeqa
where ${\rm const.} \approx 1$ is a numerical constant depending on the 
definition of the Kondo temperature, which will be explained shortly
\footnote{For simplicity we use a convention, where $k_B$ is unity.}.
The parameters $J$ and $D$ appear in the scaling equations via the 
Kondo temperature only, so that the low temperature thermodynamic 
properties of the model are universal functions of $T/T_{K}$. 

In the regime where the perturbation theory is valid, i.e., 
$T \gg T_{K}$ the zero field impurity susceptibility is
\beqa
\chi_{\rm imp} (T) = \frac{g \mu_B}{4 T} [ \, 1 - 
\frac{1}{\ln (T/T_{K})} - \frac{\ln (\ln (T/T_{K}))}{2 \ln^2 (T/T_{K})} 
+ \dots \, ] \, .
\label{a2}
\eeqa
behaving very much a Curie law with reduced moment. The induced 
impurity magnetization at $T = 0$ is:
\beqa
M_{\rm imp} (T) = \frac{g \mu_B}{2} [\, 1 - \frac{1}{2 \ln (g \mu_B H / T_{H} ) }
- \frac{\ln \ln (g \mu_B H / T_{H} )}
{4 \ln^2 (g \mu_B H / T_{H} ) } + \ldots \, ] \, ,
\label{a3} 
\eeqa
where $T_{H}$ is proportional to $T_{K}$, see Eq.\ (\ref{a10}).
The impurity resistivity in the same high temperature limit is
given by
\beqa
R_{\rm imp} (T) = \frac{R_0}{2} [ 1 - \frac{\ln (T / T_{K})}
{ [ \ln^2 (T/T_{K}) + 3 \pi^2 / 4 ]^{1/2} } \, ] \, . 
\label{a4} 
\eeqa
It is important to remark here that the Bethe Ansatz calculations 
confirmed the high temperature perturbative results of Eqs.\
(\ref{a2}), (\ref{a3}) and (\ref{a4}). 

The low temperature limit is non-perturbative: in this limit 
the energy scale tends to zero, which corresponds to the
strong coupling limit $J \rightarrow \infty$. In the low 
temperature limit the impurity spin is compensated by the 
conduction electrons and the impurity susceptibility is finite
corresponding to a Pauli contribution. At $T \rightarrow 0$
the susceptibility is
\beqa
\chi_{\rm imp} (0) = \frac{ (g \mu_B)^2 w }{4 T_{K}} 
\, ,
\label{a5} 
\eeqa
where $w = 0.4107$ is known as the Wilson constant, with exact
value $\exp ({\rm C} + 1/4) / \pi^{3/2}$, where ${\rm C}$ = 0.5772 is 
Euler's constant. The specific heat
in the same limit is 
\beqa
\gamma_{\rm imp} (0) = \frac{ \pi^2 w }{6 T_{K}} 
\, ,
\label{a6} 
\eeqa

The definition of $T_{K}$ in Eqs.\ (\ref{a5}) and (\ref{a6})
corresponds to the Wilson (1975) definition and it was chosen 
such that there are no terms of order $1 / \ln^2 (T/T_{K})$
in the high temperature expansion of the impurity susceptibility
from Eq.\ (\ref{a2}). The exact solution community followed
Wilson's definition of the Kondo temperature based on Eq.\
(\ref{a5}), or Eq.\ (\ref{a6}). For example, Eq.\ (\ref{a5})
in the Bethe Ansatz approach is
\beqa
\chi_{\rm imp} (0) = \frac{ (g \mu_B)^2 e^{\pi / c} }{4 D^{\prime}} 
\, ,
\label{a7} 
\eeqa
where $D^{\prime} = \pi N_e / L$, is the conduction electron 
charge density, with $N_e$ being the number of conduction electrons, 
$L$ is the length of the chain (effective one dimensional lattice 
that the Kondo model is mapped into) and $c = \tan J$. Accordingly,
the Bethe Ansatz Kondo temperature can be identified to be
\beqa
T^{BA}_{K} = D^{\prime} \:  \exp(-\pi / c) \, .
\label{a8} 
\eeqa
Following Andrei, Furuya and Lowenstein (1983) the exact ratio of
the two Kondo temperatures is
\beqa
T_{K} = w T^{BA}_{K} \, ,
\label{a9} 
\eeqa
where the Wilson constant is given after Eq.\ (\ref{a5}). From 
Eq.\ (\ref{a9}) it can be shown that $T_{H}$ defined through
Eq.\ (\ref{a3}) is 
\beqa
T_{H} = \frac{2 T_{K}}{w {\sqrt{e \pi}}} \, .
\label{a10} 
\eeqa

Another important quantity used in Kondo models is the so-called 
Wilson ratio for the impurity. Which defined as the ratio of the 
zero temperature specific heat and paramagnetic susceptibility,
has the value
\beqa
R = \frac{4 \pi^2}{3 (g \mu_B)^2} \: 
\frac{ \chi_{\rm imp} }{ \gamma_{\rm imp} } \, .
\label{a11} 
\eeqa
For a non-interacting system $R = 1$, while it can be seen that 
Eq.\ (\ref{a11}), using Eqs.\ (\ref{a5}) and (\ref{a6}), is 
$R = 2$. This shows that in the low temperature limit the
magnetically screened impurity scatters the conduction electrons
like a non-magnetic impurity with a resonance at the Fermi level
with width of order $T_{K}$. This explains why even in the
presence of an external magnetic field, the Wilson ration is
unchanged: 
\beqa
R = \frac{4 \pi^2}{3 (g \mu_B)^2} \: 
\frac{ \chi_{\rm imp} (H) }{ \gamma_{\rm imp} (H)} \: = \: 2
\, .
\label{a12} 
\eeqa
This ratio can only be calculated by the Bethe Ansatz approach
(Tsvelick and Wiegmann 1983). It's value is of great 
importance for heavy fermion compounds, where higher values of
$R$ (Hess, Riseborough and Smith 1993) signals a non-Fermi 
liquid behaviour.

%%%%%%%%%%%%%%%%%%%%%%%%%%%%%%%%%%%%%%%%%%%
%% appendix 3
%%%%%%%%%%%%%%%%%%%%%%%%%%%%%%%%%%%%%%%%%%%%

\cleardoublepage
\chapter{Exact Unitary Transformation}
\label{ch4sw}

In this appendix the results of section \ref{sectionSW} are derived following
Chan and Gul\'{a}csi (2003). This derivation is tailored for the 1D
periodic Anderson model. A similar infinite order canonical transformation 
can be applied to the single impurity Anderson model. Details of this
derivation are given by Chan and Gul\'{a}csi (2001a,2004). Interestingly,
hamiltonians with similar structure to the Anderson model, such as 
the two band Hubbard model, also allow an infinite order canonical 
transformation, for details, see Chan and Gul\'{a}csi (2000,2001b,2002). 

The construction of the canonical transformation is given in 
section \ref{ch4sw-one}, where the first five orders of the 
transformation will be calculated. Here it will be shown that
there is a pattern in the coefficients  of the first five
odd orders. This pattern will be proven to exist for any 
order by induction in section \ref{ch4sw-two}. The exact 
value of all the coefficients up to infinite order is derived
in section \ref{ch4sw-three}, allowing the evaluation of the
exact transformed hamoltonian in section \ref{ch4sw-four}.

\section{\label{ch4sw-one}Canonical Transformation} 

The hamiltonian of the 1D periodic Anderson model, see Eq.\ (\ref{4.12a}),  
re-written in the real space is:
\begin{equation}
H_{\PAM} = H_{0} + H_{V} \, ,
\label{eq:H(i)}
\end{equation}
with
\begin{eqnarray}
H_{0} & = & t \sum_{i, \sigma} (c^{\dagger}_{c i+1 \sigma}
c^{}_{c i \sigma} +  {\rm h.c.}) 
- \mu \sum_{i, \sigma} c^{\dagger}_{c i \sigma} c^{}_{c i \sigma}
\nonumber \\
&& +  \varepsilon_{f} \sum_{i, \sigma}
c^{\dagger}_{f i \sigma} c^{}_{f i \sigma} 
+  U \sum_{i} n^{}_{f i -\sigma} n^{}_{f i \sigma} \, ,
\label{eq:H(0)} \\
H_{V} & = & V \sum_{i \sigma} ( c^{\dagger}_{f i \sigma}
c^{}_{c i \sigma} + {\rm h.c} ) \, ,
\label{eq:H(V)}
\end{eqnarray} 
where the chemical potential of the conduction electrons $\mu$ has
been added to Eq.\ (\ref{4.12a}), so that the obtained results
will be more general, and $\varepsilon (k)$ from Eq.\ (\ref{4.12a})
is: $\varepsilon (k) = 2 t \cos k$. 

The easiest way to solve this hamiltonian is by using perturbation
theory (for details see, Hewson (1993)), e.g., with $U=0$ and use 
perturbation theory to find the solution for small $U$. Similarly, 
it is possible to solve the Hamiltonian with $V=0$ and use perturbation 
theory to find the solution when $V$ is small. As presented in 
section \ref{old4.2.2}, Schrieffer and Wolff (1966), chose the second option,
namely perturbing $H_{\PAM}$ around $V = 0$, up to second order in $V$. 
However, the novelty in their method was that they used a canonical 
transformation. They canonically transformed the hamiltonian
to a form which has eliminated the hybridization term $H_{V}$. The 
transformation was, however, calculated only up to the first order, 
and, as such, the solution is valid only for a small parameter range
\footnote{$n$th order canonical transformation corresponds to
$2 n$th order perturbation theory in $H_{V}$ (Yosida and Yoshimori 1973).}.
Nevertheless, the novelty in applying canonical transformation
is that if one can keep track of the coefficients in each order of the 
transformation and persistently carry out the transformation to higher
orders, one can extend the parameter range of the solution to an extent
beyond the reach of perturbation methods. This is what has been 
achieved by Chan and Gul\'{a}csi (2001a, 2003, 2004), and is presented
hereafter. 
 
The unitary transformation of Eqs.\ (\ref{eq:H(0)}) and
(\ref{eq:H(V)} can be written as 
\begin{equation}
\tilde{H} = e^{S} H e^{-S} = H_{0} + [S,H_{V}] / 2
+ [S,[S,H_{V}]] / 3 + \ldots
\end{equation}
in which the parameter $S$ in the transformation is determined by 
the condition 
\begin{equation}
H_{V} + [S, H_{0}] = 0, 
\label{eq:Scondition}
\end{equation}
and in the case of Eq.\ (\ref{eq:H(i)}) is explicitly given by
\begin{equation}
S  =  \sum_{i \sigma} V ( A + Z c^{\dagger}_{f i -\sigma}
c^{}_{f i -\sigma} ) 
( c^{\dagger}_{f i \sigma} c^{}_{c i \sigma} -
c^{\dagger}_{c i \sigma} c^{}_{f i \sigma}) \; ,
\label{eq:rS}
\end{equation}
with
$A = 1 / (-2t + \mu + \varepsilon_{f})$ and
$Z = 1 / (-2t + \mu + \varepsilon_{f} + U) - A$, identical to
Eq.\ (\ref{sw4}). 
One can easily verify that this expression of $S$ satisfies 
(\ref{eq:Scondition}), by using a continuum representation of
the conduction electrons, $c_{c i \pm 1 \sigma} \approx 
c_{c i \sigma} \pm \partial_{x} c_{c x \sigma} {}_{\vert x \rightarrow i}$,
as in the field theory bosonization of 1D models 
(see eg, Emery 1979, Haldane 1981, Fradkin 1991, Voit 1994, 
Stone 1994, Gul\'{a}csi 1997). For a proof of the 
completeness of the Bose representation, see also
section \ref{old2.3}. 

The first order of the transformation is identical to the Schrieffer
and Wolff (1966) result from Eq.\ (\ref{sw5}), see also Eq.\ (\ref{eq:nthorder})
and Table \ref{tab:1-3OrderResult}. 

It is straightforward to continue the transformation up to fifth order 
(beyond this stage the number of terms obtained in each order is
well over several hundreds of millions (Chan and Gul\'{a}csi 2001a,2004)). 
The result for the first ($n = 1$), 
third ($n = 3$) and fifth ($n = 5$) order terms can be written
in a compact form observing that 
the transformed hamiltonian depend on the following commutation: 
\begin{equation}
[[S,H_{V}]]_{n} \: = \:
[\overbrace{S,[S,[S,\ldots,[S}^{\mbox{$n$ times}},H_{V}] \ldots] \; ,
\label{eq:nthcomm}
\end{equation}
This commutation was found to have the following form over the first three
odd values of $n$,
\begin{eqnarray}
&& [[S,H_{V}]]_{n} \:
= \: \sum_{i, \sigma} \biggl[
J_{n} (c^{\dagger}_{c i \sigma} c^{}_{c i -\sigma}
c^{\dagger}_{f i -\sigma} c^{}_{f i \sigma}
- n^{}_{c i \sigma} n^{}_{f i -\sigma})
\nonumber \\
&& + P_{n} ( c^{\dagger}_{c i \sigma} c^{\dagger}_{c i -\sigma}
c^{}_{f i \sigma} c^{}_{f i -\sigma}  +
c^{\dagger}_{f i \sigma} c^{\dagger}_{f i -\sigma}
c^{}_{c i \sigma} c^{}_{c i -\sigma})
+ G_{n} ( n^{}_{f i \sigma} - n^{}_{c i \sigma})
+ I_{n} n^{}_{f i -\sigma} n^{}_{f i \sigma}
\nonumber \\
&& + M_{n} n^{}_{f i -\sigma} n^{}_{c i -\sigma}
( n^{}_{f i \sigma} - n^{}_{c i \sigma})
+ K_{n} n^{}_{c i -\sigma} n^{}_{c i \sigma} \biggr] \; ,
\label{eq:nthorder}
\end{eqnarray}
It can be observed, that the canonical transformation generates, 
besides the terms which renormalize the starting Hamiltonian, three 
new effective interactions, $J$, $P$, and $K$, and a higher order, 
triplet creating term, $M$. $J$ is the well-known Kondo
coupling, as it appears in Eqs.\ (\ref{4.11}) and (\ref{4.13}). 
$P$ a Josephson type two particle intersite
tunnelling and $K$ an effective on-site Coulomb repulsion
for the conduction electrons.
The values of the coefficients $J$, $P$, $G$ and $I$, $K$, $M$ for the first 
three odd $n$ are summarized in Table \ref{tab:1-3OrderResult}. 

\begin{table}
\centering
\begin{tabular}{|c|c|c|c|c|c|c|} 
\hline 
$n$ & $J$ & $P$ & $G$ & $I$ & $K$ & $M$ \\ \hline \hline
1 & 2$Z$ & -$Z$ & 2$A$ & 2$Z$ & 0 & 0 \\ \hline
3 &  - 32 $A^2 Z$  & - 8 $A^2 Z$ & - 8 $A^3$ & -28 $A^2 Z$ & -4 $A^2 Z$ 
  &  16 $A Z^2$   \\
  & - 32 $A Z^2$  & - 8 $A Z^2$ & & - 36 $A Z^2$ & + 4 $A Z^2$ 
  & + 8 $Z^3$ \\
  &  - 16 $Z^3$    & + 2 $Z^3$ & & - 16 $Z^3$   &    &     \\ \hline
5 & 512 $A^4 Z$  & 224 $A^4 Z$ & 32 $A^5$  & 336 $A^4 Z$ 
  & 176 $A^4 Z$  & - 352 $A^3 Z^2$ \\
  & + 1024 $A^3 Z^2$ & + 448 $A^3 Z^2$ & & + 848 $A^3 Z^2$ 
  & + 176 $A^3 Z^2$ & - 528 $A^2 Z^3$ \\
  & + 1024 $A^2 Z^3$ & + 328 $A^2 Z^3$ & & + 936 $A^2 Z^3$ 
  & + 88 $A^2 Z^3$ & - 368 $A Z^4$  \\
  & + 512 $A Z^4$  & + 104 $A Z^4$   & & + 520 $A Z^4$  
  & - 8 $A Z^4$    & - 96 $Z^5$ \\
  & + 128 $Z^5$      & - 4 $Z^5$   & & + 128 $Z^5$   &    & \\
\hline
\end{tabular}
\caption{The coefficients of the transformed hamiltonian in the first,
third and fifth order.} 
\label{tab:1-3OrderResult}
\end{table}

It is not by accident that the first three odd order commutators
Eq.\ (\ref{eq:nthcomm}) is given by a close form Eq.\ (\ref{eq:nthorder}). 
In fact, it can be proven by using induction that the form of 
the $n$ odd order commutation Eq.\ (\ref{eq:nthcomm}) is given by 
Eq.\ (\ref{eq:nthorder}) in general.

Other than the form of the commutation, one can also find   
a pattern in the coefficients  of the common terms among the
first, third and fifth orders, from Table \ref{tab:1-3OrderResult}. 
This pattern has been verified true up to eleventh order (Chan
and Gul\'{a}csi 2001a, 2003, 2004) and can be proven valid for any order 
by the same induction. 

\section{\label{ch4sw-two}Proof by Induction} 

The $n = 1$ case is the well known Schrieffer 
and Wolff (1966) result which represents the first rows of Table 
\ref{tab:1-3OrderResult}, notice that $J_{1}$ is identical to
Eq.\ (\ref{4.14}). By visual inspection, 
it can be observed that there is a pattern in the coefficients  of the 
common terms among the first, third and fifth orders. This pattern can 
be proven to exist for any order by induction, if the same pattern remains 
after commuting Eq.\ (\ref{eq:nthorder}) with $S$ twice. 

Two different indices are introduced here to
differentiate the order of the commutation $n$ from the recurrence
of the coefficients $J_{m}$, $P_{m}$, $\ldots$ over odd orders. The
mapping of the two sequences can be written as $n = 2 m + 1$ for
odd order $n$. Now, assuming that Eq.\ (\ref{eq:nthorder}) is true for
any $m$, we calculate its commutation with $S$ for $[[S,H_{V}]]_{n+1}$,
which yields:
\begin{eqnarray}
&& [[S,H_{V}]]_{n+1} \: = \:
\sum_{i, \sigma} \biggl[ - 2 V J_{m} ( A + Z n^{}_{f i -\sigma} )
\: (n^{}_{f i -\sigma} - n^{}_{c i -\sigma} )
\nonumber \\
&&- 2 V P_{m} ( A + Z n^{}_{c i -\sigma} ) \:
(n^{}_{f i -\sigma} - n^{}_{c i -\sigma} )
- 2 V G_{m} ( A + Z n^{}_{f i -\sigma} )
\nonumber \\
&& - 2 V M_{m} ( A + Z ) n^{}_{f i -\sigma} n^{}_{c i -\sigma}
- 2 V I_{m} ( A + Z n^{}_{f i -\sigma}) n^{}_{f i -\sigma}
\nonumber \\
&& + 2 V K_{m} ( A + Z n^{}_{f i -\sigma} ) n^{}_{c i -\sigma}
\biggr] \: \biggl(
c^{\dagger}_{c i \sigma} c^{}_{f i \sigma}
+ c^{\dagger}_{f i \sigma} c^{}_{c i \sigma} \biggr) \; .
\label{eq:n+1thorder}
\end{eqnarray}

To show that Eq.\ (\ref{eq:nthorder}) is true for any $m$, 
Eq.\ (\ref{eq:n+1thorder}) has to be commuted with $S$ again, 
obtaining:
\begin{eqnarray}
&& [[S,H_{V}]]_{n+2} \: = \: \sum_{i, \sigma}
\biggl[ J_{m+1} ( c^{\dagger}_{c i \sigma} c^{}_{c i -\sigma}
c^{\dagger}_{f i -\sigma} c^{}_{f i \sigma} -
n^{}_{c i\sigma} n^{}_{f i -\sigma} )
\nonumber \\
&&+ P_{m+1} ( c^{\dagger}_{c i \sigma} c^{\dagger}_{c i -\sigma}
c^{}_{f i \sigma} c^{}_{f i -\sigma} +
c^{\dagger}_{f i \sigma} c^{\dagger}_{f i -\sigma}
c^{}_{c i \sigma} c^{}_{c i -\sigma} )
+ G_{m+1} ( n^{}_{c i \sigma} - n^{}_{c i \sigma} )
\nonumber \\
&& + I_{m+1} n^{}_{f i -\sigma} n^{}_{f i \sigma}
+ K_{m+1} n^{}_{c i -\sigma} n^{}_{c i \sigma}
+ M_{m+1} n^{}_{f i -\sigma} n^{}_{c i -\sigma}
( n^{}_{f i \sigma} - n^{}_{c i \sigma} ) \biggr] \; ,
\label{eq:n+2thorder}
\end{eqnarray}
where
\begin{eqnarray}
J_{m+1} &=& -J_{m} 4V^{2} ( (A + Z)^{2} + A^{2})
- P_{m} 8V^{2} (A^{2} + A Z)
\nonumber \\
&& - G_{m} 4V^{2}(2 A Z + Z^{2})
- I_{m} 4V^{2}(A + Z)^{2} - K_{m} 4V^{2} A^{2} \; ,
\label{eq:Jn+1} \\
P_{m+1} &=& - J_{m} 4V^{2} (A^{2} + A Z) - P_{m} 2V^{2}
( (A + Z)^{2} + A^{2})
\nonumber \\
&& - (I_{m} + K_{m}) 2V^{2} (A + Z) A \; ,
\label{eq:Pn+1} \\
I_{m+1} &=& -J_{m} 4V^{2} (A + Z)^{2} - P_{m} 4V^{2} (A^{2} + A Z)
\nonumber \\
&& - G_{m} 4V^{2} (2 A Z + Z^{2}) - I_{m} 4V^{2} (A + Z)^{2} \; ,
\label{eq:In+1} \\
K_{m+1} &=& - J_{m} 4V^{2} A^{2} - P_{m} 4V^{2} (A^{2} + A Z)
- K_{m} 4V^{2} A^{2} \; ,
\label{eq:Kn+1} \\
M_{m+1} &=& + K_{m} 4V^{2} (2 A Z + Z^{2}) - M_{m} 4V^{2} (A + Z)^{2}
\nonumber \\
&& + J_{m} 4V^{2} (2 A Z + Z^{2}) \; ,
\label{eq:Mn+1} \\
G_{m+1} &=& - G_{m} 4V^{2} A^{2} \; .
\label{eq:Gn+1}
\end{eqnarray}
Notice that Eq.\ (\ref{eq:n+2thorder}) does not contain any new term
that is not in $[[S,H_{V}]]_{n}$. By mathematical  induction, it can be
concluded that the form of the $n$th commutation of $H_{V}$ with $S$ is 
closed and is always given by Eq.\ (\ref{eq:nthorder}).

\section{\label{ch4sw-three}Evaluating the Coefficients} 

The recursive equations (\ref{eq:Jn+1}) -( \ref{eq:Gn+1}) from the last
section can be solved simultaneously to
give the odd order coefficients of the transformed hamiltonian,
$J_{m}$, $P_{m}$, $I_{m}$, $K_{m}$ and $M_{m}$.

These recursive equations can be summarized into an matrix form in which
\begin{equation}
\left( \begin{array}{c}
J_{m+1}    \\
2 P_{m+1}  \\
I_{m+1}    \\
K_{m+1}    \end{array} \right)
\: = \: - 4 V^{2} {\bf M} \cdot
\left( \begin{array}{c}
J_{m}  \\
P_{m}  \\
I_{m}  \\
K_{m}  \end{array} \right)
- 4 V^{2}
\left( \begin{array}{c}
\alpha^{2}-\beta^{2} \\
0                    \\
\alpha^{2}-\beta^{2} \\
0                    \end{array} \right)
G_{m} \; , 
\label{eq:MatrixEq}
\end{equation}
where $\alpha = A + Z$ and $\beta = A$. ${\bf M}$ is a matrix given by:
\begin{equation}
{\bf M} \: = \: \left( \begin{array}{cccc}
\alpha^{2}+\beta^{2} & 2\alpha\beta & \alpha^{2} & \beta^{2} \\
2\alpha\beta & \alpha^{2}+\beta^{2} & \alpha\beta & \alpha\beta \\
\alpha^{2} & \alpha\beta & \alpha^{2} & 0 \\
\beta^{2} & \alpha\beta & 0 & \beta^{2}
\end{array} \right) \; .
\end{equation}

Not only is the matrix ${\bf M}$ symmetric, but its determinant is
also zero. It is, in fact, easy to see this since the first row
of the matrix is equal to the sum of the third and the fourth, while the
second row can be expressed as the linear combination of the last two.
Explicitly,
\begin{eqnarray}
J_{m+1} &=& I_{m+1} + K_{m+1} \; ,
\label{eq:J=I+K} \\
2P_{m+1} &=& \frac{\beta}{\alpha} (I_{m+1} + G_{m+1}) +
\frac{\alpha}{\beta} (K_{m+1} - G_{m+1}) \; .
\label{eq:2P=I+K+G}
\end{eqnarray}

The expression of $I_{m+1}$ and $K_{m+1}$ in terms of $J_{m+1}$ and
$P_{m+1}$ can be deduced directly from Eq.\ (\ref{eq:MatrixEq}), 
together with the general solution of $G_{m}$.
\begin{eqnarray}
I_{m+1} \: &=& \: \frac{\alpha\beta}{\beta^{2} - \alpha^{2}}
(2P_{m+1} - \frac{\alpha}{\beta} J_{m+1}) - G_{m+1} \; ,
\label{eq:In+1PJG} \\
K_{m+1} \: &=& \: \frac{\alpha\beta}{\alpha^{2} - \beta^{2}}
(2P_{m+1} - \frac{\beta}{\alpha} J_{m+1}) + G_{m+1} \; ,
\label{eq:Kn+1PJG} \\
G_{m} \: &=& \: (-2V^{2})^{m} (2\beta^{2})^{m} G_{0} \; .
\label{eq:Gn}
\end{eqnarray}
Substituting these equations back into Eqs.\ (\ref{eq:Jn+1})
and (\ref{eq:Pn+1}) yields the surprisingly simple result:
\begin{eqnarray}
J_{m+1} \: &=& \: -2^{3}V^{2} (\alpha^{2} + \beta^{2}) J_{m} \; ,
\label{eq:Jn+1Jn} \\
P_{m+1} \: &=& \: -2V^{2} (3 \alpha\beta J_{m} + (\alpha^{2} +
\beta^{2}) P_{m}) \; .
\label{eq:Pn+1JnPn}
\end{eqnarray}
In terms of $J_{0}$ and $P_{0}$ they can be written as:
\begin{eqnarray}
J_{m} \: &=& \: (-2^{3}V^{2})^{m} (\alpha^{2} + \beta^{2})^{m} J_{0} \; ,
\label{eq:Jn}\\
P_{m} \: &=& \: (-2V^{2})^{m}(\alpha^{2} + \beta^{2})^{m-1}
\nonumber\\
&&\cdot \biggl[ (\alpha^{2} + \beta^{2}) P_{0} +
\alpha \beta (4^{m} - 1) J_{0} \biggr] \; .
\label{eq:Pn}
\end{eqnarray}

Using Eqs.\ (\ref{eq:J=I+K}) and (\ref{eq:2P=I+K+G}), $K_{m+1}$ can
be written in terms of $P_{m}$, $J_{m}$, $G_{m}$ and subsequently
be solved by using Eqs.\ (\ref{eq:Gn}), (\ref{eq:Jn}) and (\ref{eq:Pn}):
\begin{equation}
K_{m} \: = \: (-2V^{2})^{m} (\alpha^{2} + \beta^{2})^{m-1} \biggl[
( 4^{m}\beta^{2} - \frac{2\alpha^{2}\beta^{2}}{\alpha^{2}-\beta^{2}})
J_{0} + 2 \alpha\beta \frac{\alpha^{2}+\beta^{2}}{\alpha^{2}-\beta^{2}}
P_{0} \biggr] + G_{m} \; .
\end{equation}

$I_{m}$ can now be evaluated from $K_{m}$ using
Eq.\ (\ref{eq:J=I+K}):
\begin{equation}
I_{m} \: = \: (-2V^{2})^{m} (\alpha^{2} + \beta^{2})^{m-1} \biggl[
(4^{m} \alpha^{2}+ \frac{2\alpha^{2}\beta^{2}}{\alpha^{2}-\beta^{2}})
J_{0} - 2 \alpha\beta \frac{\alpha^{2}+\beta^{2}}{\alpha^{2}-\beta^{2}}
P_{0} \biggr] -  G_{m} \; .
\label{eq:In}
\end{equation}

$M_{m}$ is slightly more complicated since it depends on not only
$M_{m-1}$, but also  $K_{m-1}$ and $J_{m-1}$. However, it can still 
be summed after some algebraic manipulations:
\begin{eqnarray}
M_{m} \: &=& \: -[-2^{3}V^{2}(\alpha^{2}+\beta^{2})]^{m-1} \:
(\alpha^{2}-\beta^{2}) \: J_{0}
\nonumber \\
&& - [-2V^{2}(\alpha^{2}+\beta^{2})]^{m} \:
\frac{4\alpha\beta}{\alpha^{2}-\beta^{2}} \:
(\frac{\alpha\beta J_{0}}{\alpha^{2}+\beta^{2}} - P_{0})
\nonumber \\
&& + (-4V^{2}\alpha^{2})^{m} \:
(\frac{\alpha^{2}+\beta^{2}}{\alpha^{2}-\beta^{2}}J_{0}-
\frac{4\alpha\beta}{\alpha^{2}-\beta^{2}}P_{0}
\nonumber \\
&& - G_{0} + M_{0}) + (-4V^{2}\beta^{2})^{m} \: G_{0} \; .
\label{eq:Mn}
\end{eqnarray}

These expressions can be further simplified (Chan and Gul\'{a}csi 2003)
by using the first order, ie, Schrieffer-Wolff results, which gives: 
\begin{eqnarray}
J_{m} &=& (-2^{3}V^{2})^{m} \: (\alpha^{2}+\beta^{2})^{m} \:
2 (\alpha-\beta) V^{2} \; ,
\label{eq:Jm} \\
P_{m} &=& (-2V^{2})^{m} \: (\alpha^{2}+\beta^{2})^{m-1} \:
[2^{2m+1}\alpha\beta - (\alpha+\beta)^{2}] \:
(\alpha-\beta) V^{2} \; ,
\label{eq:Pm} \\
G_{m} &=& (-2V^{2})^{m} \: (2\beta^{2})^{m} \: 2 \beta V^{2} \; ,
\label{eq:Gm} \\
K_{m} &=& \{ (-2V^{2})^{m} \: (\alpha^{2} + \beta^{2})^{m-1} \:
[4^{m}\beta(\alpha-\beta)
\nonumber \\
&& - \alpha(\alpha+\beta)]  + (-2V^{2})^{m} \:
(2\beta^{2})^{m} \} \: 2 \beta V^{2} \; ,
\label{eq:Km} \\
I_{m} &=& (-2^{3}V^{2})^{m} \: (\alpha^{2}+\beta^{2})^{m-1} \:
2 \alpha^{2} (\alpha-\beta) V^{2}
\nonumber \\
&& + (-2V^{2})^{m} \: (\alpha^{2} + \beta^{2})^{m-1} \:
2 \alpha \beta (\alpha+\beta) V^{2}
\nonumber \\
&& - (-2V^{2})^{m} \: (2\beta^{2})^{m} \: 2 \beta V^{2} \; ,
\label{eq:Im} \\
M_{m} &=& -(-2^{3}V^{2})^{m} \: (\alpha^{2}+\beta^{2})^{m-1} \:
2 (\alpha-\beta) (\alpha^{2}-\beta^{2}) V^{2}
\nonumber \\
&& - (-2V^{2})^{m} \: (\alpha^{2}+\beta^{2})^{m-1} \:
4 \alpha \beta (\alpha+\beta) V^{2}
\nonumber \\
&& + (-4\alpha^{2}V^{2})^{m} \: 2 \alpha V^{2} +
(-4\beta^{2}V^{2})^{m} \: 2 \beta V^{2} \; .
\label{eq:Mm}
\end{eqnarray}

The even order coefficients can also be deduced from the odd order
coefficients, after rearranging the $n+1$th commutation result 
Eq.\ (\ref{eq:n+1thorder}) to the form:
\begin{equation}
[[S,H_{V}]]_{n+1} \: = \: \sum_{i, \sigma}
(R_{m} + S_{m}n^{f}_{i, -\sigma} + T_{m}n^{c}_{i, -\sigma}
+ Q_{m}n^{f}_{i, -\sigma}n^{c}_{i, -\sigma}) \:
( c^{\dagger}_{i, \sigma} f^{}_{i, \sigma}
+ f^{\dagger}_{i, \sigma} c^{}_{i, \sigma}) \; ,
\label{eq:n+1order2}
\end{equation}
where
\begin{eqnarray}
R_{m} &=& -2VG_{m}A \; ,
\nonumber \\
S_{m} &=& -2V[(J_{m}+P_{m})A+I_{m}(A+Z) + (J_{m}+G_{m})Z] \; ,
\nonumber \\
T_{m} &=& 2V[(J_{m}+P_{m}+K_{m})A + P_{m}Z] \; ,
\nonumber \\
Q_{m} &=& 2V[(J_{m}+K_{m}-P_{m})Z - M_{m}(A+Z)] \; .
\label{eq:RSTQm}
\end{eqnarray}
The general solutions of $J_{m}$, $P_{m}$, $G_{m}$,
$I_{m}$, $K_{m}$, $M_{m}$ from Eqs.\ (\ref{eq:Jm}-\ref{eq:Mm}) 
can be used to find the expression of the even order coefficients:
\begin{eqnarray}
R_{m} &=& -(-4V^{2}\beta^{2})^{m} \: 4 \beta^{2} V^{3} \; ,
\label{eq:Rm} \\
S_{m} &=& -[-2^{3}V^{2}(\alpha^{2}+\beta^{2})]^{m} \:
8 \alpha (\alpha-\beta) V^{3}
\nonumber \\
&& -[-2V^{2}(\alpha^{2}+\beta^{2})]^{m} \:
2 \beta (\alpha+\beta) V^{3}
\nonumber \\
&& +(-4V^{2}\beta^{2})^{m} \: 4 \beta^{2} V^{3} \; ,
\label{eq:Sm} \\
T_{m} &=& [-2^{3}V^{2}(\alpha^{2}+\beta^{2})]^{m} \:
8 \beta (\alpha-\beta) V^{3}
\nonumber \\
&& -[-2V^{2}(\alpha^{2}+\beta^{2})]^{m} \:
2 \alpha (\alpha+\beta) V^{3}
\nonumber \\
&& +(-4V^{2}\beta^{2})^{m} \: 4 \beta^{2} V^{3} \; ,
\label{eq:Tm} \\
Q_{m} &=& [-2^3V^{2}(\alpha^{2}+\beta^{2})]^{m} \:
8 (\alpha-\beta)^{2} V^{3}
\nonumber \\
&& + [-2V^{2}(\alpha^{2}+\beta^{2})]^{m} \:
2 (\alpha+\beta)^{2} V^{3}
\nonumber \\
&& -(-4V^{2}\alpha^{2})^{m} \: 4 \alpha^{2} V^{3}
- (-4V^{2}\beta^{2})^{m} \: 4\beta^{2} V^{3} \; .
\label{eq:Qm}
\end{eqnarray}

The coefficients $R_{m},S_{m},Q_{m}$ and $T_{m}$ are calculated from a 
direct commutation of the odd order result from Eq.\ (\ref{eq:nthorder}) 
with $S$, in which the coefficients $J_{m}$, $P_{m}$, $G_{m}$,
$I_{m}$, $K_{m}$, $M_{m}$ given by Eqs.\ (\ref{eq:Jm}-\ref{eq:Mm}) 

In a nutshell, following Chan and Gul\'{a}csi (2001a,2003,2004) the result 
of the commutation of $H_{V}$ with $S$ to any order has been obtained, 
and hence have all the information needed to re-build the hamiltonian 
after the transformation.

\section{\label{ch4sw-four}The Transformed Hamiltonian} 

Using these general expression for the $n$th and the $n+1$th commutation
of $S$ with $H_{V}$, the exact infinite order transformation of the
hamiltonian can be calculated. The transformed hamiltonian
comprises $H_{0}$, the sum of the odd order commutations of $S$
with $H_{V}$ and the sum of the even order commutations:
\begin{equation}
\tilde{H}{}_{\PAM} = H_{0} + H_{odd} + H_{even} \, ,
\label{transham}
\end{equation}
where
\begin{eqnarray}
H_{odd} \: &=& \: \sum_{m=0}^{\infty}
\biggl[ \frac{1}{(2m+1)!} - \frac{1}{(2m+2)!} \biggr] \:
[[S,H_{V}]]_{2m+1} \; ,
\label{eq:oddsum} \\
H_{even} \: &=& \: \sum_{m=0}^{\infty}
\biggl[ \frac{1}{(2m+2)!} - \frac{1}{(2m+3)!} \biggr] \:
[[S,H_{V}]]_{2m+2} \; .
\label{eq:evensum}
\end{eqnarray}

Using the formulas
\begin{equation}
\sum^{\infty}_{n=0} \biggl[ \frac{1}{(2n+1)!} -
\frac{1}{(2n+2)!} \biggr] \gamma(-\xi^{2})^{n}  \: =  \: \gamma
\biggl( \frac{\sin \xi}{\xi} + \frac{\cos \xi -1}{\xi^{2}} \biggr) \; ,
\label{eq:sumfx}
\end{equation}
and
\begin{equation}
\sum^{\infty}_{n=0} \biggl[ \frac{1}{(2n+2)!} -
\frac{1}{(2n+3)!} \biggr] \gamma (-\xi^{2})^{n} \: =  \: \gamma
\biggl( \frac{\sin \xi}{\xi^{3}} - \frac{\cos \xi}{\xi^{2}}\biggr) \; ,
\label{eq:sumgx}
\end{equation}
where $\gamma$ and $\xi$ are any real or complex variables. It is 
straightforward to sum the odd order hamiltonian to get:
\begin{eqnarray}
&& H_{odd} \: = \: \sum_{i, \sigma} J (c^{\dagger}_{c i \sigma}
c^{}_{c i -\sigma} c^{\dagger}_{f i -\sigma} c^{}_{f i \sigma} -
n^{}_{c i \sigma} n^{}_{f i -\sigma})
\nonumber \\
&&+ P ( c^{\dagger}_{c i \sigma} c^{\dagger}_{c i -\sigma}
c^{}_{f i \sigma} c^{}_{f i -\sigma}  +
c^{\dagger}_{f i \sigma} c^{\dagger}_{f i -\sigma}
c^{}_{c i \sigma} c^{}_{c i -\sigma})
+ G ( n^{}_{f i \sigma} - n^{}_{c i \sigma})
\nonumber \\
&& + M n^{}_{f i -\sigma} n^{}_{c i -\sigma}
( n^{}_{f i \sigma} - n^{}_{c i \sigma})
+ I n^{}_{f i -\sigma} n^{}_{f i \sigma}
+ K n^{}_{c i, -\sigma} n^{}_{c i \sigma} \; .
\label{eq:sumoddH}
\end{eqnarray}
where $J$, $P$, $G$, $I$, $K$ and $M$ are the summation of the
corresponding $J_{m}$, $P_{m}$, $G_{m}$, $I_{m}$, $K_{m}$ and $M_{m}$
over infinite number of $m$. If  we define $\theta =
\sqrt{2V^{2}(\alpha^{2}+\beta^{2})}$,  $\theta_{\beta} = 2V\beta$,
$\theta_{\alpha} = 2V\alpha$ and $F(x) = \sin x /x + (\cos x - 1)/x^2$
then the exact values of the coupling constants from
Eq.\ (\ref{eq:sumoddH}) are:

\begin{eqnarray}
J &=&  2 (\alpha - \beta) V^{2} F(2 \theta) \; ,
\label{eq:J} \\
P &=& 2 \alpha \beta (\alpha - \beta) V^{2} F(2 \theta)
- (\alpha - \beta) V^{2}
\frac{(\alpha + \beta)^{2}}{\alpha^{2} + \beta^{2}} F(\theta) \; ,
\label{eq:P} \\
G &=& 2 \beta V^{2} F(\theta_{\beta}) \; ,
\label{eq:G} \\
K &=& 2 \beta^{2} V^{2} \frac{\alpha - \beta}{\alpha^{2} + \beta^{2}}
F(2 \theta) - 2 \alpha \beta V^{2}
\frac{\alpha + \beta}{\alpha^{2} + \beta^{2}} F(\theta) + G \; ,
\label{eq:K} \\
I &=& 2 \alpha^{2} V^{2} \frac{\alpha - \beta}{\alpha^{2} + \beta^{2}}
F(2 \theta) + 2 \alpha \beta V^{2}
\frac{\alpha + \beta}{\alpha^{2} + \beta^{2}} F(\theta) - G \; ,
\label{eq:I} \\
M &=& -2 (\alpha - \beta)
\frac{\alpha^{2} - \beta^{2}}{\alpha^{2} + \beta^{2}} V^{2} F(2 \theta)
- 4 \alpha \beta \frac{\alpha + \beta}{\alpha^{2} + \beta^{2}}
V^{2} F(\theta)
\nonumber\\
&& + 2 \alpha V^{2} F(\theta_{\alpha}) +
2 \beta V^{2} F(\theta_{\beta}) \; .
\label{eq:M}
\end{eqnarray}

In the symmetric case where $t=0, \epsilon_{f} = - U / 2$, we
have $Z = -2A$, $\theta = \theta_{\beta} = -\theta_{\alpha} = 2AV$,
$\alpha = -A$ and $\beta = A$. The summation becomes:
\begin{eqnarray}
J_{sym} &=& -4 A V^{2} F(4 A V) \; ,
\label{eq:Jsym} \\
P_{sym} &=& 2 A V^{2} F(4 A V) = -\frac{1}{2} J_{sym} \; ,
\label{eq:Psym} \\
G_{sym} &=& 2 A V^{2} F(2 A V) \; ,
\label{eq:Gsym} \\
K_{sym} &=& A V^{2}(1 - \cos 2 A V) [F(2 A V)
+ \frac{\sin 2 A V}{2 A V}] \; ,
\label{eq:Ksym} \\
I_{sym} &=& -2 A V^{2} (F(4 A V) + F(2 A V)) \; ,
\label{eq:Isym} \\
M_{sym} &=& 0 \; .
\label{eq:Msym}
\end{eqnarray}

Similarly, the even order hamiltonian can be evaluated by substitution
\begin{equation}
H_{even} \: = \: \sum_{i, \sigma} (R + S n^{}_{f i -\sigma} +
T n^{}_{c i -\sigma} + Q n^{}_{f i -\sigma} n^{}_{c i -\sigma} ) \:
(c^{\dagger}_{c i \sigma} c^{}_{f i \sigma} +
c^{\dagger}_{f i \sigma} c^{}_{f i \sigma}) \; ,
\label{eq:sumeven}
\end{equation}
where $R$, $S$, $T$ and $Q$ are the summation of the corresponding
$R_{m}$, $S_{m}$, $Q_{m}$ and $T_{m}$ over infinite number of $m$.
Using the same notations $\theta$, $\theta_{\beta}$, $\theta_{\alpha}$
as in the odd order coefficients, and $F'(x) = \sin x / x^3 -
\cos x / x^2$, we obtain:

\begin{eqnarray}
R &=& -4 \beta^{2} V^{3} F'(\theta_{\beta})  \; ,
\label{eq:R} \\
S &=& -8 (\alpha - \beta) \alpha V^{3} F'(2 \theta)
- 2 (\alpha + \beta) \beta V^{3} F' (\theta)  - R  \; ,
\label{eq:S} \\
T  &=& 8 \beta (\alpha - \beta) V^{3} F' (2 \theta)
- 2 \alpha (\alpha + \beta) V^{3} F' (\theta) - R  \; ,
\label{eq:T} \\
Q  &=& 8 (\alpha - \beta)^{2} V^{3} F' (2 \theta)
+ 2 (\alpha + \beta)^{2} V^{3} F' (\theta)
\nonumber\\
&&- 4 \beta V^{3} F' (\theta_{\beta}) - 4 \beta V^{3}
F' (\theta_{\alpha}) \; .
\label{eq:Q}
\end{eqnarray}

For the symmetric case ($\epsilon_{f} = - U / 2$),
we have $Z = -2A$, $\theta = \theta_{\beta} = - \theta_{\alpha}
= 2AV$, $\alpha = -A$ and $\beta = A$, both the odd and
even order hamiltonian coefficients simplify considerably.
In the symmetric case where $t=0, \epsilon_{f} = - U / 2$, we
have $Z = -2A$, $\theta = \theta_{\beta} = -\theta_{\alpha} = 2AV$,
$\alpha = -A$ and $\beta = A$. The summation becomes
\begin{eqnarray}
R_{sym} &=&  -4 A^{2} V^{3} F' (2 A V) \; , \\
S_{sym} &=&  -16 A^{2} V^{3} F' (4 A V) + R_{sym} \; , \\
T_{sym} &=&  -16 A^{2} V^{3} F'(4 A V) + R_{sym} \; , \\
Q_{sym} &=&  32 A^{2} V^{3} F' (4 A V) + 2 R_{sym} \; .
\end{eqnarray}

%%%%%%%%%%%%%%%%%%%%%%%%%%%%%%%%%%%%%%%%%%%
%% appendix 4
%%%%%%%%%%%%%%%%%%%%%%%%%%%%%%%%%%%%%%%%%%%%

\cleardoublepage
\chapter{\label{ch1one}Introduction to Many-Electron Systems}

For completeness, and in order to fix notation, 
it is useful to present a brief account of the 
used formalism to describe many-particle systems. 
Bases of single-electron states are given in 
section \ref{old1.1.1} for the two systems of central interest in 
this review: (i) Continuum system: electrons confined to move in 
1D, but otherwise free. (ii) Lattice system: 
electrons confined to move along a 1D chain 
of ions.\footnote{The formalism is thus written for 1D systems. 
The generalization to higher dimensions is straightforward, and 
usually involves only a replacement of scalar spatial and 
momentum variables $x$ and $k$ by vectors ${\bf r}$ and 
${\bf k}$ respectively (Nozi\`{e}res, 1964).} 
The construction of the 
many-particle state-space out of single-particle states 
is given in section \ref{old1.1.2}. This includes both many-fermion and 
many-boson systems, and systems with a variable number of 
particles. Section \ref{old1.1.3} discusses  
operators on the many-particle states. 
Creation and annihilation operators are defined, 
and their fundamental commutation relations are given. 
Creation and annihilation operators are highly convenient 
in that they account automatically for the correct 
symmetry of many-boson and many-fermion states, 
while at the same 
time admitting of a simple physical interpretation. They 
are used to construct all the many-particle operators 
subsequently required. 

\section{\label{old1.1.1}Single-electron states}

{\bf Single-electron states in a continuum}: Consider an 
electron confined to move in one spatial dimension along a 
length $L$. Since contact effects at the 
endpoints of the system are not of concern here, 
periodic boundary conditions are imposed. The state-space for an 
electron confined to move in this system is a two-component 
Hilbert space of Lebesgue square-integrable functions 
(Jauch 1968). The functions are complex-valued, are defined on 1D 
Euclidean space, and satisfy the periodic boundary 
conditions. An orthonormal basis of  
single-electron states for the system is given by the set
$\{|k\sig\rangle\}$ of normalized simultaneous eigenstates 
of momentum $k$ along the direction of motion, 
and spin-component $\sig$ along a specified axis 
in 3D space (conventionally denoted  
the $z$-axis). These properties of  
the single-electron state-space are here taken as 
axiomatic, with the reader referred to Jauch (1968),
for an extensive discussion. 

A realization of the state-space may be given using the position 
representation. The wavefunction $\Phi_{k\sig}({x})$ 
corresponding to the basis element $|{k}\sig\rangle$ satisfies 
the momentum eigenvalue equation $-i\partial_{x} 
\Phi_{k\sig}(x) = k\Phi_{k\sig}(x)$, where 
$\partial_{x}$ means $d/dx$. (Units are chosen throughout 
so that $\hbar = 1$.) Normalized 
solutions of the eigenvalue equation are 
\beqa
\Phi_{{k}\sig}({x}) = L^{-1/2} 
\, e^{i{kx}}\,\chi_{\sig}\, .
\label{1.1}
\eeqa
The periodic boundary conditions restrict the allowed 
momenta to values $k = 2\pi m/L$, with $m$ any integer. The Pauli 
spinors $\chi^{\dg}_{\uparrow} = (1,0)$ and 
$\chi^{\dg}_{\downarrow} = (0,1)$ describe spin $\sig = 
\uparrow \downarrow = \pm 1$, respectively, along the 
$z$-axis.\footnote{The labels $\uparrow, +1$ and $\downarrow, 
-1$ will be used interchangeably for $\sig$; in formulas $\sig$ 
always denotes $\pm 1$ as the spin is $\pm \hbar/2$, 
respectively, along the $z$-axis.}

The momentum/spin basis elements $|k\sig\rangle$ are also 
eigenstates of the single-electron kinetic energy operator, 
or non-interacting single-electron hamiltonian $H_{0}^{(1)}$. In 
the position representation, 
\beqa 
H_{0}^{(1)} = -\frac{1}{2m_{e}} \partial_{x}^{2}\, ,
\label{1.1a}
\eeqa
where $m_{e}$ is the bare electron mass. $\Phi_{k\sig}(x)$ is an 
eigenstate of Eq.\ (\ref{1.1a}) with kinetic energy, or 
dispersion, $\varepsilon(k) = k^{2}/2m_{e}$. 

A second orthonormal basis of single-electron states $\{|{x} \sig 
\rangle\}$, which relate to position in the continuum system, 
may be obtained 
by Fourier decomposing the momentum states: 
\beqa
|{x}\sig\rangle = L^{-1/2} \sum_{k} 
e^{-i{kx}} |{k} \sig \rangle .
\label{1.2}
\eeqa
The wavefunction $\Phi_{{x}\sig}({x}')$ 
corresponding to the basis element 
$|{x}\sig\rangle$ may be obtained from Eqs.\ (\ref{1.1}) and 
(\ref{1.2}):
\beqa
\Phi_{{x}\sig}({x}') = 
L^{-1}\sum_{k} e^{-ik(x-x')}\, \chi_{\sig}\, .
\label{1.2a}
\eeqa
This simplifies in a thermodynamically large system;
as $L \rightarrow \infty$,  
\beqa 
\sum_{k} f(k) \rightarrow 
(L/2\pi)\int_{-\infty}^{\infty} dk\, f(k)\,  . 
\label{1.3}
\eeqa
Since the Dirac $\delta$-function has the representation 
\beqa
\delta(x) = \frac{1}{2\pi}\int_{-\infty}^{\infty}dk\, 
e^{-ikx} \, ,
\label{1.3a}
\eeqa
it follows from Eqs.\ (\ref{1.2a}) and (\ref{1.3}) that 
\beqa
\Phi_{{x}\sig}({x}') \rightarrow 
\frac{1}{2\pi}\int_{-\infty}^{\infty}dk\, e^{-ik(x-x')}\, 
\chi_{\sig}  = \delta({x} - {x}') \,\chi_{\sig}\, , \quad \quad
{\rm as} \,\,\, L \rightarrow \infty\, .
\label{1.4} 
\eeqa
It is important to understand the significance of the 
limit $L \rightarrow \infty$ in Eq.\ (\ref{1.4}), for 
otherwise it appears as though the denumerable basis 
$\{ |{k}\sig\rangle\}$ for finite $L$, with wavefunctions as in 
Eq.\ (\ref{1.1}), may be transformed into a 
non-denumerable basis with wavefunctions as in 
Eq.\ (\ref{1.4}). This point has some relevance to the 
development of the bosonization formalism in chapter \ref{ch2}: 
In an influential paper, Haldane (1981) constructs a  
{\it periodic} Dirac $\delta$-function with period $L$
(cf. Eq.\ (3.4) of Haldane (1981)). This 
construction is formally correct, but is vacuous as the period 
$L = \infty$ is required, as in Eq.\ (\ref{1.3}), in order 
to obtain the Dirac $\delta$-function in the first place. 
For finite $L$ it is not possible to have both a denumerable 
momentum basis, and a non-denumerable position basis, for  
in the first case the corresponding Hilbert space is separable, 
while in the second case it is not. Haldane uses the periodic 
Dirac $\delta$-function construction 
to motivate an interpretation of a length 
$\alpha$ which appears in the bosonization formulae. 

{\bf Single-electron states in a lattice}: 
Consider now a chain of $N$ identical ions, 
with periodic boundary conditions again imposed . 
The ions are located at lattice sites $ja$, where $a$ is 
the lattice spacing and $j$ is an integer. An orthonormal 
basis of states for an electron moving along the chain  
of ions is given by the set of Bloch states  
$\{|n{k}\sig\rangle\}$, where $n$ is a discrete index labelling 
the band, and where ${k}$ is the crystal momentum. (The reader 
is referred to Haug (1972) for a detailed discussion.) 
Bloch states are eigenstates of the 
hamiltonian $H_{0}^{(1)}$ for a single electron 
moving in the periodic potential of the chain of 
ions. In the position representation, 
\beqa
H_{0}^{(1)} = -\frac{1}{2m_{e}} \partial_{x}^{2} + U(x) \, ,
\label{1.4a}
\eeqa
where $U(x)$ is the periodic potential of the lattice of ions.
Normalized Bloch wavefunctions are given by (Huag 1972)
\beqa
\Phi_{n{k}\sig}({x}) = 
N^{-1/2}\, e^{i{kx}}
\, u_{n{k}}({x}) \, \chi_{\sig}\, ,
\label{1.5}
\eeqa
where the Bloch function $u_{nk}$ satisfies $u_{nk}(x + ja) = 
u_{nk}(x)$ for all $j$, and 
is normalized in the unit cell of length $a$ 
centred on a lattice site. Periodic boundary conditions 
restrict the allowed crystal momenta to values
${k} = (2 \pi m/L)$  with integers $m$. The length of the 
lattice system is $L = Na$. To prevent multiple counting of 
states, the allowed crystal momenta $k$
are further restricted to the first 
Brillouin zone (FBZ) in the lattice (Huag 1972), $-\pi /a 
< k \leq \pi /a$, so that the integers $m$   
satisfy $-N/2 < m \leq N/2$. The Bloch states are eigenstates of 
$H_{0}^{(1)}$ with band energies $\varepsilon_{n}(k)$, which 
depend on the form of $U(x)$. 

A second orthonormal basis of Wannier states 
$\{|n{j}\sig\rangle\}$ are obtained by Fourier 
decomposing the Bloch states:  
\beqa
|n{j}\sig\rangle = N^{-1/2}\sum_{ {k} \in 
{\rm FBZ}} e^{-i{kja}}|n{k}\sig\rangle\, .
\label{1.6}
\eeqa
The Wannier states represent electrons centred on the 
ion at site $ja$, with the degree of localization depending on 
the form of the Bloch functions $u_{nk}(x)$. 
$|n{j}\sig\rangle$ are the lattice equivalent of 
the continuum electron states $|{x}\sig\rangle$ of Eq.\ 
(\ref{1.2}). Using Eqs.\ (\ref{1.5}) and (\ref{1.6}),  
the wavefunction corresponding to the Wannier state 
$|n{j}\sig\rangle$ is given by 
\beqa
\Phi_{n\sig}(x-ja) = N^{-1}\sum_{k \in {\rm FBZ}} e^{ik(x-ja)}\, 
u_{nk}(x)\, \chi_{\sig} \, , 
\label{1.6a}
\eeqa
which is the lattice equivalent of Eq.\ (\ref{1.2a}).

\section{\label{old1.1.2}Many-particle states}

{\bf Distinguishable particles}: 
Consider a system consisting of $N_{0}$ distinguishable 
particles. If 
$|\psi_{1}\rangle, |\psi_{2}\rangle, \ldots, 
|\psi_{N_{0}}\rangle$ 
are single-particle states for the 
system, then the ordered $N_{0}$-tuple
\beqa
|\psi^{(N_{0})}\rangle = 
|\psi_{1}\rangle \otimes |\psi_{2}\rangle \otimes \cdots 
\otimes |\psi_{N_{0}}\rangle 
\label{1.7} 
\eeqa
is the $N_{0}$-particle state in which particle 1 is in the 
state $|\psi_{1}\rangle$, particle 2 is in the state 
$|\psi_{2}\rangle$,  $\ldots$, and particle 
$N_{0}$ is in the state $|\psi_{N_{0}}\rangle$. 
The product $\otimes$ in Eq.\ (\ref{1.7}) is the 
Hilbert space tensor product (Prugove\v{c}ki 1981), 
and the many-particle state 
$|\psi^{(N_{0})}\rangle$ is an element in the tensor product 
of the $N_{0}$ corresponding single-particle Hilbert 
spaces. It is convenient to leave the tensor product 
implicit, so that Eq.\ (\ref{1.7}) becomes simply 
$|\psi^{(N_{0})}\rangle = 
|\psi_{1}\rangle |\psi_{2}\rangle \cdots 
|\psi_{N_{0}}\rangle$. The tensor product defines 
the inner product of two 
$N_{0}$-particle states 
$|\psi^{(N_{0})}\rangle = 
|\psi_{1}\rangle |\psi_{2}\rangle \cdots 
|\psi_{N_{0}}\rangle$ and  
$|\phi^{(N_{0})}\rangle = |\phi_{1}\rangle |\phi_{2}\rangle 
\cdots |\phi_{N_{0}}\rangle$ by Prugove\v{c}ki (1981)
\beqa 
\langle \phi^{(N_{0})} | \psi^{(N_{0})} \rangle = 
\langle \phi_{1} |\psi_{1} \rangle
\langle \phi_{2} |\psi_{2} \rangle \cdots 
\langle \phi_{N_{0}} |\psi_{N_{0}} \rangle \, , 
\label{1.8}
\eeqa
where the inner products on the right are from the 
single-particle Hilbert spaces.

{\bf Indistinguishable particles}: 
Consider now a system consisting of $N_{0}$ indistinguishable 
particles of either the Fermi or the Bose variety. It is   
taken as axiomatic here that the physically realisable 
many-fermion states 
for the system are antisymmetric with regard to the labelling of 
the particles, while the many-boson states are symmetric.  
(The reader is referred to Gross, Runge and Heinonen (1991), 
for a more detailed discussion.) 
Explicitly, if $|\psi_{1}\rangle, \ldots, 
|\psi_{N_{0}}\rangle$ are single-fermion or single-boson states 
for the system, then the physically realisable many-fermion or 
many-boson states for the system, in which a particle is in the 
state $|\psi_{1}\rangle$, and $\ldots$, and a particle is in the 
state $|\psi_{N_{0}}\rangle$, are of the form 
\beqa
|\psi_{1}, \psi_{2}, \ldots, \psi_{N_{0}} \rangle 
= \left(1/\sqrt{N_{0}!}\right) \sum_{P}\zeta^{P}
|\psi_{P(1)}\rangle |\psi_{P(2)}\rangle \cdots 
|\psi_{P(N_{0})}\rangle \, ,
\label{1.9}
\eeqa
where the sum is over the $N_{0}!$ permutations $P$ of 
$N_{0}$ things, and where $\zeta^{P} = +1$ for bosons, and 
$\zeta^{P} = \pm 1$ for fermions, depending as the 
permutation $P$ is even or odd respectively (Feynman 1972). 
The factor $1/\sqrt{N_{0}!}$ is chosen for 
convenience of normalization. Note the special designation 
$|\psi_{1}, \psi_{2}, \ldots, \psi_{N_{0}} \rangle$     
for many-particle states with the correct symmetry.

Eq.\ (\ref{1.9}) contains within it Pauli's exclusion principle, 
that no more than one fermion may occupy each single-particle 
state. For if, in Eq.\ (\ref{1.9}), two of the single-fermion 
states were identical, say $|\psi_{1}\rangle 
= |\psi_{2}\rangle$, then the 
permutation which interchanges the particle labels $1$ and $2$ 
will reproduce the same state, but with an opposite sign.  
A straightforward extension of this argument to all permutations 
establishes that the fully antisymmetrized many-fermion state 
will vanish identically. This property is fundamental to 
many-fermion systems, and distinguishes them from 
many-boson systems, for which no such restriction applies. 

Using Eq.\ (\ref{1.9}), it may be verified that the 
inner product between 
two properly symmetrized $N_{0}$-particle states,
$|\psi_{1}, \psi_{2}, \ldots, \psi_{N_{0}} \rangle$ and   
$|\phi_{1}, \phi_{2}, \ldots, \phi_{N_{0}} \rangle$, takes 
the form of a determinant:
\begin{eqnarray}
&& \langle \phi_{1}, \phi_{2}, \ldots, \phi_{N_{0}}
|\psi_{1}, \psi_{2}, \ldots, \psi_{N_{0}} \rangle 
\nonumber \\
&& =  
\left| \begin{array}{cccc}
\langle \phi_{1}|\psi_{1}\rangle & 
\langle \phi_{1}|\psi_{2}\rangle & 
\cdots & \langle \phi_{1}|\psi_{N_{0}}\rangle \\
\langle \phi_{2}|\psi_{1}\rangle & 
\langle \phi_{2}|\psi_{2}\rangle & 
\cdots & \langle \phi_{2}|\psi_{N_{0}}\rangle  \\
\vdots & \vdots & \ddots & \vdots \\
\langle \phi_{N_{0}}|\psi_{1}\rangle & 
\langle \phi_{N_{0}}|\psi_{2}\rangle & 
\cdots & \langle \phi_{N_{0}}|\psi_{N_{0}}\rangle 
\end{array} \right|_{\zeta} 
\label{1.10}
\end{eqnarray}
For fermions $\zeta = -1$ and the standard determinant 
is signified. For bosons $\zeta = +1$ and the 
permanent (Feynman 1972) is signified; this is the 
standard determinant, but with $+$ signs only attached  
to all the usual determinental terms. 

{\bf Multi-particle states}: To describe systems in 
which the particle numbers vary, and in particular in order 
to properly define creation and annihilation operators, 
it is necessary to define a multi-particle space.  
This is the direct sum (Prugove\v{c}ki 1981) of all the 
different many-particle spaces. 
A general member of this space is written 
\beqa
|\psi\rangle = |\psi^{(0)}\rangle \oplus |\psi^{(1)}\rangle  
\oplus \ldots \oplus |\psi^{(N_{0})}\rangle \oplus \ldots 
\label{1.11} 
\eeqa
where $|\psi^{(n)}\rangle$, $n = 0,1,\ldots$, denotes an 
$n$-particle state, and 
where $\oplus$ denotes the direct sum. (The state 
$|\psi^{(0)}\rangle$ describes the system with no 
particles present. It is a tensor of rank zero, i.e.\  
a complex number.) The inner product between 
two multi-particle states is the obvious one: States with 
different particle numbers are orthogonal, so that
\beqa
\langle\phi |\psi\rangle = \sum_{n=0}^{\infty} 
\langle \phi^{(n)}|\psi^{(n)}\rangle \, . 
\label{1.12}
\eeqa

\section{\label{old1.1.3}Many-particle operators}

Operators on physically realisable many-particle states 
are conveniently written in terms of creation and annihilation 
operators. These operators act on multi-particle states 
whose elements $|\psi^{(n)}\rangle$, as in Eq.\ (\ref{1.11}), 
have the required symmetry (cf.\ Eq.\ (\ref{1.9})) 
depending on whether the system 
contains fermions or bosons. Creation and annihilation 
operators have the advantage that the 
states acted on also retain the correct symmetry. This simplifies 
manipulations by avoiding direct reference to lengthy expansions, 
such as those in Eqs.\ (\ref{1.9}) and (\ref{1.10}).
 
{\bf Creation and annihilation operators}:
To define creation and annihilation operators, 
let $|\phi\rangle$ be a single-particle state for a 
system. The creation operator $a^{\dg}_{\phi}$ for this  
state places a particle in the system in the state 
$|\phi\rangle$, and properly symmetrizes the resulting 
many-particle state. $a^{\dg}_{\phi}$ is defined by 
\beqa
a^{\dg}_{\phi}|\psi_{1}, \psi_{2}, \ldots, \psi_{N_{0}}\rangle 
= |\phi, \psi_{1}, \psi_{2}, \ldots, \psi_{N_{0}}\rangle\, ,
\label{1.13}
\eeqa
where $|\psi_{1}, \psi_{2}, \ldots, \psi_{N_{0}}\rangle$ is 
an $N_{0}$-particle state as defined in Eq.\ 
(\ref{1.9}); it is symmetric or antisymmetric 
in the particle labelling as the 
system contains bosons or fermions, respectively. Note 
that $a^{\dg}_{\phi}$ is an operator on the multi-particle
space, and generally maps an $N_{0}$-particle state to an 
$N_{0} + 1$-particle state. The exception to this rule is a 
many-fermion 
state in which $|\phi\rangle$ is already occupied. In this case,  
in accordance with Pauli's exclusion principle 
(cf.\ section \ref{old1.1.2}),
$a^{\dg}_{\phi}$ maps the many-fermion state 
to zero. Creation operators are linear by definition, and 
are linear functionals of their arguments; if 
$|\phi\rangle 
= c_{1}|\phi_{1}\rangle + c_{2}|\phi_{2}\rangle$, 
then 
\beqa
a^{\dg}_{\phi} = c_{1}a^{\dg}_{\phi_{1}} 
+ c_{2}a^{\dg}_{\phi_{2}} \, ,
\label{1.14}
\eeqa
 where $c_{1}$ and $c_{2}$ are 
complex numbers. 

The annihilation operator $a_{\phi}$ for the single-particle 
state $|\phi\rangle$ is defined as the hermitian conjugate 
(h.c.), or adjoint, of the creation operator $a^{\dg}_{\phi}$. 
Using Eqs.\ (\ref{1.10}) and (\ref{1.13}), it may 
be verified that (Feynman 1972) 
\beqa
a_{\phi}|\psi_{1}, \ldots, \psi_{N_{0}}\rangle 
= \sum_{n=1}^{N_{0}} \zeta^{n-1} 
\langle\phi | \psi_{n}\rangle 
|\psi_{1}, \ldots ({\rm no}\, \psi_{n}) \ldots, 
\psi_{N_{0}}\rangle ,
\label{1.15}
\eeqa
where $\zeta = \pm 1$ as the system contains bosons or 
fermions, respectively.
The annihilation operator thus removes the particles in the 
states $|\psi_{n}\rangle$ one at a time, with an appropriate 
sign to maintain the proper symmetry, and with a weight 
depending on the overlap of $|\psi_{n}\rangle$ with 
$|\phi\rangle$. From Eq.\ (\ref{1.14}), the annihilation 
operator is a conjugated linear functional of its argument: 
\beqa
a^{}_{\phi} = c^{*}_{1}a^{}_{\phi_{1}} 
+ c^{*}_{2}a^{}_{\phi_{2}} \, ,
\label{1.16}
\eeqa
where $|\phi\rangle 
= c_{1}|\phi_{1}\rangle + c_{2}|\phi_{2}\rangle$, and where 
$c_{1}$ and $c_{2}$ are complex numbers with conjugates 
$c_{1}^{*}$ and $c_{2}^{*}$ respectively. 

Using Eqs.\ (\ref{1.13}) and (\ref{1.15}), it is 
straightforward to verify (Feynman 1972) that for 
single-particle states $|\phi_{1}\rangle$ and 
$|\phi_{2}\rangle$, boson creation and annihilation 
operators satisfy the commutation relations
\beqa
\left[a^{}_{\phi_{1}}, a^{}_{\phi_{2}}\right] =
\left[a^{\dg}_{\phi_{1}}, a^{\dg}_{\phi_{2}}\right]  
= 0 \, ,  \quad \, \,
\left[a^{}_{\phi_{1}}, a^{\dg}_{\phi_{2}}\right]  
= \langle\phi_{1}|\phi_{2}\rangle \, ,
\label{1.17}
\eeqa
where the commutator $[A, B] = AB - BA$. In a similar fashion 
one can show that fermion creation and annihilation operators 
satisfy the anticommutation relations 
\beqa
\left\{a^{}_{\phi_{1}}, a^{}_{\phi_{2}}\right\} =
\left\{a^{\dg}_{\phi_{1}}, a^{\dg}_{\phi_{2}}\right\}  
= 0 \, ,  \quad \, \, 
\left\{a^{}_{\phi_{1}}, a^{\dg}_{\phi_{2}}\right\}  
= \langle\phi_{1}|\phi_{2}\rangle \, ,
\label{1.18}
\eeqa
where the anticommutator $\{A, B\} = AB + BA$. 

{\bf Other many-particle operators}: The construction of 
general many-particle operators out of creation and 
annihilation operators begins from the decomposition 
of an operator into single-particle projectors: 
If $O^{(1)}$ is an operator on single-particle states, such as a 
number operator, then there is always  
a decomposition 
\beqa
O^{(1)} = \sum_{\alpha,\beta}O_{\alpha \beta}| 
\alpha\rangle\langle\beta |
\label{1.19}
\eeqa
where $|\alpha\rangle$ and $|\beta\rangle$ are elements in an 
orthonormal basis of 
single-particle states, and where $O_{\alpha \beta} = 
\langle\alpha |O^{(1)}|\beta\rangle$. 
The corresponding many-particle operator $O$, 
representing the sum of $O^{(1)}$ over each occupied state, is 
then given by (Feynman 1972, Negele and Orland 1988)  
\beqa
O = \sum_{\alpha,\beta}O_{\alpha \beta} \, 
a^{\dg}_{\alpha} a^{}_{\beta}\, .
\label{1.20}
\eeqa
Similarly, if $O^{(2)}$ is a two-particle operator, such as an 
electron-electron interaction, then the corresponding 
many-particle operator $O$, representing the sum of $O^{(2)}$ 
over all pairs of particles, is given by 
\beqa
O = \frac{1}{2}\sum_{\alpha, \beta, \gamma, \delta} 
O_{\alpha \beta \gamma \delta} \, a^{\dg}_{\alpha}a^{\dg}_{\beta} 
a^{}_{\gamma}a^{}_{\delta}\, , 
\label{1.21}
\eeqa
where the matrix element 
$O_{\alpha \beta \gamma \delta} = \langle \alpha \beta|O^{(2)}| 
\gamma \delta \rangle$ for single-particle states 
$|\alpha\rangle, \ldots, |\delta\rangle$ in an 
orthonormal basis. The factor $1/2$ in Eq.\ (\ref{1.21}) is to 
redress the counting of each pair of particles twice.
(See Negele and Orland (1988) for a detailed derivation, 
together with extensions to operators involving more than two 
particles.) 

{\bf Some notation}: Creation and annihilation operators 
for the various single-electron states given in section 
\ref{old1.1.1} will be denoted by $c$ with the appropriate 
subscript. Thus $c^{}_{k\sig}$ annihilates a continuum electron 
in the momentum/spin state $|k\sig\rangle$, $c^{\dg}_{nj\sig}$ 
creates a lattice electron of spin $\sig$ at the lattice 
site $ja$ 
in the band $n$, and so on. There is an exception to this rule, 
dictated by convention. For continuum electrons in the position 
basis $\{|x\sig\rangle\}$ (cf.\ Eq.\ (\ref{1.2})), 
the corresponding creation and 
annihilation operators are denoted by $\psi^{\dg}_{\sig}(x)$ 
and $\psi^{}_{\sig}(x)$, respectively, and are called the Fermi 
fields. 

The relationships between the creation and 
annihilation operators for the different bases given in 
section \ref{old1.1.1} are as follows. 
From Eqs.\ (\ref{1.2}) and (\ref{1.14}), 
\beqa
\psi^{\dg}_{\sig}(x) = L^{-1/2}\sum_{k}e^{-ikx}\, 
c^{\dg}_{k\sig}\, .
\label{1.22}
\eeqa
This may be inverted using the representation of the 
Kronecker delta
\beqa
\delta_{k, k'} = L^{-1}\int_{L}dx\, e^{i(k' - k)x}
\label{1.23}
\eeqa
to give 
\beqa
c^{\dg}_{k\sig} = L^{-1/2}\int_{L}dx\, e^{ikx}\, 
\psi^{\dg}_{\sig}(x)\, . 
\label{1.24}
\eeqa
Similar expressions obtain for the lattice operators: From Eqs.\ 
(\ref{1.6}) and (\ref{1.14}),  
\beqa
c^{\dg}_{nj\sig} = N^{-1/2}\sum_{k \in {\rm FBZ}}e^{-ikja}\, 
c^{\dg}_{nk\sig}\, .
\label{1.25}
\eeqa
This may be inverted using the lattice equivalent of 
Eq.\ (\ref{1.23}),  
\beqa
\delta_{k, k'} = N^{-1}\sum_{j} e^{i(k' - k)ja}\, ,
\label{1.26}
\eeqa
where $k, k' \in {\rm FBZ}$ (Huag 1972).
Eqs.\ (\ref{1.25}) and (\ref{1.26}) give
\beqa
c^{\dg}_{nk\sig} = N^{-1/2}\sum_{j}e^{ikja}\, 
c^{\dg}_{nj\sig}\, .
\label{1.27}
\eeqa
These relationships will be used extensively throughout this 
review.

In chapter \ref{ch2} it was made clear that one of the special 
features of 1D Fermi systems is that they may be written in 
terms of bosonic excitations. The letter $b$ will be reserved to 
denote bosonic creation and annihilation operators, 
so that $b^{\dg}_{k\sig}$ creates a bosonic density 
fluctuation of wave vector $k$ out of electrons of spin $\sig$.  
(See section \ref{old2.2.1} for more details). 

%%%%%%%%%%%%%%%%%%%%%%%%%%%%%%%%%%%%%%%%%%%
%% appendix 5 
%%%%%%%%%%%%%%%%%%%%%%%%%%%%%%%%%%%%%%%%%%%%

\cleardoublepage
\chapter{\label{ch1two}Properties of Many-Electron Systems}

The low-temperature properties of many-electron systems are 
dominated by the antisymmetry requirement, Eq.\ (\ref{1.9}), 
on the physically realisable many-electron states. Some 
consequences of antisymmetry 
will now be presented as a general introduction to the low-energy 
properties of many-electron systems. 
Section \ref{old1.2.1} considers non-interacting many-electron 
systems, and defines the important 
concept of the Fermi sea. Section \ref{old1.2.2} 
considers interactions between the electrons, and 
gives some standard examples of interaction terms.  
It is common in 1D many-electron systems to
replace the Fermi sea by Dirac seas. This constitutes a field 
theory approximation to the condensed matter system of interest. 
The field theory construction is outlined in section \ref{old1.2.3}.

\section{\label{old1.2.1}Non-interacting Fermi systems: The Fermi gas}

Consider a system consisting of non-interacting electrons. 
In this case the only energy in the system is kinetic, and comes  
from the motion of the electrons through the system. 
The total energy of a non-interacting continuum 
system is obtained as follows: Count the number of electrons 
occupying each momentum/spin state $|k\sig\rangle$, and multiply 
the occupation number by the kinetic energy, or dispersion, 
$\varepsilon (k) = k^{2}/2m_{e}$ corresponding to that state
(cf. Eq.\ (\ref{1.1a})).  The total energy for the 
system is obtained by adding the contributions from each 
momentum/spin state. Using Eq.\ (\ref{1.20}), the operator for 
the number of electrons in the 
state $|k\sig\rangle$ is given by $c^{\dg}_{k\sig}c^{}_{k\sig}$.  
The total energy operator for the non-interacting continuum 
system, that is the non-interacting many-electron 
hamiltonian $H_{0}$, is thus 
\beqa
H_{0} = \sum_{k, \sig} \varepsilon(k) \, 
c^{\dg}_{k\sig}c^{}_{k\sig} \, . 
\label{1.2.1}
\eeqa
To determine the ground-state of the non-interacting 
system, recall that the antisymmetry requirement  
of many-fermion states prevents more than 
one electron occupying each single-electron state $|k\sig\rangle$.
(Pauli's exclusion principle.) The ground-state of $H_{0}$ for a 
fixed number of electrons is thus 
obtained by placing exactly one electron in each momentum/spin 
state $|k\sig\rangle$, beginning with the lowest energy 
states at $k = 0$, and continuing through 
increasing $|k|$ values until no electrons remain. This 
construction gives the non-interacting ground-state, denoted 
$|0\rangle$. If the 
system contains $N_{e}$ electrons, then simple counting arguments 
show that $|0\rangle$ consists of all momentum/spin states 
$|k\sig\rangle$ occupied for $|k| < k_{F} = N_{e} \pi/2L$ 
(in 1D\footnote{The 2D and 3D Fermi momenta take a different 
form: $k_{F} = (3\pi^{2}N_{e}/L^{3})^{1/3}$ is the 3D form in a 
box of volume $L^{3}$; $k_{F} = (2\pi N_{e}/L^{2})^{1/2}$ is the 
2D form on a square of area $L^{2}$. In all cases the 
ground-state consists of a Fermi sea with momentum/spin states 
$|{\bf k}\sig\rangle$ with $|k| < k_{F}$ occupied, and those with 
$|k| > k_{F}$ empty. Based on analogy with the 3D case, in all 
dimensions the set of points with $|k| = k_{F}$ is called the 
Fermi surface.}), and all other states empty. 
$k_{F}$ is called the Fermi 
momentum, and is of the order of an inverse angstrom for 
(3D) metallic elements. For simplicity, it will always be 
assumed that $|0\rangle$ is non-degenerate, so that $N_{e} = 
2(2m_{\rm max} + 1)$ with $m_{\rm max}$ a (large) positive 
integer. In this case $k_{F} = 2\pi (m_{\rm max} + 1/2)/L$ 
falls between the highest occupied $|k|$ state, and the lowest 
empty $|k|$ state. The range of occupied states $-k_{F} < k < 
k_{F}$ in $|0\rangle$ is called the Fermi sea, and is the 
natural object out of which to build low-energy excitations due 
to weakly-interacting electrons. 
The existence of the Fermi sea is a direct consequence of the 
antisymmetry of many-fermion states. 

For lattice systems the situation is similar. The Bloch states 
$|nk\sig\rangle$ are eigenstates of the single-electron 
hamiltonian $H_{0}^{(1)}$ of Eq.\ (\ref{1.4a}), with band 
energies $\varepsilon_{n}(k)$ depending on the periodic 
potential $U(x)$ of the lattice of ions. Using Eq.\ 
(\ref{1.20}), and following the 
derivation of Eq.\ (\ref{1.2.1}), the non-interacting hamiltonian 
$H_{0}$ for many electrons moving in the lattice is given by 
\beqa
H_{0} = \sum_{n, k \in {\rm FBZ}, \sig} \varepsilon_{n}(k) \, 
c^{\dg}_{nk\sig}c^{}_{nk\sig} \, . 
\label{1.2.2}
\eeqa
To determine the dispersion $\varepsilon_{n}(k)$ 
in a special case, consider writing $H_{0}$ 
in terms of Wannier states. Using Eq.\ (\ref{1.27}), and writing 
$H_{0}^{(1)}$ in position representation as in Eq.\ 
(\ref{1.4a}), $H_{0}$ takes the form
\beqa
H_{0} &=& \sum_{n, j, l, \sig} t_{njl}\,  
c^{\dg}_{nj\sig}c^{}_{nl\sig} \, ,
\nonumber \\
t_{njl} &=& \int_{L}dx\, \Phi^{*}_{n}(x-ja)\left[ 
-\frac{1}{2m_{e}}\partial_{x}^{2} + U(x)\right]\Phi_{n}(x-la)\, ,
\label{1.2.3}
\eeqa
where $\Phi_{n}(x - ja)$ is the Wannier state wavefunction, 
as in Eq.\ (\ref{1.6a}), but without the Pauli spinor. $t_{njl}$ 
is the amplitude for an electron in band $n$ to hop from site 
$l$ to site $j$. Since hopping only occurs between states in the 
same band, it suffices to restrict attention to a single band, 
and the band index is suppressed. If the Wannier states 
$|j\sig\rangle$ in the band are largely localized at their 
lattice sites $j$, then the hopping amplitudes $t_{jl}$ will be 
appreciable only for sites $j$ close to $l$. In the simplest 
approximation, the amplitudes are appreciable only if $j$ and 
$l$ are nearest neighbouring sites in the lattice. In this case 
the non-interacting hamiltonian $H_{0}$ for the band reduces to 
\beqa
H_{0} = -t \sum_{j, \sig} \left( 
c^{\dg}_{j, \sig}c^{}_{j+1\sig} + {\rm h.c.} \right) \, ,
\label{1.2.3a}
\eeqa
to an additive constant, where h.c.\ denotes hermitian 
conjugate, and where the nearest-neighbour hopping
\beqa
t = \int_{L}dx\, \Phi^{*}(x)\left[ 
\frac{1}{2m_{e}}\partial_{x}^{2} - U(x)\right]\Phi(x-a)\, .
\label{1.2.4}
\eeqa
Throughout this review the hopping amplitude $t$ will be assumed 
to be positive.\footnote{For negative $t$ the roles of 
electrons and holes are interchanged.} Using Eq.\ (\ref{1.25}) 
to invert the above procedure, and to write Eq.\ (\ref{1.2.3a}) 
in terms of Bloch states, the dispersion for a band 
with nearest-neighbour hopping is found to be given by 
\beqa
\varepsilon(k) = N^{-1}\sum_{j,l}t_{jl}\, 
e^{ik(j-l)a} = -2t\cos(ka)\, .  
\label{1.2.5}
\eeqa
The non-interacting ground-state $|0\rangle$ for a band with 
nearest-neighbour hopping thus takes the same form as 
for the continuum electron system, and has the same Fermi 
momentum $k_{F} = N_{e}\pi / 2L$, but with a modified 
ground-state energy.

\section{\label{old1.2.2}Electron-electron interactions}

A general electron-electron interaction may be defined as follows:
If an electron of spin $\sig$ is located at $x$, 
and an electron of spin $\sig'$ is located at 
$x'$, then there is a cost in energy given by 
\beqa
V_{\sig, \sig'}(x - x') &=& V_{\sig, \sig'}(|x - x'|)
\nonumber \\
&=& V_{\parallel}(x-x')\,\delta_{\sig, \sig'} 
+ V_{\perp}(x-x')\,\delta_{\sig, -\sig'} \, .
\label{1.2.6} 
\eeqa
For electrostatic Coulomb repulsion, for example, the 
interaction energy is 
$V_{\parallel}(x-x') = V_{\perp}(x-x') =e^{2}/|x - x'|$, 
where $-e$ is the electron charge. The operator for the 
electron-electron interaction Eq.\ (\ref{1.2.6}) 
in a continuum many-electron system may be  
written in terms of creation and annihilation operators using 
Eq.\ (\ref{1.21}): 
\beqa
V &=& \frac{1}{2}\sum_{\sig, \sig'} \int_{L} dx \int_{L} dx'\,  
\psi^{\dg}_{\sig}(x) \psi^{\dg}_{\sig'}(x') 
V_{\sig, \sig'}(x - x')\psi^{}_{\sig'}(x')\psi^{}_{\sig}(x) 
\nonumber \\
&=&  \frac{1}{2L} \sum_{k,k',q} \sum_{\sig, \sig'}
V_{\sig, \sig'}(q)\, c^{\dg}_{k+q\sig} c^{\dg}_{k'-q\sig'}
c^{}_{k'\sig'}c^{}_{k\sig} \, ,
\label{1.2.7}
\eeqa
where $V_{\sig, \sig'}(q)$ is the Fourier transform of 
$V_{\sig, \sig'}(x)$:
\beqa
V_{\sig, \sig'}(q) = V_{\sig, \sig'}(-q) = 
\int_{L}dx\, V_{\sig, \sig'}(x)\,e^{iqx}\, .
\label{1.2.8}
\eeqa
The second form in Eqs.\ (\ref{1.2.7}) is obtained using 
Eqs.\ (\ref{1.22}) and (\ref{1.23}).    

The effect of electron-electron interactions on the 
non-interacting Fermi sea may be read from the second of 
Eqs.\ (\ref{1.2.7}): An electron in the state $|k\sig\rangle$ 
interacts with an electron in the state $|k'\sig'\rangle$. The 
electrons scatter into states $|k+q\sig\rangle$ and 
$|k'-q\sig'\rangle$ with an amplitude given by 
$V_{\sig, \sig'}(q)$. Note that the scattering conserves both 
momentum and spin. The scattering processes 
perturb the Fermi sea of non-interacting electrons. For 
weak interactions at low temperatures, 
electrons are scattered from states   
just below $k_{F}$ into states just above $k_{F}$, and 
a weakly interacting system consists of a Fermi sea, together 
with a population of particle-hole pairs. The holes are 
unoccupied momentum states in the Fermi sea just below the 
Fermi surface at $|k| = k_{F}$, and the particles are electrons 
in excited momentum states just above $k_{F}$. The particle-hole 
pairs are the elementary fermionic excitations, and may be used 
to construct any number conserving interacting state.    

The 1D state-space greatly restricts the allowed low-energy 
particle-hole excitations. In 2D and 3D systems, the  
low-energy particle-hole excitations may have momentum transfers 
$q$ anywhere in the range $0 \leq |q| \lesssim 2k_{F}$, since in 
2D and 3D the Fermi surface $|k| = k_{F}$ is continuous. In 
1D the Fermi surface consists of just the two points 
$\pm k_{F}$. The low-energy scattering processes in 1D systems 
must thus have momentum transfers either
close to zero, called forward scattering, or close to 
$2k_{F}$, called backscattering. In forward scattering, an 
electron just inside the Fermi sea near $rk_{F}$, $r = \pm$, is 
scattered to a state just outside the Fermi sea near the same 
Fermi point $rk_{F}$. The momentum transfer $q \approx 0$ in 
this process. In backscattering, an electron is scattered from 
just inside the Fermi sea near one Fermi point $rk_{F}$, to 
just outside the Fermi sea near the opposite Fermi point at 
$-rk_{F}$. The momentum transfer is backscattering processes is 
$q \approx \pm 2k_{F}$, and corresponds physically to an 
electron at the Fermi surface rebounding from a hard wall. 
Low-energy scattering 
processes in 1D are discussed in detail in S\'{o}lyom's 
review (S\'{o}lyom 1979).  

Turning now to lattice systems, from Eq.\ (\ref{1.21}) the 
operator for the electron-electron interaction of Eq.\ 
(\ref{1.2.6}) in a single band of a lattice system is 
\beqa
V &=& \frac{1}{2}\sum_{i, j, l, m} \sum_{\sig, \sig'} 
V_{\sig, \sig'}(i, j, l, m) \, 
c^{\dg}_{i\sig}c^{\dg}_{j\sig'}c^{}_{l\sig'}c^{}_{m\sig}\, . 
\label{1.2.9}
\eeqa
The matrix element is given by 
\beqa
V_{\sig, \sig'}(i, j, l, m) &=& \int_{L}dx\int_{L}dx'\, 
\Phi^{*}_{\sig}(x-ia)\Phi^{*}_{\sig'}(x'-ja)
\nonumber \\
&& \quad \times V_{\sig, \sig'}(x-x')\, 
\Phi^{}_{\sig'}(x'-la)\Phi^{}_{\sig}(x-ma) \, 
\label{1.2.9a}
\eeqa
where $\Phi_{\sig}(x-ja)$ are Wannier state wavefunctions, 
Eq.\ (\ref{1.6a}), and where the band index has been suppressed. 
The generalization of Eq.\ (\ref{1.2.9}) to more than one band 
is straightforward (cf.\ the two band example of Eqs.\ 
(\ref{4.1}) and (\ref{4.2}) in chapter 4). 
The low-energy excitations 
out of the Fermi sea for an interacting 1D lattice system are 
the same as those for a continuum system as discussed above, but 
include one further interaction. As well as forward and 
backscattering, the 1D lattice system supports low-energy 
Umklapp scattering if the number of electrons equals the number 
of lattice sites, $N_{e} = N$ (called half-filling). 
In Umklapp scattering, two electrons near one Fermi point at 
$rk_{F}$ scatter into two states near the opposite Fermi point 
at $-rk_{F}$. This `double backscattering' interaction conserves 
momentum because at half-filling the momentum transfer 
$4k_{F} = 2\pi /a$ is a reciprocal lattice vector, and 
belongs to the equivalence class of zero crystal momentum.

There is a simplification 
of the interaction Eq.\ (\ref{1.2.9}) which leads to an 
important model of strongly-correlated electron systems. 
If the Wannier states are strongly localized 
at their lattice sites, the matrix element 
$V_{\sig, \sig'}(i, j, l, m)$ will be non-negligible only if the 
interacting electrons are at the same site: $i = j = l = m$. 
For example, for $3d$-electrons in transition metals, 
the on-site matrix 
elements are around 20eV, while the nearest-neighbour matrix 
elements are down to roughly 2 or 3eV (Hubbard 1963, 1964), and may 
be neglected to a good approximation.
The operator for on-site electron-electron interactions in a 
single band, called the Hubbard interaction, 
may be obtained by considering only on-site terms in 
Eq.\ (\ref{1.2.9}). This gives 
\beqa
V_{\Hub} &=& U\sum_{j}n_{j\uparrow}n_{j\downarrow}\, , \quad 
n_{j\sig} = c^{\dg}_{j\sig}c^{}_{j\sig} \, ,
\nonumber \\
U &=& \int_{L}dx\int_{L}dx'\, 
|\Phi(x)|^{2}\, V_{\perp}(x-x')\, |\Phi(x')|^{2}\, , 
\label{1.2.10}
\eeqa
where $\Phi(x)$ is the $j = 0$ Wannier state wavefunction, 
Eq.\ (\ref{1.6a}), without the Pauli spinor. Note that for a 
repulsive interaction between the electrons, 
as in Coulomb repulsion, $V_{\sig, \sig'}(x - x') > 0$ and the 
Hubbard interaction parameter $U > 0$. The on-site 
interaction Eq.\ (\ref{1.2.10}), together with nearest-neighbour 
electron hopping Eq.\ (\ref{1.2.3a}), is called the 
Hubbard model, and has the hamiltonian
\beqa
H_{\Hub} = -t\sum_{j, \sig}\left( c^{\dg}_{j\sig}c^{}_{j+1\sig} 
+ {\rm h.c.} \right) + U\sum_{j}n_{j\uparrow}n_{j\downarrow}\, .
\label{1.2.11}
\eeqa
The 1D Hubbard model is solved using bosonization in section 
\ref{old3.2}.

\section{\label{old1.2.3}Field theory approximation}

For weak interactions at low temperatures, 1D many-electron 
continuum systems may be described 
by particle-hole excitations close to the two Fermi points at 
$\pm k_{F}$. It seems reasonable in this situation 
to treat the electrons in the states near each Fermi 
point as belonging to a different species; those close to 
$+k_{F}$ are referred to as right-moving electrons, and those 
close to $-k_{F}$ are referred to as left-moving electrons. 
If the electron dispersion is approximated by its linear 
expansion about the Fermi points, which is generally valid for 
weak-interactions, then the different electron 
species have different dispersion relations. The right-movers 
(labelled by $r = +$) have 
\beqa
\varepsilon_{+}(k) = \varepsilon(k_{F}) + v_{F}(k - k_{F})\, , 
\label{1.2.12}
\eeqa
and the left-movers (labelled by $r= -$) have 
\beqa
\varepsilon_{-}(k) = \varepsilon(k_{F}) + v_{F}(-k - k_{F})\, , 
\label{1.2.13}
\eeqa
where the Fermi velocity $v_{F} = d\varepsilon(k_{F})/dk$. 

For condensed matter systems the division of the electrons 
into different species is matter of nomenclature only.  
In principle, the dispersion relations above should be 
accompanied by a bandwidth cut-off $k_{0}$, which satisfies  
$0 < k_{0} \leq k_{F}$ (S\'{o}lyom 1979), and then  
Eq.\ (\ref{1.2.12}) holds for $k_{F} - k_{0} < k 
< k_{F} + k_{0}$, and Eq.\ (\ref{1.2.13}) holds for 
$-k_{F} - k_{0} < k < -k_{F} + k_{0}$. This is not the case 
in field theory; the relativistic Dirac hamiltonian for 
spinless massless fermions moving in 1D generates two 
species of fermions which satisfy the dispersion relations 
Eqs.\ (\ref{1.2.12}) and (\ref{1.2.13}) exactly for all 
values of $k$. (Luttinger 1963).
In order to make use of the powerful methods of field theory 
in the condensed matter system, it is common to ignore the 
physical restrictions on the ranges of validity of the 
dispersion relations Eqs.\ (\ref{1.2.12}) and (\ref{1.2.13}),  
and to assume that the relations hold for all $k$ values 
in the condensed matter system as well. This turns the Fermi sea 
into two Dirac seas, one for right-moving electrons and one for 
left-moving electrons, and constitutes a field theory 
approximation to the condensed matter system of interest.  

The rationale behind the field theory approximation 
is that the electrons deep in the Fermi sea in the states 
$|k| \ll k_{F}$ remain largely unaffected by weak-interactions, 
and are irrelevant in the description of 
low-energy macroscopic properties. It should therefore be 
possible to choose any convenient form for 
the states far from the Fermi surface without altering the 
low-energy behaviour. This is illustrated by Haldane (1981). The
field theory approximation does however give rise to problems. 
The problems do not arise from the introduction
of unphysical states far from the Fermi surface, 
but from the fact that an 
infinite number of such states are introduced, and that many of 
the central results of field theory rely crucially on this; 
any large but finite number of states will not suffice. 
As a consequence, the field theory approximation is in a 
certain sense uncontrolled;  there are aspects of the formal 
manipulations which make no sense from a condensed 
matter perspective, and there are terms in the final results 
for which {\it ad hoc} interpretations have to be made. These 
issues were discussed at length in chapters 2 and 3.

%%%%%%%%%%%%%%%%%%%%%%%%%%%%%%%%%%%%%%%%%%%
%% appendix 6
%%%%%%%%%%%%%%%%%%%%%%%%%%%%%%%%%%%%%%%%%%%%

\cleardoublepage
\chapter{\label{appc}Non-Abelian bosonization}

Non-Abelian bosonization in the field theoretical approach 
has been introduced by Witten (1984), see also Stone (1994)
for several high energy applications. For the 1D Kondo lattice, 
Le Hur (1998) studied the finite temperature fixed points
of the model using non-Abelian bosonization in both the
half filled case and close to half filling. The following,
however is a straightforward non-Abelian generalization of 
the Abelian bosonization performed in chapters \ref{ch2} 
and \ref{ch5}. The purpose of this section is to prove that
the double-exchange interaction, see section \ref{old5.1.3}, 
is also present in the non-Abelian limit. 

The hamiltonian of the 1D Kondo lattice from Eq.\ (\ref{4.11}) 
may be written
\begin{equation}
H^{\alpha}_{\KL} = - t \; \frac{1}{3} \sum_{\alpha = x, y, z} 
\sum_{\langle ij \rangle} \sum_{\sig = \uparrow, \downarrow} 
c^{\dg}_{c i \sig_{\alpha}} c^{}_{c j \sig_{\alpha}} 
+ \; J \sum_{\alpha = x, y, z}  \sum_{j} 
{\uparrow}_{j \alpha} \, ( n_{c j \uparrow_{\alpha}} - 
n_{c j \downarrow_{\alpha}} )  \, , 
\label{appc.1}
\end{equation}
corresponding to all components $\alpha = x, y, z$ of the spin
$\sig_{\alpha} = \uparrow_{\alpha}, \downarrow_{\alpha} $. The
conduction band can be bosonized in the low energy, long 
wavelength subspace with states $k_{0}$, as in the non-Abelian limit
(cf.\ section \ref{old2.2}), with $|k| < 3 k_{0}/2$
full and $|k| > \pi/a - 3 k_{0}/2$ empty, and spanned by number 
fluctuations $\rho(k)$ with $|k| < k_{0}$ (see Eq.\ (\ref{tom})
of section \ref{old2.2.1}). 

Bosonization proceeds as in section
\ref{old2.1}, but in the present case from 12 fundamental
number fluctuations operators 
\begin{equation}
\rho_{r \sig_{\alpha}}(k) \; = \sum_{0 < r \overline{k} < \pi/a}
 c^{\dg}_{\overline{k} - \frac{k}{2} \sig_{\alpha}}
c^{}_{\overline{k} + \frac{k}{2} \sig_{\alpha}} \, ,  
\label{appc.2}
\end{equation}
where $r = R, L$, as before. 
For a comparison with the non-Abelian limit, see Eq.\ (\ref{2.1.4}). 
Eq.\ (\ref{appc.2}) defines {\it two} charge and {\it six} spin 
number fluctuation operators:
\beqa
\rho_{r} (k) &=& \rho_{r \uparrow_{\alpha}} (k) 
+ \rho_{r \downarrow_{\alpha}} (k) \, ,
\nonumber \\
\sig_{r \alpha} (k) &=& \rho_{r \uparrow_{\alpha}} (k) 
- \rho_{r \downarrow_{\alpha}} (k) \, ,
\label{appc.3}
\eeqa
where for the charge fluctuation operators $\alpha = x$, or 
$\alpha = y$, or $\alpha = z$, does not matter, while for the
spin fluctuation operators $\alpha = x, y, z$, as previously defined.
The commutation relations (for the non-Abelian limit, see 
section \ref{old2.2.2}) for $|k|, |k'| < k_{0}$ are
\beqa
\left[ \rho_{r}(k), \rho_{r'}(k') \right] &=& 
\delta_{r,r'} \, \delta_{k,-k'} \, \frac{rkL}{\pi} \, ,
\nonumber \\
\left[ \sig_{r \alpha}(k), \rho_{r' \beta}(k') \right] &=& 
\delta_{r,r'} \, \delta_{\alpha, \beta} \, \delta_{k,-k'} \, \frac{rkL}{\pi}
\, + \, 2 i \varepsilon_{\alpha \beta \gamma} \sig_{r \gamma} (k + k') \, ,
\nonumber \\
\left[ \rho_{r}(k), \sig_{r' \alpha}(k') \right] &=& 0 \, ,
\label{appc.4}
\eeqa
where $\varepsilon_{\alpha \beta \gamma}$ is the totally antisymmetric
3rd rank unit tensor. 

The bosonized Kondo lattice hamiltonian, corresponding to Eq.\ (\ref{bklm}),
turns out to be
\beqa
H^{\alpha}_{\KL} &=& \frac{ \pi v_{F}}{L} \sum_{0 < k < k_{0}} 
[ \rho_{R}(-k) \rho_{R}(k) + \rho_{L}(k)\rho_{L}(-k) 
\nonumber \\
&& + \frac{1}{3} \left( \vec{\sig}_{R}(-k) {\bf \cdot} \vec{\sig}_{R}(k)
+ \vec{\sig}_{L}(k) {\bf \cdot} \vec{\sig}_{L}(-k) \right) ]
\label{appc5.1} \\
&& + \frac{J}{N} \sum_{j, \; 0 < |k| < k_{0}} \vec{\uparrow}_{j} {\bf \cdot}
\left[ \vec{\sig}_{R}(k) + \vec{\sig}_{L}(k) \right] e^{i a k_j} \, .
\label{appc5.2}
\eeqa
Here, the first term of (\ref{appc5.1}) corresponds to Eq.\ (\ref{Ho}), 
the kinetic energy,
while the remaining hamiltonian term, (\ref{appc5.2}) is the Abelian 
bosonized form of the Kondo coupling. Obviously, in obtaining Eqs. 
(\ref{appc5.1}) and (\ref{appc5.2}) the strongly fluctuating $2 k_{F}$
terms have been neglected. 

Following section \ref{old5.1.2}, a canonical transformation can be
performed also in the Abelian case, where, instead of Eq.\ (\ref{5.1.1}),
it is used
\beqa
{\rm S}^{\alpha} = i \, \frac{3 J}{v_{F} N} \sum_{j, \; 0 < |k| < k_{0}}
\vec{\uparrow}_{j} \, {\bf \cdot} \,  
\left[ \vec{\sig}_{R}(k) - \vec{\sig}_{L}(k) \right] \, 
\frac{e^{i a k_j}}{k} \, .
\label{appc6} 
\eeqa
In section \ref{old5.1.2} the canonical transformation has been 
carried out exactly up to infinite order, however this is not possible for 
Eqs.\ (\ref{appc5.1}) and (\ref{appc5.2}) (The methods described
in section \ref{sectionSW}, see also Appendix \ref{ch4sw}, cannot
be applied in this case.) Thus, the first order transformation of
\beqa
\tilde{H}{}^{\alpha}_{\KL} \, = \, e^{-{\rm S}^{\alpha}} \, 
H^{\alpha}_{\KL} \, e^{{\rm S}^{\alpha}} 
\label{appc7}
\eeqa
is only presented here: 
\beqa
&& \tilde{H}{}^{\alpha}_{\KL} 
= \frac{ \pi v_{F}}{L} \sum_{0 < k < k_{0}} 
[ \rho_{R}(-k) \rho_{R}(k) + \rho_{L}(k)\rho_{L}(-k) 
\nonumber \\
&& + \frac{1}{3} \left( \vec{\sig}_{R}(-k) {\bf \cdot} \vec{\sig}_{R}(k)
+ \vec{\sig}_{L}(k) {\bf \cdot} \vec{\sig}_{L}(-k) \right) ]
\label{appc8.1} \\
&& - \frac{3}{v_F} \, \left( \frac{a J}{\pi} \right)^{2} \, 
\sum_{j, j'} \, \frac{ [\sin a (j - j') k_{0}] }{a (j - j')} \, 
\vec{\uparrow}_{j} \, {\bf \cdot} \, \vec{\uparrow}_{j'} 
\label{appc8.2} \\
&& + i \, \frac{3}{v_F} \, \left( \frac{J}{N} \right)^{2} \, 
\sum_{j, \; 0 < |k| < k_{0}} \, e^{i a k_j} 
\nonumber \\
&& \, \left\{ \sum_{j', \; 0 < |k'| < k_{0}} \, \frac{e^{i a k'{}_{j'}}}{k'} \, 
\, \vec{\uparrow}_{j} \, {\bf \cdot} \, \left[ \vec{\uparrow}_{j'} {\bf \times}
\left( \vec{\sig}_{R}(k + k') - \vec{\sig}_{L}(k + k') \right) \right] \right.
\nonumber \\
&& + \left. \frac{1}{2} \sum_{0 < |k'| < k_{0}} \, \frac{e^{i a k'{}_j}}{k'} \,
\left[ \vec{\uparrow}_{j} {\bf \times}
\left( \vec{\sig}_{R}(k') - \vec{\sig}_{L}(k') \right) \right] {\bf \cdot} 
\left( \vec{\sig}_{R}(k) - \vec{\sig}_{L}(k) \right) \right\}
\label{appc8.3}
\eeqa
It can be seen that (\ref{appc8.1}) is identical to (\ref{appc5.1}),
which is due to the fact that the canonical transformation have been
performed only to first order. 

Compared the term (\ref{appc8.2}) to the Abelian result from 
Eq.\ (\ref{tbklm}), it can be seen that even in the non-Abelian
approach a double-exchange ordering, for details see section
\ref{old5.1.3}, does appear between the
conduction electrons and impurity spins. The term (\ref{appc8.2}) 
couples impurity  spins $\uparrow_{\alpha \, j}$ to their 
perpendicular spin number fluctuations $\sig_{\beta} (k)$
and $\sig_{\gamma} (k)$ in the conduction band. 
However, the most important fact is that the Abelian results from 
chapters \ref{ch5} and \ref{ch6} can be extended to a full, non-Abelian
description. 

%%%%%%%%%%%%%%%%%%%%%%%%%%%%%%%%%%%%%%%%%%%
%% appendix 7
%%%%%%%%%%%%%%%%%%%%%%%%%%%%%%%%%%%%%%%%%%%%

\cleardoublepage
\chapter{\label{appa}The Luttinger Model Bosonization}

The Luttinger model is obtained from the continuum   
system described in section \ref{old2.1} by independently extending 
the linearized branches near the right ($+k_{F}$) and 
left ($-k_{F}$) Fermi points through all $k$ values 
(S\'{o}lyom 1979). This introduces two Dirac seas of 
right- and left-moving electrons, labelled by $r = \pm$ 
respectively, and forms a field theory approximation to 
the realistic system (cf.\ section \ref{old1.2.3}). In this appendix, 
the derivation of the bosonization formalism for the 
Luttinger model is outlined. In the main, the appendix follows 
the paper of Haldane (1981), and the review by Voit (1994). 

{\bf System}: The non-interacting 
(kinetic energy) hamiltonian for the Luttinger model is 
given by (Haldane 1981, Voit 1994)
\beqa
H_{0} = \sum_{r, k, \sig} v_{F}(rk - k_{F})\, 
:c^{\dg}_{rk\sig}c^{}_{rk\sig}: \, ,
\label{A.1}
\eeqa
where $c^{\dg}_{rk\sig}$ creates an $r$-electron 
of spin $\sig$ with momentum $k = 2\pi m/L$, $m$ 
any integer, and where the Fermi velocity 
$v_{F}$ and Fermi momentum $k_{F}$ are taken from the 
realistic system of interest. The non-interacting 
ground-state $|0\rangle$ of the Luttinger model 
consists of right-moving electrons  
in $k$ states from $-\infty$ to $k_{F}$, and 
left-moving electrons in $k$ states 
from $-k_{F}$ to $\infty$. To avoid direct reference 
to the Dirac seas, which contain an infinite number of 
electrons in (non-physical) negative energy states, or 
in other words to exclude unbounded operators, it is 
necessary in the Luttinger model to introduce a 
normal-ordering convention, designated by colons, which 
subtracts non-interacting ground-state 
expectation values. Thus 
\beqa
:c^{\dg}_{rk\sig}c^{}_{rk\sig}:\, \, 
= c^{\dg}_{rk\sig}c^{}_{rk\sig}    
- \langle 0|c^{\dg}_{rk\sig}c^{}_{rk\sig}|0\rangle \, , 
\label{A.2}
\eeqa
and ensures that the eigenvalues of $H_{0}$ are finite. 

{\bf Notations}: 
Density fluctuation operators may be defined in the 
Luttinger model by (Voit 1994)
\beqa
\rho_{r\sig}(k) = \sum_{k'} 
:c^{\dg}_{rk'\sig}c^{}_{rk'+k\sig}:  
= \sum_{k'}\left(c^{\dg}_{rk'\sig}c^{}_{rk'+k\sig} - 
\delta_{k,0}\langle 0|c^{\dg}_{rk'\sig}
c^{}_{rk'\sig}|0\rangle\right) \, .
\label{A.3}
\eeqa
These are the components in a Fourier expansion 
(cf.\ Eq.\ (\ref{2.1.2})) 
of the real-space density of $r$-electrons of spin $\sig$. 
The $k=0$ normal-ordered number operators $\rho_{r\sig}(0)$ 
are denoted by $N_{r\sig}$, as in Eq.\ (\ref{2.1.5}), and 
charge and spin density fluctuations are defined in terms 
of $\rho_{r\sig}(k)$ as in Eq.\ (\ref{2.3.5}).

{\bf Two Theorems}: 
It is readily verified that all the commutation relations 
between the operators $\rho_{r\sig}(k)$ vanish identically, 
except for those in which the normal-ordering 
convention is non-trivial:
\beqa
[\rho_{r\sig}(k), \rho_{r'\sig'}(k')] &=&
\delta_{r,r'}\delta_{k,-k'}\delta_{\sig,\sig'}\sum_{k''} 
\left(\langle 0|c^{\dg}_{rk''\sig}c^{}_{rk''\sig}|0\rangle
- \langle 0|c^{\dg}_{rk''+k\sig}
c^{}_{rk''+k\sig}|0\rangle\right) 
\nonumber \\
&=& \delta_{r,r'}\delta_{k,-k'}\delta_{\sig,\sig'}
\frac{rkL}{2\pi} \, .
\label{A.4}
\eeqa
This is analogous to Tomonaga's result Eq.\ (\ref{tom}) 
for the condensed matter system, but applies for 
fluctuations $\rho_{r\sig}(k)$ with $|k|$ arbitrarily large. 

The proof of the completeness of the states 
generated by the fluctuations $\rho_{r\sig}(k)$ 
proceeds as in section \ref{old2.2.3}, with a result equivalent 
to Eq.\ (\ref{2.2.7}), but applies without restrictions 
(Haldane 1981); $Z_{b} = Z_{f}$ rigorously 
at all temperatures $\beta^{-1}$.

{\bf Bose representation}: In contrast to the derivations 
of section \ref{old2.3}, only one independent Bose representation 
needs to be derived for the Luttinger model; 
the representation for the Fermi fields $\psi^{}_{r\sig}(x)$. 
The derivation begins from the commutation relation
\beqa
[\rho_{r\sig}(k), \psi_{r'\sig'}(x)] = 
-\delta_{r,r'}\delta_{\sig,\sig'}e^{-ikx}
\psi_{r\sig}(x) \, ,
\label{A.5}
\eeqa
which holds exactly for all $k$. This contrasts with 
the result in chapter \ref{ch2}, Eq.\ (\ref{2.3.10}), which holds 
only for asymptotically long-wavelength density fluctuations. 
Fermionic ladder operators are defined by 
\beqa
U^{\dg}_{r\sig} = L^{-1/2} \int_{L}dx\,\, e^{-irk_{F}x}
e^{-i\Phi^{\dg}_{r\sig}(x)}\psi_{r\sig}(x)
e^{-i\Phi_{r\sig}(x)}\, ,
\nonumber \\
\Phi_{r\sig}(x) = r\frac{\pi x}{L} N_{r\sig} 
- ir \lim_{\alpha \rightarrow 0} 
\sum_{rk > 0}\frac{2\pi}{kL}\rho_{r\sig}(k) 
e^{ikx - \alpha |k|/2}\, , 
\label{A.5a}
\eeqa
and commute with all density fluctuations $\rho_{r\sig}(k)$ 
with $k \neq 0$. As in section \ref{old2.3.2}, the ladder operator 
expression may be inverted to give an exact Bose 
representation for the Fermi fields; 
\beqa
\psi_{r\sig}(x) & = & 
\lim_{\alpha \rightarrow 0} 
\frac{1}{\sqrt{2\pi \alpha}}e^{ir(k_{F} + \pi/L)x} 
e^{i\Psi_{r\sig}(x)}U^{\dg}_{r\sig} \, ,
\nonumber \\
\Psi_{r\sig}(x)  
& = & \{\theta_{\rho}(x) + r\phi_{\rho}(x)
+ \sig[\theta_{\sig}(x) + r\phi_{\sig}(x)]\}/2  \, ,
\label{A.7}
\eeqa 
where the Bose fields are defined by 
\beqa
\phi_{\nu}(x) & = &  \frac{\pi x}{L}(N^{\nu}_{+} + N^{\nu}_{-}) 
- i\lim_{\alpha \rightarrow 0}\sum_{k \neq 0} \frac{\pi}{kL}
[\nu_{+}(k) + \nu_{-}(k)]e^{ikx - \alpha |k|/2} \, ,
\nonumber \\ 
\theta_{\nu}(x) & = & \frac{\pi x}{L}(N^{\nu}_{+} - N^{\nu}_{-}) 
- i\lim_{\alpha \rightarrow 0}\sum_{k \neq 0} \frac{\pi}{kL}
[\nu_{+}(k) - \nu_{-}(k)]e^{ikx - \alpha |k|/2} \, .
\label{A.8}
\eeqa
The Bose fields correspond to those used in chapter \ref{ch2}, 
Eq.\ (\ref{2.3.6}), but with an exponential cut-off function 
and $\alpha \rightarrow 0$.

{\bf Working with the Bose representation}: Since any operator 
may be expressed in terms of the Fermi fields, and since 
Eq.\ (\ref{A.7}) is an exact representation, the bosonization 
formalism for the Luttinger model is complete; there is no need 
to separately derive representations for other Fermi operators, 
in contrast to the derivation of chapter \ref{ch2}. However, it is 
important to proceed carefully when using Eq.\ (\ref{A.7}) to 
determine representations for some of the Fermi bilinears.  
As an example, consider the density operator $\rho_{r\sig}(x) 
= \psi^{\dg}_{r\sig}(x)\psi^{}_{r\sig}(x)$.\footnote{Similar 
considerations apply in calculating the representation for 
the non-interacting hamiltonian.} A naive application of 
Eq.\ (\ref{A.7}) gives $\rho_{r\sig}(x) = \lim_{\alpha 
\rightarrow 0} 1/2\pi \alpha$. This is incorrect since it 
violates the normal-ordering convention; without 
normal-ordering the density $\rho_{r\sig}(x)$ is infinite due 
the Dirac seas, and normal-ordering is non-trivial for the 
Luttinger model densities. With due regard for normal-ordering, 
and taking the limit $\alpha \rightarrow 0$, which excludes  
further unwanted terms, the Fourier decomposition of 
$\rho_{r\sig}(x)$ is recovered;
\beqa
\rho_{r\sig}(x)  =  L^{-1}\sum_{k} \rho_{r\sig}(k)
\, e^{ikx}\, .
\label{A.9}
\eeqa

%%%%%%%%%%%%%%%%%%%%%%%%%%%%%%%%%%%%%%%%%%%
%% appendix 8
%%%%%%%%%%%%%%%%%%%%%%%%%%%%%%%%%%%%%%%%%%%%

\cleardoublepage
\chapter{\label{appb}The Jordan-Wigner transformation}

A derivation of the Jordan-Wigner transformation has been included
for the benefit of chapter \ref{ch5}. The transformation is used in 
section \ref{old5.1.2}, but most importantly in section \ref{old5.2.1},
where it offers an alternative way to determine the critical transition 
line of the ferromagnetic-paramagnetic transition. 

Consider a chain of $N$ spins one-half ${\bf S}_{j}$, 
$j = 1,\ldots,N$. The components satisfy generic angular 
momentum commutation relations
\beqa
\left[S_{j}^{\alpha}, S_{l}^{\beta}\right] = 
i\,\delta_{j,l}\,\varepsilon_{\alpha \beta \gamma}\,
S_{j}^{\gamma},
\label{JW1}
\eeqa
where $\alpha, \beta, \gamma$ label choices from the components 
$x,y,z$, and where $\varepsilon_{\alpha \beta \gamma}$ is the 
totally antisymmetric third rank unit tensor.  Spin one-half 
is signalled by the normalization 
\beqa
\left(S_{j}^{x}\right)^{2} = 
\left(S_{j}^{y}\right)^{2} = 
\left(S_{j}^{z}\right)^{2} = 1/4\,, 
\label{JW2}
\eeqa
and the closure property 
\beqa
S^{x}_{j}S^{y}_{j} + S^{y}_{j}S^{x}_{j} = 0\, , 
\quad {\rm etc.}\,{\rm cycl.}\, ,
\label{JW3}
\eeqa
which may be verified by going to a particular representation, 
as for example in terms of the Pauli matrices. 

The raising and lowering operators
\beqa
S^{\pm}_{j} \equiv S^{x}_{j} \pm iS^{y}_{j}
\label{JW4}
\eeqa
partially resemble bosons in that they commute at different 
sites (from Eq.\ (\ref{JW1})), and partially resemble fermions 
in that they anticommute at the same site (from 
Eqs.\ (\ref{JW2}) and (\ref{JW3})). 
This mixed behaviour is inconvenient. For 
example, it prevents any direct diagonalization of the 
quadratic forms $S^{+}_{j}S^{-}_{j+1}$ (cf. Eq.\ (\ref{JW8a})). 
The problem may be circumvented by defining new 
operators out of $S^{\pm}_{j}$ which contain an additional 
{\it non-local} factor:
\beqa
c^{}_{j} &\equiv& \exp\left\{i\pi\sum_{l=1}^{j-1}S^{+}_{l} 
S^{-}_{l}\right\}\, S^{-}_{j}, 
\nonumber \\
c^{\dg}_{j} &=& S^{+}_{j}\,
\exp\left\{-i\pi\sum_{l=1}^{j-1}S^{+}_{l} 
S^{-}_{l}\right\}. 
\label{JW5}
\eeqa
These operators satisfy the anticommutation relations of 
spinless fermions
\beqa
\left\{c^{}_{j}, c^{}_{l}\right\} = 
\left\{c^{\dg}_{j}, c^{\dg}_{l}\right\} = 0, \quad \quad
\left\{c^{}_{j}, c^{\dg}_{l}\right\} = \delta_{j,l}\, ,
\label{JW6}
\eeqa
and are called {\it Jordan-Wigner fermions}. The  
anticommutation relations of Eq.\ (\ref{JW6}) 
may be verified by first noting that 
\beqa
S^{\pm}_{j}S^{\mp}_{j} = 1/2 \pm S^{z}_{j}, \quad \quad
\exp\left(\pm i\pi S^{z}_{j}\right) = \pm 2iS^{z}_{j}\, .
\nonumber
\eeqa
The Jordan-Wigner transformation Eq.\ (\ref{JW5}) 
may now be written 
\beqa
c_{j} = \prod_{l=1}^{j-1}\left(-2S_{l}^{z}\right)\, S^{-}_{j}\, ,
\quad \quad
c^{\dg}_{j} = S^{+}_{j}
\prod_{l=1}^{j-1}\left(-2S_{l}^{z}\right)\, . 
\label{JW6a}
\eeqa
Since $(-2S^{z}_{j})^{2} = 1$, it follows that 
\beqa
\left\{c^{}_{j}, c^{\dg}_{j}\right\} = S^{-}_{j}S^{+}_{j} 
+ S^{+}_{j}S^{-}_{j} = 1.
\nonumber 
\eeqa
The other two on-site anticommutators follow immediately from 
$(S^{+}_{j})^{2} = (S^{-}_{j})^{2} = 0$. To verify 
anticommutation at different sites, consider first $j < l$. 
Eq.\ (\ref{JW6a}) gives   
\beqa
\left\{c^{}_{j}, c^{\dg}_{l}\right\} = 
-2\left(S^{-}_{j}S^{z}_{j} + S^{z}_{j}S^{-}_{j}\right)
\prod_{m=j+1}^{l-1}\left(-2S^{z}_{m}\right)\,S^{+}_{l}, 
\label{JW7}
\eeqa
and this vanishes since 
$S^{\pm}_{j}S^{z}_{j} = -S^{z}_{j}S^{\pm}_{j}$. The remaining 
anticommutators for $j \neq l$ take the same basic form as 
Eq.\ (\ref{JW7}), and similarly vanish. 

The reason that the Jordan-Wigner fermions 
$c^{}_{j}, c^{\dg}_{l}$ anticommute at 
different sites, whereas the original operators 
$S^{-}_{j}, S^{+}_{l}$ commute, is because 
the non-local factor 
\beqa
K_{<}(j) = \exp\left\{i\pi\sum_{l=1}^{j-1}S^{+}_{l} 
S^{-}_{l}\right\}
\label{B.9}
\eeqa
provides an extra minus sign (cf. Eq.\ (\ref{JW7})). $K_{<}(j)$ 
is called the {\it disorder} or 
{\it soliton} term, 
and its non-locality greatly restricts the utility of the 
Jordan-Wigner transformation: there is no gain in simplicity if 
the transformed hamiltonian explicitly contains $K_{<}(j)$ 
or related non-local objects. The transformation 
is most useful in spin systems with 
nearest-neighbour interactions between at most two 
components, and such that there are no applied 
fields in the directions of the interacting components.
The transverse-field Ising chain 
\beqa
H = -\sum_{j}\left\{ {\cal J}_{j}S^{x}_{j}S^{x}_{j+1} 
+ h_{j}S^{z}_{j}\right\} 
\label{JW8} 
\eeqa
fortunately falls into this category. To write Eq.\ (\ref{JW8}) 
in terms of Jordan-Wigner fermions, first note that 
$S^{z}_{j} = S^{+}_{j}S^{-}_{j} - 1/2 = c^{\dg}_{j}c^{}_{j} 
- 1/2$. To transcribe the interaction term, note that 
\beqa
S^{x}_{j}S^{x}_{j+1} = \frac{1}{4} \left(
S^{+}_{j}S^{+}_{j+1} + S^{+}_{j}S^{-}_{j+1} +
S^{-}_{j}S^{+}_{j+1} + S^{-}_{j}S^{-}_{j+1} \right).
\label{JW8a}
\eeqa
From Eq.\ (\ref{JW6a}), 
\beqa
c^{\dg}_{j}c^{}_{j+1} = -2S^{+}_{j}S^{z}_{j}S^{-}_{j+1} = 
S^{+}_{j}S^{-}_{j+1}\, , 
\label{JW9}
\eeqa
where the last equality follows from $S^{\pm}_{j}S^{z}_{j} = 
\mp S^{\pm}_{j}/2$. The simple form of Eq.\ (\ref{JW9}) 
relies crucially on the interactions being between 
nearest-neighbours only. Similarly, 
\beqa
c^{}_{j}c^{\dg}_{j+1} = -S^{-}_{j}S^{+}_{j+1}, \quad 
c^{\dg}_{j}c^{\dg}_{j+1} = S^{+}_{j}S^{+}_{j+1}, \quad 
c^{}_{j}c^{}_{j+1} = -S^{-}_{j}S^{-}_{j+1}\, .
\nonumber
\eeqa
In the limit as $N \rightarrow \infty$, the transverse-field 
Ising hamiltonian Eq.\ (\ref{JW8}) 
now reads
\beqa
H = -\sum_{j}\left\{ h_{j}c^{\dg}_{j}c^{}_{j} + 
\frac{{\cal J}_{j}}{4}\left(c^{\dg}_{j} - c^{}_{j}\right)
\left(c^{\dg}_{j+1}+ c^{}_{j+1}\right) \right\} + {\rm const.} 
\label{JW10}
\eeqa
where ${\rm const.} = \sum_{j}h_{j}/2$. The transformed 
hamiltonian is a quadratic form in Fermi operators, and may 
be diagonalized via a Bogoliubov unitary transformation.

%%%%%%%%%%%%%%%%%%%%%%%%%%%%%%%%%%%%%%%%%%%
%% references
%%%%%%%%%%%%%%%%%%%%%%%%%%%%%%%%%%%%%%%%%%%%

\cleardoublepage
\references
\chapter{\label{refe}References}
\parindent0pt

\begin{description}

\item[]
\item[]Abdalla, E., Abdalla, M. C. B., and Rothe, K. D., 1991, {\em 
          Non-perturbative methods in 2 
          dimensional quantum field theory}, World Scientific, Singapore. 
\item[]Aeppli, G. and Fisk, Z., 1992, Comments Cond. Mat. Phys. {\bf 16}, 155.
\item[]Apostol, M., 1983, J. Phys. C: Solid State Phys. {\bf 16}, 5937. 
\item[]Aristov, D. N., 1997, Phys. Rev. B {\bf 55}, 8064. 
\item[]Anderson, P. W., 1961, Phys. Rev. {\bf 124}, 41. 
\item[]Anderson, P. W., 1967, Phys. Rev. {\bf 164}, 352.
\item[]Anderson, P. W., 1970, J. Phys. C {\bf 3}, 2439.
\item[]Anderson, P. W., 1987, Science {\bf 235}, 1196. 
\item[]Anderson, P. W. and Hasegawa, H., 1955, Phys. Rev. {\bf 100}, 675.
\item[]Anderson, P. W. and Yuval, G., 1969, Phys. Rev. Lett. {\bf 23}, 89.
\item[]Anderson, P. W., Yuval, G. and Hammann, D. R., 1970, Phys. Rev. 
          B{\bf 1}, 4464. 
\item[]Andrei, N., 1980, Phys. Rev. Lett. {\bf 45}, 379.
\item[]Andrei, N, Furuya, K. and Lowenstein, J. H., 1983, Rev. Mod. Phys. 
         {\bf 55}, 331. 
\item[]Ashcroft, N. W. and Mermin, N. D., 1976, {\em Solid State Physics}, 
        Saunders, Philadelphia.
\item[]Assad, F. F., 1999, Phys. Rev. Lett. {\bf 83}, 796. 

\item[]
\item[]Bares, P. A. and Blatter, G., 1990, Phys. Rev. Lett. {\bf 64}, 2567.
\item[]Batista, C. D., Eroles, J., Avignon, M. and Alascio, B., 1998, Phys. 
          Rev. B {\bf 58}, 14689.
\item[]Batista, C. D., Eroles, J., Avignon, M. and Alascio, B., 2000, Phys. 
          Rev. B {\bf 62}, 15047.
\item[]Batlogg, B., Ott, H. R. and Wachter, P., 1979, Phys. Rev. Lett. 
          {\bf 42}, 278. 
\item[]Baxter, R. J., 1982, {\em Exactly solvable models in statistical
           mechanics}, Academic Press, London. 
\item[]Baym, G. and Pethick, C., 1978, in {\em The Physics of Liquid and 
         Solid Helium}, Part II, K. H. Bennemann and J. B. Ketterson  
         (Eds.), Wiley, New York. 
\item[]Belavin, A. A., Polyakov, A. M., and Zamolodchikov, A. B., 1984,
         Nucl. Phys. B {\bf 241}, 333. 
\item[]Bethe, H. A., 1931, Z. Phys. {\bf 71}, 205. 
\item[]Biedenharn, L. C. and Louck, J. D., 1981, {\em Angular Momentum 
          in Quantum Physics}, Addison-Wesley, Massachusetts.
\item[]Blatt, J. M. and Weisskopf, V. F., 1979, {\em Theoretical
           Nuclear Physics}, Springer-Verlag, New York. 
\item[]Bowen, G. and Gul\'{a}csi, M., 2001, Phil. Mag. B{\bf 81}, 1409. 
\item[]Br\'{e}zin, E. and Zinn-Justin, J., 1989, {\em Fields, Strings 
         and Critical Phenomena}, Elsevier Science Publishers, Amsterdam. 
\item[]Bray, J. and Chui, S., 1987, Phys. Rev. B {\bf 36}, 8600.
\item[]Brazovskii, S.,Matveenko, S., and Nozi\`{e}res, P., 1994, J. Phys. I 
        France {\bf 4}, 571.

\item[]
\item[]Capponi, A. and Assad, F. F., 2001, Phys. Rev. B {\bf 63}, 155114.
\item[]Caprara, S. and Rosengren, A., 1997, Europhys. Lett. {\bf 39}, 55. 
\item[]Caron, L. G. and Pratt, G. W., 1968, Rev. Mod. Phys. {\bf 40}, 802.
\item[]Chao, K. A., Spalek, J. and Oles, A. M., 1977, J. Phys. C: Solid 
         State Phys. {\bf 10}, L271. 
\item[]Chan, R. and Gul\'{a}csi, M., 2000, J. Supercond. {\bf 13}, 917. 
\item[]Chan, R. and Gul\'{a}csi, M., 2001a, Phil. Mag. Lett. {\bf 81}, 673.
\item[]Chan, R. and Gul\'{a}csi, M., 2001b, J. Supercond. {\bf 14}, 651. 
\item[]Chan, R. and Gul\'{a}csi, M., 2002, Phil. Mag. Lett. {\bf 82}, 671. 
\item[]Chan, R. and Gul\'{a}csi, M., 2003, unpublished, 
         preprint cond-mat/0308405. 
\item[]Chan, R. and Gul\'{a}csi, M., 2004, Phil. Mag. {\bf 84}, 1265. 
\item[]Chen, Y., Chen, H., Yuan, Q. and Zhang, Y., 1999, J. Phys.: Condens.
         Matter {\bf 11}, 5623. 
\item[]Coleman, S., 1975, Phys. Rev. D{\bf 11}, 2088.
\item[]Coleman, P., Georges, A. and Tsvelick, A., 1997, J. Phys.: 
          Condens. Matter {\bf 79}, 345. 
\item[]Coll, C. F., 1974, Phys. Rev. B {\bf 9}, 2150. 
\item[]Coqblin, B. and Schrieffer, J. R., 1969, Phys. Rev. {\bf 185}, 847.

\item[]
\item[]Dagotto, E., 1994, Rev. Mod. Phys. {\bf 66}, 763. 
\item[]Dagotto, E., {\it et al.}, 1998, Phys. Rev. B {\bf 58}, 6414. 
\item[]Daul, S., 2000, Euro. Phys. J. B{\bf 14}, 649. 
\item[]Daul, S. and Noack, R. M., 1998, Phys. Rev. B{\bf 58}, 5, 2635.
\item[]Doniach, S., 1977, Physica {\bf 91B}, 231. 
\item[]Dwight, H. B., 1961, {\em Tables of Integrals and other Mathematical 
         Data}, 4th Edn., Macmillan, New York.

\item[]
\item[]Emery, V. J., 1979, in {\it Highly Conducting One-dimensional Solids}, 
         J. T. Devreese, R. P. Evrard and V. E. van Doren (Eds.), 
	  Plenum, New York.
\item[]Emery, V. J. and Kivelson, S., 1992, Phys. Rev. B {\bf 46}, 10812.
\item[]Emery, V. J., Luther, A. and Peschel, I., 1976, Phys. Rev. 
         B {\bf 13}, 1272.

\item[]
\item[]Fetter, A. L. and Walecka, J. D, 1971, {\it Quantum Theory of
          Many-Partcile Systems}, McGraw-Hill, New York. 
\item[]Feynman, R. P., 1972, {\em Statistical Mechanics}, 
         Benjamin, Reading (especially chapter 6). 
\item[]Finkel'stein, A. M. and Larkin, A. I., 1993, Phys. Rev. B 
         {\bf 47}, 10461.
\item[]Fisher, D. S., 1992, Phys. Rev. Lett. {\bf 69}, 534. 
\item[]Fisher, D. S., 1995, Phys. Rev. B {\bf 51}, 6411. 
\item[]Fisk, Z., {\it et al.}, 1995, Physica B {\bf 206-207}, 798. 
\item[]Fradkin, E., 1991, {\em Field Theories of Condensed Matter Systems}, 
          1991, Addison-Wesley, Massachusetts, Chap. 4. 
\item[]Fradkin, E. and Hirsch, J. E., 1983, Phys. Rev. B{\bf 27}, 1680. 
\item[]Fr\"{o}hlich, H. and Nabarro, F. R. N., 1940, Proc. Roy. Soc. 
         A {\bf 175}, 382.
\item[]Fulde, P, 1993, {\em Electron Correlations in Molecules and Solids}, 
          Springer-Verlag, Heidelberg. 
\item[]Fujimoto, S. and Kawakami, N., 1997, J. Phys. Soc. Japan {\bf 66}, 2157. 
\item[]Fye, R. M. and Scalapino, D. J., 1990, Phys. Rev. Lett. {\bf 65}, 3177. 
\item[]Fye, R. M. and Scalapino, D. J., 1991, Phys. Rev. B {\bf 44}, 7486. 

\item[]
\item[]Garcia, D. J., {\it et al.}, 2002, Phys. Rev. B {\bf 65}, 134444. 
\item[]Glazek, S. D. and Wilson, K. G., 1993, Phys. Rev. D{\bf 48}, 5863. 
\item[]Goodenough, J. B., 1955, Phys. Rev. {\bf 100}, 564.
\item[]Gradshteyn, I. S. and Ryzhik, I. M., 1965,{\em  Table of Integrals, 
       Series, and Products}, 4th Edn., Academic Press,  New York.
\item[]Griffiths, R. B., 1969, Phys. Rev. Lett. {\bf 23}, 17.
\item[]Gross, E. K. U., Runge, E. and Heinonen, O., 1991, {\em Many-Particle 
         Theory}, Hilger, Bristol.
\item[]Guerrero, M. and Yu, C. C., 1995, Phys. Rev. B {\bf 51}, 10301. 
\item[]Gul\'{a}csi, M., 1997a, Phil. Mag. B{\bf 76}, 731. 
\item[]Gul\'{a}csi, M., 1997b, in {\em Recent Progress in Many-Body T
         heories}, Neilson, D. and Bishop, R. F. (Eds.), 
	 World Scientific, Singapore, p. 485.
\item[]Gul\'{a}csi, M. and Bedell, K. S., 1994, Phys. Rev. Lett. 
          {\bf 72}, 2765.  
\item[]Gul\'{a}csi, M., Bussmann-Holder, A. and Bishop, A. R., 2003, 
          unpublished, preprint cond-mat/0307069.
\item[]Gul\'{a}csi, M., Bussmann-Holder, A. and Bishop, A. R., 2004,
          Jour. Supercond. {\bf 17}, 167. 
\item[]Gul\'{a}csi, M., McCulloch, I. P., Juozapavicius, A. and 
           Rosengren, A., 2003, unpublished, preprint cond-mat/0304351.
\item[]Gul\'{a}csi, M., McCulloch, I. P., Juozapavicius, A. and 
           Rosengren, A., 2004, Phys. Rev. B {\bf 69}, 174425. 
\item[]Gul\'{a}csi, Zs. and Gul\'{a}csi, M., 1998, Adv. Phys. {\bf 47}, 1. 
\item[]Gutzwiller, M. C., 1963, Phys. Rev. Lett. {\bf 10}, 159. 

\item[]
\item[]Ha, Y. K., 1984, Phys. Rev. D{\bf 29}, 1744. 
\item[]Haldane, F. D. M., 1981, J. Phys. C: Solid State Phys. {\bf 14}, 2585. 
\item[]Harris, A. B. and Lange, R. V., 1967, Phys. Rev. {\bf 157}, 295. 
\item[]Haug, A., 1972, {\em Theoretical Solid State Physics},  Vol. 1, 
          Pergamon Press, Oxford. 
\item[]Heidenreich, R., Schroer, B., Seiler, R. and Uhlenbrock, D., 1975,
          Phys. Lett. A{\bf 54}, 119. 
\item[]Hess, D. W., Riseborough, P. S. and Smith, J. L., 1993, 
         Encyclopedia of Applied Physics, 7, 435, VCH Publishers, New York. 
\item[]Hewson, A. C., 1993, {\em The Kondo Problem to Heavy Fermions}, 
          Cambridge University Press, Cambridge. 
\item[]Hirsch, J. E., 1985, Phys. Rev. Lett. {\bf 54}, 1317.
\item[]Hirsch, J. E. and Fradkin, E., 1982, Phys. Rev. Lett. {\bf 49}, 402.
\item[]Hirsch, J. E. and Fradkin, E., 1983, Phys. Rev. B{\bf 27} 4302. 
\item[]Hirsch, J. E., 1984, Phys. Rev. B {\bf 30}, 5383.
\item[]Holstein, T., 1959, Ann. Phys. (N.Y.) {\bf 8}, 325; 343.
\item[]Holstein, T., 1961, Ann. Phys. (N.Y.) {\bf 16}, 407. 
\item[]Honner, G. and Gul\'{a}csi, M., 1997a, Phys. Rev. Lett. {\bf 78}, 2180.
\item[]Honner, G. and Gul\'{a}csi, M., 1997b, Phil. Mag. B{\bf 76}, 849. 
\item[]Honner, G. and Gul\'{a}csi, M., 1997c, Z. Phys. B{\bf 104}, 733.
\item[]Honner, G. and Gul\'{a}csi, M., 1998a, J. Magn. Magn. Matter. 
        {\bf 184}, 307. 
\item[]Honner, G. and Gul\'{a}csi, M., 1998b, Phys. Rev. B{\bf 58}, 2662. 
\item[]Honner, G. and Gul\'{a}csi, M., 1999, J. Supercond. {\bf 12}, 237.
\item[]Honner, G. and Gul\'{a}csi, M., 2002, unpublished. 
\item[]Horsch, P., Jaklic, J. and Mack, F., 1999, Phys. Rev. B {\bf 59},
        R14149.
\item[]Huang, K., 1987, {\em Statistical Mechanics}, 2nd Edn., Wiley, New York.
\item[]Hubbard, J., 1963, Proc. Roy. Soc. A {\bf 276}, 238.
\item[]Hubbard, J., 1964, Proc. Roy. Soc. A {\bf 277}, 237.

\item[]
\item[]Itoyama, H., McCoy, B. M., and Perk, J. H. H., 1990, Int. Jour.
          Mod. Phys. B {\bf 4}, 295.
\item[]Itzykson, C. and Drouffe, J. M., 1989, {\em Statistical field
           theory}, vol. 2, Cambridge University Press, Cambridge. 

\item[]
\item[]Jauch, J. M., 1968, {\em Foundations of Quantum Mechanics}, 
         Addison-Wesley, Reading. 
\item[]Jin, S., {\it et al.}, 1994, Science {\bf 264}, 413.
\item[]Jonker, G. H. and Van Santen, J. H., 1950, Physica {\bf 16}, 337. 
\item[]Jullien, R., Fields, J. N. and Doniach, S., 1977, Phys. Rev. B 
         {\bf 16}, 4889. 
\item[]Juozapavicius, A., Caprara, S. and Rosengren, A., 1997, Phys. 
         Rev. B {\bf 56}, 11097. 
\item[]Juozapavicius, A., McCulloch, I. P., Gul\'{a}csi, M. and Rosengren, 
         A., 2002, Phil. Mag. B{\bf 82}, 1211. 
\item[]Jurecka, C. and Brening, W., 2001, Phys. Rev. B {\bf 64}, 92406. 

\item[]
\item[]Kasuya, T., 1956, Prog. Theor. Phys. {\bf 16}, 45. 
\item[]Kehrein, S. K. and Mielke, A., 1996, Ann. Phys. {\bf 252}, 1. 
\item[]Kittel, C., 1968, in {\em Solid State Physics}, F. Seitz, D. Turnbull, 
        and H. Ehrenreich (Eds.), Vol. 22, Academic Press, New York.
\item[]Kohn, W., 1964, Phys. Rev. {\bf 133}, A171. 
\item[]Koller, W., Pr\"{u}ll, A., Evertz, H. G. and von der Linden, W., 2003,
          Phys. Rev. B {\bf 67}, 174418. 
\item[]Kolley, E., Kolley, W. and Tietz, R., 1992, J. Phys. Cond. Matter. 
         {\bf 4}, 3517. 
\item[]Kondo, J., 1964, Prog. Theor. Phys. {\bf 32}, 37.
\item[]Korepin, V. E., Bogoliubov, N. M. and Izergin, A. G., 1993, 
         {\em{ Quantum Inverse Scattering method and Correlation Functions}}, 
         Cambridge University Press, Cambridge. 
\item[]Krishna-murthy, H. R., Wilkins, J. W. and Wilson, K. G., 1980, 
          Phys. Rev. B {\bf 21}, 1003. 
\item[]Kruis, H. V., McCulloch, I. P., Nussinov, Z. and Zaanen, J., 2002,
          unpublished, preprint cond-mat/0209493. 
\item[]Kubo, K. and Ohata, N., 1972, J. Phys. Soc. Jpn {\bf 63}, 3214.

\item[]
\item[]Lacroix, C., 1985, Solid State Commun. {\bf 54}, 991.
\item[]Lacroix, C. and Cyrot, M., 1979, Phys. Rev. B {\bf 20}, 1969. 
\item[]Landau, L. D. and Lifshitz, E. M., 1965, {\em Quantum  Mechanics}, 
          Addison-Wesley, Reading.
\item[]Langer, W., Plischke, M. and Mattis, D., 1969, Phys. Rev. Lett.
          {\bf 23}, 1448.
\item[]Lavagna, M. and P\'{e}pin, C., 2000, Phys. Rev. B {\bf 62}, 6450. 
\item[]Le Hur, K., 1997, Phys. Rev. B {\bf 56}, 14058. 
\item[]Le Hur, K., 1998, Phys. Rev. B {\bf 58}, 10261.
\item[]Lebedev, N. N., 1965, {\em Special Functions and their Applications}, 
        Prentice-Hall, Englewood Cliffs. 
\item[]Lee, P. A., Rice, T. M., Serene, J. W., Sham, L. J. and Wilkins, 
          J. W., 1986, Comm. Cond. Mat. Phys. {\bf 12}, 99. 
\item[]Lieb, E., Schultz, T. and Mattis, D., 1961, Ann. Phys. (N.Y.) 
         {\bf 16}, 407. 
\item[]Lieb, E. H. and Wu, F. Y., 1968, Phys. Rev. Lett. {\bf 20}, 1445. 
\item[]Linden, von der W., 1992, Phys. Rep. {\bf 220}, 53. 
\item[]Luther, A. and Emery, V. J., 1974, Phys. Rev. Lett. {\bf 33}, 589.
\item[]Luther, A. and Peschel, I., 1974, Phys. Rev. B {\bf 9}, 2911. 
\item[]Luther, A., and Peschel, I., 1975, Phys. Rev. B{\bf 12}, 3908.
\item[]Luttinger, J. M., 1963, J. Math. Phys. {\bf 4}, 1154.

\item[]
\item[]Mahan, G. D., 1990, {\em Many-particle physics}, 2nd Edn., 
         Plenum, New York. 
\item[]Mahan, G. D., 1993, {\em An encyclopedia of exactly solved
          models in one dimension}, World Scientific, Singapore. 
\item[]Makhankov, V. G., 1989, {\it Soliton Phenomenology}, Mathematics 
        and Its Applications (Soviet Series) Vol. 33, Kluwer, Dordrecht. 
\item[]Mandelstam, S., 1975, Phys. Rev. D{\bf 11}, 3026. 
\item[]Mattis, D. C., 1974, J. Math. Phys. {\bf 15}, 609.
\item[]Mattis, D. C. and Lieb, E. H., 1965, J. Math. Phys.  {\bf 6}, 304.
\item[]Matveenko, S. and Brazovskii, S., 1994, Sov. Phys. JETP {\bf 78}, 892.
\item[]McCulloch, I. P. and Gul\'{a}csi, M., 2000, Aust. J. Phys. 
         {\bf 53}, 597. 
\item[]McCulloch, I. P. and Gul\'{a}csi, M., 2001, Phil. Mag. Lett. 
         {\bf 81}, 447. 
\item[]McCulloch, I. P. and Gul\'{a}csi, M., 2002, Europhys. Lett. 
         {\bf 57}, 852. 
\item[]McCulloch, I. P., Gul\'{a}csi, M., Caprara, S., Juozapavicius, A. 
         and Rosengren, A., 1999, J. Low Temp. phys. {\bf 117}, 323. 
\item[]McCulloch, I. P., Bishop, A. R. and Gul\'{a}csi, M., 2001,
         Phil. Mag. B {\bf 81}, 1603. 
\item[]McCulloch, I. P., Juozapavicius, A., Rosengren, A. and Gul\'{a}csi, 
         M., 2001, Phil. Mag. Lett. {\bf 81}, 869. 
\item[]McCulloch, I. P., Juozapavicius, A., Rosengren, A. and Gul\'{a}csi, 
         M., 2002, Phys. Rev. B{\bf 65}, 52410. 
\item[]Millis, A. J., Littlewood, P. B. and Shraiman, B. I., 1995, Phys. Rev. 
         Lett. {\bf 74}, 5144. 
\item[]Moukouri, S. and Caron, L. G., 1995, Phys. Rev. B {\bf 52}, R15723. 
\item[]Moukouri, S., Chen, L. and Caron, L. G., 1996, Phys. Rev. B 
         {\bf 53}, R488. 

\item[]
\item[]Nagaoka, Y., 1966, Phys. Rev. {\bf 147}, 392. 
\item[]Negele, J. W. and Orland, H., 1988, {\em Quantum Many-Particle 
          Systems}, Addison-Wesley, New York. 
\item[]Newns, D. M. and Hewson, A. C., 1980, J. Phys. F{\bf 10}, 2429. 
\item[]Nishino, T., Hikihara, T., Okunishi, K. and Hiedida, Y., 1999, 
          Int. J. Mod. Phys. B.{\bf 13}, 1.
\item[]Noack, R. M. and White, S. R., 1993, Phys. Rev. B{\bf 47}, 9243. 
\item[]Novais, E., Miranda, E., Castro Neto, A. H. and Cabrera, 2002a,
           Phys. Rev. Lett. {\bf 88}, 217201. 
\item[]Novais, E., Miranda, E., Castro Neto, A. H. and Cabrera, 2002b, 
           Phys. Rev. B {\bf 66}, 174409. 
\item[]Nozi\`{e}res, P., 1964, {\em Interacting Fermi Systems}, 
          Benjamin, New York. 
\item[]Nozi\`{e}res, P. and Blandin, A., 1980, J. Physique {\bf 41}, 193.

\item[]
\item[]Ogata, M. and Shiba, H., 1990, Phys. Rev. B {\bf 41}, 2326.
\item[]Ogawa, M. Y., {\it et al.}, 1987, J. Am. Chem. Soc.{\bf 109}, 1115. 
\item[]Ogawa, T., Kanda, K. and Matsubara, T., 1975, Prog. Theor. Phys.
           {\bf 53}, 614.
\item[]Oshikawa, M., 2000, Phys. Rev. Lett. {\bf 84}, 3370. 
\item[]Overhauser, A. W., 1965, Physics {\bf 1}, 307.
\item[]{\"O}stlund, S. and Rommer, S., 1995, Phys. Rev. Lett. {\bf 75}, 3537.

\item[]
\item[]Peschel, I., Wang, X., Kaulke, M. and Hallberg, K., 1999, 
          {\em Density Matrix Renormalization}, Lecture Notes in Physics 
	  Vol. 528, Springer, Berlin. 
\item[]P\'{e}pin, C. and Lavagna, M., 1999, Phys. Rev. B {\bf 59}, 2591.
\item[]Pfeuty, P., 1970, Ann. Phys. (N.Y.) {\bf 57}, 79. 
\item[]Pfeuty, P., 1979, Phys. Lett {\bf 72A}, 245. 
\item[]Prugove\v{c}ki, E., 1981, {\em Quantum Mechanics in  Hilbert Space}, 
         Academic Press, New York. 

\item[]
\item[]Roth, L., 1966, Phys. Rev. {\bf 149}, 306. 
\item[]Ruderman, M. A. and Kittel, C., 1954, Phys. Rev. {\bf 96}, 99.

\item[]
\item[]Sakamoto, H. and Kubo, K., 1996, J. Phys. Soc. Jpn. {\bf 65}, 3732.
\item[]Satija, I. I., 1990, Phys. Rev. B {\bf 41}, 7235.
\item[]Satija, I. I., 1994, Phys. Rev. B {\bf 49}, 3391.
\item[]Satija, I. I. and Doria, M. M., 1989, Phys. Rev. B 
         {\bf 39}, 9757. 
\item[]Schick, M., 1968, Phys. Rev. {\bf 166}, 404. 
\item[]Schlottmann, P., 1987, Phys. Rev. B {\bf 36}, 5177.
\item[]Schlottmann, P., 1992, Phys. Rev. B {\bf 46}, 998. 
\item[]Schotte, K. D. and Schotte, U., 1969, Phys. Rev. {\bf 182}, 479.
\item[]Sch\"{o}nhammer, K. and Meden, V., 1996, Am. J. Phys. {\bf 64}, 1168. 
\item[]Schrieffer, J. R. and Wolff, P. A., 1966, Phys. Rev.{\bf 149}, 491.
\item[]Schulz, H. J., 1990, Phys. Rev. Lett. {\bf 64}, 2831.
\item[]Schulz, H. J., 1991, Int. J. Mod. Phys. B {\bf 5}, 57.
\item[]Searle, C. W. and S. T. Wang, S. T., 1970, Can. J. Phys. 
         {\bf 48}, 2023. 
\item[]Shankar, R., 1995, Act. Phys. Pol. B {\bf 26}, 1835.
\item[]Shen, S-Q., 1996, Phys. Rev. B {\bf 53}, 14252. 
\item[]Shi, Z. P., Singh, R.R. P. Gelfand, M. P., and Wang, Z., 1995, 
          Phys. Rev. B {\bf 51}, 15630. 
\item[]Shibata, N., Ishii, C. and Ueda, K., 1995, Phys. Rev. B {\bf 51}, 3626.
\item[]Shibata, N. and Tsunetsugu, H., 1999, J. Phys. Soc. Japan {\bf 68}, 3138.
\item[]Shibata, N., Nishino, T., Ueda, K. and Ishii, C., 1996, Phys. Rev. 
         B {\bf 53}, R8828.
\item[]Shibata, N. and Ueda, K., 1999, J. Phys. Condes. Matt. {\bf 11}, R1. 
\item[]Sierra, G. and Nishino, T., 1997, Nucl. Phys. B{\bf 456}, 505. 
\item[]Sigrist, M., Tsunetsugu, H. and Ueda, K., 1991, Phys. Rev. Lett. 
         {\bf 67}, 2211. 
\item[]Sigrist, M., Tsunetsugu, H., Ueda, K. and Rice, T. M., 1992b, Phys. 
         Rev. B {\bf 46}, 13838. 
\item[]Sigrist, M., Ueda, K. and Tsunetsugu, H., 1992a, Phys. Rev. B 
         {\bf 46}, 175.
\item[]Sikkema, A. E. Affleck and I. White, S. R., 1997, Phys. Rev. Lett.
          {\bf 79}, 929. 
\item[]Sinjukow, P. and Nolting, W., 2002, unpublished, preprint 
         cond-mat/0206270.
\item[]Sokoloff, J. B., 1970, Phys. Rev. B {\bf 1}, 1144.
\item[]S\'{o}lyom, J., 1979, Adv. Phys. {\bf 28}, 201.
\item[]S{\o}rensen, E. S. and Affleck, I., 1996, Phys. Rev. B {\bf 53}, 9153.
\item[]Stewart, G. R., 1984, Rev. Mod. Phys. {\bf 56}, 755. 
\item[]Stone, M. (Ed.), 1994, {\em Bosonization}, World Scientific, Singapore.
\item[]Strong, S. P. and Millis, A. J., 1994, Phys. Rev. B {\bf 50}, 9911. 
\item[]Su, W. P., Schrieffer, J. R. and Heeger, A. H., 1980a, Phys. Rev.
           Lett. {\bf 42}, 1698. 
\item[]Su, W. P., Schrieffer, J. R. and Heeger, A. H., 1980b, 
           Phys. Rev. B{\bf 22}, 2099.
\item[]Sutherland, B., 1974, Phys. Rev. B {\bf 12}, 3795. 
\item[]Suzuki, T., 1993, Physica B {\bf 186-188}, 347. 

\item[]
\item[]Tokura, Y., {\it et al.}, 1996, J. Appl. Phys.  {\bf 79}, 5288. 
\item[]Tomonaga, S., 1950, Prog. Theor. Phys. {\bf 5}, 544.
\item[]Troyer, M. and W\"{u}rtz, D., 1993, Phys. Rev. B {\bf 47}, 2886. 
\item[]Toulouse, G., 1969, C. R. Acad. Sci.{\bf 268}, 1200.
\item[]Tsunetsugu, H., 1997, Phys. Rev. B {\bf 55}, 3042.
\item[]Tsunetsugu, H., Hatsugai, Y., Ueda, K. and Sigrist, M., 1992, Phys. 
        Rev. B {\bf 46}, 3175.
\item[]Tsunetsugu, H., Sigrist, M. and Ueda, K., 1993, Phys. Rev. B 
         {\bf 47}, 8345. 
\item[]Tsunetsugu, H., Sigrist, M. and Ueda, K., 1997, Rev. Mod. Phys. 
         {\bf 69},  809. 
\item[]Tsvelik, A. M., 1994, Phys. Rev. Lett. {\bf 72}, 1048. 
\item[]Tsvelick, A. M. and Wiegmann, P. B., 1983, Adv. Phys. {\bf 32}, 453.

\item[]
\item[]Van Santen, J. H. and Jonker, G. H., 1950, Physica {\bf 16}, 599. 
\item[]Van Vleck, J. H., 1962, Rev. Mod. Phys. {\bf 34}, 681.
\item[]Varma, C. M., 1976, Rev. Mod. Phys. {\bf 48}, 219.
\item[]Varma, C. M., 1979, Solid State Commun. {\bf 30}, 537.
\item[]Varma, C. M., 1984, in {\it Moment Formation in Solids}, 
          W. J. L. Buyers (Ed.), Plenum, New York. 
\item[]Varma, C. M., 1994, Phys. Rev. B {\bf 50}, 9952.
\item[]Visscher, P. B., 1974, Phys. Rev. B {\bf 10}, 943.
\item[]Voit, J., 1994, Rep. Prog. Phys. {\bf 57}, 977. 
\item[]von Delft, J. and Sch{\"o}ller, H., 1998, Ann. Phys. {\bf 4}, 225.

\item[]
\item[]Wagner, W., 1986, {\em Unitary Transformations in Solid State Physics}, 
          Modern Problems in Condensed Matter Sciences Vol. 15, Elsevier, 
	  Amsterdam. 
\item[]Wang, Z., Li, X. P., Lee, D. H., 1993, Phys. Rev. B {\bf 47}, 11935.
\item[]Wang, Z., Li, X. P., Lee, D. H., 1994, Physica B {\bf 199} - 
          {\bf 200}, 463. 
\item[]White, S. R., 1992, Phys. Rev. Lett. {\bf 69}, 2863.
\item[]White, S. R., 1993, Phys. Rev. B {\bf 48}, 10345.
\item[]White, S. R., 1998, Phys. Rep. {\bf 301}, 187.
\item[]White, S. R. and Affleck, I., 1996, Phys. Rev. B {\bf 54},  9862. 
\item[]White, S. R. and Scalapino, D. J., 1998, Phys. Rev. Lett. {\bf 80}, 1272. 
\item[]Wiegmann, P. B., 1980, Sov. Phys. JETP Lett. {\bf 31}, 392.
\item[]Wiegmann, P. B. and Finkel'shtein, A. M., 1978, Zh. Eksp. Teor. 
          Fiz. {\bf 75}, 204. 
\item[]Wilson, K. G., 1975, Rev. Mod. Phys. {\bf 47}, 773.
\item[]Witten, E., 1984, Commun. Math. Phys. {\bf 92}, 455.

\item[]
\item[]Xavier, J. C., Novais, E. and Miranda, E., 2002, Phys. Rev. 
          B {\bf 65}, 214406.
\item[]Xavier, J. C., Pereira, R. G., Miranda, E. and Affleck, I., 2003, 
          Phys. Rev. Lett. {\bf 90}, 247204. 

\item[]
\item[]Yamanaka, M., Oshikawa, M. and Affleck,I., 1997,  Phys.
           Rev. Lett. {\bf 79}, 1110.
\item[]Yafet, Y., 1987, Phys. Rev. B {\bf 36}, 3948. 
\item[]Yanagisawa, T. and Harigaya, K., 1994, Phys. Rev. B {\bf 50}, 9577. 
\item[]Yanagisawa, T. and Shimoi, M., 1996, Int. J. Mod. Phys. B {\bf 10}, 3383.
\item[]Yosida, K., 1957, Phys. Rev. {\bf 106}, 893.
\item[]Yosida, K., 1980, {\em Functional Analysis}, Springer-Verlag, Berlin. 
\item[]Yosida, K. and Yoshimori, A., 1973, in {\it Magnetism}, Vol. 5, 
         G. T. Rado and H. Suhl (Eds.), Academic Press, New York. 
\item[]Yu, C. C. and White, S. R., 1993, Phys. Rev. Lett. {\bf 71}, 3866.
\item[]Yunoki, S., {\it et al.}, 1998, Phys. Rev. Lett. {\bf 80}, 845. 
\item[]Yuval, G. and Anderson, P. W., 1970, Phys. Rev. B{\bf 1}, 1522.

\item[]
\item[]Zaanen, J. and Ole\'{s}, A. M., 1988, Phys. Rev. B {\bf 37}, 9423. 
\item[]Zachar, O., Kivelson, S. E. and V. J. Emery, V. J., 1996, Phys. Rev. 
         Lett. {\bf 77}, 1342. 
\item[]Zang, J., R\"{o}der, H., Bishop, A. R. and Trugman, S. A., 1997, 
          J. Phys.: Condens. Matter {\bf 9}, L157.
\item[]Zener, C., 1951, Phys. Rev. {\bf 82}, 403.
\item[]Zhang, G. M. and Yu, L., 2000, Phys. Rev. B {\bf 62}, 76.
\item[]Zheng, W. and Oitmaa, J., 2003, Phys. Rev. B {\bf 67}, 214406. 
\item[]Zhou, L-J. and Q.-Q. Zheng, Q.-Q., 1992, J. Magn. Magn. Matter. 
         {\bf 109}, 237.  

\end{description}

%%%%%%%%%%%%%%%%%%%%%%%%%%%%%%%%%%%%%%%%%%%
%% figures
%%%%%%%%%%%%%%%%%%%%%%%%%%%%%%%%%%%%%%%%%%%%

\cleardoublepage
\figures
\chapter{\label{fig}Figures}

%%\newpage
\hspace*{\fill}
\vfill

%%\newpage
\hspace*{\fill}
\vfill

%%\newpage
\hspace*{\fill}
\vfill

%%\newpage
\hspace*{\fill}
\vfill

%%\newpage
\hspace*{\fill}
\vfill

%%\newpage
\hspace*{\fill}
\vfill

%%\newpage
\hspace*{\fill}
\vfill

%%\newpage
\hspace*{\fill}
\vfill

%%\newpage
\hspace*{\fill}
\vfill

%%\newpage
\hspace*{\fill}
\vfill

%%\newpage
\hspace*{\fill}
\vfill

%%\newpage
\hspace*{\fill}
\vfill

%%\newpage
\hspace*{\fill}
\vfill

%%\newpage
\hspace*{\fill}
\vfill

%%\newpage
\hspace*{\fill}
\vfill

%%\newpage
\hspace*{\fill}
\vfill

%%\newpage
\hspace*{\fill}
\vfill

%%\newpage
\hspace*{\fill}
\vfill

%%\newpage
\hspace*{\fill}
\vfill

\end{document}